# Modèle théorique de la contraction musculaire squelettique :
## ressorts entropiques, processus stochastiques et mécanique classique

Sylvain LOUVET ♦

15/11/2013

## Résumé

Le modèle analyse la fibre musculaire comme un système matériel déformable dont les réponses aux sollicitations expérimentales sont calculées à partir des lois de la mécanique Newtonienne. Les nanomoteurs biologiques générant force et mouvement sont les têtes de myosine de classe II (tetM) qui transforment l'énergie chimique en énergie mécanique par l'hydrolyse d'une molécule d'ATP. Cet échange énergétique se caractérise par une modification de la conformation de la tetM, nommée *working stroke* (WS). Mécaniquement, la tetM est modélisée par trois segments rigides articulés entre eux (S1a, S1b et S2). Une fois S1a fixé au filament d'actine (filA), le WS se produit d'une manière similaire à celle d'un ressort angulaire localisé dans S1a exerçant son action sur un levier, fonction attribuée à S1b, S2 servant de tige de liaison entre S1b et le filament de myosine (filM). L'hypothèse principale du modèle précise que la position angulaire de S1b, lorsque que le WS entame une action mécaniquement efficace, varie aléatoirement entre deux valeurs constantes afin d'assurer à S2 son rôle de tige de liaison rigide. En conséquence, le déclenchement du WS dépend du positionnement initial de la tetM par rapport à la molécule d'actine (molA) à laquelle elle est liée. A l'appui des théorèmes de la mécanique du solide, associés à des raisonnements statistiques, le modèle prédit les quatre phases transitoires qui suivent le raccourcissement d'une fibre musculaire après une perturbation selon un échelon de longueur ou de force. Le modèle pointe, d'une part, l'importance de la viscosité durant la première phase de ces deux types d'échelon et, d'autre part, l'interdépendance des 6 autres phases vis-à-vis d'un état bimodal après disparition des actions dues à la viscosité. Le modèle interprète le comportement d'une fibre musculaire selon une perturbation en escalier. Le modèle reste compatible avec les données expérimentales lors de variations de température ou de diverses modifications structurelles.

♦Laboratoire de Mécanique des Solides, UMR 6610, Chasseneuil Futuroscope, France

Email: sylvain.louvet@univ-nantes.fr

A theoretical model of skeletal muscular contraction :
entropic springs, stochastic processes and classical mechanics

## Abstract

The model analyzes the muscle fiber as a deformable material system whose responses to experimental stresses are calculated from the laws of Newtonian mechanics. Biological nanomotors generating force and motion are myosin II heads (tetM) that convert chemical energy into mechanical energy by hydrolysis of one ATP molecule. This energy exchange is characterized therein by changing the conformation of the tetM, the working stroke (WS). Mechanically, the tetM is modeled by three rigid segments articulated between them (S1a, S1b and S2). Once S1a fixed on the actin filament (filA), the WS occurs similarly to that of an angular spring located in S1a exerting its action on a lever, function assigned to S1b, S2 serving as a connecting rod between S1b and the myosin filament (filM). The main assumption of the model states that the S1b angular position, when the WS starts a mechanically effective action, varies randomly between two constant values in order to ensure for S2 its role of rigid rod. Consequently the beginning of the WS depends on the initial positioning of the tetM from the actin molecule (molA) to which it is bounded. By applying the theorems of solid mechanics with statistical calculations, the model predicts the four transient phases for the shortening of a muscle fiber according to a force or length step. The model shows, on the one hand, the importance of the viscosity during the first phase of these two types of step and secondly, the interdependence of the six other phases towards a bimodal state after disappearance of viscosity actions. The model provides an interpretation for the behavior of the fiber with shortening staircase. The model is consistent with experimental data during temperature changes or various structural alterations.

# Introduction

Depuis plus d'un siècle, la fibre musculaire est l'objet d'innombrables expériences dont les résultats apparemment déterministes puisque reproductibles restent inexplicables dans leur globalité.

Parmi beaucoup de modèles proposés, deux émergent :

- le modèle du « ***lever arm*** », modèle où le mouvement de S1 serait généré par l'énergie mécanique d'un ressort linéaire situé dans S2 (Huxley 1957; Huxley 1974), puis depuis les travaux de I. Rayment et de ses coauteurs (Rayment, Holden et al. 1993; Rayment, Rypniewski et al. 1993) par l'énergie mécanique d'un ressort angulaire situé dans S1a et exerçant son action sur S1b (Holmes 1997 ; Geeves and Holmes 1999 ; Highsmith 1999 ; Huxley 2000 ).
- le modèle du « ***brownian ratchet*** » ou « ***brownian motor*** », modèle élaboré à partir d'un concept[1] développé par Richard Feynman (Feynman, Leighton et al. 1963), où le mouvement de S1 proviendrait de forces créées par les fluctuations thermiques (Astumian and Bier 1996 ; Yanagida, Kitamura et al. 2000 ; Ait-Haddou and Herzog 2002 ; Karatzaferi, Chinn et al. 2004).

Le principal modèle de la contraction musculaire demeure le modèle de A.F. Huxley (Huxley 1957; Huxley and Simmons 1971; Huxley 1974) dont l'équation de base combine la vitesse de raccourcissement de la fibre musculaire avec les vitesses d'attachement et de détachement des têtes de myosine (tetM), elles-mêmes fonctions des vitesses des réactions chimiques associées aux différentes étapes de l'hydrolyse de l'ATP lors d'un pont d'actomyosine (Lymn and Taylor 1971). La majorité des autres modèles publiés suivent cette voie conceptuelle (Cooke, White et al. 1994; Piazzesi and Lombardi 1995; Smith and Geeves 1995; Duke 1999; Nielsen 2002; Piazzesi, Reconditi et al. 2002; Lan and Sun 2005; Chin, Yue et al. 2006; Offer and Ranatunga 2013 ).

Certains modèles concilient les 2 thèses dominantes en conjecturant que S1a et/ou S1b seraient susceptibles d'osciller entre 2 ou plusieurs états stables (Huxley and Simmons 1971; Piazzesi and Lombardi 1995; Cooke 1997; Duke 1999; Huxley 2000 ; Houdusse and Sweeney 2001; Nielsen 2002 ; Offer and Ranatunga 2013 ).

Si ces modèles fournissent des équations prédictives pour une ou plusieurs phases transitoires survenant lors d'un raccourcissement d'une fibre musculaire selon un échelon en longueur ou en force, aucun d'entre eux n'est à même de prédire l'ensemble de toutes les phases.

Tout en nous inspirant des modèles précédents, nous suggérons un autre type d'approche en présentant la fibre musculaire comme un système matériel déformable auquel les lois de la mécanique du solide sont appliquées. Nous supposons que le moment moteur d'une tetM résulte de l'élasticité entropique, suivant la voie ouverte par R. Cooke (Cooke 1997) ou B. Nielsen (Nielsen 2002), mais le comportement, notamment collectif, s'apparente à celui d'un ressort mécanique.

[1] *En pratique, la roue à rochet ou cliquet brownien de Feynman tourne indifféremment dans les 2 sens et reste donc en moyenne immobile sur une longue durée; le modèle du mouvement brownien rectifié («biased brownian motion model») a été introduit afin que la tetM se déplace dans une direction privilégiée, i.e. vers la ligne M.*

## Méthode

L'article est organisé en 18 chapitres comme suit :

Le 1er chapitre cerne une série d'évènements subis par une tête de myosine (tetM) dans un cadre probabiliste, géométrique et mécanique où sont présentées les 14 hypothèses que nécessitent le modèle, chacune de ces 14 hypothèses étant reformulée contextuellement dans les chapitres suivants.

Le chapitre 2 propose une modélisation de la fibre musculaire et de ces différents composants. Puis seront étudiées la cinétique d'une tetM (chapitre 3) et celle d'un collectif de tetM dans un demi-sarcomère (chapitre 4).

Le chapitre 5 fournit les équations générales s'appliquant à une myofibrille.

Le chapitre 6 examine le comportement d'une fibre musculaire en conditions isométriques maximales à l'appui des équations apportées par les chapitres précédents.

Les chapitres 7 à 10 sont consacrés à la modélisation des phases 1 à 4 d'un échelon en longueur.

Le chapitre 11 analyse le comportement d'une fibre musculaire selon une perturbation en escalier

Les chapitres 12 à 15 sont dévolus à la modélisation des phases 1 à 4 d'un échelon en force.

Le chapitre 16 développe les calculs des valeurs minimale et maximale du nombre de WS.

Le chapitre 17 caractérise la signature de la fibre musculaire en contraction à partir d'une relation bimodale.

Le 18ème chapitre indique les limites du modèle.

Une discussion clôt la plupart des chapitres.

Le modèle repose sur un travail de lectures et de réflexions. De nombreux articles scientifiques ont retenu notre attention. Parmi ceux-ci, sept publications servent de base expérimentale[1] à ce modèle théorique; nous les citons par ordre de parution :

**Ford, L. E., A. F. Huxley, et al. (1977)**. "Tension responses to sudden length change in stimulated frog muscle fibres near slack length." J Physiol 269(2): 441-515.

**Edman, K. A., C. Reggiani, et al. (1988)**. "Maximum velocity of shortening related to myosin isoform composition in frog skeletal muscle fibres." J Physiol 395: 679-94.

**Piazzesi, G. and V. Lombardi (1995)**. "A cross-bridge model that is able to explain mechanical and energetic properties of shortening muscle." Biophys J 68(5): 1966-79.

**Linari, M., V. Lombardi, et al. (1997)**. "Cross-bridge kinetics studied with staircase shortening in single fibres from frog skeletal muscle." J Muscle Res Cell Motil 18(1): 91-101.

**Piazzesi, G., L. Lucii, et al. (2002)**. "The size and the speed of the working stroke of muscle myosin and its dependence on the force." J Physiol 545(Pt 1): 145-51

**Huxley, H., M. Reconditi, et al. (2006)**. "X-ray interference studies of crossbridge action in muscle contraction: evidence from muscles during steady shortening." J Mol Biol 363(4): 762-72.

**Reconditi, M., E. Brunello, et al. (2011)**. "Motion of myosin head domains during activation and force development in skeletal muscle." Proc Natl Acad Sci U S A 108(17): 7236-40.

[1] *La plupart des expérimentations référencées ont été réalisées sur le muscle de grenouille (Rana temporaria, Rana esculenta, Rana pipiens).*

## Liste des Abréviations

*( Certains sigles ont été anglicisés car communément employés dans la communauté scientifique )*

| Sigle | Définition | Valeur |
|---|---|---|
| **A** | Liaison encastrée entre le filA et S1a durant le WS<br>Point caractérisant le sitA d'1 molA | |
| $\mathbf{a_X}$ | Coefficient entrant dans la formule du moment moteur d'origine entropique en fonction de la position linéaire relative de A par rapport à D | 52 |
| $\mathbf{a_\theta}$ | Coefficient entrant dans la formule du moment moteur d'origine entropique en fonction de la position angulaire $\theta$ de S1b | 316.9 |
| **AL** | Bras de levier (*Arm Lever*) | |
| **b** | N° itératif d'1 S1 effectuant 1 WS dans 1 hs | de **1** à $N_{s,R_L}^{WS}$ |
| **B** | Liaison rotule entre S1a et S1b<br>Liaison pivot moteur entre S1a et S1b durant 1 WS | |
| **C** | Liaison rotule entre S1b et S2 | |
| **CB** | Pont formé par l'union d'1 S1 à 1 sitA ou pont d'actomyosine (*Crossbridge*) | |
| **D** | Liaison rotule entre S2 et le filM | |
| **d1** | domaine 1 du raccourcissement d'un hs, défini par $[0;d_{2sitA}]$ | de 0 à 5.4 nm |
| **d2** | domaine 2 du raccourcissement d'un hs, défini par $[d_{2sitA};2d_{2sitA}]$ | de 5.4 à 10.8 nm |
| **DE** | Evt Détachement (*Detachment*) d'une molA par S1a<br>Dans le modèle, état de S1a non lié | |
| $\mathbf{d_{AB}}$ | distance fixe entre AB selon OY durant 1 WS | 4 nm |
| $\mathbf{d_{filAM}}$ | distance entre un filA et un filM voisins (centre à centre) selon OY | 24.8 nm |
| $\mathbf{d_{2sitA}}$ | distance entre 2 sitA | 5.4 nm |
| **e** | N° itératif de l'échelon d'une perturbation en escalier | |
| **filA** | filament d'actine dans 1 hs | |
| **filM** | filament de myosine dans 1 hs | |
| **fmI** | fibre musculaire idéalisée | |
| $\mathbf{I_{M3}}$ | intensité d'une des réflexions d'un faisceau de rayons X par l'ensemble quasi-cristallin que constitue la fmI ; considéré comme un bon indicateur du nombre de tetM en cours de WS | |
| **j** | N° itératif de l'échelon d'une perturbation en force | de 0 à N, N entier |
| **G** | bijection entre $pT^{(j)}$ et $\Delta X^{(j)}$ | |
| **hs** | demi sarcomère (*half-sarcomer* ) | |
| **hsL** | demi sarcomère gauche (*half-sarcomer in the Left*) | |
| **hsR** | demi sarcomère droit (*half-sarcomer in the Right*) | |
| **k** | N° itératif de l'échelon d'une perturbation en longueur | de 0 à N, N entier |
| $\mathbf{L_{S1b}}$ | Longueur de S1b | 10 nm |
| $\mathbf{L_{S2}}$ | Longueur de S2 | 30 nm |
| $\mathbf{L_{hs}}$ | Longueur d'1 hs | de 0.625 µm à 1.825 µm |
| $\mathbf{L0^{hs}}$ | Longueur de référence d'1 hs à T0 | 1.125 µm |
| $\mathcal{M}_B$ | Moment moteur d'un ressort entropique appliqué en B selon OZ | |
| $\mathcal{M}_B^{lin}$ | Moment moteur d'un ressort mécanique appliqué en B selon OZ | |
| $\mathcal{M}0$ | Moment moyen d'origine entropique des S1 initiant leurs WS et dont les positions angulaires $\theta$ sont uniformément distribuées sur $\delta_\theta$ | |
| $\mathcal{M}0^{lin}$ | Moment moyen d'origine mécanique des S1 initiant leurs WS et dont les positions angulaires $\theta$ sont uniformément distribuées sur $\delta_\theta$ | |
| $\mathcal{M}0^*$ | Moment moyen d'origine entropique des S1 initiant leurs WS et dont les positions angulaires $\theta$ sont uniformément distribuées sur $\Delta\theta_{WS}^{Max}$ | $\mathcal{M}0^*/\mathcal{M}0 = 0.75$ |
| $\mathcal{M}1$ | Valeur algébrique du moment moteur entropique correspondant à $\theta 1$ | |
| $\mathcal{M}2$ | Valeur algébrique du moment moteur entropique correspondant à $\theta 2$ | |
| $\mathbf{M_S}$ | disque M ou ligne M situé au centre du sarcomère n°s | |
| $M_{S,R_L} + filM$ | système mécanique rigide composé de la moitié du disque M avec ses filM associés à droite ou à gauche | |
| $M_S + 2 \cdot filM$ | système mécanique rigide composé du disque M avec ses filM associés à droite et à gauche | |
| **molA** | molécule d'actine, brique d'un filA | |
| **molM** | molécule de myosine classe II, composée de 2S1+S2 | |
| **N** | Projection orthogonale de toute force F selon OY | |
| $N_{myof}$ | Nombre de myofibrilles de la fmI | |
| $N_S$ | Nombre de sarcomères d'une myofibrille | |
| $N^{WS}_{s,R_L}$ | Nombre de S1 réalisant 1 WS dans hsR ou hsL du sarcomère n° s | |
| **p1** , **p2** , **p3** , **p4** | phases 1 ,2, 3 ou 4 classiques et transitoires d'une perturbation en force ou en longueur appliquée à la fmI après une mise en tension isométrique max | |
| **pT** | Tension relative de la fmI égale au rapport de la tension instantanée de la fmI par rapport à la tension isométrique de la fmI tétanisée $= T / T0^{fmI}$ | |
| $pT^{(k)}$ ; $pT^{(j)}$ | Tension relative appliquée à la fmI selon OX pour l'échelon en longuer n° k, ou en force n° j | |
| **pT0** | Tension relative maximale isométrique de la fmI où les positions angulaires des $\Lambda 0$ S1b suivent une distribution uniforme centrée sur $\theta 0$ avec une dispersion égale à $\delta_\theta$ | 1 |
| **pT0*** | Tension relative maximale isométrique de la fmI où les positions angulaires des $\Lambda 0$ S1b suivent une distribution uniforme centrée sur $\theta 0*$ avec une dispersion égale à $\Delta\theta^{Max}_{WS}$ | 0.75 |
| **pT1** | Tension relative de la fmI à la fin de p1 d'un échelon de longueur ou de force | |
| $pT1_{z1}$ | Tension relative de la fmI dans z1 à la fin de p1 d'un échelon de longueur ou de force , | |
| $pT1_{z2}$ | Tension relative de la fmI dans z2 à la fin de p1 d'un échelon de longueur ou de force | |
| $pT1/2$ | Tension relative de la fmI après disparition des forces dues à la viscosité au début de p2 d'un échelon de longueur ou de force | |
| $pT1/2^{A}_{z2}$ | Tension relative de la fmI dans z2 après disparition des forces dues à la viscosité au début de p2 d'un échelon de longueur<br>A pour Amorti (*Amortized*) | |
| $pT1/2^{E}_{z2}$ | Tension relative de la fmI dans z2 après disparition des forces dues à la viscosité au début de p2 d'un échelon de longueur<br>E pour Exagéré (*Exaggerated*) | |
| **pT2** | Tension relative de la fmI à la fin de p2 d'un échelon de longueur | |
| $pT2_{SBfast}$ | contribution à la remontée de la tension relative de $pT1/2^{E}_{z2}$ jusqu'à **pT2** | |
| **pT2DE** | Tension relative de la fmI à la fin de p2 d'un échelon de longueur ou de force calculée avec $pT1/2^{A}_{z2}$ | |
| $r_{filA}$ | Rayon d'1 filA | 3.5 nm |
| $r_{filM}$ | Rayon d'1 filM | 7.5 nm |
| $R^{WS}_{R_L}(\theta,\beta)$ | = cosθ + sinθ tanφ cosβ | |
| $R_{WS}$ | valeur moyenne de $R^{WS}_{R_L}(\theta,\beta)$ durant $\Delta t_{WS}$ considérée comme constante si $\|\beta\| \leq 42°$ | 0.96 (varie de 0.9 à 1.1 selon $L_{S1b}$ , $L_{S2}$ , $d_{AB}$ et $d_{filAM}$ ) |
| **r1** | région 1 du raccourcissement d'un hs, défini par $\left[0;\Delta X^{min}_{WS}\right]$ | de 0 à 3.2 nm |
| **r2** | région 2 du raccourcissement d'un hs, défini par $\left[\Delta X^{min}_{WS};d_{2sitA}\right]$ | de 3.2 à 5.4 nm |
| **r3** | région 3 du raccourcissement d'un hs, défini par $\left[d_{2sitA};2\cdot d_{2sitA}\right]$ | de 5.4 à 10.8 nm |
| **res** | Terme de l'éq. (2.7) considéré comme négligeable | |
| **RS** | *Recovery Stroke* : chemin menant de l'état DE à l'état WB | |
| **s** | N° itératif du sarcomère d'1 myofibrille | de 1 à Ns |
| $s,R_L$ | hsR ou hsL du sarcomère N° s | |
| $S_{R_L}^{WS}(\theta,\beta)$ | = cosθ + sinθ tanφ / cosβ | |
| $S_{WS}$ | valeur moyenne de $S_{R_L}^{WS}(\theta,\beta)$ durant $\Delta t_{WS}$ considérée comme constante si \|β\|≤42° | 0.96 (varie de 0.9 à 1.1 selon $L_{S1b}$ , $L_{S2}$ , $d_{AB}$ et $d_{filAM}$ ) |
| **SB** | Liaison forte (*Strong Binding*) de S1a avec 1 molA<br>Dans le modèle, l'état SB concerne la préparation du WS pendant laquelle aucun effort de traction n'est exercé sur le filM auquel est relié la tetM. | |
| $SB_{fast}$ | Evt composé de l'état SB suivi de l'évt « initiation du WS » avec apparition d'une force de traction exercée sur le filM auquel est relié la tetM.<br>Opérationnel uniquement en conditions isométriques | |
| $SB_{slow}$ | Evt composé de l'état SB suivi de l'évt « initiation du WS » avec apparition d'une force de traction exercée sur le filM auquel est relié la tetM Opérationnelles uniquement lors d'un raccourcissement continu | |
| **S1** | Sous-fragment 2 composé de S1a et S1b | |
| **S1a** | domaine catalytique ou moteur d'1 S1 modélisé mécaniquement par un cylindre | |
| **S1b** | domaine composé de chaines légères (ELC et RLC) ou levier d'1 S1 modélisé mécaniquement par une barre cylindrique | |
| **S2** | Sous-fragment 2 modélisé mécaniquement par une tige cylindrique | |
| **sitA** | site de liaison forte d'1 S1 avec 1 molA, situé sur la molA | |
| **t** | temps instantané | |
| $t_{stop_p1}^{1}$ | Temps de la fin de la phase 1 classique | 0.15 ms |
| $t_{stop_p1}^{2}$ | Temps de la fin de la phase 1 du modèle, lorsque les actions dues à la viscosité ont cessé ; appartient à la phase 2 classique | 0.25 ms |
| $t_{stop_p2}$ | Temps de la fin de la phase 2 | ~ 3 ms |
| **T** | Projection orthogonale de toute force F selon OX | |
| **T°** | Température exprimée en degré Kelvin (°K) ou degré Celsius (°C) | |
| $T^{fmI}$ | Tension appliquée à la fmI selon OX | |
| $T^{(k)}$ ; $T^{(j)}$ | Tension appliquée à la fmI selon OX pour l'échelon en longueur n° k, ou en force n° j | |
| $T^{myof}$ | Tension appliqué à chacune des $N_{myof}$ myofibrilles selon OX | $T^{myof} = T^{fmI} / N_{myof}$ |
| $T0^{hsL}$ | somme des actions exercées par les tetM du hsL initiant leur WS | |
| $T0^{fmI}$ | Tension maximale isométrique de la fmI où les positions angulaires des $\Lambda 0$ S1b suivent une distribution uniforme centrée sur $\theta 0$ avec une dispersion égale à $\delta_\theta$ | |
| **T1** | Tension minimale atteinte par la fmI à la fin de p1 après un raccourcissement par un échelon de longueur ou de force n° k | |
| $T_{stop_p1}^{1\ (k)}$ | Valeur de $T^{(k)}$ à $t_{stop_p1}^{1}$ pour l'échelon en longueur n° k | $T_{stop_p1}^{1\ (k)} = T1$ |
| $T_{stop_p1}^{2\ (k)}$ | Valeur de $T^{(k)}$ à $t_{stop_p1}^{2}$ pour l'échelon en longueur n° k | |
| **tetM** | tête de myosine II, composée de S1a+S1b+S2 | |
| $u^{(j)}$ | module de la vitesse relative constante de chaque M par rapport à Z durant la phase 4 d'une perturbation par un échelon de force n° j | |
| $u_{p1}^{(k)}$ | module de la vitesse relative constante de chaque M par rapport à Z durant la phase 1 d'une perturbation par un échelon de force n° k | |
| $u_{Ms,R}$ | vitesse relative de Ms par rapport à Zs-1 selon OX | |
| $u_{Ms,L}$ | vitesse relative de Ms par rapport à Zs selon OX | |
| **V** | Vitesse de raccourcissement de la fmI selon OX | |
| $V^{(j)}$ | Vitesse de raccourcissement de la fmI selon OX à l'échelon de force n° j | |
| $V_{p2}^{(j)}$ | Vitesse de raccourcissement de la fmI durant p2 d'un échelon de force n° j | |
| $V_{start_p2}^{(j)}$ | Vitesse de raccourcissement de la fmI OX au début de p2 d'un échelon de force n° j | |
| $V_{p1}^{(k)}$ | Vitesse de raccourcissement de la fmI durant p1 d'un échelon de longueur ou de force n° k | |
| $V^{max}$ | Vitesse maximale de raccourcissement de la fmI obtenue lors de p4 pour un échelon de force inférieur ou égal à $0.1 \cdot T0^{fmI}$ | |
| **WB** | Liaison faible (*Weak Binding*) de S1a avec 1 molA | |
| **WS** | *Working Stroke*<br>Dans le modèle, l'état ou l'évt WS signifie liaison forte de S1a avec moment moteur exercé sur S1b et force de traction exercée sur le filM auquel est relié la tetM | |
| **X** | Abscisse, où OX est un axe parallèle aux axes longitudinaux des filM, des filA, de la myofibrille et de la fmI | |
| $\mathbf{X0^*}$ | Abscisse sur OX caractérisant le déplacement linéaire relatif moyen correspondant au déplacement angulaire $\theta 0^*$ | 1.5 nm |
| **X1** | Abscisse sur OX définissant le déplacement linéaire relatif du disqM par rapport au disqZ correspondant au déplacement angulaire $\theta 1$ | -3.3 nm |
| **X2** | Abscisse sur OX définissant le déplacement linéaire relatif du disqM par rapport au disqZ correspondant au déplacement angulaire $\theta 2$ | 3.5 nm |
| **X3** | Abscisse sur OX définissant le déplacement linéaire relatif du disqM par rapport au disqZ correspondant au déplacement angulaire $\theta_{stopWS}$ | 6.7 nm |
| **Y** | Ordonnée, où OY est un axe quelconque perpendiculaire à OX appartenant au plan transversal du sarcomère, de la fmI ou de la myofibrille | |
| **Y°** | Ordonnée, où OY° est un axe perpendiculaire à OX et parallèle à l'axe qui passe par les centres du filA et du filM | |
| **Z** | 3ème coordonnée, où OZ est l'axe complémentaire, perpendiculaire au plan OXY | |
| **Z°** | 3ème coordonnée, où OZ° est l'axe complémentaire, perpendiculaire au plan OXY° | |
| $Z_s$ | disque Z ou ligne Z situé à gauche de Ms dans le sarcomère N°s | |
| $Z_{s,R_L} + \text{filA}$ | système mécanique rigide composé de la moitié du disque Z n° s avec ses filA associés à droite ou à gauche | |
| $Z_S + 2 \cdot \text{filA}$ | Système mécanique rigide composé du disque Z n° s avec ses filA associés à droite et à gauche | |
| **z1** | Zone 1 du raccourcissement d'un hs, défini par $\left[0;\Delta X_{WS}^{min}\right]$ | de 0 à 3.2 nm |
| **z2** | Zone 2 du raccourcissement d'un hs défini par $\left[\Delta X_{WS}^{min};\Delta X_{WS}^{Max}\right]$ | de 3.2 à 10 nm |
| $\beta$ | Angle entre le plan OXY dans le quel s'effectue le WS par rapport au plan de référence OXY° | de -42° à +42° |
| $\beta^{Max}$ | Module de la valeur maximale de $\beta$ | 42° |
| $\delta_X$ | $= \lvert X2 - X1 \rvert = \Delta X_{WS}^{Max} - \Delta X_{WS}^{min}$ | 6.8 nm |
| $\delta_\theta$ | $= \lvert \theta 2 - \theta 1 \rvert$ | 41° |
| $\Delta L_{p1}^{(k)}$ | longueur de raccourcissement de la fmI durant la phase 1 de l'échelon en longueur n° k | |
| $\Delta t_{DE}^{stepF}$ | Durée moyenne de survenue de l'évt DE concernant les phases qui suivent une perturbation par un échelon de force | 8 ms |
| $\Delta t_{DE}^{stepL}$ | Durée moyenne de survenue de l'évt DE concernant les phases qui suivent une perturbation par un échelon de longueur | <8 ms |
| $\Delta t_{p1}$ ou $\Delta t_{p1}^{F}$ | Durée de la phase 1 classique pour un échelon en longueur ou en force | 0.15 ms |
| $\Delta t_{RS}$ | Durée moyenne de RS | 30 ms |
| $\Delta t_{SBfast}$ | Durée moyenne de l'évt SBfast | 0.7 ms |
| $\Delta t_{SBslowt}$ | Durée moyenne de l'évt SBslowt | de 4 à 8 ms (conditions standard) |
| $\Delta t_{WS}^{(b)}$ | Durée du WS concernant le S1 n° b | |
| $\Delta X_{p1}^{(j)}$ | Variation de longueur d'un hs à la fin de p1 pour l'échelon de forcer n° j | |
| $\Delta X_{p1}^{(k)}$ | Variation de longueur d'un hs à la fin de p1 pour l'échelon de longueur n° k | |
| $\Delta X_{p1}^{T=0}$ | Raccourcissement maximal d'uns hs à la fin de p1 pour lequel la tension s'annule (T=0) | 4.45 nm |
| $\Delta X_{p2}^{T=0}$ | Raccourcissement maximal d'uns hs à la fin de p2 pour lequel la tension s'annule (T=0) | 10.7 nm |
| $\Delta X_{p2_vico=0}$ | Raccourcissement d'uns hs entre le début de p2 et l'instant où les forces de viscosité ont cessé leur action | |
| $\Delta X_{p2_amort}$ | Raccourcissement d'uns hs entre l'instant où les forces de viscosité ont cessé leur action et la fin de p2 | |
| $\Delta X^{(b)}_{s,R_L}$ | déplacement linéaire relatif de $D^{(b)}_{s,L}$ relativement à $A^{(b)}_{s,R}$ dans hsL et de $D^{(b)}_{s,R}$ relativement à $A^{(b)}_{s-1,L}$ dans hsR au cours d'un WS. | |
| $\Delta X_{WS}$ | Pas d'1 tetM selon OX lors du WS | ~ 7 nm en moyenne |
| $\Delta X^{Max}_{WS}$ | Valeur maximale de $\Delta X_{WS}$ | 10 nm |
| $\Delta X^{min}_{WS}$ | Valeur minimale de $\Delta X_{WS}$ | 3.2 nm |
| $\Delta\theta^{(b)}_{WS,R_L}$ | déplacement angulaire du S1b n° b dans hsL ou hsR au cours d'un WS. | |
| $\Delta\theta_{WS}$ | Variation angulaire de θ durant WS | |
| $\Delta\theta^{Max}_{WS}$ | Valeur maximale de $\Delta\theta_{WS}$ | 60° |
| $\Delta\theta^{(j)}_{WS}$ | intervalle de répartition des positions angulaires des $\Lambda^{(j)}_{WS}$ S1b dans 1 hs pour l'échelon de force n° j | |
| $\varphi$ | Angle entre S2 et OX dans le plan OXY° | de +15° à +10° dans hsR<br>de -15° à -10° dans hsL |
| $\chi^{fmI}_{z1}$ | Raideur de la fmI considérée comme un ressort mécanique linéaire durant p1 dans z1 avec $\chi^{fmI}_{z1} = \chi^{hs}_{z1}/2 \cdot Ns$ | |
| $\chi^{fmI}_{z2}$ | Raideur de la fmI considérée comme un ressort mécanique linéaire durant p1 dans z2 avec $\chi^{fmI}_{z2} = \chi^{hs}_{z2}/2 \cdot Ns$ | |
| $\chi^{hs}_{z1}$ | Raideur d'un hs considéré comme un ressort mécanique linéaire durant p1 dans z1 | 0.16 $nm^{-1}$ |
| $\chi^{hs}_{z2}$ | Raideur d'un hs considéré comme un ressort mécanique linéaire durant p1 dans z2 | 0.072 $nm^{-1}$ |
| $\Lambda_{TOT}$ | Nombre total de S1 dans 1 hs | de 50 000 à 300 000<br>selon typologie |
| $\Lambda_{CB}$ | Nombre de CB par hs | |
| $\Lambda^{(k)}_{WS}$ ; $\Lambda^{(j)}_{WS}$ | Nombre de WS par hs à l'échelon de longueur n°k ;<br>à l'échelon de force n° j | |
| $\Lambda 0$ | Nombre de WS par hs à T0<br>Nombre maximal de WS par hs | |
| $\nu^{hs}$ | coefficient de proportionnalité des forces de viscosité s'exerçant sur $(Z_{S,R_L} + filA)$ et $(M_{S,R_L} + filM)$ | |
| $\nu^{visc}_{p1}$ | module du coefficient de linéarité des forces de viscosité durant p1 | |
| $\theta$ | Angle entre S1b et CY donnant la position angulaire de S1b durant le WS | de +20° à -40° dans hsR<br>de -20° à +40° dans hsL |
| $\theta 0$ | Position angulaire moyenne de S1b avec le modèle de ressort entropique | -5° |
| $\theta 0^{lin}$ | Position angulaire moyenne de S1b avec le modèle de ressort mécanique | -4°30'' |
| $\theta 1$ | Borne inférieure de $\theta^{(b)}_{startW}$ | +25° dans hsR<br>-25° dans hsL |
| $\theta 2$ | Borne supérieure de $\theta^{(b)}_{startW}$ | -16° dans hsR<br>+16° dans hsL |
| $\theta^{(b)}_{startWS}$ | Position angulaire de θ au démarrage du WS du S1b de la tetM n° b | $\in [\theta 1; \theta 2]$ |
| $\theta_{stopWS}$ | Position angulaire de θ à l'arrêt du WS considérée comme une constante | +35° dans hsR<br>-35° dans hsL |
| $\theta 0$ | Position angulaire moyenne de la distribution uniforme des valeurs de θ en conditions isométriques | -5° dans hsR<br>+5° dans hsL |
| $\theta 0^*$ | Position angulaire moyenne des valeurs de θ répartis uniformément sur $\Delta\theta^{Max}_{WS}$ | +4° dans hsR<br>-4° dans hsL |

## Liste des Figures

# 1 Evènements auxquels est astreinte une tetM lors du raccourcissement d’une fibre musculaire

## 1.1 Cadre probabiliste, géométrique et mécanique

### *1.1.1 Préambule*

Nous renvoyons à la lecture d’articles de référence (Hanson and Huxley 1953; Huxley 1957; Huxley 1969; Squire, Al-Khayat et al. 2005) pour une description complète d’une fibre musculaire, d’une myofibrille et d’un sarcomère. Le comportement d’une fibre musculaire en contraction isométrique selon diverses conditions expérimentales (Gordon, Huxley et al. 1966) est supposé acquis. Nous présumons aussi connues les 4 phases de la réponse mécanique d’une fibre musculaire, qui après une stimulation tétanique, subit un raccourcissement transitoire, soit par un échelon de longueur (« *length step* »), soit par un échelon de force (« *force or load step* »); ces 4 phases ont été décrites en fonction des deux types de perturbation par A.F. Huxley (Huxley 1974).

Chaque sarcomère d’une myofibrille se décompose en 2 demi-sarcomères (Fig. 2 et 3E; chap. 2). Un demi-sarcomère sera noté **hs** *(« half-sarcomer* ») ; nous distinguerons le demi-sarcomère droit noté **hsR** («*half-sarcomer on the Right* ») et le demi-sarcomère gauche noté **hsL** («*half-sarcomer on the Left* »).

La tête de myosine[1] de type II (tetM) est modélisée par 3 segments rigides articulés entre eux, S1a, S1b et S2 (Fig. 2 et 3A ; chap. 2), suivant la description donnée par I. Rayment et ses coauteurs (Rayment, Rypniewski et al. 1993): soit, d’une part, le sous-fragment 1 (S1) formé par le domaine catalytique ou moteur (S1a) et le domaine composé de chaines légères ou levier (S1b), et d’autre part le sous-fragment 2 (S2) qui sert de tige de liaison entre S1 et le filament de myosine[1] (filM) .

Dans la suite de l’article, tous les sites de fixation des molécules d’actine présentes au sein de la fibre musculaire seront considérés comme disponibles pour un éventuel CB (« *cross-bridge* »; pont d’actomyosine).

Les évènements que subit une tetM sont formés d’états et de transitions d’états (Fig. 1). La série d’évènements étudiés dans les paragraphes suivants ne se veut pas une description exhaustive du « *cross-bridge cycle* » ou du « cycle de Lymn_Taylor », *i.e.* du cycle d’interactions chimico-mécaniques des tetM avec les filaments d’actine (filA) dont beaucoup de mécanismes restent à découvrir. Nous nous contentons de rapporter des évènements qui font l’objet d’un large consensus et pour lesquels nous émettons des hypothèses dans un cadre probabiliste, géométrique ou mécanique. Nous nous gardons de toute interprétation de nature biochimique.

---

[1] *Une molécule de myosine (molM) est composée de 2 S1, chacun relié au filM par S2 (Fig. 3A; chap. 2),*
*i.e. 1 molM = 2 S1a+ 2 S1b+ S2*
*1 tetM est composée d’un S1 et de S2, i.e. 1 tetM = 1 S1a+ 1 S1b+ S2*
*1 filM comporte environ 100 molécules de myosine (i.e, 200 S1) par hs.*
*1 filament d’actine (filA) est entouré par 3 filM, et d’après le système hexagonal, un nombre maximal d’environ 100 S1 par hs ont la possibilité spatiale de se lier au filA.*

**Fig. 1 : Graphe des transitions d'états concernant une tetM**

Les cercles bruns représentent les états d'une tetM
Les flèches vertes qui relient les états sont les transitions ou changements d'états subis par une tetM

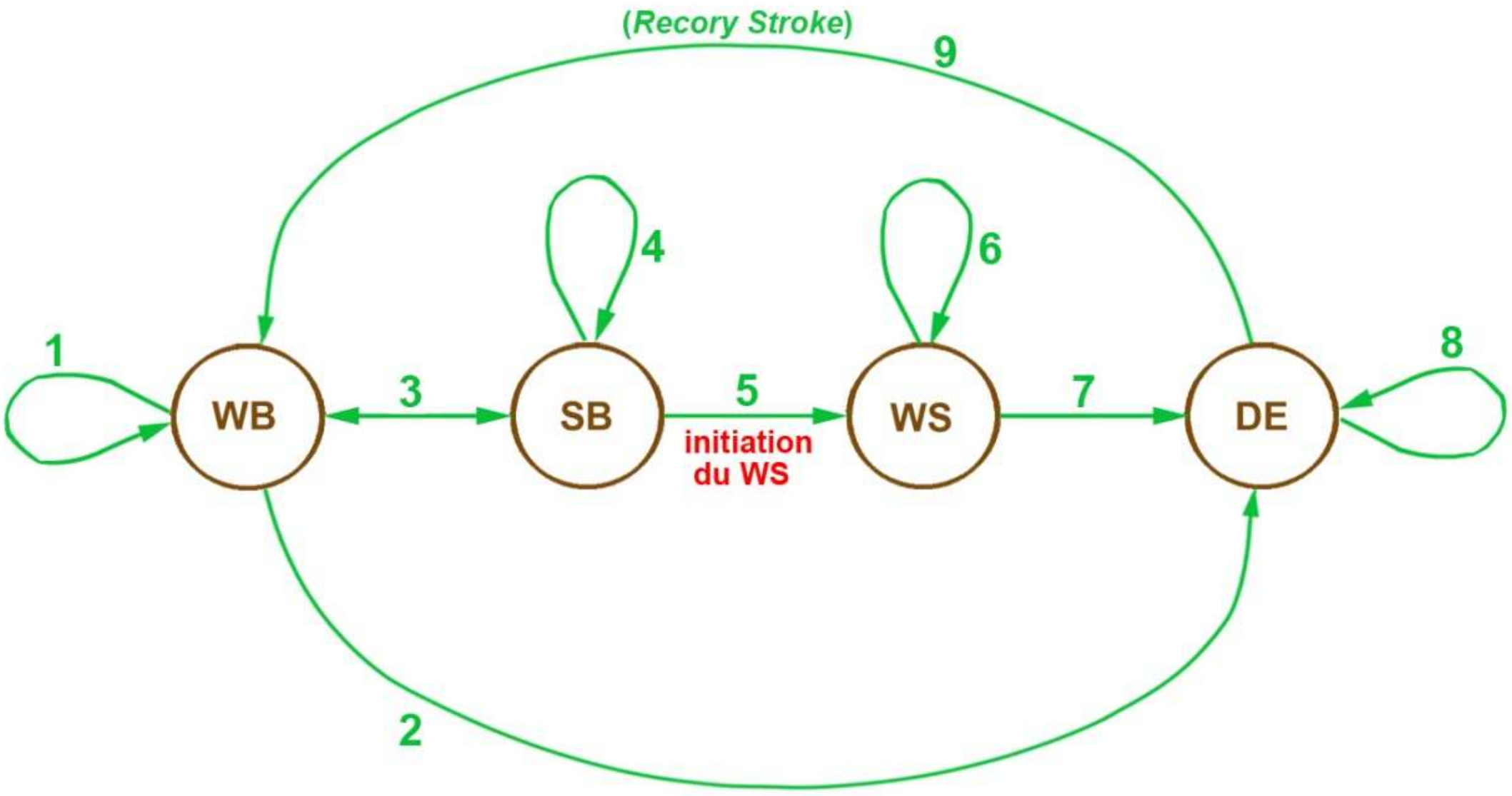

**WB** *(Weak Binding)* = liaison faible de S1a
**SB** *(Strong Binding)* = liaison forte de S1a et/ou déplacement de S1b et/ou moment moteur sans effort de traction exercé sur le filM
**WS** *(Working Stroke)* = liaison forte de S1a et moment moteur de S1b avec effort de traction exercé sur le filM
**DE** *(Detachment)* = S1a non lié

### *1.1.2 Probabilité de survenue d'un évènement aléatoire*

Soit un évènement aléatoire A concernant une tetM ; on définit la v.a. [1] $\boldsymbol{\varepsilon_A}$ comme le temps de survenue de A dans l'intervalle $[\mathbf{0};\mathbf{t}]$. La fonction de densité de $\boldsymbol{\varepsilon_A}$, notée $\mathbf{f_A}$, est classiquement la loi exponentielle qui décrit la survenue d'un événement unique aléatoire, telle que:

$$\mathbf{f_A(t) = \frac{1}{\Delta t_A} \cdot e^{-\frac{t}{\Delta t_A}}}$$

avec $\mathbf{t}$, le temps instantané

et $\mathbf{\Delta t_A = E(\boldsymbol{\varepsilon_A})}$, *i.e.* la durée moyenne de survenue de A

Selon l'éq. (A.16) de l'Annexe A, la probabilité de réalisation de A dans l'intervalle $[\mathbf{0};\mathbf{t}]$ est donnée par :

$$\mathbf{p_A = P(\boldsymbol{\varepsilon_A} \le t) = \int_0^t \frac{1}{\Delta t_A} \cdot e^{-\frac{u}{\Delta t_A}} \cdot du = 1 - e^{-\frac{t}{\Delta t_A}}} \qquad (1.1)$$

[1] *v.a. signifie variable aléatoire*

### *1.1.3 Probabilité de survenue d'une succession d'évènements aléatoires*

Pour déterminer la probabilité d'une suite d'évènements concernant 1 tetM, nous allons recourir à différentes méthodes :

**Méthode 1 où les durées moyennes de survenue des événements successifs sont connues**:
Considérons $\mathbf{E1}$ et $\mathbf{E2}$ deux événements aléatoires successifs dont les temps de survenue sont 2 v.a. suivant chacune une loi exponentielle de paramètres respectifs $1/\Delta t_1$ et $1/\Delta t_2$. Par définition d'une loi exponentielle, les durées de réalisation de $\mathbf{E1}$ et $\mathbf{E2}$ sont 2 v.a. indépendantes, qui suivent les 2 mêmes lois exponentielles de paramètres respectifs $1/\Delta t_1$ et $1/\Delta t_2$.
Avec (A.1) et (A.10) de l'Annexe A, la probabilité de réalisation d'un évènement noté $\mathbf{E_1 \rightarrow E2}$, succession des évènements E1 et E2, dont le temps de survenue est la somme des durées de réalisation de $\mathbf{E1}$ et $\mathbf{E2}$, se formule :

$$p_{E1\rightarrow E2} = 1 - \frac{\Delta t_1 \cdot e^{-\frac{t}{\Delta t_1}} - \Delta t_2 \cdot e^{-\frac{t}{\Delta t_2}}}{\Delta t_1 - \Delta t_2} \qquad (1.2)$$

avec $\Delta t_1 \neq \Delta t2$

La densité déterminée avec (A.13) à l'Annexe A permet le calcul de la probabilité de survenue de 3 évènements aléatoires successifs.

**Méthode 2 où au moins une des durées moyenne de survenue d'un des événements successifs est inconnue**:
L'ensemble de la chaine d'évènements est modélisé comme un seul évènement aléatoire G qui globalise toute la chaîne. La probabilité de réalisation de G s'identifie à celle donné par (1.1), soit :

$$p_G = \int_0^t \frac{1}{\Delta t_G} \cdot e^{-\frac{u}{\Delta t_G}} \cdot du = 1 - e^{-\frac{t - \Delta t_{P_G}}{\Delta t_G}} \qquad (1.3)$$

avec $\Delta t_G$, la durée moyenne de survenue de G qui sera appréciée empiriquement.
et $\Delta t_{P_G}$, le décalage temporel de mise en place de l'événement final de la chaine G, qui tient compte de la durée moyenne de la succession des évènements le précédant, succession notée $\{P_G\}$.
$\Delta t_{P_G}$ sera aussi apprécié empiriquement.

## 1.2 Déplacements browniens de S1 et S2 avant et après le CB

Dans un liquide (par ex, le liquide intracellulaire d'une fibre musculaire), une grosse molécule de la taille d'une dizaine de nanomètres (par ex, S1 ou S2) subit les multiples chocs[1] des petites molécules environnantes qui composent le liquide. On observe que la trajectoire de la grosse molécule suit une trajectoire erratique, *i.e.* se déplace suivant un mouvement brownien. L'agitation brownienne correspond à la fluctuation du nombre d'états microscopiques de la grosse molécule et donc à la fluctuation de son entropie. En 1910, A. Einstein a calculé que cette fluctuation était égale à $\mathbf{k_B/2}$, ce qui correspond à une fluctuation d'énergie à chaque choc, notée $\mathbf{\delta E_{th}}$ (**th** pour thermique), telle que :

$$\mathbf{\delta E_{th} = k_B \cdot T°/2} \tag{1.4}$$

où $\mathbf{k_B = 1.38 \cdot 10^{-23}\ J.°K^{-1}}$ est la constante de Boltzmann.

[1] *avec un temps moyen entre 2 chocs consécutifs de l'ordre de $10^{-9}$ à $10^{-11}$ s, temps fonction de divers paramètres dont la température.*

et $\mathbf{T°}$ est la température du liquide exprimée en °K.

Soit pour $\mathbf{0°C \leq T° \leq 30°C}$ :

$$\mathbf{\delta E_{th} \approx 2 \cdot 10^{-21}\ J} \tag{1.5}$$

Ces fluctuations d'énergie génèrent des forces appelées usuellement forces thermiques. En conséquence, les tetM (et tous les éléments constitutifs d'une fibre musculaire) se mouvant dans le liquide intracellulaire sont soumises en permanence à de telles forces thermiques.

Lorsque la tetM n'est pas liée fortement au filA (état DE; Fig. 1), *i.e.* avant et après le CB, S2 tout en restant lié au filM d'une part, et S1 relié à S2 d'autre part, suivent des trajectoires aléatoires.

## 1.3 Evénement WB : liaison faible de la tetM due à la force d'attraction de la molA

Chaque molécule d'actine (molA) présente un site de fixation (Geeves and Holmes 1999) pour une tetM que nous nommerons sitA.

Les molA constituant un filA forment une double hélice ; $\mathbf{d_{2sitA}}$, la distance entre 2 sitA voisins sur la même hélice, est selon les auteurs comprise entre 5.3 et 5.5 nm (Piazzesi and Lombardi 1995; Geeves and Holmes 1999 ; Nielsen 2002 ; Squire, Al-Khayat et al. 2005 ). Par la suite, nous retiendrons la valeur médiane suivante :

$$\mathbf{d_{2sitA} = 5.4\,nm} \tag{1.6}$$

Avant le CB, S1a et S1b forment un segment rigide (Rayment, Holden et al. 1993). Si, poussé par S2 selon l'axe longitudinal du filM, S1 se rapproche suffisamment d'une molécule d'actine (molA) du filA tout en étant attiré par une force d'origine électrique, un CB se forme. et la nature de la liaison entre la tetM et la molA est dite « faible » (Yu and Brenner 1989 ; Rayment, Holden et al. 1993 ; Holmes, Angert et al. 2003 ; Geeves and Holmes 2005; Squire and Knupp 2005).

La valeur de force de la liaison faible doit :

- être supérieure à $\mathbf{3 \cdot \delta E_{th}}$ [1] sinon il n'y a pas attraction (rappelons que S1 et S2 subissent en permanence l'action des forces thermiques).
- permettre la rupture de cette liaison (principe d'une liaison faible) pour qu'aléatoirement plusieurs chocs libèrent simultanément la tetM.

---

[1] *Les chocs engendrent des déplacements d'1 tetM selon les 3 axes de l'espace; il faut donc multiplier par 3 d'après le théorème de l'équipartition de l'énergie.*

On peut donc raisonnablement formuler que l'énergie de la liaison faible notée $\delta E_{WB}$ (WB pour « *Weak Binding* », *i.e.* liaison faible) doit varier entre 2 et 4 fois $(3 \cdot \delta E_{th})$, soit :

$$6 \cdot \delta E_{th} \leq \delta E_{WB} \leq 12 \cdot \delta E_{th} \qquad (1.7)$$

La force d'attraction de la liaison faible notée $F_{WB}$ permet alors le déplacement entre 2 sitA voisins avant ou après sur la même hélice du filA, telle qu'approximativement[1] d'après (1.7) :

$$\frac{6 \cdot \delta E_{th}}{d_{2sitA}} \leq F_{WB} \leq \frac{12 \cdot \delta E_{th}}{d_{2sitA}} \qquad (1.8)$$

Soit avec les valeurs fournies par (1.5) et (1.6) :

$$2\,pN \leq F_{WB} \leq 5\,pN \qquad (1.9)$$

Durant la liaison faible, S1 continue à subir les chocs des molécules du liquide intercellulaire et suit une trajectoire brownienne dans un volume au dessus de la molA circonscrit par la force de rappel de la liaison faible du CB (état WB et chemin 1; Fig. 1). Plusieurs enchainements d'évènements peuvent advenir; nous en distinguons trois :

- la tetM démarre un WS (chemins 3 et 4 avec état SB, puis chemin 5 vers état WS) : voir paragraphes suivants.
- la tetM rompt la liaison faible (chemin 2 vers état DE), soit sous l'action de poussée de S1 par S2 lors d'un raccourcissement de la fibre musculaire, soit de manière aléatoire sous l'action des forces thermiques, soit sous l'action conjointe des 2.
- la tetM demeure en WB (chemin 1)

## 1.4 Evénement SB : préparation du WS suivie de son initiation

La succession des évènements menant de l'état WB à l'état WS est une boîte noire. Le modèle initial du « swinging lever arm » (Rayment, Holden et al. 1993) semblait indiquer que les 3 occurrences, « liaison forte », « début du mouvement de rotation de S1b » et « démarrage du couple moteur », étaient quasi-simultanées, mais des études ultérieures ont suggéré que les survenues de ces 3 évènements pouvaient être découplées (Geeves and Holmes 1999; Houdusse, Kalabokis et al. 1999; Houdusse and Sweeney 2001; Geeves and Holmes 2005; Caremani, Dantzig et al. 2008; Lowey and Trybus 2010; Prochniewicz, Guhathakurta et al. 2013).

A l'appui de ces considérations, nous présentons les 3 évènements suivants comme conditions nécessaires au démarrage du WS :

- soit $\mathbf{OX}$, un axe parallèle aux axes longitudinaux des filM et des filA (Fig. 2, 3B et 3E ; chap. 2), et $\mathbf{OY^o}$ l'axe perpendiculaire à $\mathbf{OX}$ passant par les centres des filA et filM associés par la liaison faible de la tetM (Fig. 3B et 3E). S1 se positionne dans le plan $\mathbf{OXY}$ formant selon OX un angle $\beta$ avec le plan $\mathbf{OXY^o}$ de telle manière que S2 devienne une tige de liaison rigide entre S1b et le filM (Fig. 3B et 3D).

[1] *Le calcul du travail repose sur la formule W=F·L valable quand F est cste, ce qui n'est évidemment pas le cas pour une force d'attraction en $1/r^n$ ; les inégalités apportent un ordre de grandeur moyen.*

- S1 rencontre un emplacement précis à la surface de la molA qui est le site de la liaison forte (sitA). La liaison faible entre S1a et la molA du filA se transforme en liaison « forte » de type encastrement; S1 et S2 constituent alors une chaine déformable fermée composée de 3 segments rigides (terminologie mécanique du CB de notre modèle durant le WS).
- un mécanisme biologique logé dans S1a agit sur S1b de manière analogue au comportement d'un ressort angulaire agissant sur un levier.

Une fois ces 3 évènements réunis, le WS démarre, *i.e.* la liaison entre S1a et S1b étant une liaison pivot moteur et la liaison entre S1b et S2 étant une liaison rigide, la tetM exerce une force de traction sur le filM. La mise en place des ces 3 évènements, dont nous ne connaissons ni l'ordre temporel, ni les mécanismes, est nommée, peut-être improprement, SB (« *strong binding* »).

La position angulaire de S1b dans le plan $\mathbf{OXY}$ au démarrage du WS est notée $\boldsymbol{\theta}_{\mathbf{startWS}}$ (Fig. 3C ; chap. 2). Nous postulons (**Hypothèse 1**) que $\boldsymbol{\theta}_{\mathbf{startWS}}$ varie, dans le plan $\mathbf{OXY}$, à l'intérieur d'un intervalle borné par 2 constantes, nommées $\boldsymbol{\theta 1}$ et $\boldsymbol{\theta 2}$ pour tout S1 de la fibre musculaire :

$$\boldsymbol{\theta}_{\mathbf{startWS}} \in [\boldsymbol{\theta 1};\boldsymbol{\theta 2}] \tag{1.10}$$

L'**Hypothèse 1** contient implicitement la réunion des 3 conditions d'obtention d'un WS.

Nous postulons que pour un intervalle de temps égal ou supérieur à la milliseconde, la distribution des $\boldsymbol{\theta}_{\mathbf{startWS}}$ est uniforme dans l'intervalle $[\boldsymbol{\theta 1};\boldsymbol{\theta 2}]$ dans chaque hs de la fibre musculaire (**Hypothèse 2**).

L'Hyp. 2 d'uniformité a été constatée par M. Reconditi et ses coauteurs (Reconditi, Linari et al. 2004) et confirmée par des chercheurs de la même équipe (Piazzesi, Reconditi et al. 2007 ; Reconditi, Brunello et al. 2011) ou travaillant en collaboration (Huxley, Reconditi et al. 2006).

L'évènement global SB (chemins 3 et 4 vers état SB, puis chemin 5 avant état WS ; Fig. 1) peut comporter d'autres évènements que les 3 qui concourent à l'initiation du WS. Par exemple, un mouvement de rotation de S1b peut s'effectuer de manière aléatoire à cause des forces thermiques, ou de manière déterministe par un moment moteur, mais en supposant que S2 est courbée du fait de sa position actuelle, S2 n'étant pas un solide indéformable ne remplit pas son rôle de tige de liaison et aucune force motrice n'est exercée sur le filM : le WS ne pourra être effectif que lorsque S2 se rigidifiera dans une position plus avancée. Il s'agit d'un scénario éventuel parmi beaucoup d'autres.

Deux catégories d'évènement SB, auxquelles les Hyp. 1 et 2 s'appliquent, sont présentées.

### *1.4.1 Evènement $SB_{fast}$ en conditions isométriques*

La variable aléatoire $\varepsilon_{SBfast}$ est définie comme la durée mise par S1 pour que, quasi simultanément, S1b atteigne la position géométrique de déclenchement de WS postulée avec l'Hyp. 1 et que S1a se fixe fortement au sitA.

Avec (1.3), la probabilité pour que l'évènement global $SB_{fast}$ advienne dans l'intervalle $[0; t]$ est égale à :

$$p_{SBfast} = P(\varepsilon_{SBfast} \leq t) = 1 - e^{-\frac{t}{\Delta t_{SBfast}}} \tag{1.11}$$

avec $\Delta t_{SBfast} = E(\varepsilon_{SBfast})$, *i.e.* la durée moyenne du déplacement de S1 pour atteindre la position adéquate relativement à l'évt $SB_{fast}$

Dans (1.11) par rapport à la formulation de (1.3), il n'y a pas de décalage temporel, soit :

$$\Delta t_{p_{SBfast}} = 0 \tag{1.12}$$

On note que $\Delta t_{SBfast}$ est une constante homogène à un temps et dépendante des conditions expérimentales (température, pH, …etc.).

Nous postulons (**Hypothèse 3**) que $\Delta t_{SBfast}$ est de l'ordre de la milliseconde tel que :

$$\Delta t_{SBfast} = 0.7\ ms \tag{1.13}$$

Avec la valeur de $\Delta t_{SBfast}$ apportée par (1.13), le plateau de l'exponentielle mise en équation dans (1.11) commence à 3 ms environ, valeur qui définira le début de la phase 2 d'une perturbation par un échelon de longueur (voir chap. 8).

Nous formulons les conditions de survenue de l'évènement $SB_{fast}$ avec l'**Hypothèse 4** : l'événement $SB_{fast}$ est réalisé lorsque la longueur de la fibre musculaire reste constante et cesse dès que la fibre musculaire se raccourcit continument. Autrement dit, en conditions isométriques, tous les S1 se comportent en mode $SB_{fast}$ avant d'initier un WS.

### *1.4.2 Evènement $SB_{slow}$ lors du raccourcissement continu de la fibre musculaire*

La variable aléatoire $\varepsilon_{SBslow}$ est définie comme la durée mise par S1 pour que, de manière non immédiate, S1a se fixe fortement au sitA et que S1b atteigne la position géométrique d'initiation du WS selon l'Hyp. 1.

Avec (1.3), la probabilité pour que $SB_{slow}$ advienne dans l'intervalle $[0;t]$ est égale à :

$$p_{SBslow} = P(\varepsilon_{SBslow} \leq t) = 1 - e^{-\frac{t}{\Delta t_{SBslow}}} \quad (1.14)$$

où $\Delta t_{SBslow} = E(\varepsilon_{SBslow})$, est la durée moyenne de réalisation de l'évt $\varepsilon_{SBslow}$

avec $\Delta t_{SBslow} > \Delta t_{SBfast}$ (1.15)

Dans (1.14) par rapport à la formulation de (1.3), on ne prend pas en compte de décalage temporel, soit :

$$\Delta t_{P_{SBslow}} = 0 \quad (1.16)$$

$\Delta t_{SBslow}$ est une constante homogène à un temps et dépendante des conditions expérimentales (typologie musculaire, température, pH, …etc.). $\Delta t_{SBslow}$ est un paramètre important du modèle car sa valeur est le facteur déterminant de la vitesse de raccourcissement continu de la fibre musculaire, notamment lors de la phase 4 d'une perturbation par un échelon de force.

Nous postulons (**Hypothèse 5**) que $\Delta t_{SBslow}$ est égal à plusieurs millisecondes, soit aux températures expérimentales proches de 0 °C :

$$4ms \leq \Delta t_{SBslow} \leq 8\ ms \quad (1.17)$$

**Hypothèse 6** : l'événement $SB_{slow}$ survient en condition de raccourcissement continu, i.e. lorsque la fibre musculaire se raccourcit continument, tous les S1 se comportent en mode $SB_{slow}$ avant d'initier un WS.

**<u>Précision apportée :</u>**

L'état SB concerne la préparation du WS pendant laquelle aucun effort de traction n'est exercé sur le filM auquel est reliée la tetM. Le retour vers l'état WB est possible avec chemin 3.

Les 2 évts $SB_{fast}$ et $SB_{slow}$ comportent l'état SB avec chemins 3 et 4, suivi de l'évt « initiation du WS » représenté sur la Fig. 1 par le chemin 5, *i.e.* la fin de ces 2 évts est caractérisée par la réalisation des conditions imposées à l'Hyp. 1 et par l'apparition d'une force de traction exercée sur le filM auquel est relié la tetM. Le retour vers l'état WB est impossible.

.

## 1.5 Evénement WS : déroulement du WS

Par définition, un WS (état WS et chemin 6; Fig. 1) est mécaniquement efficace lorsqu'un moment moteur est appliqué sur S1b et qu'au cours de leurs déplacements les 3 segments S1a, S1b et S2 restant rigides, un effort de traction est exercé sur le filM auquel est relié la tetM par l'entremise de S2.

Une nouvelle hypothèse est émise : durant la totalité du WS la trajectoire de S1b est une trajectoire déterministe[1] qui s'effectue entièrement dans le même plan OXY défini au paragraphe précédent (Fig. 3B et 3D; chap. 2); en conséquence $\beta$ , l'angle entre $\mathbf{OXY}$ et $\mathbf{OXY^0}$ selon OX est une constante durant le WS **(Hypothèse 7)**.

A l'échelle nanométrique les quantités d'accélération linéaires et angulaires, d'origines gravitationnelle ou inertielle, sont négligeables; idem pour la poussée d'Archimède. A la formulation précédente de l'Hyp. 7, on ajoute que les forces de viscosité s'exerçant sur une molécule de myosine[2] (molM) sont aussi négligeables **(Hypothèse 7+)**.

Un complément qui sera justifié au chapitre 3 est apporté à l'Hyp. 7 : la valeur maximale de $|\beta|$ notée $\beta^{\mathbf{Max}}$ est égale à :

$$\beta^{\mathbf{Max}} = 42° \quad (1.18)$$

### *Conséquences des hypothèses 1 et 7 :*

**Corolaire 1 :** seul un des 2 S1 reliés à S2 (Fig. 3A ; chap. 2) a la possibilité de se lier à un sitA pour y effectuer un WS. En effet la structure à double hélice du filA et le fait qu'il n'y ait qu'un seul S1 par sitA rendent géométriquement impossible le cas où les 2 S1 d'une molM effectueraient leur WS en partie ou en totalité ensemble; car dans ce cas, au moins un des 2 S1 ne suivrait pas les modalités imposées par les Hyp. 1 et 7. Cette incidence est corroborée par d'autres travaux (Juanhuix, Bordas et al. 2001; Piazzesi, Reconditi et al. 2002).

**Corolaire 2 :** d'après la disposition hélicoïdale des sitA (Fig. 1 et 2), l'Hyp. 7 avec (1.18) et la condition dictée par (1.10) impliquent que seuls 4 sitA[3] parmi les 13/6 sitA d'un motif d'un filA sont accessibles pour 1 molM. Ainsi, par symétrie par rapport au plan OXY°, et selon le corolaire 1, il n'y a au maximum que 2 sitA accessibles par S1 parmi les 13/6 sitA d'un motif d'un filA.
Dans la structure hexagonale transversale propre au sarcomère du muscle squelettique, chaque filA est entouré de 3 filM ; ainsi la valeur de $\beta^{\mathbf{Max}}$ de (1.18) autorise la cohabitation de 6 S1 à la même abscisse X puisque l'inégalité suivante est vérifiée :

$$\beta^{\mathbf{Max}} < \frac{360°}{6} = 60°$$

---

[1] *Les perturbations créées par les forces thermiques sont considérées comme négligeables.*

[2] *Mais à l'échelle de la myofibrille, les forces de viscosité doivent être prises en compte (voir Hyp. 9).*

[3] *L'angle selon OX formé par 4 sitA consécutifs est égal à : $3 \cdot (360° / 13) \approx 83° \approx 2 \cdot \beta^{Max}$*

Selon l'Hyp. 2 d'uniformité et le résultat obtenu (une seule tetM[1] sur les 2 a la possibilité de se fixer sur 1 des 4 sitA possibles à chaque motif 13/6 du filA), on déduit la probabilité maximale de WS, notée $\mathbf{p_{WS}^{Max}}$, égale au rapport de la distance entre 4 sitA sur la longueur du motif 13/6 (36 nm), soit avec (1.6) :

$$\mathbf{p_{WS}^{Max} = \frac{1}{2} \cdot \frac{3 \cdot 5.4}{36} \approx 22.5\%} \qquad (1.19)$$

Ce résultat est compatible avec les valeurs de la littérature qui varient entre 15 et 30% (Ford, Huxley et al. 1985; Linari, Dobbie et al. 1998 ; Corrie, Brandmeier et al. 1999; Hopkins, Sabido-David et al. 2002 ; Huxley, Reconditi et al. 2006a ; Piazzesi, Reconditi et al. 2007; Reconditi, Brunello et al. 2011).

Nous formulons une hypothèse supplémentaire (**Hypothèse 8**) : le moment actif $\mathcal{M}$ produit par le domaine moteur est d'origine entropique et est modélisé durant le WS par une équation de type hyperbolique, tel que :

$$\mathcal{M} = \mathbf{A \cdot \frac{k_B \cdot T^\circ}{(\theta - \theta_0)} + B} \qquad (1.20)$$

où $\boldsymbol{\theta}$ est l'angle formé par S1b avec la verticale dans le plan $\mathbf{OXY}$ (Fig. 3C et 3D)

et où $\mathbf{A}$, $\mathbf{B}$ et $\boldsymbol{\theta_0}$ sont 3 constantes qui seront définies au chapitre 3.

**Hypothèse 9** : les caractéristiques mécaniques du disque Z (avec les filA répartis à droite et à gauche et avec les autres protéines associées) et du disque M (avec les filM répartis à droite et à gauche et avec les autres protéines associées) sont similaires. Les masses des tetM dans les états WB, SB et DE (Fig. 1) sont intégrées à la masse du disque M et on admet que les masses des tetM dans l'état WS ne perturbent pas cette égalité. On postule ainsi que les masses et les constantes de proportionnalité entre force de viscosité[2] et vitesse de déplacement de ces 2 systèmes, (disqM+2·filM) et (disqZ+2·filA), sont égales.

La fin du WS nécessite une autre hypothèse (**Hypothèse 10**) : la position angulaire de fin de WS[3] de S1b dans le plan $\mathbf{OXY}$ est une constante notée $\boldsymbol{\theta_{stopWS}}$ pour tout S1.

**Corolaire 3 :** à l'appui des Hyp. 1, 7 et 10 et sur la base de données de la littérature, nous déduisons que le pas d'une tetM durant le WS ($\mathbf{\Delta X_{WS}}$) varie entre une valeur minimale ($\mathbf{\Delta X_{WS}^{min}}$) et une valeur maximale ($\mathbf{\Delta X_{WS}^{Max}}$), tel que (voir chap. 3) :

$$\mathbf{\left(\Delta X_{WS}^{min} \approx 3.2\,nm\right) \leq \Delta X_{WS} \leq \left(\Delta X_{WS}^{Max} \approx 10\,nm\right)} \qquad (1.21)$$

[1] *La probabilité, que les 2 S1 d'une molM se lient chacun à la suite sur l'un des 4 sitA voisins propres à chacun, est faible.*

[2] *A l'échelle de la myofibrille dans laquelle sont présentes une trentaine de protéines, notamment les protéines de maintien, les forces de viscosité ne sont pas négligeables pour les disques Z et M pendant la phase 1 (voir chap. 7 et 10, Annexe D).*

[3] *A l'exception notable de la phase 4 relative aux échelons de force proches de la tension isométrique maximale ou aux faibles valeurs, pour lesquels est postulée la possibilité d'un détachement précoce (voir Hyp. 13 et chap. 15).*

**<u>Conséquences</u> : d**'après l'Hyp. 2 de distribution uniforme, les valeurs de l'inégalité (1.21) fournissent celle du pas moyen, noté $\overline{\Delta X_{WS}}$ :

$$\overline{\Delta X_{WS}} = \frac{\Delta X_{WS}^{Max} + \Delta X_{WS}^{min}}{2} = 6.6 \text{ nm} \tag{1.22}$$

Valeur conforme à celles de différents modèles proposés dans la littérature (voir Fig. 4.C dans (Piazzesi, Reconditi et al. 2007)).

**Hypothèse 11** : aux basses températures proches de 0°C, le nombre des S1 qui appartiennent aux hs situés au milieu de la fibre musculaire et qui ont effectué leurs WS complets au cours des 20 nm de mise en place de la tension isométrique est négligeable.

## 1.6 Evénement DE : détachement de S1 du filA en cours ou à la fin du WS

Le détachement de la tetM après le WS (chemin 7 vers état DE; Fig. 1) est une succession d'évènements ; ce processus est, notamment, conditionné par l'arrivée d'une molécule d'ATP (Yu and Brenner 1989 ; Rayment, Holden et al. 1993; Geeves and Holmes 1999; Geeves and Holmes 2005). L'enchaînement des évènements concourant au détachement d'un S1a du filA reste à être élucidé (Kuhner and Fischer 2011).

Dans notre modèle, l'évènement DE (« *DEtachment*») est apprécié comme un évènement aléatoire global dont la probabilité de réalisation est donnée selon (1.3) par :

$$p_{DE} = P(\varepsilon_{DE} \leq t) = 1 - e^{-\frac{t - \Delta t_{P_{DE}}}{\Delta t_{DE}}} \tag{1.23}$$

avec $\Delta t_{DE} = E(\varepsilon_{DE})$, *i.e.* la durée moyenne de détachement de S1

$\Delta t_{DE}$ et $\Delta t_{P_{DE}}$ sont des constantes dépendantes des conditions expérimentales (température, pH, etc) et de la concentration intracellulaire en ATP (Cooke and Bialek 1979; Nyitrai, Rossi et al. 2006 ).

Le modèle distingue deux cas :

### *1.6.1 Conditions isométriques*

**Hypothèse 12** : les valeurs de $\Delta t_{DE}$ et $\Delta t_{P_{DE}}$ sont prises égales à plusieurs millisecondes telles que[1] :

$$\Delta t_{DE}^{stepL} \approx 8 \text{ ms} \tag{1.24}$$

$$\Delta t_{P_{DE}}^{stepL} \approx 3 \text{ ms} \tag{1.25}$$

Valeur compatible avec les données correspondantes de la littérature (Piazzesi and Lombardi 1995 ; Kuhner and Fischer 2011).

[1] *La valeur de 3 ms pour $\Delta t_{P_{DE}}$ correspond à la fin de la phase 2 d'un échelon de longueur (voir chap.8.)*

**Complément de l'hypothèse 12** : une tetM en cours de WS se détache lorsque S1b a atteint la position finale commune $\theta_{stopWS}$.

En conditions isométriques, le démarrage de WS est conditionné par les Hyp 1, 3 et 4 et la fin du WS par l'Hyp 12 et son complément.

#### *1.6.2 Raccourcissements continus*

**Hypothèse 13** : les valeurs de $\Delta t_{DE}$ et $\Delta t_{P_{DE}}$ sont inférieures aux précédentes telles que :

$$\Delta t_{DE}^{stepF} < 8 \text{ ms} \qquad (1.26)$$

$$\Delta t_{P_{DE}}^{stepF} < 3 \text{ ms} \qquad (1.27)$$

**Complément de l'hypothèse 13** : une tetM en cours de WS a la possibilité de se détacher avant que S1b n'atteigne la position finale $\theta_{stopWS}$

Lors d'un raccourcissement continu, le démarrage de WS est conditionné par les Hyp. 1, 5 et 6 et la fin du WS par l'Hyp 13 et son complément.

## 1.7 Evénement RS : « *Recovery Stroke* »

S1 étant détaché, l'événement intitulé « ***Recovery Stroke*** » concerne classiquement le rétablissement de la conformation entre S1a et S1b qui a précédé la mise en place du WS, *i.e.* la rotation de S1b opposée en direction et égale en module à la rotation effectuée par S1b durant le WS (Yu and Brenner 1989 ; Rayment, Holden et al. 1993; Geeves and Holmes 1999 ; Houdusse, Szent-Gyorgyi et al. 2000; Geeves and Holmes 2005), *i.e.* le passage de l'étape 3 à l'étape 4 du cycle de Lymn-Taylor (Koppole, Smith et al. 2007) .

L'événement RS est différentié de l'évènement DE puisqu'il a été observé sans que la tetM ne soit attachée au filA et donc ne se libère (Sugi, Minoda et al. 2008).

Par définition, **RS** est une succession d'évènements entre l'état DE vers l'état WB avec le chemin 9, comprenant possiblement l'état WB et les chemins 1, 2, 3 et 8. Dans notre modèle, **RS** est considérée comme un évènement aléatoire global. La probabilité de réalisation de RS est donnée avec (1.3) par :

$$p_{RS} = 1 - e^{-\frac{t - \Delta t_{P_{RS}}}{\Delta t_{RS}}} \qquad (1.28)$$

avec $\Delta t_{RS} = E(\varepsilon_{RS})$, *i.e.* la durée moyenne du RS

$\Delta t_{RS}$ est une constante dépendante des conditions expérimentales (température, pH, etc).

$\Delta t_{RS}$ est de l'ordre de plusieurs dizaine de millisecondes (**Hypothèse 14**), tel qu'empiriquement[1] :

$$\Delta t_{RS} \approx 30\ \mathrm{ms} \qquad (1.29)$$

$$\Delta t_{P_{RS}} \approx 3\ \mathrm{ms} \qquad (1.30)$$

La valeur de $\Delta t_{RS}$ est comprise entre 25 ms (Piazzesi, Reconditi et al. 1999) et 50ms (Huxley 2000).

Le mécanisme mis en œuvre lors du RS pourrait être d'origine entropique (Baumketner and Nesmelov 2011).

---

[1] *La valeur de 3 ms $\Delta t_{PRS}$ correspond à la fin de la phase 2 (voir chap.8)*

# 2 Modélisation mécanique

## <u>2.1 Cinématique d'une fibre musculaire idéalisée soumise à une force T et se raccourcissant à vitesse constante V</u>

La fibre musculaire idéalisée ($\mathbf{fmI}$) est formée de $\mathbf{N_{myof}}$ myofibrilles similaires réparties de manière parfaitement homogène dans le plan $\mathbf{OYZ}$ de la Fig. 2. Chacune des $\mathbf{N_{myof}}$ myofibrilles est constituée d'un nombre identique de sarcomères ($\mathbf{Ns}$) qui sont alignés en série selon l'axe longitudinal $\mathbf{OX}$ (Fig. 2). Lors la phase de raccourcissement, l'extrémité droite de la fmI reste fixe et une force $\mathbf{T}$[1] constante est appliquée à son extrémité gauche[2]; chaque myofibrille se raccourcit à la même vitesse $\mathbf{V}$, la vitesse de raccourcissement de la $\mathbf{fmI}$. De par l'homogénéité de la répartition, la force qui s'exerce sur chaque myofibrille ($\mathbf{T^{myof}}$) est telle que (Fig. 2) :

$$\mathbf{T^{myof}} = \frac{\mathbf{T}}{\mathbf{N_{myof}}} \qquad (2.1)$$

Ainsi l'étude d'une myofibrille suffit pour comprendre le comportement mécanique de la $\mathbf{fmI}$.

**Fig. 2 : Cinématique d'une myofibrille constituée de $N_S$ sarcomères alignés en série selon OX**

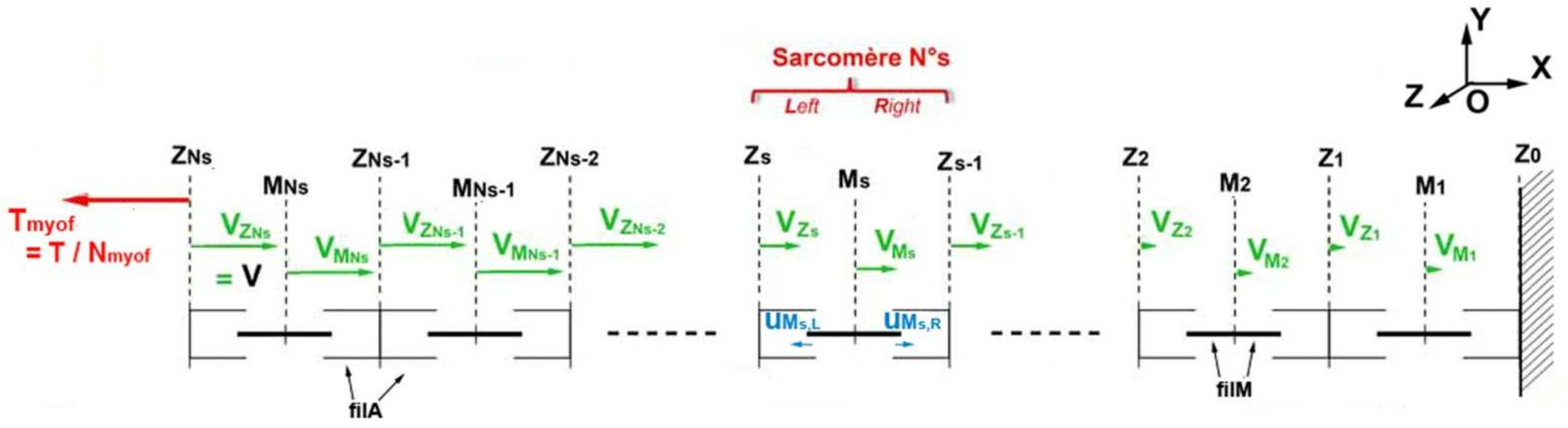

On définit le repère galiléen $\mathbf{OXYZ}$ ainsi :

- $\mathbf{O}$ est un point fixe quelconque du laboratoire
- $\mathbf{OX}$ est un axe parallèle aux axes longitudinaux des $\mathbf{filM}$ et des $\mathbf{filA}$ (Fig. 2, 3B et 3E)
- $\mathbf{OY}$ est un axe quelconque perpendiculaire à $\mathbf{OX}$ appartenant à un plan de coupe transversale de la myofibrille (Fig. 2, 3B, 3D et 3E)
- $\mathbf{OZ}$ est l'axe complémentaire, perpendiculaire au plan $\mathbf{OXY}$ (Fig. 2, 3B, 3D et 3E)

---

[1] ***T** s'écrit ici en valeur algébrique. Si T est le module, il faudrait écrire **–T** dans le plan OXY. **T** est inférieure ou égale à **T0**$^{fmI}$, la tension isométrique de la fmI.*

[2] *La ligne Z0 est toujours fixe avec la précision que la fmI est préparée de telle manière que ses structures tendineuses ne jouent aucun rôle mécaniquement.*

.

Dans le plan $\mathbf{OXY}$ d'une myofibrille (Fig. 2), chaque sarcomère, numéroté de $\mathbf{1}$ à $\mathbf{Ns}$ (de droite à gauche) est compris entre deux lignes Z (disques Z dans $\mathbf{OXYZ}$) et est centré sur la ligne M (disque M dans $\mathbf{OXYZ}$) (Fig. 2) grâce aux protéines de maintien, notamment la titine.

Soit le sarcomère n° s compris entre les lignes $\mathbf{Z_s}$ et $\mathbf{Z_{s-1}}$ et approximativement centré sur la ligne $\mathbf{M_S}$. Les lignes $\mathbf{Z_s}$, $\mathbf{Z_{s-1}}$ et $\mathbf{M_s}$ sont uniquement animées de mouvements de translation linéaire selon OX ; aussi dans le repère galiléen OXYZ du laboratoire, la loi de composition des vitesses donne la vitesse absolue de la ligne $\mathbf{Z_s}$ par itération (Fig. 2 et 3E), en partant, de droite à gauche:

$$\mathbf{V_{Zs}} = \frac{\mathbf{dOZ_s}}{\mathbf{dt}} = \sum_{\mathbf{j=1}}^{\mathbf{s}} \left( \frac{\mathbf{dZ_{j-1}M_j}}{\mathbf{dt}} + \frac{\mathbf{dM_jZ_j}}{\mathbf{dt}} \right) = \sum_{\mathbf{j=1}}^{\mathbf{s}} \left( \mathbf{u_{Mj,R}} - \mathbf{u_{Mj,L}} \right) \quad (2.\ 2)$$

avec $\mathbf{u_{Mi,L}} = \frac{\mathbf{dZ_jM_j}}{\mathbf{dt}}$ $<$ 0 : vitesse relative de $\mathbf{M_j}$ par rapport à $\mathbf{Z_j}$ (2. 3)

$\mathbf{u_{Mj,R}} = \frac{\mathbf{dZ_{j-1}M_j}}{\mathbf{dt}}$ $>$ 0 : vitesse relative de $\mathbf{M_j}$ par rapport à $\mathbf{Z_{j-1}}$ (2. 4)

Soit pour le cas $\mathbf{s = Ns}$ :

$$\mathbf{V} = \mathbf{V_{ZNs}} = \sum_{\mathbf{j=1}}^{\mathbf{Ns}} \left( \mathbf{u_{Mj,R}} - \mathbf{u_{Mj,L}} \right) \quad (2.\ 5)$$

De même la vitesse absolue du disque $\mathbf{M_s}$ est :

$$\mathbf{V_{Ms}} = \frac{\mathbf{dOM_s}}{\mathbf{dt}} = \sum_{\mathbf{j=1}}^{\mathbf{s-1}} \left( \mathbf{u_{Mj,R}} - \mathbf{u_{Mj,L}} \right) + \mathbf{u_{Ms,R}} = \mathbf{V_{Zs-1}} + \mathbf{u_{Ms,R}} \quad (2.\ 6)$$

Considérons le cas particulier où la fmI se raccourcit à vitesse $\mathbf{V}$ constante. Si dans ce cas, certains disques M associés aux filaments de myosine répartis symétriquement par rapport à $\mathbf{M}$, *i.e.* $(\mathbf{M + 2 \cdot filM})$ ou certains disques Z associés aux filaments d'actine répartis symétriquement par rapport à $\mathbf{Z}$, *i.e.* $(\mathbf{Z + 2 \cdot filA})$ (Fig. 2), ensembles de segments rigides indéformables, présentaient des accélérations, alors celles-ci devraient être exactement compensées à chaque instant t par les décélérations d'autres ensembles $(\mathbf{M + 2 \cdot filM})$ ou $(\mathbf{Z + 2 \cdot filA})$ sans qu'aucune fluctuation importante de $\mathbf{V}$ n'apparaisse[1]. Comme le cycle d'un CB est à l'échelle de la milliseconde (Ford, Huxley et al. 1974; Lombardi, Piazzesi et al. 1992), ce cas mécanique apparait impossible à réaliser sur des durées de plusieurs centaines de millisecondes. D'après les équations (2.2) à (2.6), il s'en déduit par récurrence que les vitesses absolues $\mathbf{V_{Zs}}$ et $\mathbf{V_{Ms}}$ ainsi que les vitesses relatives $\mathbf{u_{Ms,L}}$ et $\mathbf{u_{Ms,R}}$ sont toutes constantes pendant le raccourcissement de la fmI réalisé à $\mathbf{V}$ cste; cette assertion ne signifie pas que $\mathbf{u_{Ms,L}}$ et $\mathbf{u_{Ms,R}}$ sont égales (voir Chapitre 7 et Annexe D).

[1] *Comme pour le cas d'oscillations observables (Huxley, A. F. (1974).*

## 2.2 Modèles mécaniques d'un filament d'actine et d'un filament de myosine

Les **filA** sont modélisés par des tiges cylindriques rigides alignées selon l'axe longitudinal $\mathbf{OX}$ de la **fmI** et sont répartis symétriquement par rapport à la ligne Z. Le sarcomère N°s est compris entre les 2 disques $\mathbf{Z_s}$ et $\mathbf{Z_{s-1}}$ (Fig. 2 et 3E). Les sites de liaison forte des molécules d'actine (sitA) situés dans la partie gauche du sarcomère (hsL) appartiennent aux filA placées à droite de $\mathbf{Z_s}$ (Fig. 3E) et sont nommés $\mathbf{A_{s,R}}$ (R pour « *Right* »; à droite de $\mathbf{Z_s}$ ). Les sitA situés dans la partie droite du sarcomère (hsR) appartiennent aux filA placées à gauche de $\mathbf{Z_{s-1}}$ (Fig. 3E) et sont nommés $\mathbf{A_{s-1,L}}$ (L pour « *Left* »; à gauche de $\mathbf{Z_{s-1}}$ ).

Les filM sont modélisés par des tiges cylindriques rigides alignées selon l'axe longitudinal OX de la fmI et sont répartis symétriquement par rapport au disque M (Fig. 2 et 3E).

## 2.3 Définition et conditions d'un WS « mécaniquement efficace »

Par définition, un WS est mécaniquement efficace lorsqu'au cours de leurs déplacements les 3 segments S1a, S1b et S2 restent rigides avec une longueur constante, formant ainsi un système polyarticulé auquel peuvent s'appliquer les lois de la mécanique. Lors d'un WS, S1a étant lié fortement à une molA est solidaire du filA auquel appartient la molA.

On caractérise l'axe $\mathbf{OY^o}$ comme l'axe perpendiculaire à OX passant les centres du filA et filM associés par la tetM durant le WS (Fig. 3D et 3E).

Rappel de l' **Hypothèse 7** : durant le raccourcissement du hs, la tetM poursuit son WS de telle manière que S1b effectue entièrement son mouvement, selon une trajectoire déterministe donc paramétrable, dans le plan OXY qui forme avec le plan $\mathbf{OXY^o}$ un angle constant $\boldsymbol{\beta}$ durant la totalité du WS (Fig. 3B, 3C et 3D).

Ce postulat repose sur la symétrie des 2 S1 par rapport au plan $\mathbf{OXY^o}$ auquel appartient le point D qui relie les 2 S1 au filM par l'entremise de S2 (Fig. 3A et 3D). Pour que le WS de chacun[1] des 2 S1 puisse se dérouler dans OXYZ de façon équivalente selon les 2 angles $\boldsymbol{\beta}$ et $-\boldsymbol{\beta}$ , la liaison S1a/sitA doit entrainer chaque S1b selon un mouvement symétrique par rapport à $\mathbf{OX}$ et seule l'Hyp. 7 autorise cette condition corroborée par diverses observations (Hopkins, Sabido-David et al. 2002).

## 2.4 Modèle mécanique d'une tête de myosine liée fortement au filament d'actine

Nous allons maintenant modéliser dans hsR et dans hsL la liaison forte, qui correspond mécaniquement à une chaine fermée que forme la tetM d'un filM fixée sur le sitA d'un filA ; cette chaîne est constituée de 3 segments rigides (Rayment, Rypniewski et al. 1993): le domaine moteur S1a, le levier S1b et la tige S2 (Fig. 3A).

S1a est modélisé par un cylindre dont la hauteur $\mathbf{L_{S1a}}$ est du même ordre de grandeur que son diamètre (approximativement le diamètre ou la diagonale d'1 molA selon le modèle géométrique choisi pour représenter la molA). S1a est lié au filA par une liaison encastrement (points $\mathbf{A_{s-1,L}}$ et $\mathbf{A_{s,R}}$ ; Fig. 3D et 3E) et à S1b par une liaison rotule (points $\mathbf{B_{s,R}}$ et $\mathbf{B_{s,L}}$ ; Fig. 3C et 3D).

[1] *Un seul S1 sur les 2 effectue réellement un WS d'après le corollaire 1 du paragraphe 1.5. Dans ce paragraphe sont examinées les conditions géométriques de faisabilité d'un WS.*

**Fig. 3 : Cinématique d'une tetM au cours d'un WS dans hsR et hsL**

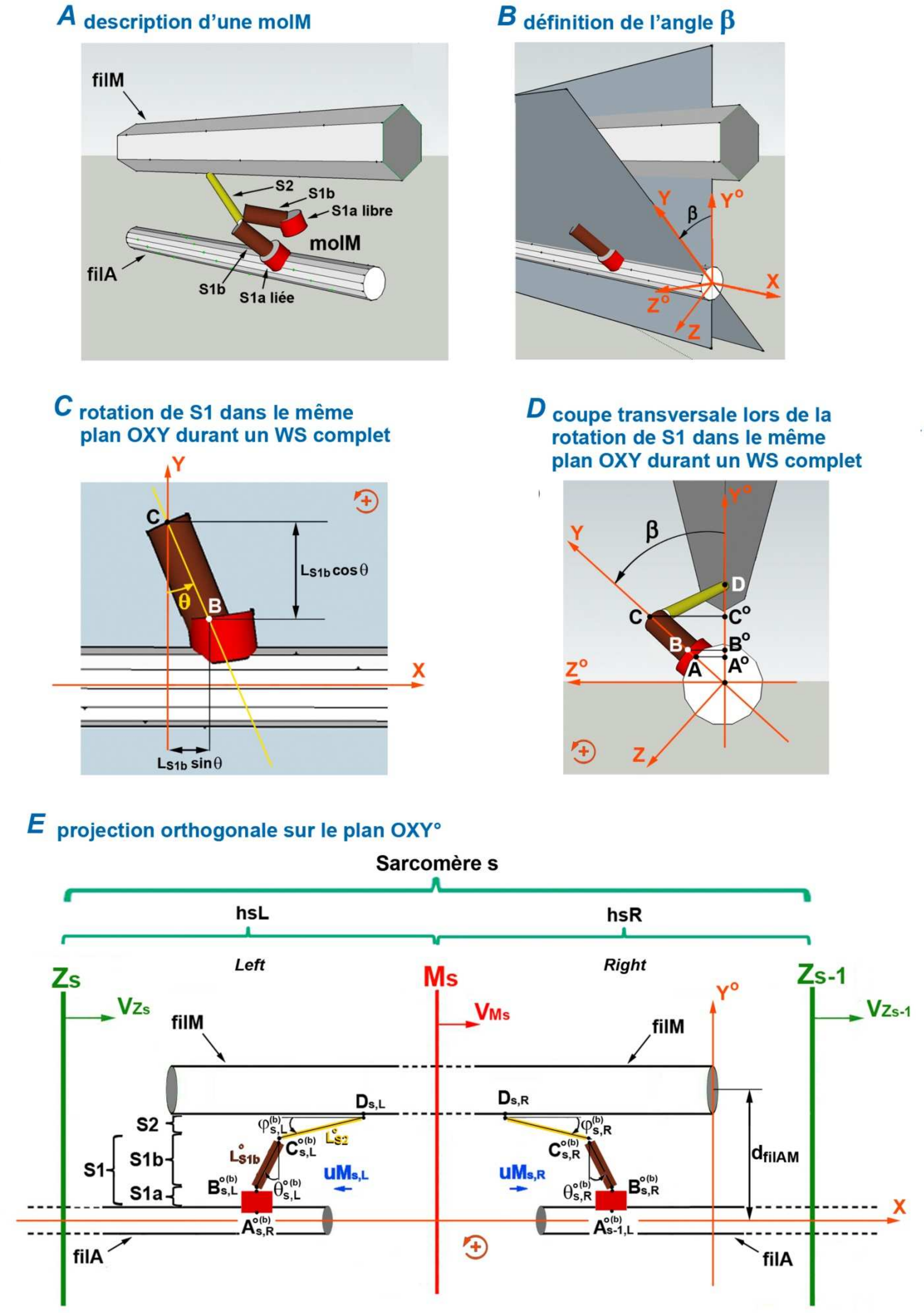

S1b est modélisé par une barre cylindrique (de longueur $\mathbf{L_{S1b}}$) et est relié à S2 par une liaison rotule (points et $\mathbf{C_{s,R}}$ et $\mathbf{C_{s,L}}$ ; Fig. 3C et 3D).

S2 est modélisé par une tige cylindrique (de longueur $\mathbf{L_{S2}}$) et est relié à filM par une liaison rotule (points $\mathbf{D_{s,R}}$ et $\mathbf{D_{s,L}}$ ; Fig. 3A, 3D et 3E).

## 2.5 Cinématique d'une tête de myosine durant WS à V cste

Dans le hsR (hsL) du sarcomère n°s, il y a $\mathbf{N_{s,R}^{WS}}$ ($\mathbf{N_{s,L}^{WS}}$) **S1** en cours de WS, respectivement.
Appelons **b** l'indice qui identifie chaque **S1** en cours de WS, tel que **b** varie de 1 à $\mathbf{N_{s,R}^{WS}}$ (de 1 à $\mathbf{N_{s,L}^{WS}}$) dans hsR (hsL).

D'après l'Hyp. 7, le plan **OXY** spécifique à chaque **S1** n° **b** dans hsR ou hsL forme un angle constant $\boldsymbol{\beta}$ avec le plan $\mathbf{OXY^o}$ durant l'intégralité du WS (Fig. 3D).

**S1b** est représenté par le segment **CB** dans le plan **OXY** tel que **CB** forme un angle instantané $\boldsymbol{\theta_{s,R}^{(b)}}$ ($\boldsymbol{\theta_{s,L}^{(b)}}$) avec l'axe parallèle à **OY** passant par $\mathbf{C_{s,R}^{(b)}}$ ($\mathbf{C_{s,L}^{(b)}}$), respectivement (Fig. 3C).
$\mathbf{C^oB^o}$, qui est la projection orthogonale de **CB** selon $\mathbf{OZ^o}$ sur le plan $\mathbf{OXY^o}$ (Fig. 3D), forme ainsi un angle instantané $\boldsymbol{\theta_{s,R}^{o(b)}}$ ($\boldsymbol{\theta_{s,L}^{o(b)}}$) avec l'axe parallèle à $\mathbf{OY^o}$ passant par $\mathbf{C_{s,R}^{o(b)}}$ ($\mathbf{C_{s,L}^{o(b)}}$), respectivement (Fig. 3E)

**S2** est représenté par le segment **DC** (Fig. 3D). $\mathbf{DC^o}$, qui est la projection orthogonale de **DC** selon $\mathbf{OZ^o}$ sur le plan $\mathbf{OXY^o}$, forme un angle $\boldsymbol{\varphi_{s,R}^{(b)}}$ ($\boldsymbol{\varphi_{s,L}^{(b)}}$) avec l'axe parallèle à **OX** passant par $\mathbf{D_{s,R}}$ ($\mathbf{D_{s,L}}$), respectivement (Fig. 3E).

Des calculs[1] trigonométriques[2] menés à l'Annexe B conduisent à[3] :

$$\mathbf{u_{Ms,R_L}} = -\dot{\boldsymbol{\theta}}_{\mathbf{s,R_L}}^{\mathbf{(b)}} \cdot \mathbf{L_{S1b}} \cdot \left[ \left( \cos\theta_{s,R_L}^{(b)} + \sin\theta_{s,R_L}^{(b)} \cdot \tan\varphi_{s,R_L}^{(b)} \cdot \cos\beta_{s,R_L}^{(b)} \right) + res_{s,R_L}^{(b)} \right] \quad (2.7)$$

où (b) est l'indice de S1 dans hsR ou hsL

1 *Les calculs du paragraphe semblent impliquer que les WS respectifs des 2 S1 dans hsR et dans hsL sont synchrones; les équations restent valables avec 2 WS décalés dans le temps.*

2 *Si le S1 de hsR et celui de hsL n'appartiennent pas au même plan au moment de leur WS respectif, une rotation selon OX d'un des 2 plans par rapport à l'autre suffit pour les faire coïncider dans le même plan OXY de la Fig. 2.*

3 *Le sigle R_L signifie R ou L (« Right or Left »; droit ou gauche)*
*Ex :* $\boldsymbol{\theta_{s,R_L}} \equiv \boldsymbol{\theta_{s,R}}$ ou $\boldsymbol{\theta_{s,L}}$

Comme $\varphi^{(b)}_{s,R_L}$ dépend de $\theta^{(b)}_{s,R_L}$ et $\beta^{(b)}_{s,R_L}$ (voir équations (B.10) et (B.14) de l'Annexe B), on pose :

$$\mathbf{R}(\beta^{(b)},\theta^{(b)}) = \left(\cos\theta^{(b)} + \sin\theta^{(b)} \cdot \tan\varphi^{(b)} \cdot \cos\beta^{(b)}\right) \tag{2.8}$$

$$\mathbf{AL}^{(b)} = \mathbf{L}_{S1b} \cdot \left[\mathbf{R}(\beta^{(b)},\theta^{(b)}) + \mathbf{res}^{(b)}\right] \tag{2.9}$$

Et (2.7) se réécrit :

$$\mathbf{u}_{Ms,R_L} = -\dot{\theta}^{(b)}_{s,R_L} \cdot \mathbf{AL}^{(b)}_{s,R_L} \tag{2. 10}$$

L'équation (2.10) souligne la relation entre $\mathbf{u}_{Ms,R_L}$, la vitesse linéaire relative de raccourcissement d'un demi-sarcomère, et $\dot{\theta}^{(b)}_{s,R_L}$, la vitesse angulaire d'une tête de myosine durant son WS. On note que dans ce modèle théorique le bras de levier $\mathbf{AL}^{(b)}_{s,R_L}$ (AL pour *arm lever*) dépend de 2 facteurs, $\mathbf{R}(\theta^{(b)}_{s,R_L},\beta^{(b)}_{s,R_L})$ et $\mathbf{res}^{(b)}_{s,R_L}$, dont il faut étudier le comportement sur l'intervalle $\lfloor\theta 1;\theta_{stopWS}\rfloor$.

## 3 Dynamique d’une tête de myosine durant un WS à V= cste

Fig. 4 : Dynamique d’une tetM au cours d’un WS dans hsR et hsL

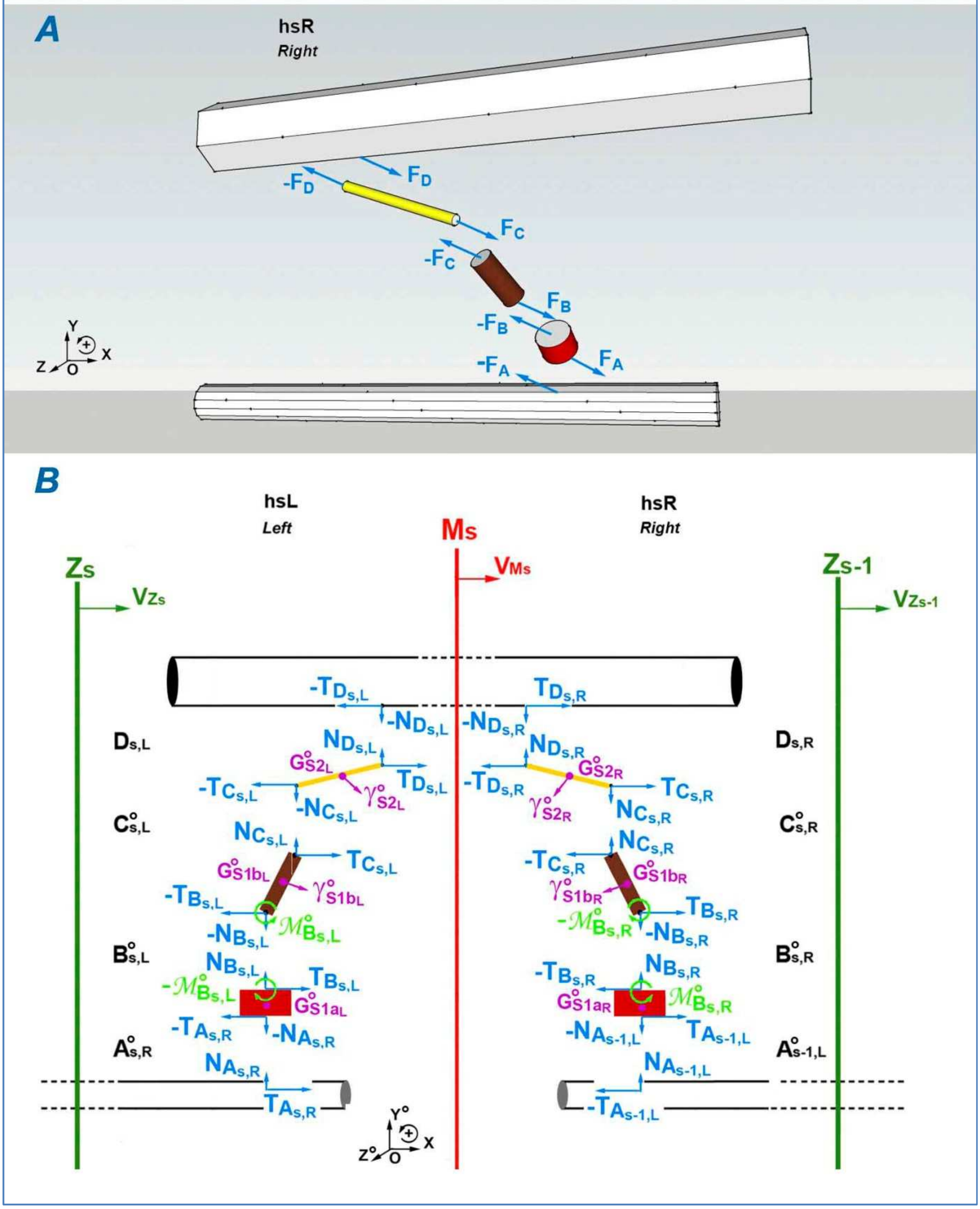

## 3.1 Equations

Rappel de l'**Hypothèse 7+** : à l'échelle nanométrique les quantités d'accélérations linéaire et angulaire, d'origines gravitationnelle ou inertielle, sont négligeables; idem pour la poussée d'Archimède. Nous retenons l'hypothèse que les forces de viscosité s'exerçant sur une molM sont aussi négligeables.

Si ces conditions sont vérifiées, les seules actions en présence dans le système mécanique proposé par ce modèle sont les forces et moments de liaison.

Les forces exercées sur S1a, S1b et S2 à chacune des 4 liaisons A, B, C et D dans hsR et hsL sont nommées F (Fig. 4A) avec pour composantes les projections orthogonales selon OX et OY° , T et N, respectivement (Fig. 4B).

Durant le WS, les 2 moments pivot moteur appliqués en $\mathbf{B}_{s,R}$ et $\mathbf{B}_{s,L}$ selon OZ sont nommés $\mathcal{M}_{\mathbf{B}_{s,R}}$ et $\mathcal{M}_{\mathbf{B}_{s,L}}$ avec pour composantes les projections orthogonales selon OZ°, $\mathcal{M}^{\mathbf{o}}_{\mathbf{B}_{s,R}}$ et $\mathcal{M}^{\mathbf{o}}_{\mathbf{B}_{s,L}}$ , respectivement (Fig. 4B).

Enfin à tout instant t durant le WS, aucun moment moteur n'est appliqué aux liaisons rotule en C et en D (moments nuls).

A V cste, les théorèmes généraux de la mécanique du solide appliqués aux 2 S1 respectifs de hsR et de hsL, dans le repère galiléen[1] OXY°Z° (Fig. 4B)[2], donnent (Annexe C) :

Pour hsR[3]

$$\mathbf{T}_{\mathbf{A}^{(b)}_{s-1,L}} = \mathbf{T}_{\mathbf{D}^{(b)}_{s,R}} \tag{3.1}$$

$$\mathcal{M}_{\mathbf{B}^{(b)}_{s,R}} = \mathbf{L}_{S1b} \cdot \mathbf{T}_{\mathbf{A}^{(b)}_{s-1,L}} \cdot \mathbf{S}(\theta^{(b)}_{s,R}, \beta^{(b)}_{s,R}) \tag{3.2}$$

où (b) est l'indice identifiant la tête de myosine n° b localisée dans hsR

Pour hsL[3]

$$\mathbf{T}_{\mathbf{A}^{(b)}_{s,R}} = \mathbf{T}_{\mathbf{D}^{(b)}_{s,L}} \tag{3.3}$$

$$\mathcal{M}_{\mathbf{B}^{(b)}_{s,L}} = \mathbf{L}_{S1b} \cdot \mathbf{T}_{\mathbf{A}^{(b)}_{s,R}} \cdot \mathbf{S}(\theta^{(b)}_{s,L}, \beta^{(b)}_{s,L}) \tag{3.4}$$

où (b) est l'indice identifiant la tête de myosine n° b localisée dans hsL

avec :

$$\mathbf{S}(\theta^{(b)}_{s,R_L}, \beta^{(b)}_{s,R_L}) = \cos\theta^{(b)}_{s,R_L} + \frac{\sin\theta^{(b)}_{s,R_L} \cdot \tan\varphi^{(b)}_{s,R_L}}{\cos\beta^{(b)}_{s,R_L}} \tag{3.5}$$

---

[1] *OXY°Z° peut être placé en Ms ou en Zs (s= 1, 2,…, Ns) puisqu'à V cste tous les Ms et Zs se déplacent à vitesse constante (voir paragraphe 1.2) les uns par rapport aux autres et par rapport au repère du laboratoire, condition constitutive d'un repère galiléen.*

[2] *Les moments des actions et réactions de liaison entre S1a et filA en $A_{s,R}$ et en $A_{s-1,L}$ n'apparaissent pas sur la Fig. 4B pour cause d'encombrement.*

[3] *Les moments et les forces sont exprimés en module, mais les angles sont en valeur algébrique.*

## 3.2 Dispersion angulaire en début de WS et constance angulaire en fin de WS

Rappel de l'**Hypothèse 1** : $\theta_{startWS}^{(b)}$ est la position angulaire de début de WS dans le plan **OXY** de la tetM n° b localisée dans hsR ou hsL qui permet à S2 d'assurer son rôle de liaison mécanique avec une longueur fixe. $\theta_{startWS}^{(b)}$ varie dans un intervalle angulaire borné par 2 constantes ( $\theta 1$ et $\theta 2$ ) pour tout S1 de la fmI tel que:

dans hsR : $\theta 1 \geq \theta_{startWS}^{(b)} \geq \theta 2$ (3.6a)

dans hsL : $\theta 1 \leq \theta_{startWS}^{(b)} \leq \theta 2$ (3.6b)

On note $\delta_\theta$, la dispersion angulaire de début de WS, telle que :

$$\delta_\theta = |\theta 1 - \theta 2| \tag{3.7}$$

Rappel de l'**Hypothèse 10** : $\theta_{stopWS}^{(b)}$ , la position angulaire de fin de WS dans le plan OXY, est une constante notée $\theta_{stopWS}$ pour tout S1 de la fmI.

On note $\Delta\theta_{WS}^{Max}$ la variation angulaire maximale au cours d'un WS, telle que:

$$\Delta\theta_{WS}^{Max} = |\theta 1 - \theta_{stopWS}| \tag{3.8}$$

## 3.3 Applications

Nous complétons **l'hypothèse 1** en précisant les valeurs de $\theta 1$ et $\theta 2$ :

dans hsR : $\theta 1 = +25°$ $\theta 2 = -16°$ (3.9a)

dans hsL : $\theta 1 = -25°$ $\theta 2 = +16°$ (3.9b)

Ce choix a été arrêté pour que divers paramètres demeurent aussi constants que possible durant le WS (voir Fig. 5 et Table 1); cette proposition peut sembler arbitraire mais les valeurs restent compatibles avec les données de la littérature (voir paragraphe 6.4.2.2). Avec ces données, la dispersion angulaire de départ est égale à :

$$\delta_\theta = |\theta 1 - \theta 2| = 41° \tag{3.9c}$$

La valeur de $\Delta\theta_{WS}^{Max}$ est prise à (Geeves and Holmes 1999; Fischer, Windshugel et al. 2005; Huxley, Reconditi et al. 2006; Llinas, Pylypenko et al. 2012 ):

$$\Delta\theta_{WS}^{Max} = |\theta 1 - \theta_{stopWS}| = 60° \tag{3.10}$$

Et avec (3.10) on déduit de l'hyp. 10 et (3.8) :

$$\forall b, \quad \left|\theta_{stopWS}^{(b)}\right| = \left|\theta_{stopWS}\right| = 35° \tag{3.11}$$

**Table. 1** : **Evolutions de divers paramètres caractérisant la cinématique d'une tête de myosine durant un WS selon 5 valeurs de $\beta$**

| $\beta$ | $\cos\beta$ | $\overline{R(\beta)}^{\oplus}$ | $\overline{S(\theta,\beta)}$ | $\|res\|^{\oplus}$ | $\overline{AL}$ | $\Delta X_{WS}^{Max1\mp}$ (nm) | $\Delta X_{WS}^{Max2\mp}$ (nm) |
|---|---|---|---|---|---|---|---|
| **0°** | 1 | 0.96<br>*~ constant* | 0.96<br>*~ constant* | 0<br>*constant* | 0.96 $L_{S1b}$<br>*~ constant* | 10 | 10.05 |
| **27.7°** | 0.89 | 0.96<br>*~ constant* | 0.96<br>*~ constant* | ≤ 0.05<br>*~constant* | 0.96 $L_{S1b}$<br>*~constant* | 9.9 | 10.05 |
| **42°** | 0.74 | 0.96<br>*~ constant* | 0.97<br>*~ constant* | ≤ 0.11<br>*~ constant* | 0.96 $L_{S1b}$<br>*~ constant* | 9.9 | 10.05 |
| **55.4°** | 0.57 | 0.97<br>*~ constant* | 1<br>*Non constant* | ≤ 0.19<br>*Non constant* | 0.97 $L_{S1b}$<br>*~ constant* | 9.7 | 10.1 |
| **70°** | 0.34 | 0.97<br>*~ constant* | 1.08<br>*Non constant* | ≤ 0.27<br>*Non constant* | 0.97 $L_{S1b}$<br>*~ constant* | 9.6 | 10.1 |

$^{\oplus}\overline{R(\beta)}$ *et* $|res|$ *sont des termes issus des équations (2.7) et (2.8) du chapitre 2*

$^{\mp}\Delta X_{WS}^{Max1}$ *est calculé d'après (B.1) et (B.10) de l'Annexe B, et* $\Delta X_{WS}^{Max2}$ *avec (3.10), (3.12) et (3.20)*

Les 13 sitA d'un motif d'un filA forment une double hélice avec un pas angulaire de 27.7° (=360°/13). 3 cas vont être analysés qui correspondent aux 3 WS d'un S1 effectués sur 3 sitA successifs positionnés sur le même filA tels que : $\beta^{(b)}$ = 0°, $\beta^{(b)}$ = 27.7° et $\beta^{(b)}$ = 55.4°

Dans les exemples de la Fig. 5, on observe pour 4 valeurs de $\beta$ les évolutions (en fonction de $\theta$ variant entre $\theta 1$ = -25° et $\theta_{stopWS}$ = +35°) des paramètres suivants :

- $R(\theta,\beta)$, $res$ et $[R(\theta,\beta)+res]$ (Fig. 5A, 5D, 5G et 5J), calculés d'après les équations (2.8), (B9) et (B.13).

- $\dot{\theta}$ (Fig. 5B, 5E, 5H et 5K) calculé à partir de l'équation (B9)

- $S(\theta,\beta)$ (Fig. 5C, 5F, 5I et 5L) d'après (3.5) comparativement à $R(\theta,\beta)$

Avec les données classiques ($L_{S1b}$ = 10 nm, $L_{S2}$ = 30 nm, $d_{filAM}$[1]= 24.8 nm, $d_{AB}$ = 4 nm, $r_{filM}$ = 7.5 nm et $r_{filA}$ = 3.5 nm ), les évolutions moyennes de divers paramètres caractérisant la cinématique d'une tête de myosine selon 5 positions angulaires $\beta$ d'un sitA sur un filA sont résumées dans la Table 1.

---

[1] *Calculé en prenant la distance entre les centres de 2 filM égale à 43 nm (Squire and. Knupp 2005), et en considérant que le centre du filA se trouve confondu avec l'orthocentre du triangle équilatéral formé par les 3 filM voisins.*

**Fig. 5 : Evolution de 5 paramètres pour 4 valeurs de $\beta$ dans un hsL entre $\theta 1 = -25°$ et $\theta_{stopWS} = +35°$**

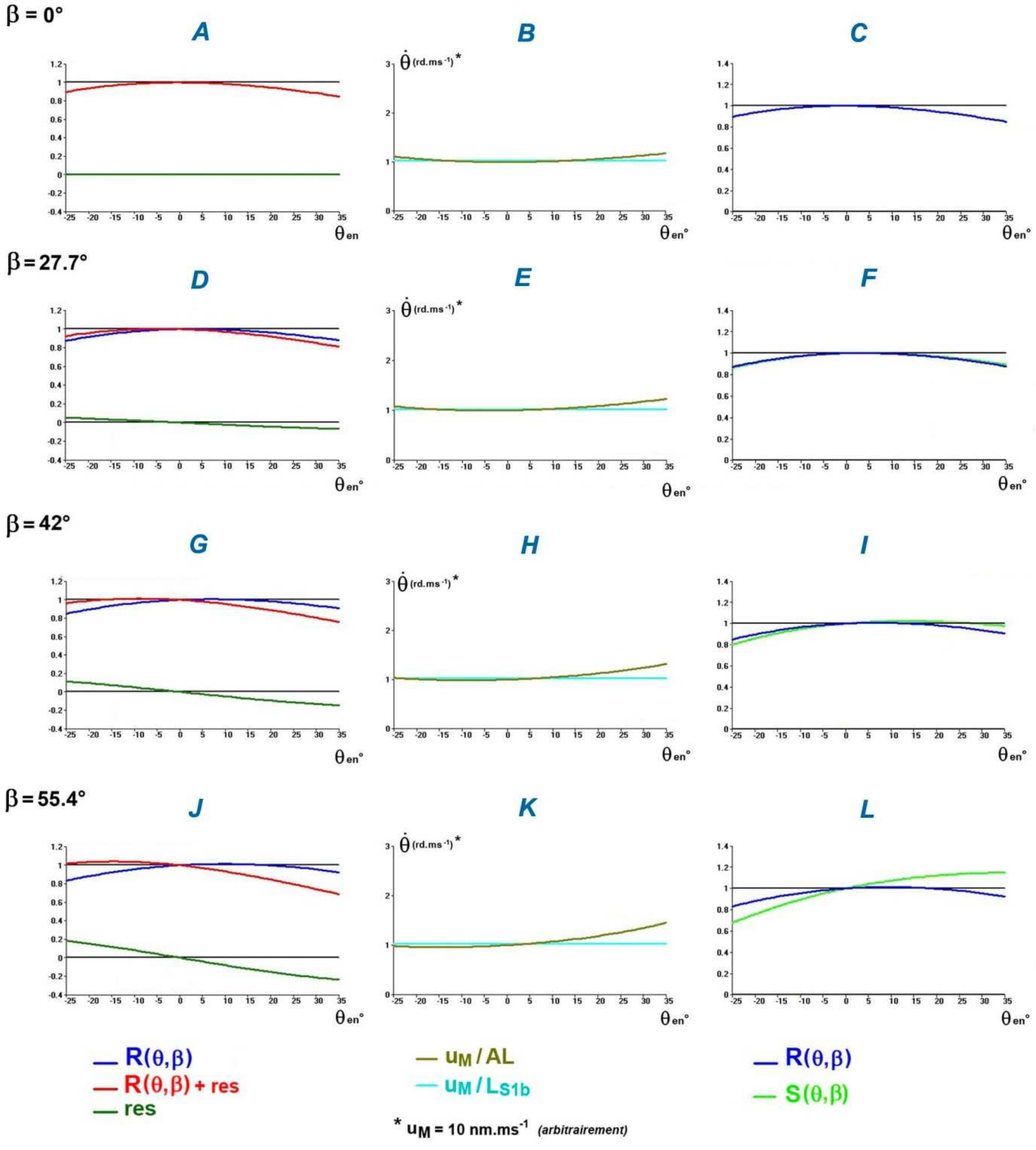

Durant le WS et pour $|\beta| \le 42°$, on note (Fig. 5 et Table 1):

- $\mathbf{R(\theta,\beta)}$ est à peu près constant autour d'une valeur moyenne $\mathbf{R_{WS}}$ :

$$\mathbf{R(\theta^{(b)}_{s,R_L},\beta^{(b)}_{s,R_L}) \approx R_{WS} = \frac{\int_{\theta 1}^{\theta_{stopWS}} (\cos\theta + \sin\theta \cdot \tan\varphi \cdot \cos\beta) \cdot d\theta}{\Delta\theta_{WS}} \approx 0.96} \quad (3.12)$$

- $\mathbf{S(\theta,\beta)}$ est à peu près constant autour d'une valeur moyenne $\mathbf{S_{WS}}$ :

$$\mathbf{S(\theta^{(b)}_{s,R_L},\beta^{(b)}_{s,R_L}) \approx S_{WS} = \frac{\int_{\theta 1}^{\theta_{stopWS}} \frac{(\cos\theta + \sin\theta \cdot \tan\varphi)}{\cos\beta} \cdot d\theta}{\Delta\theta_{WS}} \approx 0.96} \quad (3.13)$$

- (3.12) et (3.13) autorisent (Fig. 5C, 5F et 5I) :

$$\mathbf{R_{WS} \approx S_{WS}} \quad (3.14)$$

- L'évolution du dernier terme de l'équation (1.7), $\mathbf{res^{(b)}_{s,R_L}}$, est tel que $|\mathbf{res}|$ reste inférieur à 0.11 durant le WS (Fig. 5A, 5D et 5G ; Table 1) ; en le négligeant et avec (3.12), (2.7) se reformule :

$$\mathbf{u_{Ms,R_L} = -\dot{\theta}^{(b)}_{s,R_L} \cdot L_{S1b} \cdot R_{WS}}$$

  Dans ce cas, puisqu'il a été démontré au paragraphe 2.1 qu'à V=cste, $\mathbf{u_{Ms,R}}$ et $\mathbf{u_{Ms,L}}$ restaient constants, il est permis de conclure qu'à V cste, $\mathbf{\dot{\theta}^{(b)}_{s,R_L}}$ prend une valeur constante ( $\mathbf{\dot{\theta}_{s,R_L}}$), quelque soit la valeur de $|\beta^{(b)}| \le 42°$ (Fig. 5B, 5E et 5H) au cours du WS[1] :

$$\mathbf{u_{Ms,R_L} = -\dot{\theta}_{s,R_L} \cdot L_{S1b} \cdot R_{WS}} \quad (3.15)$$

- On nomme $\mathbf{\Delta X^{(b)}_{s,R_L}}$ le déplacement linéaire relatif de $\mathbf{D^{(b)}_{s,L}}$ relativement à $\mathbf{A^{(b)}_{s,R}}$ dans hsL et de $\mathbf{D^{(b)}_{s,R}}$ relativement à $\mathbf{A^{(b)}_{s-1,L}}$ dans hsR au cours d'un WS. Par intégration de (3.15), ou simplement par définition d'une vitesse, on déduit que $\mathbf{\Delta X^{(b)}_{s,R_L}}$ est relié au déplacement angulaire de S1b noté $\mathbf{\Delta\theta^{(b)}_{Ws,R_L}}$ par :

$$\mathbf{\Delta X^{(b)}_{s,R_L} = \Delta\theta^{(b)}_{Ws,R_L} \cdot L_{S1b} \cdot R_{WS}} \quad (3.16)$$

  avec $\mathbf{\Delta\theta^{(b)}_{s,R_L} = \left|\theta^{(b)}_{startWS} - \theta^{(b)}_{s,R_L}\right|}$ (3.17)

---

[1] *On remarque qu'à V cste, les points appartenant aux solides S1b (par exemple, $C_{s,R}$ ou $C_{s,L}$) se déplacent à vitesses linéaire et angulaire constantes et suivent approximativement durant 1 WS des arcs de spirales d'Archimède dans le plan OXY.*

***Cas particuliers*** :

Les éq. (3.16) et (3.17) fournissent la valeur du pas d'une tetM ( $\Delta X_{WS}^{(b)}$ ) durant un WS dans hsR ou hsL :

$$\Delta X_{WS}^{(b)} = \Delta\theta_{WS}^{(b)} \cdot L_{S1b} \cdot R_{WS} \qquad (3.18)$$

avec $$\Delta\theta_{WS}^{(b)} = \left|\theta_{startWS}^{(b)} - \theta_{stopWS}\right| \qquad (3.19)$$

On vérifie que le pas d'une tetM varie entre 2 constantes (Fig. 8; chap. 4) :

$$\Delta X_{WS}^{Max} \approx L_{S1b} \cdot R_{WS} \cdot \Delta\theta_{WS}^{Max} \qquad (3.20)$$

$$\Delta X_{WS}^{min} \approx L_{S1b} \cdot R_{WS} \cdot \Delta\theta_{WS}^{min} \qquad (3.21)$$

avec $$\Delta\theta_{WS}^{min} = \Delta\theta_{WS}^{Max} - \delta_\theta \qquad (3.22)$$

tel que :

$$\Delta X_{WS}^{min} \le \Delta X_{WS}^{(b)} \le \Delta X_{WS}^{Max} \qquad (3.23a)$$

Soit avec les données précédentes fournies par (3.9a), (3.9b), (3.10) et (3.12), l'inégalité (3.23a) devient :

$$3.2\,nm \le \Delta X_{WS}^{(b)} \le 10\,nm \qquad (3.23b)$$

Enfin on vérifie avec ce qui précède et (3.16) :

$$0 \le \Delta X_{s,R_L}^{(b)} \le \Delta X_{WS}^{(b)} \le \Delta X_{WS}^{Max} \qquad (3.24)$$

- On déduit de (3.15), (3.20), (3.21) et (3.23a), que la durée mécanique du WS ( $\Delta t_{WS}^{(b)}$ ) est une variable comprise entre 2 constantes pour chaque valeur de la vitesse angulaire ou linéaire:

$$\Delta t_{WS}^{min} \le \Delta t_{WS}^{(b)} \le \Delta t_{WS}^{Max} \qquad (3.25)$$

avec : $$\Delta t_{WS}^{Max} = \frac{\Delta X_{WS}^{Max}}{\left|uM_{s,R_L}\right|} = \frac{\Delta\theta_{WS}^{Max}}{\left|\dot{\theta}_{s,R_L}\right|} \qquad (3.26)$$

$$\Delta t_{WS}^{min} = \frac{\Delta X_{WS}^{min}}{\left|uM_{s,R_L}\right|} = \frac{\Delta\theta_{WS}^{min}}{\left|\dot{\theta}_{s,R_L}\right|} \qquad (3.27)$$

- Pour les valeurs angulaires de $\beta = 55.4° = 2x27.7°$ et de $\beta = 70°$ (moins de 3 fois 27.7°), on observe des évolutions très divergentes (Table 1 et Fig. 5).

A l'issue de ces observations, on conclut que la valeur maximale de $\left|\beta^{(b)}\right|$ est égale à 42° (complément de l'**Hyp. 7**) et sera notée $\beta^{Max}$ avec l'inégalité : $\left|\beta^{(b)}\right| \le \beta^{Max}$

Les hypothèses précédentes autorisent l'application de toutes les équations précédentes à tous les S1 de la fmI lors de leur WS.

## 3.5 Robustesse du modèle mécanique

### *3.5.1 Influence de la distance inter-filamentaire*

Dans le modèle soumis, l'homogénéité de la fmI suppose que la distance entre le milieu du diamètre du filA et le milieu du diamètre du filM, $\mathbf{d_{filAM}}$, est une constante (voir Annexe C). Or le travail de A.M. Gordon, A.F. Huxley et F.J. Julian (Gordon, Huxley et al. 1966) souligne que cette hypothèse ne peut être vérifiée que pour une gamme réduite de longueur du hs ($\mathbf{L_{hs}}$), telle que :

$$\mathbf{1\ \mu m \le L_{hs} \le 1.825\ \mu m} \tag{3.29}$$

En deçà de la borne minimale, il y recouvrement partiel des filA des hs droit et gauche, et de facto $\mathbf{d_{filAM}}$ augmente.

D'autres expérimentations ont été conduites où l'espace interstitiel entre les filA et les filM varie en diminuant ou augmentant selon divers critères physiologiques (Edman and Hwang 1977; Edman 1988).

Nous avons testé l'évolution des paramètres issus des équations précédentes en reprenant les mêmes données du paragraphe 3.4 pour $\mathbf{d_{filAM} = 24.8 - 7 = 17.8\ nm}$ et pour $\mathbf{d_{filAM} = 24.8 + 7 = 31.8\ nm}$, où la valeur de 7 nm correspond à la valeur approximative du diamètre d'un filA. Les valeurs sont récapitulées dans la Table 2 et les cinétiques des différents paramètres apparaissent dans la Fig. 6.

On observe :

- La stabilité des divers paramètres est respectée grosso modo, excepté $\mathbf{\overline{S(\theta,\beta)}}$ et $\mathbf{|res|}$ aux conditions $\mathbf{d_{filAM} = 31.8\ nm}$ et $\mathbf{\beta = 42°}$
- Le pas maximal de S1, $\mathbf{\Delta X_{WS}^{Max}}$, diminue légèrement pour $\mathbf{d_{filAM} = 17.8\ nm}$ et augmente légèrement pour $\mathbf{d_{filAM} = 31.8\ nm}$

### *3.5.2 Influence de la longueur du segment rigide nommé S2*

Notre modèle n'apporte pas d'explication sur les modalités dont la tetM s'organise structurellement pour rendre S2 rigide (voir chapitre 18). Il est possible que la longueur de S2 se révèle être une variable d'ajustement; il est donc impératif de vérifier comment évolue les paramètres selon différentes valeurs de $\mathbf{L_{S2}}$. Les valeurs sont récapitulées dans la Table 3.

On note :

- La stabilité des divers paramètres est respectée grosso modo et les cinétiques sont pratiquement superposables à celles de la Fig. 5 aux mêmes valeurs de $\mathbf{\beta}$, excepté $\mathbf{|res|}$ pour $\mathbf{L_{S2} = 20\ nm}$ et $\mathbf{\beta = 42°}$
- La stabilité se dégrade logiquement quand $\mathbf{L_{S2}}$ diminue et s'améliore quand $\mathbf{L_{S2}}$ augmente.
- Les égalités (3.12), (3.13) et (3.14) sont vérifiées pour ces 2 valeurs de $\mathbf{L_{S2}}$ validant conséquemment les équations ou égalités qui les suivent.

**Fig. 6 : Evolution de 5 paramètres pour 2 valeurs de $d_{filAM}$ et pour 3 valeurs de $\beta$ dans un hsL entre $\theta 1$ = -25° et $\theta_{stopWS}$ = +35°**

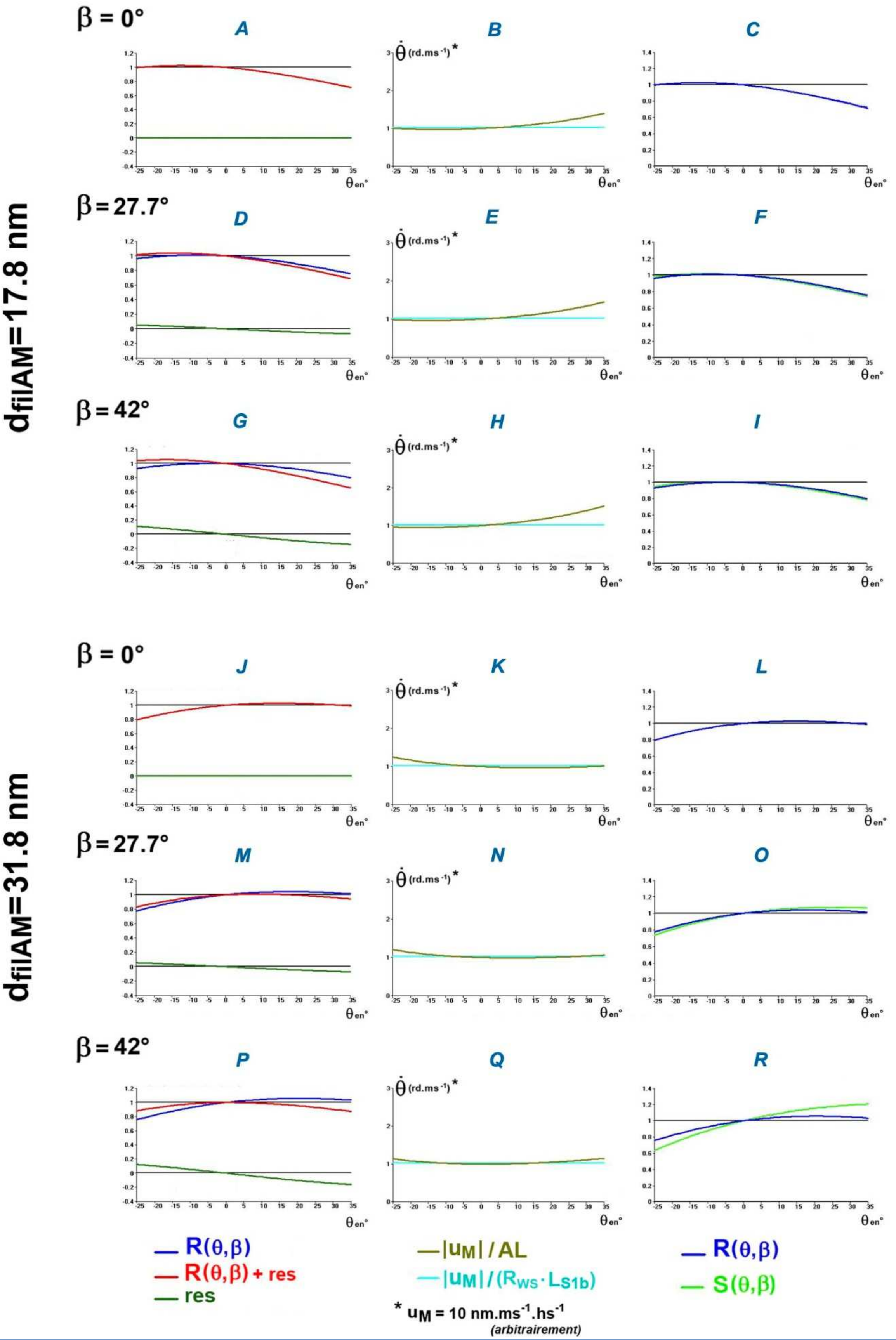

**Table. 2** : **Evolutions de divers paramètres caractérisant la cinématique d'une tête de myosine durant un WS selon 2 valeurs de $d_{filAM}$ et 3 valeurs de $\beta$**

| $d_{filAM}$ (nm) | $\beta$ | $\cos\beta$ | $\overline{R(\theta,\beta)}$ | $\overline{S(\theta,\beta)}$ | $\lvert res \rvert$ | $\overline{AL}$ | $\Delta X_{WS}^{Max1\mp}$ (nm) | $\Delta X_{WS}^{Max2\mp}$ (nm) |
|---|---|---|---|---|---|---|---|---|
| **17.8** | **0°** | 1 | 0.94 *~ constant* | 0.94 *~ constant* | 0 *constant* | 0.94 $L_{S1b}$ *~ constant* | 9.8 | 9.8 |
| | **27.7°** | 0.89 | 0.94 *~ constant* | 0.94 *~ constant* | ≤ 0.05 *~constant* | 0.94 $L_{S1b}$ *~constant* | 9.8 | 9.85 |
| | **42°** | 0.74 | 0.95 *~ constant* | 0.95 *~ constant* | ≤ 0.11 *~ constant* | 0.95 $L_{S1b}$ *~ constant* | 9.7 | 9.9 |
| **31.8** | **0°** | 1 | 0.99 *~ constant* | 0.98 *~ constant* | 0 *constant* | 0.99 $L_{S1b}$ *~ constant* | 10.2 | 10.3 |
| | **27.7°** | 0.87 | 0.98 *~ constant* | 0.99 *~ constant* | ≤ 0.06 *~constant* | 0.98 $L_{S1b}$ *~constant* | 10.1 | 10.3 |
| | **42°** | 0.74 | 0.98 *~ constant* | 1.01 *Non constant* | ≤ 0.125 *~ constant* | 0.98 $L_{S1b}$ *~ constant* | 10 | 10.3 |

$^{\mp}\Delta X_{WS}^{Max1}$ *est calculé d'après (B.1) et (B.10) de l'annexe B, et* $\Delta X_{WS}^{Max2}$ *avec (3.10), (3.12) et (3.20)*

**Table. 3** : **Evolutions de divers paramètres caractérisant la cinématique d'une tête de myosine durant un WS selon 2 valeurs de $L_{S2}$ et 3 valeurs de $\beta$**

| $L_{S2}$ (nm) | $\beta$ | $\cos\beta$ | $\overline{R(\theta,\beta)}$ | $\overline{S(\theta,\beta)}$ | $\lvert res \rvert$ | $\overline{AL}$ | $\Delta X_{WS}^{Max1\mp}$ (nm) | $\Delta X_{WS}^{Max2\mp}$ (nm) |
|---|---|---|---|---|---|---|---|---|
| **20** | **0°** | 1 | 0.96 *~ constant* | 0.96 *~ constant* | 0 *constant* | 0.96 $L_{S1b}$ *~ constant* | 10 | 10.05 |
| | **27.7°** | 0.89 | 0.96 *~ constant* | 0.97 *~ constant* | ≤ 0.08 *~constant* | 0.96 $L_{S1b}$ *~constant* | 9.9 | 10.05 |
| | **42°** | 0.74 | 0.97 *~ constant* | 0.97 *~ constant* | ≤ 0.2 *Non constant* | 0.97 $L_{S1b}$ *~ constant* | 9.8 | 10.1 |
| **40** | **0°** | 1 | 0.95 *~ constant* | 0.95 *~ constant* | 0 *constant* | 0.95 $L_{S1b}$ *~ constant* | 9.9 | 10 |
| | **27.7°** | 0.89 | 0.96 *~ constant* | 0.96 *~ constant* | ≤ 0.04 *~constant* | 0.96 $L_{S1b}$ *~constant* | 9.9 | 10.05 |
| | **42°** | 0.74 | 0.96 *~ constant* | 0.97 *~ constant* | ≤ 0.08 *~ constant* | 0.96 $L_{S1b}$ *~ constant* | 9.9 | 10.05 |

$^{\mp}\Delta X_{WS}^{Max1}$ *est calculé d'après (B.1) et (B.10) de l'annexe B, et* $\Delta X_{WS}^{Max2}$ *avec (3.10), (3.12) et (3.20)*

## 3.6 Evolution temporelle de l'angulation de S1b durant un WS

A V cste, pour n'importe quel S1 n° b effectuant un WS dans le hsR ou le hsL du sarcomère n° s, $s \in [1,2,...,N_S]$, on observe à l'appui des Hyp. 1 et 7 et d'après les égalités des équations (3.7), (3.8), (3.9a), (3.9b), (3.10) et (3.15) que $\theta$ varie linéairement et algébriquement de $\theta^{(b)}_{startWS}$ à $\theta_{stopWS}$ (Fig. 7A et 7B) :

$$\theta = \theta_{stopWS} + \left(\theta^{(b)}_{startWS} - \theta_{stopWS}\right) \cdot \left(1 - \frac{t}{\Delta t^{(b)}_{WS}}\right) \quad (3.30)$$

## 3.7 Comportement mécanique d'un ressort d'origine entropique durant le WS

Le moment moteur angulaire exercé par S1a sur S1b au point B durant le WS de la S1 n°b dans hsR ou hsL est noté $\mathcal{M}^{(b)}_B$ .

Rappel de l'**Hypothèse 8** : $\mathcal{M}^{(b)}_B$ , d'origine entropique, est modélisé par une équation de type hyperbolique. L'équation se formule comme suit :

$$\mathcal{M}^{(b)}_B = \mathcal{M}_B(\theta) = \frac{\mathcal{M}1}{\Delta\theta^{Max}_{WS}} \cdot (\theta 1 + a_\theta) \cdot \left(\frac{\theta_{stopWS} + a_\theta}{\theta + a_\theta} - 1\right) \quad (3.31)$$

où $a_\theta$ est une constante valant $+a_\theta$ dans hsL et $-a_\theta$ dans hsR

et $\mathcal{M}1$ la valeur algébrique du moment angulaire correspondant à la position angulaire $\theta 1$

avec $|\mathcal{M}1| = \left|\mathcal{M}1^{0°C}\right| + c \cdot k_B \cdot T°$

où $T°$ est la température expérimentale de la fmI exprimée en °K

Dans (3.31), $\theta$ et $\mathcal{M}_B$ sont exprimés en valeur algébrique.

On note (Fig. 7C et 7D) :

$$\mathcal{M}_B(\theta_{stopWS}) = 0 \quad (3.32)$$

$$\mathcal{M}_B(\theta 2) = \mathcal{M}2 \quad (3.33)$$

$$\mathcal{M}_B\left(\theta^{(b)}_{startWS}\right) = \mathcal{M}^{(b)}_{startWS} \quad (3.34)$$

dans hsR : $\mathcal{M}1 \geq \mathcal{M}^{(b)}_{startWS} \geq \mathcal{M}2$ (3.35a)

dans hsL : $\mathcal{M}1 \leq \mathcal{M}^{(b)}_{startWS} \leq \mathcal{M}2$ (3.35b)

En combinant les équations (3.2), (3.4), (3.5), (3.8), (3.12), (3.13), (3.14), (3.20) avec (3.30), on obtient :

$$\forall s \in [1,2,...,N_S], \forall t \in \Delta t^{(b)}_{WS} \qquad T_{A^{(b)}_{s,R_L}} = \frac{\mathcal{M}1}{\Delta X^{Max}_{WS}} \cdot (\theta 1 + a_\theta) \cdot \left(\frac{\theta_{stopWS} + a_\theta}{\theta^{(b)} + a_\theta} - 1\right) \quad (3.36a)$$

De même avec (3.1), (3.3), (3.5), (3.8), (3.12), (3.13), (3.14), (3.20) et (3.30) impliquent :

$$\forall s \in [1,2,...,N_S], \forall t \in \Delta t^{(b)}_{WS} \qquad T_{D^{(b)}_{s,R_L}} = \frac{\mathcal{M}1}{\Delta X^{Max}_{WS}} \cdot (\theta 1 + a_\theta) \cdot \left(\frac{\theta_{stopWS} + a_\theta}{\theta^{(b)} + a_\theta} - 1\right) \quad (3.36b)$$

**Fig. 7 : Evolutions temporelles de θ et de $\mathcal{M}_B$ au cours d'un WS dans hsL et hsR**

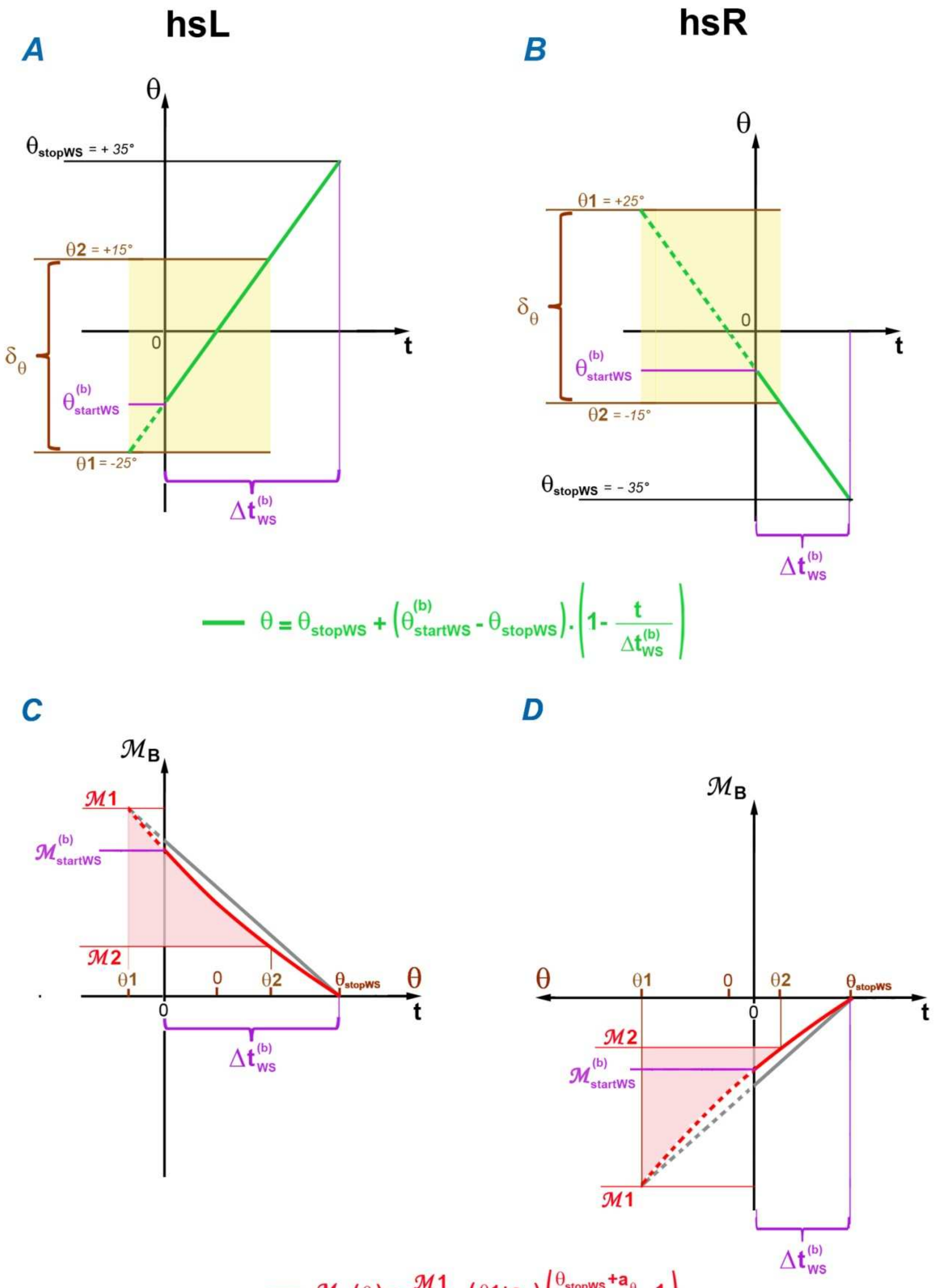

$$\theta = \theta_{stopWS} + \left(\theta^{(b)}_{startWS} - \theta_{stopWS}\right) \cdot \left(1 - \frac{t}{\Delta t^{(b)}_{WS}}\right)$$

$$\mathcal{M}_B(\theta) = \frac{\mathcal{M}1}{\Delta\theta^{Max}_{WS}} \left(\theta 1 + a_\theta\right) \left(\frac{\theta_{stopWS} + a_\theta}{\theta + a_\theta} - 1\right)$$

$$\mathcal{M}^{lin}_B(\theta) = \frac{\mathcal{M}1}{\Delta\theta^{Max}_{WS}} \left(\theta_{stopWS} - \theta\right)$$

## 3.8 Modélisation par approximation linéaire

En choisissant une valeur de $\mathbf{a}_\theta$ élevée[1], le comportement du ressort entropique est proche de celui d'un ressort mécanique (Fig .7C et 7D) dont le moment angulaire, noté $\mathcal{M}_B^{lin}$, est modélisé par l'équation linéaire classique :

$$\mathcal{M}_B^{lin} = \tau \cdot (\theta_{stopWS} - \theta) \tag{3.37}$$

où $\tau$ est la raideur angulaire de ce ressort telle que : $\tau = \frac{\mathcal{M}1}{\theta_{stopWS} - \theta 1} = \frac{|\mathcal{M}1|}{\Delta\theta_{WS}^{Max}}$

Avec (3.8), l'éq. (3.37) se réécrit algébriquement (Fig. 6a):

$$\mathcal{M}_B^{lin} = \frac{|\mathcal{M}1|}{\Delta\theta_{WS}^{Max}} \cdot (\theta_{stopWS} - \theta) \tag{3.38}$$

Ainsi (3.2), (3.4), (3.5), (3.8), (3.14), (3.18) et (3.38) induisent au même instant t ∈ $\Delta t_{WS}^{(b)}$ :

$$\forall s \in [1,2,...,N_S] \qquad T_{A^{(b)}}^{lin} = \frac{|\mathcal{M}1|}{\Delta X_{WS}^{Max}} \cdot |\theta_{stopWS} - \theta| \tag{3.39a}$$

De même (3.1), (3.3), (3.5), (3.14), (3.18) et (3.38) impliquent au même instant t ∈ $\Delta t_{WS}^{(b)}$ :

$$\forall s \in [1,2,...,N_S] \qquad T_{D^{(b)}}^{lin} = \frac{|\mathcal{M}1|}{\Delta X_{WS}^{Max}} \cdot |\theta_{stopWS} - \theta| \tag{3.39b}$$

Avec l'approximation linéaire, les équations (3.30) et (3.38) induisent :

$\mathcal{M}_B^{lin}$ varie linéairement en fonction du temps de $\mathcal{M}1$ à $\mathbf{0}$ (Fig. 7C et 7D) tel que :

$$\mathcal{M}_B^{lin} = |\mathcal{M}1| \cdot \frac{\left(\theta_{stopWS} - \theta_{startWS}^{(b)}\right)}{\Delta\theta_{WS}} \cdot \left(1 - \frac{t}{\Delta t_{WS}^{(b)}}\right) \tag{3.40}$$

On en déduit l'évolution temporelle de $T_{A^{(b)}}^{lin}$ et de $T_{D^{(b)}}^{lin}$.

Il est à noter que le modèle mécanique linéaire s'avère une approximation correcte du modèle entropique hyperbolique (Fig. 7C et 7D). Pour une valeur angulaire identique, on observe une différence maximale de l'ordre de 5 à 8% entre les valeurs des moments angulaires formulés par le modèle du ressort entropique (éq. 3.31) et par celui du ressort angulaire classique (éq. 3.38).

[1] *Dans notre application numérique, $a_\theta$ prend la valeur 316.9 (voir paragraphe 4.5)*

## 3.9 Discussion

### *3.9.1 Forces produites par l'élasticité entropique*

L'équation (3.31) de type hyperbolique est calquée sur l'équation des gaz parfaits. Nous rappelons que les N molécules d'un gaz enfermées dans une enceinte se comportent collectivement comme un ressort entropique; pour de petits déplacements (ou pour des sauts de température), le ressort entropique s'apparente à un ressort mécanique linéaire de raideur égale à la pente de la tangente à l'hyperbole au point d'équilibre.

### *3.9.2 Faits conformes au modèle*

La valeur maximale du pas d'une tetM durant un WS de 10 nm correspond à la valeur usuelle de la littérature (Geeves and Holmes 1999; Piazzesi, Reconditi et al. 2002 ; Reconditi, Linari et al. 2004; Huxley, Reconditi et al. 2006 ).

### *3.9.3 Remarques*

Pour obtenir les relations (3.36a) et (3.36b) qui relient moments et forces tangentielles, on relève l'importance de l'égalité (3.14) et son implication sur la géométrie inter-segmentaire d'une tetM en cours de WS.

Le travail moteur de S1 est toujours positif, que ce soit dans hsR ou hsL (Fig. 7).

# 4 Comportement collectif de tetM en cours de WS dans un hs

Dans ce chapitre, la fmI est supposée être, soit en conditions isométriques (V=0), soit en en raccourcissement continu à vitesse constante lente. Les équations des chap. 2 et 3 sont donc applicables en l'état.

## 4.1 Distribution uniforme des positions angulaires de début de WS

Rappel de **Hypothèse 1** : les angles $\theta_{startWS}^{(b)}$ des différents S1b indicés par la lettre b et démarrant un WS se répartissent entre 2 constantes $\theta 1$ et $\theta 2$ sur un intervalle de longueur $\delta_\theta = |\theta 2 - \theta 1|$.

Rappel de **Hypothèse 2** : sur une durée égale ou supérieure à la milliseconde, $\theta_{startWS}^{(b)}$ est une variable répartie continument[1] et aléatoirement sur $\delta_\theta$ ; par conséquence, $\theta_{startWS}^{(b)}$ suit une loi uniforme sur $\delta_\theta$ notée $f_0$ (rectangle grisé; Fig. 9A), telle que[2] :

$$f_0(\theta) = \frac{1}{\delta_\theta} \cdot \mathbb{1}_{[\theta 1;\theta 2]}(\theta) \qquad (4.1)$$

## 4.2 Tension exercée par un ensemble de tetM débutant leur WS

Soit $N_{startWS}^{hsL}$ le nombre de S1 initiant leur WS dans un intervalle de temps supérieur à la milliseconde dans un hsL de la fmI, nombre suffisamment grand pour justifier le passage au continu et le recours à l'Hyp. 2.

La somme des actions exercées par les $N_{startWS}^{hsL}$ tetM liées aux filA du hsL est nommée $T0^{hsL}$, soit avec (3.3) :

$$T0^{hsL} = \sum_{b=1}^{N_{startWS}^{hsL}} T_{D^{(b)}} = \sum_{b=1}^{N_{startWS}^{hsL}} T_{A^{(b)}}$$

Ce qui se réécrit avec (3.4), (3.5), (3.13), (3.14) et (3.20) :

$$T0^{hsL} = \frac{\Delta\theta_{WS}^{Max}}{\Delta X_{WS}^{Max}} \cdot \sum_{b=1}^{N_{startWS}^{hsL}} \mathcal{M}_B^{(b)} \qquad (4.2)$$

Considérons $\mathcal{M}0$, le moment moyen de $N_{startWS}^{hsL}$ S1 dont les positions angulaires $\theta$ sont uniformément distribuées sur $\delta_\theta$ tel que :

$$\mathcal{M}0 = \frac{\sum_{b=1}^{N_{startWS}^{hsL}} \mathcal{M}_B^{(b)}}{N_{startWS}^{hsL}} \cdot \qquad (4.3)$$

---

[1] *Le nombre de S1 en début de WS par hs est un nombre entier, mais ce nombre est toujours suffisamment grand pour autoriser le passage au continu.*
*Rappel : selon la typologie des fibres musculaires, il y a entre 50 000 et 200 000 S1 disponibles par hs (1 sur 2 par molM); ainsi 1‰ S1 commençant leur WS au même temps t donne un entier compris entre 50 et 200.*

[2] *Voir Annexe A1 pour la définition d'une fonction indicatrice.*

D'après (3.31) et (4.1), en passant au continu avec la définition d'une moyenne, on a :

$$\mathcal{M}0 = \frac{\mathcal{M}1 \cdot (\theta 1 + a_\theta)}{\Delta\theta_{WS}^{Max} \cdot \delta_\theta} \cdot \int_{\theta 1}^{\theta 2} \left( \frac{\theta_{stopWS} + a_\theta}{\theta + a_\theta} - 1 \right) \cdot d\theta$$

Soit après intégration :

$$\frac{\mathcal{M}0}{\mathcal{M}1} = \frac{(\theta 1 + a_\theta)}{\Delta\theta_{WS}^{Max}} \cdot \left[ \frac{(\theta_{stopWS} + a_\theta)}{\delta_\theta} \cdot \log\left( \frac{\theta 2 + a_\theta}{\theta 1 + a_\theta} \right) - 1 \right] \qquad (4.4)$$

On caractérise l'éq (3.31) par l'hyperbole $h_{\mathcal{M}}(\theta)$ représentée en rouge sur la Fig. 8, telle que :

$$h_{\mathcal{M}}(\theta) = \frac{\mathcal{M}_B}{\mathcal{M}1} = \frac{(\theta 1 + a_\theta)}{\Delta\theta_{WS}^{Max}} \cdot \left( \frac{\theta_{stopWS} + a_\theta}{\theta + a_\theta} - 1 \right) \qquad (4.5)$$

Et on définit la position angulaire $\theta 0$ telle que :

$$h_{\mathcal{M}}(\theta 0) = \frac{\mathcal{M}0}{\mathcal{M}1} = \frac{(\theta 1 + a_\theta)}{\Delta\theta_{WS}^{Max}} \cdot \left( \frac{\theta_{stopWS} + a_\theta}{\theta 0 + a_\theta} - 1 \right) \qquad (4.6)$$

Le point de coordonnées $\left( \theta 0, h_{\mathcal{M}}(\theta 0) = \frac{\mathcal{M}0}{\mathcal{M}1} \right)$ est le point violet de la Fig. 8.

On observe que $\theta 0$ se situe approximativement au milieu de $\delta_\theta$ .

On obtient confirmation de ce résultat en menant le même raisonnement avec le modèle linéaire.

Soit $\mathcal{M}0^{lin}$, le moment moyen de $N_{startWS}^{hsL}$ S1 dont les positions angulaires $\theta$ sont uniformément distribuées sur $\delta_\theta$ calculé avec la valeur donnée par (3.38), à savoir :

$$\mathcal{M}0^{lin} = \frac{\mathcal{M}1}{\Delta\theta_{WS}^{Max} \cdot \delta_\theta} \cdot \int_{\theta 1}^{\theta 2} (\theta_{stopWS} - \theta) \cdot d\theta$$

Soit après intégration :

$$\frac{\mathcal{M}0^{lin}}{\mathcal{M}1} = \frac{\left( \theta_{stopWS} - \theta 0^{lin} \right)}{\Delta\theta_{WS}^{Max}} \qquad (4.7)$$

avec : $$\theta 0^{lin} = \frac{\theta 1 + \theta 2}{2} \qquad (4.8)$$

Avec l'égalité (3.16), il y a correspondance entre l'échelle de la variable $\mathbf{X}$ (abscisse sur l'axe OX définissant le déplacement linéaire relatif[1] entre disqM et disqZ du hsL) et celle de la variable $\theta_{hsL}^{(b)}$ (position angulaire définissant la rotation du levier S1b de la tetM n° b initiant son WS dans hsL). En faisant coïncider l'origine des $\mathbf{X}$ avec $\theta 0$ défini par (4.6), on écrit l'équation de $\mathbf{X}$ en fonction[2] de $\theta$ :

$$\mathbf{X} = \mathbf{L_{S1b}} \cdot \mathbf{R_{WS}} \cdot (\theta - \theta 0) \quad (4.9a)$$

Ou inversement :

$$\theta = \theta 0 + \frac{\mathbf{X}}{\mathbf{L_{S1b}} \cdot \mathbf{R_{WS}}} \quad (4.9b)$$

***Cas particuliers*** **(Fig. 8)**:

$$\mathbf{X1} = \mathbf{L_{S1b}} \cdot \mathbf{R_{WS}} \cdot (\theta 1 - \theta 0) \quad (4.10a)$$

$$\mathbf{X2} = \mathbf{L_{S1b}} \cdot \mathbf{R_{WS}} \cdot (\theta 2 - \theta 0) \quad (4.10b)$$

$$\mathbf{X3} = \mathbf{L_{S1b}} \cdot \mathbf{R_{WS}} \cdot \left(\theta_{stopWS} - \theta 0\right) = \Delta \mathbf{X}_{WS}^{Max} + \mathbf{X1} \quad (4.10c)$$

$$\delta_X = \mathbf{X2} - \mathbf{X1} = \mathbf{L_{S1b}} \cdot \mathbf{R_{WS}} \cdot \delta_\theta \quad (4.10d)$$

Avec la transformation d'échelle fournie par (4.9b), (3.31) se reformule :

$$\mathcal{M}_B^{(b)} = \mathcal{M}_B(\mathbf{X}) = \frac{\mathcal{M}1}{\Delta \mathbf{X}_{WS}^{Max}} \cdot (\mathbf{X1} + \mathbf{a_X}) \cdot \left(\frac{\mathbf{X3} + \mathbf{a_X}}{\mathbf{X} + \mathbf{a_X}} - 1\right) \quad (4.11)$$

avec la constante : $\mathbf{a_X} = \mathbf{L_{S1b}} \cdot \mathbf{R_{WS}} \cdot (\theta 0 + \mathbf{a}_\theta)$ (4.12)

Avec le changement d'échelle où $\delta_X$ correspond à $\delta_\theta$, (4.4) devient :

$$\frac{\mathcal{M}0}{\mathcal{M}1} = \frac{(\mathbf{X1} + \mathbf{a_X})}{\Delta \mathbf{X}_{WS}^{Max}} \cdot \left[\frac{(\mathbf{X3} + \mathbf{a_X})}{\delta_X} \cdot \log\left(\frac{\mathbf{X2} + \mathbf{a_X}}{\mathbf{X1} + \mathbf{a_X}}\right) - 1\right] \quad (4.13)$$

De même pour le modèle linéaire, (3.38) se réécrit avec (4.9b) et (4.10c) :

$$\mathcal{M}_B^{lin}(\mathbf{X}) = \frac{\mathcal{M}1}{\Delta \mathbf{X}_{WS}^{Max}} \cdot (\mathbf{X3} - \mathbf{X}) \quad (4.14)$$

Et (4.7) s'énonce maintenant :

$$\frac{\mathcal{M}0^{lin}}{\mathcal{M}1} = \frac{\left(\mathbf{X3} - \mathbf{X0}^{lin}\right)}{\Delta \mathbf{X}_{WS}^{Max}} \quad (4.15)$$

avec : $\mathbf{X0}^{lin} = \frac{\mathbf{X1} + \mathbf{X2}}{2}$ (4.16)

---

[1] *Le déplacement rendant compte d'un raccourcissement est habituellement et logiquement calculé comme négatif. Pour faire coïncider l'évolution du déplacement avec le découlement temporel représenté de la gauche vers la droite dans le sens des temps positifs (voir Fig. 25 et 26 ; chap. 13), nous avons choisi de l'afficher contrairement à l'usage comme positif (Fig.9 et 10). Ainsi toutes les variations de longueur, notées $\Delta X$ , durant le raccourcissement d'un hs seront calculées positivement.*

[2] *rappel : $\theta$ est exprimé en radiant dans les formules de la page*

Fig. 8 : Moment relatif moyen de N tetM, chacune initiant un WS avec une position angulaire de S1b répartie aléatoirement entre θ1 et θ2

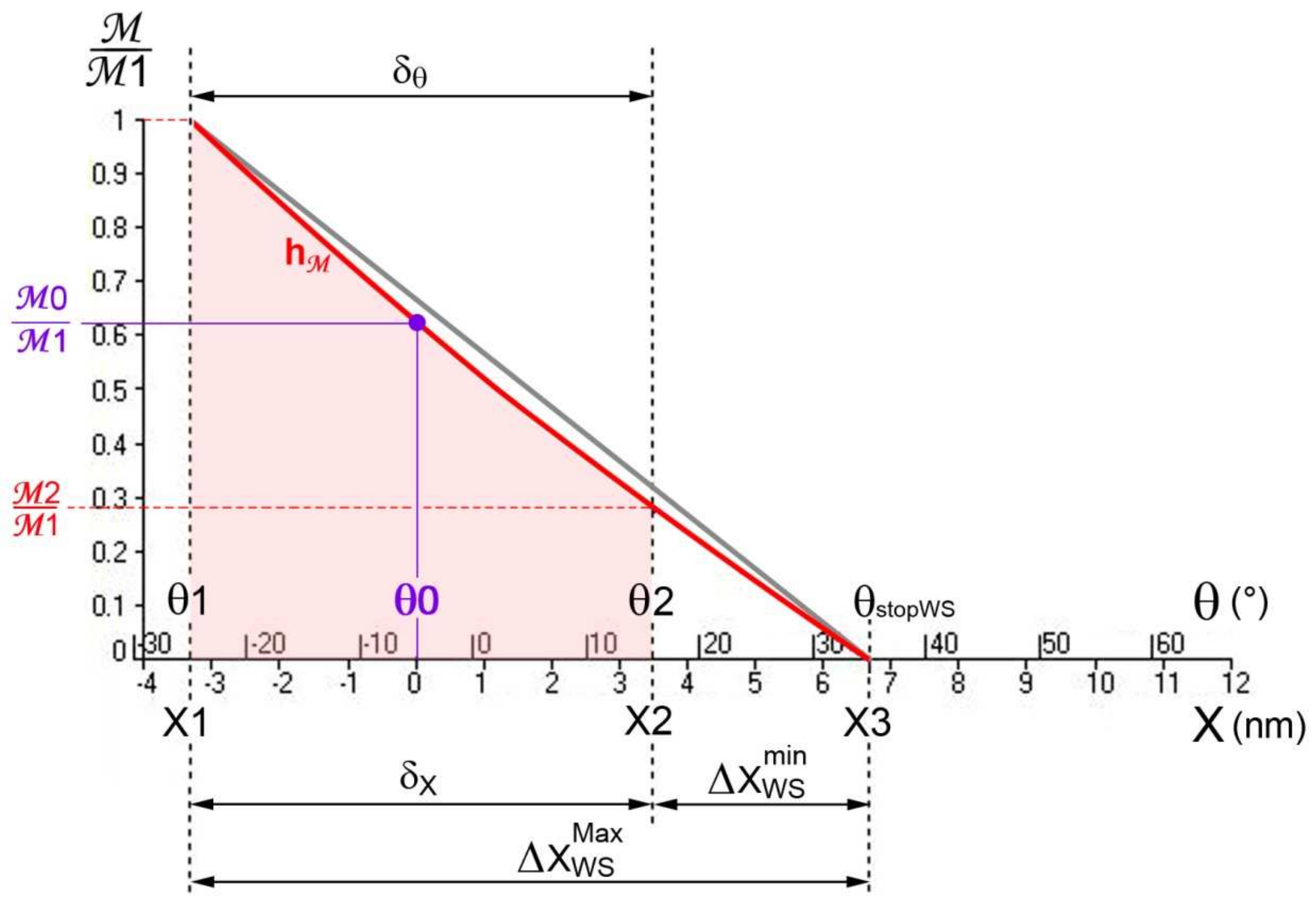

**Application numérique**[1] **(Fig. 8)**

Avec (3.9b), (3.10) et $\mathbf{a_\theta = 316.9}$ , le calcul de (4.4) donne :

$$\frac{\mathcal{M}0}{\mathcal{M}1} = \frac{(-25+316.9)}{60}\cdot\left[\frac{(35+316.9)}{41}\cdot\log\left(\frac{16+316.9}{-25+316.9}\right)-1\right]\approx 0.625 \qquad (4.17a)$$

avec $\theta 0 \approx -5°$ (4.17b)

Pour le modèle linéaire, le calcul de (4.7) mène à :

$$\frac{\mathcal{M}0^{lin}}{\mathcal{M}1} = \frac{39.5}{60} = 0.658 \qquad (4.18a)$$

avec $\theta 0^{lin} = -4°30"$ (4.18b)

Soit une différence de l'ordre de 5% entre les 2 modèles.

$$\mathbf{X1 \approx -3.3\,nm} \qquad (4.19a)$$

$$\mathbf{X2 \approx 3.5\ nm} \qquad (4.19b)$$

$$\mathbf{X3 \approx 6.7\ nm} \qquad (4.19c)$$

$$\mathbf{\delta_X \approx 6.8\,nm} \qquad (4.19d)$$

Avec (3.23b), on vérifie : $\delta_X = \Delta X_{WS}^{Max} - \Delta X_{WS}^{min} = 10\,nm - 3.2\,nm$

$$a_X = 10\cdot 0.96\cdot(-5+316.9)\cdot\frac{\pi}{180}\approx 52 \qquad (4.20)$$

---

[1] *Dans les calculs qui suivent, θ est exprimé en degré*

**En conclusion (Fig. 8, 9A et 9B) :**

La tension exercée par les $\mathbf{N^{hsL}_{startWS}}$ ($\mathbf{N^{hsR}_{startWS}}$) S1 initiant leur WS dans un hsL (hsR) de la fmI à l'instant t modélisés par un ressort entropique et formulée par (4.2) se réécrit[1] en module avec (4.3) :

$$\mathbf{T0^{hsL} = N^{hsL}_{startWS} \cdot \frac{\Delta\theta^{Max}_{WS}}{\Delta X^{Max}_{WS}} \cdot \mathcal{M}0} \qquad (4.21a)$$

$$\mathbf{T0^{hsR} = N^{hsR}_{startWS} \cdot \frac{\Delta\theta^{Max}_{WS}}{\Delta X^{Max}_{WS}} \cdot \mathcal{M}0} \qquad (4.21b)$$

où la valeur de $\mathcal{M}\mathbf{0}$ est fournie par (4.4) ou (4.13)

De manière similaire, (4.2) donne[2] avec le modèle linéaire dans hsL et hsR :

$$\mathbf{{T0^{hsL}}^{lin} = N^{hsL}_{startWS} \cdot \frac{\Delta\theta^{Max}_{WS}}{\Delta X^{Max}_{WS}} \cdot \mathcal{M}0^{lin}} \qquad (4.22a)$$

$$\mathbf{{T0^{hsR}}^{lin} = N^{hsR}_{startWS} \cdot \frac{\Delta\theta^{Max}_{WS}}{\Delta X^{Max}_{WS}} \cdot \mathcal{M}0^{lin}} \qquad (4.22b)$$

où la valeur de $\mathcal{M}\mathbf{0}^{\mathbf{lin}}$ est apportée par (4.7) ou (4.15)

## 4.3 Tension exercée par un ensemble de tetM ayant initié leur WS dans un hsL après un raccourcissement discret en longueur de la fmI

Si le hsL se raccourcit[3] à vitesse constante, il est possible avec les Hyp. 1 à 15 d'utiliser les équations des chapitres précédents. Ainsi si le hsL se raccourcit de $\mathbf{\Delta X}$ selon OX, les $\mathbf{N^{hsL}_{startWS}}$ tetM qui ont initié leur WS vont tous être déplacés du même échelon linéaire $\mathbf{\Delta X}$, correspondant d'après (3.15), (3.16) et (3.17) à la rotation effectuée à vitesse angulaire constante de chaque S1b d'un différentiel angulaire identique.

Avec (3.23a) et (3.24), deux cas de figure se présentent pour chacune des $\mathbf{N^{hsL}_{startWS}}$ tetM :

cas 1 : $\mathbf{\Delta X < \Delta X^{(b)}_{WS}}$ , *i.e.* cas où le WS de la tetM n° b est toujours en cours

cas 2 : $\mathbf{\Delta X \geq \Delta X^{(b)}_{WS}}$ , *i.e.* cas où le WS de la tetM n° b est terminé

On distingue ainsi 2 zones (Fig. 9D, 9F et 9H).

---

[1] *On aurait obtenu le même résultat en intégrant (3.36a) et (3.36b) sur* $\boldsymbol{\delta_\theta}$ *en passant au continu avec la définition d'une moyenne.*

[2] *On aurait obtenu le même résultat en intégrant (3.39a) et (3.396b) sur* $\boldsymbol{\delta_\theta}$ *en passant au continu avec la définition d'une moyenne.*

[3] *Voir la note n° 1 de la page 47*

### *4.3.1 Zone 1* : $0 \le \Delta X \le \Delta X_{WS}^{min}$

La totalité des $N_{startWS}^{hsL}$ tetM sont toujours réparties uniformément sur $\delta_X$ d'après l'Hyp. 4 et suivent par passage au continu la loi de distribution notée $f_{z1}$ telle que (rectangle grisé; Fig. 9C) :

$$f_{z1}(X) = \frac{1}{\delta_X} \cdot \mathbb{1}_{[X1+\Delta X; X2+\Delta X]}(X) \tag{4.23}$$

Avec (3.4), (3.5), (3.13), (3.14) et (3.20), la somme de toutes les actions exercées par les $N_{startWS}^{hsL}$ tetM, déplacées collectivement de l'échelon $\Delta X$ sur les filA du hsL auxquels elles sont liées, est égale à :

$$T_{z1}^{hsL}(\Delta X) = \frac{\Delta\theta_{WS}^{Max}}{\Delta X_{WS}^{Max}} \cdot \sum_{b=1}^{N_{hsL}^{startWS}} \mathcal{M}_B^{(b)}(\Delta X) \tag{4.24}$$

Par définition d'une moyenne en passant au continu sur l'intervalle $[X1+\Delta X;\, X2+\Delta X]$ de longueur $\delta_X$ en intégrant (4.11) avec (4.21a), l'éq. (4.24) mène à :

$$\frac{T_{z1}^{hsL}(\Delta X)}{T0^{hsL}} = \frac{\mathcal{M}1 \cdot (X1 + a_X)}{\mathcal{M}0 \cdot \Delta X_{WS}^{Max} \cdot \delta_X} \cdot \int_{X1+\Delta X}^{X2+\Delta X} \left( \frac{X3 + a_X}{X + a_X} - 1 \right) \cdot dX$$

Soit :

$$\frac{T_{z1}^{hsL}(\Delta X)}{T0^{hsL}} = \frac{\mathcal{M}1 \cdot (X1 + a_X)}{\mathcal{M}0 \cdot \Delta X_{WS}^{Max}} \cdot \left[ \frac{(X3 + a_X)}{\delta_X} \cdot \log\left( \frac{\Delta X + X2 + a_X}{\Delta X + X1 + a_X} \right) - 1 \right] \tag{4.25}$$

L'évolution de $T_{z1}^{hsL}(\Delta X)/T0^{hsL}$ pour $\Delta X \in \left[0; \Delta X_{WS}^{min}\right]$ est représentée sur les Fig. 9D et 10 avec un trait épais rouge; elle est approximativement linéaire (voir modèle entropique linéarisé du sous-paragraphe 4.3.4.2).

### *4.3.2 Zone 2* : $\Delta X_{WS}^{min} \le \Delta X \le \Delta X_{WS}^{Max}$

#### 4.3.2.1 Mode A (A pour Amorti; *Amortized* ) : les S1 en fin de WS n'ont aucune influence

Soit $N_{stopWS}^{hsL}(\Delta X)$, le nombre de tetM qui, parmi les $N_{startWS}^{hsL}$ tetM, ont terminé leur WS et n'exercent plus d'action sur les filA parce qu'elles sont détachées.

On postule que la distribution de ces $N_{stopWS}^{hsL}(\Delta X)$ tetM est aussi uniforme.

**Fig. 9 : Evolutions de la tension exercée par N tetM qui après avoir initié leurs WS sont déplacées d'un échelon $\Delta X$**

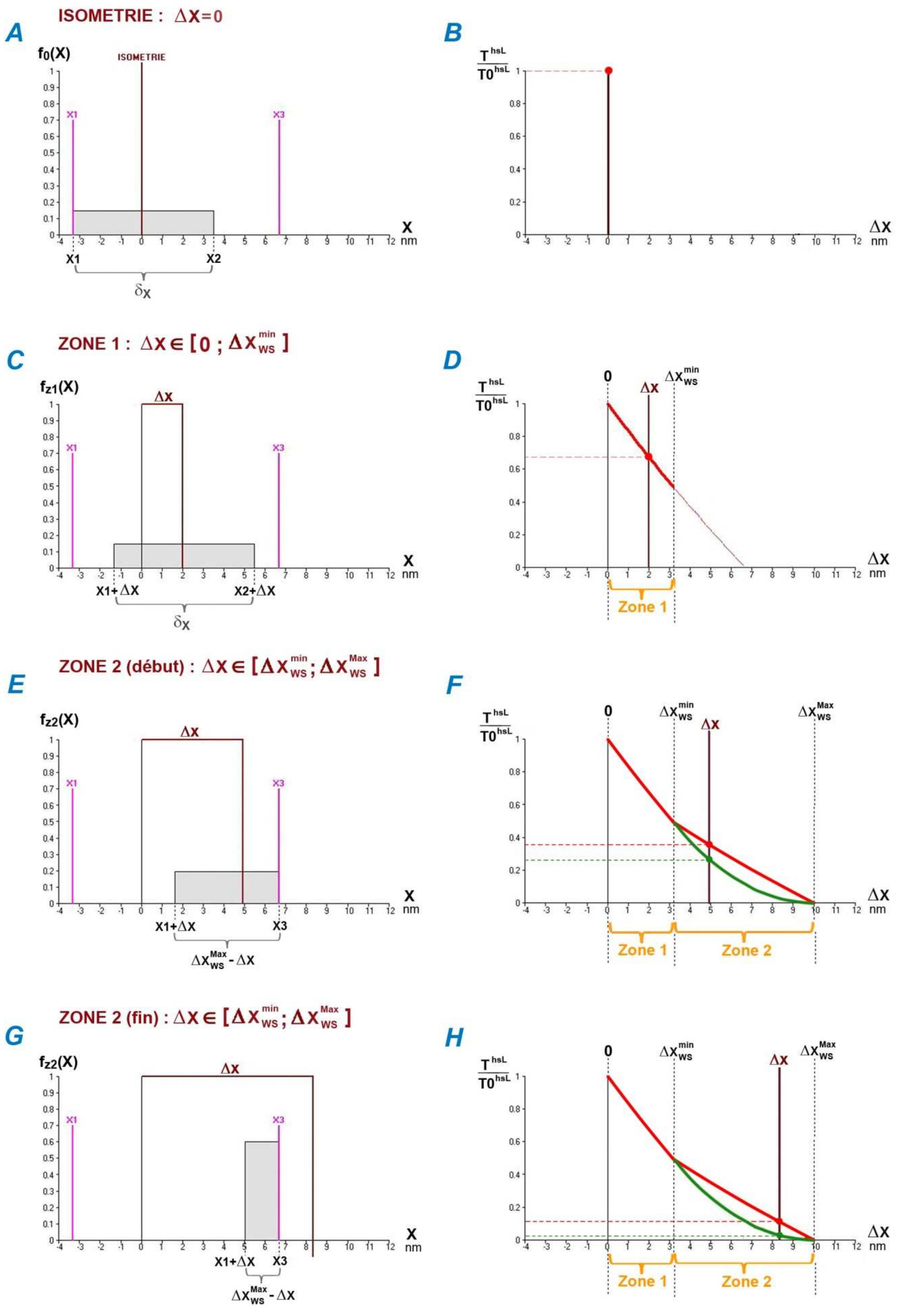

Soit les $N_{WS}^{hsL}(\Delta X)$ tetM encore en cours de WS; on vérifie :

$$N_{startWS}^{hsL} = N_{stopWS}^{hsL}(\Delta X) + N_{WS}^{hsL}(\Delta X) \tag{4.26}$$

A l'appui de l'Hyp. 2 et du postulat précédent, les $N_{WS}^{hsL}(\Delta X)$ tetM encore en cours de WS sont toujours réparties uniformément sur l'intervalle $[X1+\Delta X; X3]$ de longueur $\left(\Delta X_{WS}^{Max} - \Delta X\right)$ d'après (4.10c). Par définition d'une loi uniforme, on a :

$$N_{WS}^{hsL}(\Delta X) = N_{startWS}^{hsL} \cdot \int_{X1+\Delta X}^{X3} \frac{dX}{\delta_X} = N_{startWS}^{hsL} \cdot \frac{\left(\Delta X_{WS}^{Max} - \Delta X\right)}{\delta_X} \tag{4.27}$$

Les $N_{WS}^{hsL}(\Delta X)$ tetM toujours en cours de WS suivent, en passant au continu, la loi de distribution uniforme notée $f_{z2}$ (rectangles grisés; Fig. 9E et 9G) :

$$f_{z2}(X) = \frac{1}{\left(\Delta X_{WS}^{Max} - \Delta X\right)} \cdot \mathbb{1}_{[X1+\Delta X;X3]}(X) \tag{4.28}$$

Avec (3.4), (3.13), (3.14) et (3.20), la somme de toutes les actions exercées par les $N_{WS}^{hsL}(\Delta X)$ tetM, déplacées collectivement de l'échelon $\Delta X$ sur les filA du hsL auxquels elles sont liées, est égale à[1]:

$$T_{z2}^{hsL^A}(\Delta X) = \frac{\Delta\theta_{WS}^{Max}}{\Delta X_{WS}^{Max}} \cdot \sum_{b=1}^{N_{WS}^{hsL}(\Delta X)} \mathcal{M}_B^{(b)}(\Delta X) \tag{4.29}$$

Par définition d'une moyenne en passant au continu sur l'intervalle $[X1+\Delta X; X3]$ en intégrant (4.11) avec (4.21a), (4.27) et (4.28), l'éq. (4.29) conduit à :

$$\frac{T_{z2}^{hsL^A}(\Delta X)}{T0^{hsL}} = \frac{\mathcal{M}1 \cdot (X1 + a_X)}{\mathcal{M}0 \cdot \Delta X_{WS}^{Max} \cdot \delta_X} \cdot \int_{X1+\Delta X}^{X3} \left(\frac{X3 + a_X}{X + a_X} - 1\right) \cdot dX$$

Soit :

$$\frac{T_{z2}^{hsL^A}(\Delta X)}{T0^{hsL}} = \frac{\mathcal{M}1 \cdot (X1 + a_X) \cdot \left(\Delta X_{WS}^{Max} - \Delta X\right)}{\mathcal{M}0 \cdot \Delta X_{WS}^{Max} \cdot \delta_X} \cdot \left[\frac{(X3 + a_X)}{\left(\Delta X_{WS}^{Max} - \Delta X\right)} \cdot \log\left(\frac{X3 + a_X}{\Delta X + X1 + a_X}\right) - 1\right] \tag{4.30}$$

L'évolution de $T_{z2}^{hsL^A}(\Delta X) / T0^{hsL}$ pour $\Delta X \in \left[\Delta X_{WS}^{min}; \Delta X_{WS}^{Max}\right]$ est représentée sur les Fig. 9F, 9H et 10 par une courbe verte; au début de la zone 2, sur une étendue d'environ 1 nm, elle semble prolonger l'évolution approximativement linéaire de la zone 1, puis s'amortit de manière similaire à un arc de parabole (voir modèle linéaire).

[1] *A pour Amorti ou Amortized*

### 4.3.2.2 Mode E (E pour Exagéré; *Exaggerated* ) : les S1 en fin de WS ont une influence exagérée

L'éq. (4.30) repose sur le postulat que les $N_{stopWS}^{hsL}(\Delta X)$ tetM qui ont terminé leur WS n'exercent plus d'action sur les filA, ce qui semble suggérer que leur détachement est rapide. Or divers travaux montrent que ce détachement prend effet sur une durée qui peut se révéler égale à plusieurs ms (voir paragraphe 1.6), *i.e.* une durée très supérieure, par exemple, à celle de la phase 1 d'un échelon de longueur ou de force.

Pour tenir compte de l'influence des $N_{stopWS}^{hsL}(\Delta X)$ tetM encore attachées, il faudrait d'une part remanier le calcul de (3.31) et d'autre part connaître la distribution des $N_{stopWS}^{hsL}(\Delta X)$ tetM. Comme nous ne pouvons caractériser ces données, nous envisageons un cas extrême : les $N_{stopWS}^{hsL}(\Delta X)$ et les $N_{WS}^{hsL}(\Delta X)$ tetM , soit au total $N_{startWS}^{hsL}$ d'après (4.26), sont réparties uniformément sur l'intervalle $[X1+\Delta X; X3]$ de longueur $\left(\Delta X_{WS}^{Max} - \Delta X\right)$ *i.e*, la même fonction de densité fournie par (4.28) ; voir Fig. 9E et 9G.

Avec (3.4), (3.13), (3.14) et (3.20), la somme de toutes les actions exercées par les $N_{WS}^{hsL}(\Delta X)$ tetM, déplacées collectivement de l'échelon $\Delta X$ sur les filA du hsL auxquels elles sont liées, est égale à[1]:

$$T_{z2}^{hsL^E}(\Delta X) = \frac{\Delta\theta_{WS}^{Max}}{\Delta X_{WS}^{Max}} \cdot \sum_{b=1}^{N_{startWS}^{hsL}} \mathcal{M}_B^{(b)}(\Delta X) \qquad (4.31)$$

Par définition d'une moyenne en passant au continu sur l'intervalle $[X1+\Delta X; X3]$ en intégrant (4.11) avec (4.21a), (4.26) et (4.28), l'éq. (4.31) conduit à :

$$\frac{T_{z2}^{hsL^E}(\Delta X)}{T0^{hsL}} = \frac{\mathcal{M}1 \cdot (X1 + a_X)}{\mathcal{M}0 \cdot \Delta X_{WS}^{Max} \cdot \left(\Delta X_{WS}^{Max} - \Delta X\right)} \cdot \int_{X1+\Delta X}^{X3} \left(\frac{X3 + a_X}{X + a_X} - 1\right) \cdot dX$$

Soit :

$$\frac{T_{z2}^{hsL^E}(\Delta X)}{T0^{hsL}} = \frac{\mathcal{M}1 \cdot (X1 + a_X)}{\mathcal{M}0 \cdot \Delta X_{WS}^{Max}} \cdot \left[\frac{(X3 + a_X)}{\left(\Delta X_{WS}^{Max} - \Delta X\right)} \cdot \log\left(\frac{X3 + a_X}{\Delta X + X1 + a_X}\right) - 1\right] \qquad (4.32)$$

L'évolution de $T_{z2}^{hsL^E}(\Delta X) / T0^{hsL}$ pour $\Delta X \in \left[\Delta X_{WS}^{min}; \Delta X_{WS}^{Max}\right]$ est représentée sur les Fig. 9F, 9H et 10 par une courbe rouge et s'apparente à une évolution approximativement linéaire avec une pente 2 fois moindre que celle de la zone 1 (voir modèle entropique linéarisé du sous-paragraphe 4.3.4.3).

Sur la Fig. 10, on observe l'aspect bimodal de $T^{hsL} / T0^{hsL}$ en fonction de $\Delta X$ dans la zone 2.

---

[1] *E pour Exagéré ou Exaggerated*

**Fig. 10: Evolutions de la tension relative en fonction de l'échelon ΔX selon différentes modélisations**

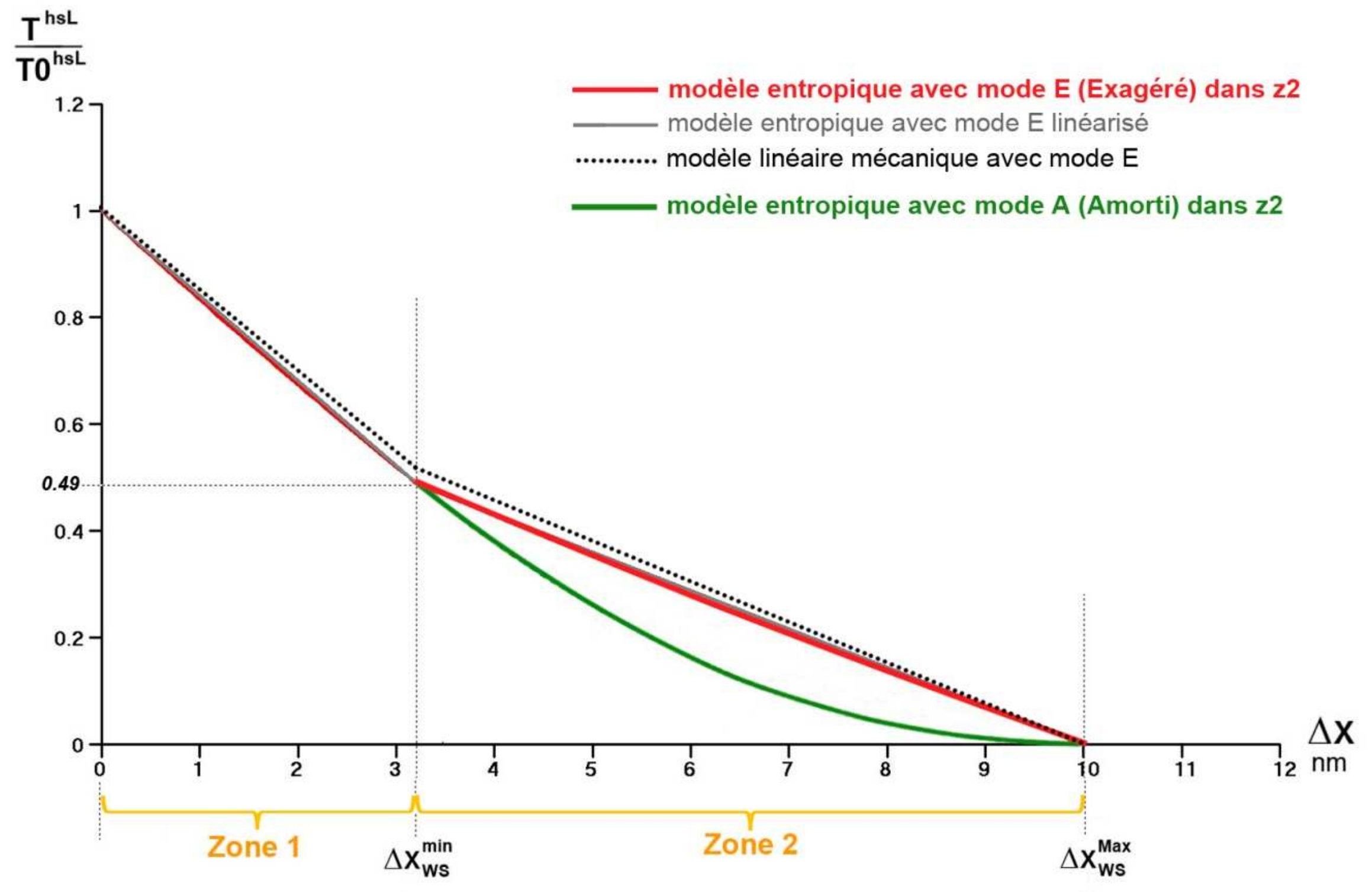

## *4.3.3 Modèle linéaire*

En suivant le même raisonnement par intégration de (4.14) sur les mêmes 2 intervalles avec (4.22a), (4.23), (4.24), (4.26) à (4.24), on obtient les équations des tensions relatives en fonction d un échelon en longueur $\Delta X$ selon le modèle linéaire :

### 4.3.3.1 Zone 1 : $0 \le \Delta X \le \Delta X_{WS}^{min}$

Soit un segment de droite (pointillés noirs; Fig. 10) :

$$\frac{T_{z1}^{hsL^{lin}}(\Delta X)}{T0^{hsL^{lin}}} = \frac{\mathcal{M}1}{\mathcal{M}0^{lin} \cdot \Delta X_{WS}^{Max}} \cdot \left[\left(X3 - X0^{lin}\right) - \Delta X\right] \qquad (4.33a)$$

où $$X0^{lin} = \frac{(X1 + X2)}{2} \qquad (4.33b)$$

avec une pente : $$p_{z1}^{lin} = -\frac{\mathcal{M}1}{\mathcal{M}0^{lin} \cdot \Delta X_{WS}^{Max}} \qquad (4.33c)$$

### 4.3.3.2 Zone 2 : $\Delta X_{WS}^{min} \le \Delta X \le \Delta X_{WS}^{Max}$

Soit un arc de parabole pour la Version A (Amorti) où les S1 en fin de WS n'ont aucune influence :

$$\frac{T_{z2}^{hsL^{lin_A}}(\Delta X)}{T0^{hsL^{lin}}} = \frac{\mathcal{M}1}{2 \cdot \mathcal{M}0^{lin} \cdot \Delta X_{WS}^{Max} \cdot \delta_X} \cdot \left(\Delta X_{WS}^{Max} - \Delta X\right)^2 \qquad (4.34a)$$

Soit un segment de droite (pointillés noirs; Fig. 10) pour la Version E (Exagéré) où les S1 en fin de WS ont une influence exagérée:

$$\frac{T_{z2}^{hsL^{lin_E}}(\Delta X)}{T0^{hsL^{lin}}} = \frac{\mathcal{M}1}{2 \cdot \mathcal{M}0^{lin} \cdot \Delta X_{WS}^{Max}} \cdot \left[ \Delta X_{WS}^{Max} - \Delta X \right] \qquad (4.34b)$$

avec une pente : $$p_{z2}^{lin_E} = -\frac{\mathcal{M}1}{2 \cdot \mathcal{M}0^{lin} \cdot \Delta X_{WS}^{Max}} = \frac{p_{z1}^{lin}}{2} \qquad (4.34c)$$

Les cinétiques de ces 3 fonctions du modèle linéaire corroborent les cinétiques précédentes du modèle hyperbolique entropique, présentées dans les Fig. 9 et 10.

## *4.3.4 Remarques*

### 4.3.4.1 Equations dans hsR

Pour hsR, d'après (4.21b) et (4.22b), on retrouve pour les 2 modèles (entropique/hyperbolique et mécanique/linéaire) les mêmes expressions relatives aux zones 1 et 2 (en prenant la valeur absolue de $\mathcal{M}1$ pour le modèle linéaire).

Rappelons que les valeurs de $T0^{hs}$ et des différentes versions de $T_{z1}^{hs}(\Delta X)$ et de $T_{z2}^{hs}(\Delta X)$ dépendent de $N_{startWS}^{hs}$, le nombre de S1 ayant initié leur WS dans hsR ou hsL à l'instant t (environ 1 ms).

### 4.3.4.2 Approximation linéaire du modèle entropique pour la Zone 1 (trait gris; Fig. 10)

Avec la remarque précédente, l'allure linéaire de la courbe donnée par (4.25) permet d'assimiler chaque hs de la fmI à un ressort linéaire de raideur $\chi_{z1}^{hs}$ tel que :

$$T_{z1}^{hs}(\Delta X) \big/ T0^{hs} \approx 1 - \Delta X \cdot \chi_{z1}^{hs} \qquad (4.35a)$$

***Application numérique :***

Avec les valeurs fournies par (4.19a) à (4.19d), l'éq. (4.25) donne pour $\Delta X_{WS}^{min} = 3.2\,nm$ (Fig. 10) :

$$\frac{T_{z1}^{hs}\left(\Delta X_{WS}^{min}\right)}{T0^{hs}} \approx 0.49 \qquad (4.35b)$$

Il s'en déduit :

$$\chi_{z1}^{hs} \approx 0.16 \;\; nm^{-1} \qquad (4.35c)$$

Pour information, le modèle linéaire avec (4.33a) mène à :

$$\chi_{z1}^{hs^{lin}} = 0.152 \;\; nm^{-1} \qquad (4.35d)$$

Soit une différence de 5% entre ces 2 valeurs.

#### 4.3.4.3 Approximation linéaire du modèle entropique pour la Zone 2 (trait gris; Fig. 10) dans le cas où les S1 parviennent rapidement en fin de WS sans avoir le temps de se détacher

Dans ce cas particulier, on se rapproche de la Version 2 pour la zone 2 et l'allure linéaire de la courbe donnée par (4.32) associée à la remarque 4.3.4.1 permet d'assimiler chaque hs de la fmI à un ressort linéaire de raideur $\chi_{z2}^{hs}$, soit :

$$\frac{T_{z2}^{hs}(\Delta X)}{T0^{hs}} \approx \chi_{z2}^{hs} \cdot \left(\Delta X_{WS}^{Max} - \Delta X\right) \tag{4.36a}$$

$$\text{avec: } \chi_{z2}^{hs} = \frac{1 - \chi_{z1}^{hs} \cdot \Delta X_{WS}^{min}}{\delta_X} \tag{4.36b}$$

Ce cas particulier prendra sens pour la phase 1 d'un raccourcissement après un échelon de longueur ou de force.

***<u>Application numérique :</u>***

La valeur de (4.35c) introduite dans (4.46b) conduit à :

$$\chi_{z2}^{hs} \approx 0.072 \ \text{nm}^{-1} \tag{4.36c}$$

Le modèle linéaire avec (4.34c) et (4.35d) mène à :

$$\chi_{z2}^{hs^{lin}} = \frac{\chi_{z1}^{hs^{lin}}}{2} = 0.076 \ \text{nm}^{-1} \tag{4.36d}$$

Soit une différence de 5% entre ces 2 valeurs.

## 4.4 Tension exercée par $N_{startWS}^{hsL}$ tetM avec une distribution uniforme des $N_{startWS}^{hsL}$ S1b dans $\Delta\theta_{WS}^{Max}$

Considérons $\mathcal{M}0^*$, le moment moyen de $N_{WS}^{hsL}$ S1 dont les positions angulaires des S1b sont uniformément distribuées sur $\Delta\theta_{WS}^{Max}$ dans un hsL.

En suivant le même raisonnement qu'au paragraphe 4.2, l'intégration de (3.31) entre $\theta 1$ et $\theta_{stopWS}$ fournit la valeur relative de $\mathcal{M}0^*$ :

$$\frac{\mathcal{M}0^*}{\mathcal{M}1} = \frac{(\theta 1 + a_\theta)}{\Delta\theta_{WS}^{Max}} \cdot \left[ \frac{(\theta_{stopWS} + a_\theta)}{\Delta\theta_{WS}^{Max}} \cdot \log\left( \frac{\theta_{stopWS} + a_\theta}{\theta 1 + a_\theta} \right) - 1 \right] \tag{4.37a}$$

Avec la transformation d'échelle fournie par (4.9b) où $\Delta X_{WS}^{Max}$ correspond à $\Delta\theta_{WS}^{Max}$, (4.37a) devient :

$$\frac{\mathcal{M}0^*}{\mathcal{M}1} = \frac{(X1 + a_X)}{\Delta X_{WS}^{Max}} \cdot \left[ \frac{(X3 + a_X)}{\Delta X_{WS}^{Max}} \cdot \log\left( \frac{X3 + a_X}{X1 + a_X} \right) - 1 \right] \tag{4.37b}$$

**Fig. 11 : Moment relatif moyen de N tetM, chacune initiant un WS avec une position angulaire de S1b répartie aléatoirement entre $\theta 1$ et $\theta_{stopWS}$**

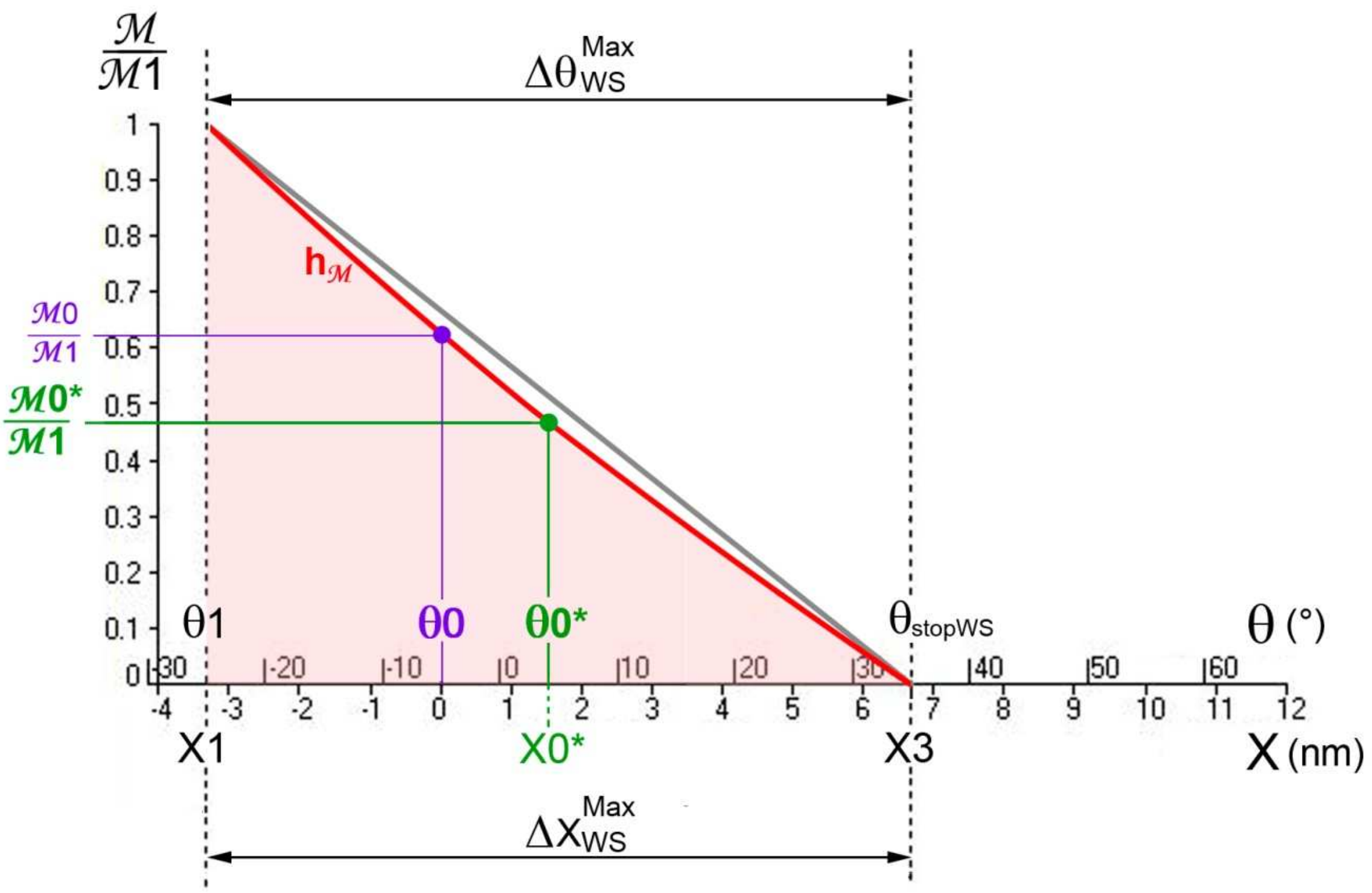

Le point de coordonnées $\left(\boldsymbol{\theta 0^*}, \mathbf{h}_{\mathcal{M}}(\boldsymbol{\theta 0^*}) = \dfrac{\mathcal{M}\mathbf{0^*}}{\mathcal{M}\mathbf{1}}\right)$ est le point vert de la Fig. 11.

Avec la transformation d'échelle, $\boldsymbol{\theta 0^*}$ correspond à un déplacement relatif du hs nommé $\mathbf{X0^*}$ (Fig. 11).

**<u>Application numérique</u> (Fig. 11)**

Avec les valeurs fournies par (4.19a) à (4.19d) et (4.20), le calcul de (4.42b) donne :

$$\frac{\mathcal{M}0^*}{\mathcal{M}1} = \frac{(-3.3+52)}{10} \cdot \left[\frac{(6.7+52)}{10} \cdot \log\left(\frac{6.7+52}{-3.3+352}\right) - 1\right] \approx 0.47 \tag{4.38}$$

soit une valeur légèrement inférieure à la moitié de $\mathcal{M}\mathbf{1}$ (le modèle linéaire donne exactement la moitié).

De (4.17a) et (4.38), on déduit :

$$\frac{\mathcal{M}0^*}{\mathcal{M}0} \approx 0.75 \tag{4.39}$$

Et les interpolations de (3.31) et (4.13) fournissent, respectivement :

$$\theta 0^* \approx 4° \tag{4.40}$$

$$X0^* \approx 1.5 \text{ nm} \tag{4.41}$$

## 4.5 Commentaires

La valeur de la constante $\mathbf{a_\theta}$ entrant dans la formulation du moment moteur d'une tetM selon l'éq. (3.31), a été posée à $\mathbf{316.9}$ afin d'obtenir le résultat fourni par (4.39).

Tout au long du chapitre, nous avons testé 2 types de modélisations, entropique (hyperbolique) et mécanique (linéaire), afin de les comparer. Ils présentent des résultats proches avec des différences de l'ordre de 5 à 10% selon les données de la littérature.
Dans les prochains chapitres, seule la modélisation de type entropique sera étudiée.

# 5 Mécanique musculaire à V cste : équations générales

**Fig. 12 : Dynamique d'une myofibrille se raccourcissant à V cste**

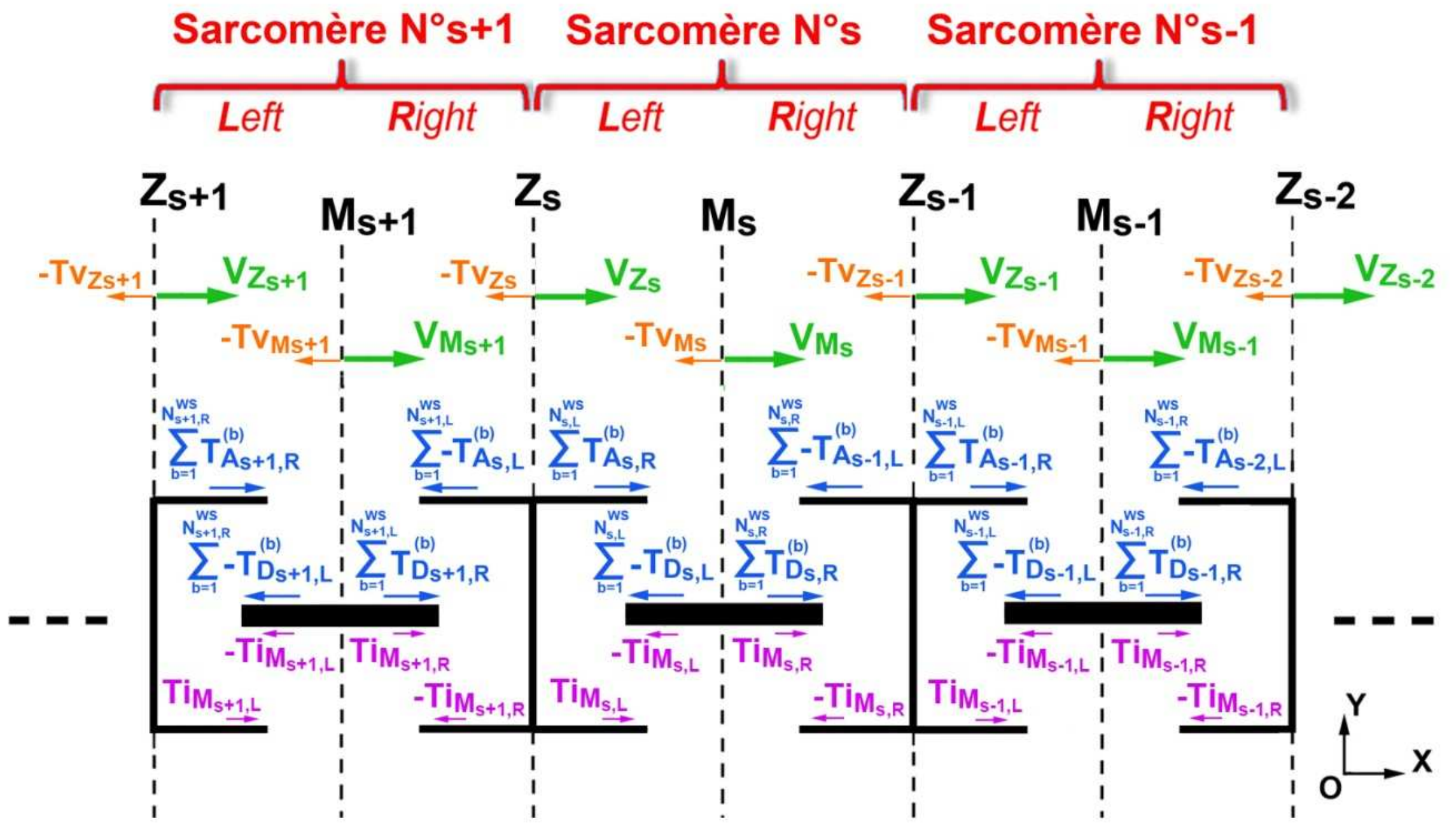

## 5.1 Bilan des forces s'exerçant à l'intérieur d'un sarcomère

Le mouvement de raccourcissement se produisant selon l'axe longitudinal OX, seules les forces tangentielles sont considérées[1].

Au chapitre précédent, les actions des tetM ont été exclusivement étudiées lorsqu'elles sont cours de WS; aussi les actions des autres tetM doivent être inventoriées selon leur séquence spatio-temporelle par rapport au filA. Ces forces seront comptabilisées comme positives si elles s'exercent dans le sens du mouvement de raccourcissement du demi-sarcomère, comme négatives dans le cas contraire, ou comme nulles. Sans présumer de l'importance de leur valeur réelle, c'est-à-dire sans considérer si elles devront être intégrées dans les calculs ou être considérées comme donnée négligeable, on recense :

- force nulle pour les S1 non liés et non à proximité d'une molA
- force positive d'attraction d'origine électrique des molA pour les S1 non liés et à proximité d'une molA
- force négative pour les S1 en cours de liaison faible, force apparentée à une force de viscosité due au mouvement brownien, force pour laquelle la vitesse de déplacement à considérer est approximativement la vitesse relative entre filA et filM du même hs, *i.e.* $\mathbf{u_{Ms,R_L}}$
- force négative pour les S1 rompant la liaison faible sans déclencher de WS

---

[1] *Les forces normales et radiales par rapport à l'axe longitudinal de la fibre musculaire s'annulent statistiquementl.*

- force négative pour les S1 à l'instant de la mise en place de la liaison forte juste avant de démarrer leur WS
- force négative pour les S1 rompant la liaison forte une fois le WS terminé

Ce sont toutes des forces d'interaction égales en module et opposées en vecteur.

Il faut aussi tenir compte de la force négative des forces de viscosité s'exerçant sur les éléments massifs du sarcomère $\left(\mathbf{M_S + 2 \cdot filM}\right)$, $\left(\mathbf{Z_{s,L} + filA}\right)$ et $\left(\mathbf{Z_{s-1,R} + filA}\right)$. Dans ce cas, la vitesse de déplacement à considérer est la vitesse absolue par rapport au référentiel du laboratoire de chaque élément constitutif de chaque sarcomère de la fmI.

## 5.2 Equations générales

Considérons le sarcomère n° s d'une myofibrille de la fmI (Fig. 2, 3E et 12).

Les actions s'appliquant sur l'ensemble rigide $\left(\mathbf{M_S + 2 \cdot filM}\right)$ sont:

**dans le hsR du sarcomère n° s**

- les actions des $\mathbf{N_{s,R}^{WS}}$ S1 en cours de WS.
- les forces d'interactions des autres S1 réunies globalement et algébriquement dans le terme générique : $\mathbf{Ti_{Ms,R}}$

**dans le hsL du sarcomère n° s**

- les actions des $\mathbf{N_{s,L}^{WS}}$ S1 en cours de WS.
- les forces d'interactions des autres S1 réunies globalement et algébriquement dans le terme générique : $\mathbf{-Ti_{Ms,L}}$

**Et les forces de viscosité** réunies globalement sous le terme générique : $\mathbf{-Tv_{Ms}}$

Appliquons le théorème de la résultante cinétique dans le repère du laboratoire OXYZ à l'ensemble rigide $\left(\mathbf{M_S + 2 \cdot filM}\right)$ se déplaçant à vitesse constante; la projection sur OX de l'ensemble des forces générées sur $\left(\mathbf{M_S + 2 \cdot filM}\right)$ donne par sommation (Fig. 12) :

$$\mathbf{0} = \left( \sum_{b=1}^{N_{s,R}^{WS}} \mathbf{T}_{\mathbf{D_{s,R}^{(b)}}} + \sum_{b=1}^{N_{s,L}^{WS}} -\mathbf{T}_{\mathbf{D_{s,L}^{(b)}}} \right) + \left( \mathbf{Ti_{M_{s,R}}} - \mathbf{Ti_{M_{s,L}}} \right) - \mathbf{Tv_{M_s}} \qquad (5.1)$$

où b est le numéro indiciel d'un S1 accomplissant un WS dont l'action provoque en $\mathbf{D_{s,R}^{(b)}}$ ($\mathbf{D_{s,L}^{(b)}}$) une force instantanée selon OX égale à $\mathbf{T}_{\mathbf{D_{s,R}^{(b)}}}$ ($-\mathbf{T}_{\mathbf{D_{s,L}^{(b)}}}$) si S1 se trouve dans hsL n° s (hsR n° s+1), respectivement.

Les actions s'appliquant sur l'ensemble rigide $\left(\mathbf{Z_S} + \mathbf{2} \cdot \mathbf{filA}\right)$ sont:

**dans le hsL du sarcomère n° s**

- les actions des $\mathbf{N_{s,L}^{WS}}$ S1 en cours de WS.
- les forces d'interactions des autres S1 réunies globalement et algébriquement dans le terme générique : $\mathbf{+Ti_{Ms,L}}$ car égales en module et opposés vectoriellement aux actions s'appliquant sur $\left(\mathbf{M_{s,L} + filM}\right)$

**dans le hsR du sarcomère n° s+1**

- les actions des $\mathbf{N_{s+1,R}^{WS}}$ S1 en cours de WS.
- les forces d'interactions des autres S1 réunies globalement et algébriquement dans le terme générique : $\mathbf{-Ti_{Ms+1,R}}$ car égales en module et opposés vectoriellement aux actions s'appliquant sur $\left(\mathbf{M_{s+1,R} + filM}\right)$

**Et les forces de viscosité** réunies globalement sous le terme générique : $\mathbf{-Tv_{Zs}}$

Appliquons le théorème de la résultante cinétique dans le repère du laboratoire OXYZ à l'ensemble rigide $\left(\mathbf{Z_S} + \mathbf{2} \cdot \mathbf{filA}\right)$ se déplaçant à vitesse constante; la projection sur OX de l'ensemble des forces générées sur $\left(\mathbf{Z_S} + \mathbf{2} \cdot \mathbf{filA}\right)$ donne par sommation (Fig. 12) :

$$\mathbf{0} = \left( \sum_{b=1}^{N_{s,L}^{WS}} \mathbf{T}_{A_{s,R}^{(b)}} + \sum_{b=1}^{N_{s+1,R}^{WS}} -\mathbf{T}_{A_{s,L}^{(b)}} \right) + \left( \mathbf{Ti}_{M_{s,L}} - \mathbf{Ti}_{M_{s+1,R}} \right) - \mathbf{Tv}_{Z_s} \qquad (5.2)$$

où b est le numéro indiciel de S1 accomplissant un WS dont l'action provoque en $\mathbf{A_{s,R}^{(b)}}$ ( $\mathbf{A_{s,L}^{(b)}}$ ) une force instantanée selon OX égale à $\mathbf{T}_{A_{s,R}^{(b)}}$ ( $-\mathbf{T}_{A_{s,L}^{(b)}}$ ) si S1 se trouve dans hsL n° s (hsR n° s+1), respectivement.

<u>**Cas particulier**</u>: pour le dernier disque $\mathbf{Z_{Ns}}$ du dernier sarcomère sur lequel s'applique directement la tension $\mathbf{T/N_{myof}}$ (Fig. 2), l'équation (5.2) donne[1]:

$$\mathbf{0} = \sum_{b=1}^{N_{Ns,L}^{WS}} \mathbf{T}_{A_{Ns,R}^{(b)}} + \frac{\mathbf{T}}{\mathbf{N_{myof}}} + \mathbf{Ti}_{M_{Ns,L}} - \mathbf{Tv}_{Z_{Ns}} \qquad (5.3)$$

Les équations (5.1), (5.2) et (5.3) sont valables à tout instant t à la condition que V soit constante.

---

[1] *Rappel : T est exprimée en valeur algébrique (T<0)*

# 6 Tension maximale isométrique

## 6.1 Définition de la tension maximale isométrique de la fmI

Les conditions d'isométrie sont au nombre de 3

***Condition 1 (rappel)*** : tous les sitA de la fmI sont disponibles

***Condition 2*** : les Ns sarcomères des $N_{myof}$ myofibrilles présentent tous une longueur identique égale à $\mathbf{2 \cdot L0^{hs}}$ (avec une variabilité de **± 10nm** pour tenir compte du raccourcissement lors de la mise ne tension) où $\mathbf{L0^{hs}}$ est la longueur d'un hs telle que la tension exercée sur la fmI soit maximale; cette valeur maximale est nommée $\mathbf{T0^{fmI}}$ . Par conséquent, tous les S1 de chaque filM sont géométriquement à une distance égale du filA voisin, soit à une distance maximale de 3 nm (le rayon d'une molA). Classiquement $\mathbf{L0^{hs}}$ vérifie (Gordon, Huxley et al. 1966) :

$$\mathbf{1\,\mu m \leq L0^{hs} \leq 1.125\,\mu m} \qquad (6.1)$$

Soit un intervalle de 125 nm qui autorise sans difficulté une variabilité de ± 10nm.

Dans le modèle, la valeur retenue est :

$$\mathbf{L0^{hs} \approx 1.1\,\mu m} \qquad (6.2)$$

***Condition 3*** : $\mathbf{T0^{fmi}}$ est appliquée à la fmI telle que $\mathbf{V = 0}$

## 6.2 Mise en place de la tension maximale isométrique de la fmI

Le travail de M. Reconditi et de ses coauteurs (Reconditi, Brunello et al. 2011) (voir aussi (Brunello, Bianco et al. 2006)) indique que $\mathbf{T0^{fmI}}$ s'établit progressivement sur une durée d'une centaine de millisecondes, alors que la longueur des hs situés au centre de la fmI se raccourcissent d'une longueur d'environ 20 nm, soit d'après (2.23a) et (2.23b) , 2 fois $\mathbf{\Delta X_{WS}^{Max}}$ . Ceci signifie qu'un certain nombre de S1 ont effectué leurs WS complets au cours de ce déplacement avant que tous les hs ne s'immobilisent à une longueur fixée.

Rappel de l' **Hypothèse 11** : aux basses températures proches de 0°C, le nombre des S1 qui ont contribué à ce raccourcissement de 20 nm et qui se trouve durant la phase isométrique sur l'intervalle angulaire $\mathbf{[\theta 2; \theta_{stopWS}]}$ est négligeable. Ainsi les S1 qui concourent à $\mathbf{T0}$ ont initié leur WS et se répartissent uniformément sur l'intervalle $\boldsymbol{\delta_\theta}$ dans chaque hsL conformément à l'Hyp. 2.

## 6.3 Calcul de la tension maximale isométrique de la fmI selon le modèle proposé

A $\mathbf{V = 0}$ , donc à $\mathbf{V}$ cste, et avec les Hyp. 1 à 15, il est possible d'utiliser les éq. (5.1) à (5.3). A l'aide des éq. (1.2) à (1.6), on vérifie que tous les (Z+2·filA) et (M+2·filM) de la fmI ont une vitesse linéaire nulle. On note que dans ces conditions, aucune des interactions mentionnées n'est impliquée et à $\mathbf{V = 0}$ , la viscosité n'intervient pas.

La durée de la mise en isométrie étant très supérieure à 1 ms, les conditions requises pour l'Hyp. 2 sont vérifiées et il est possbile d'utiliser les équations du chap. 4. Les éq. (4.21a) et (4.21b) s'appliquent à chaque disque Z de chaque myofibrille de la fmI et on obtient à partir de (5.3) :

$$\mathbf{T0}^{\mathbf{fmI}} = -\mathbf{N}_{\mathbf{myof}} \cdot \mathbf{N}_{\mathbf{Ns,L}}^{\mathbf{WS}} \cdot \mathcal{M}\mathbf{0} \cdot \frac{\Delta\theta_{\mathbf{WS}}}{\Delta \mathbf{X}_{\mathbf{WS}}^{\mathbf{Max}}} \tag{6.3}$$

En introduisant le résultat précédent dans (5.1) et (5.2), et par itération, d'une part avec (3.1) et (3.3), et, d'autre part avec (4.21a) et (4.21b), on parvient à :

$$\mathbf{N}_{\mathbf{Ns,L}}^{\mathbf{WS}} = \mathbf{N}_{\mathbf{Ns,R}}^{\mathbf{WS}} = \mathbf{N}_{\mathbf{Ns-1,L}}^{\mathbf{WS}} = \mathbf{N}_{\mathbf{Ns-1,R}}^{\mathbf{WS}} = \ldots = \mathbf{N}_{\mathbf{s,L}}^{\mathbf{WS}} = \mathbf{N}_{\mathbf{s,R}}^{\mathbf{WS}} = \ldots = \mathbf{N}_{\mathbf{1,L}}^{\mathbf{WS}} = \mathbf{N}_{\mathbf{1,R}}^{\mathbf{WS}} = \mathbf{\Lambda 0} \tag{6.4}$$

En condition isométrique maximale, la répartition des WS initiés est homogène dans chaque hs et le nombre de WS par hs est égale à la constante $\mathbf{\Lambda 0}$ figurant à la fin de (6.4).

La condition 2 donnée au paragraphe 6.1 implique que $\mathbf{\Lambda 0}$ est la valeur maximale du nombre de S1 initiant un WS par hs de la fmI.

Les éq. (6.3) et (6.4) mènent à la valeur en module de la tension isométrique maximale :

$$\left|\mathbf{T0}^{\mathbf{fmi}}\right| = \mathbf{N}_{\mathbf{myof}} \cdot \mathbf{\Lambda 0} \cdot \mathcal{M}\mathbf{0} \cdot \frac{\Delta\theta_{\mathbf{WS}}}{\Delta \mathbf{X}_{\mathbf{WS}}^{\mathbf{Max}}} \tag{6.5}$$

où la valeur de $\mathcal{M}\mathbf{0}$ est donnée par (4.4) ou (4.13)

## <u>6.4 Discussion</u>

### *6.4.2 Faits validant le modèle*

**Répartition uniforme**

**L**a dispersion des positions angulaires des S1b a été observée et décrite comme uniforme en conditions isométriques (Reconditi, Linari et al. 2004; Huxley, Reconditi et al. 2006a ; Reconditi, Brunello et al. 2011).

**Valeurs de la dispersion angulaire**

L'égalité (3.9c) mène à $\delta_\theta/2 = 20°30'$ ; cette valeur proposée par le modèle est proche des données de la littérature : « autour de 20° » dans (Hopkins, Sabido-David et al. 2002), « au moins 17° » dans (Reconditi, Linari et al. 2004), « entre 20 et 25° » dans (Huxley, Reconditi et al. 2006a).

**Valeur du déplacement moyen des S1 ayant initié un WS en conditions isométriques**

M. Reconditi a calculé un déplacement relatif moyen de $+3.27\ \mathbf{nm}$ par rapport à la position initiale de la jonction entre S1 et S2 (Reconditi, Brunello et al. 2011). Dans le modèle, la projection sur l'axe OX de la jonction entre S1 et S2 (point C ; Fig 3E) correspond à $\mathbf{X1}$ (Fig. 8 ; chap. 4), et le déplacement axial moyen est égal à $|\mathbf{X1}| = +3.3\ \mathbf{nm}$ selon le calcul effectué dans (4.19a).

### *6.4.3 Faits observés prévus par le modèle*

**Plateaux temporels isométrique et tétanique**

L'éq. (6.5) indique que la valeur de $\mathbf{T0}$ est une constante pour des conditions expérimentales fixées; elle est donc constante au cours du temps, d'où l'existence d'un plateau tétanique en condition isométrique (Gordon, Huxley et al. 1966).

**Répétitions de la mesure de $\mathbf{T0}$**

Si l'expérience est itérée à plusieurs reprises avec la même fmI pour la même valeur initiale $\mathbf{L_0^{hs}}$, le raisonnement suivi mène à la même distribution uniforme, ainsi qu'à un nombre $\mathbf{\Lambda 0}$ de WS quasi identique dans chaque hs; de ce fait on retrouve une valeur quasi identique de $\mathbf{T0}$. On doit donc observer des valeurs étalées sur un écart-type de distribution gaussienne des différentes valeurs mesurées de $\mathbf{T0}$ dans des conditions similaires pour des expériences répétées. Les travaux de K.A. Edman ont mis en exergue un tel intervalle (voir Fig. 7 de (Edman 1988)); on le note aussi sur la Fig. 6 de (Piazzesi, Reconditi et al. 2003).

**Relation entre tension isométrique et nombre de SB**

L'égalité (6.3) indique que $\mathbf{T0}$ est proportionnelle au nombre de tetM liés aux filA par hs, résultat conforme aux observations (Brunello, Bianco et al. 2006 ; Caremani, Dantzig et al. 2008).

Ainsi la tension isométrique doit augmenter avec la taille des fibres musculaires.

Avec des fibres prélevées sur le même muscle de la même espèce animale, ce phénomène est bien observé (voir Fig. 3 de (Edman, Mulieri et al. 1976)).

Selon la classification typologique des fibres musculaires squelettiques, on doit vérifier :

$$\mathbf{T0_I < T0_{IIa} < T0_{IIb}}$$

où $\mathbf{T0_I}$, $\mathbf{T0_{IIa}}$ et $\mathbf{T0_{IIb}}$ sont les tensions isométriques des fmI de type I, IIa et IIb.

Ces inégalités sont confirmées par divers articles (Rome, Sosnicki et al. 1990; Larsson and Moss 1993; Bottinelli, Canepari et al. 1996).

**Influence de $\mathbf{L0^{hs}}$, la longueur initiale du sarcomère**

Considérons les 2 longueurs initiales $\mathbf{L0^{hs} = 2.2\ \mu m}$ et $\mathbf{L0^{hs} = 3.2\ \mu m}$. La géométrie inter-filamentaire implique que le nombre de sitA est inférieur pour $\mathbf{L0^{hs} = 3.2\ \mu m}$, ce qui implique avec des conditions expérimentales identiques (mêmes type de fibres, température identique, même durée de phase 1) :

$$\mathbf{\Lambda 0_{L0^{hs}=3.2\,\mu m} < \Lambda 0_{L0^{hs}=2.2\,\mu m}}$$

où $\mathbf{\Lambda 0_{L0^{hs}=2.2\,\mu m}}$ est le nombre de WS initiés à $\mathbf{L0^{hs} = 2.2\ \mu m}$

et $\mathbf{\Lambda 0_{L0^{hs}=3.2\,\mu m}}$ est le nombre de WS initiés à $\mathbf{L0^{hs} = 3.2\ \mu m}$

Ce qui d'après (6.5) entraine :

$$\mathbf{T0}_{\mathbf{L0}^{hs}=3.2\,\mu m} < \mathbf{T0}_{\mathbf{L0}^{hs}=2.2\,\mu m} \tag{6.6}$$

où $\mathbf{T0}_{\mathbf{L0}^{hs}=2.2\,\mu m}\left(\mathbf{T0}_{\mathbf{L0}^{hs}=3.2\,\mu m}\right)$ est la tension isométrique pour les 2 longueurs initiales respectives

Résultat en accord avec les observations expérimentales (voir Fig. 3 de (Gordon, Huxley et al. 1966) et Fig. 5 de (Ford, Huxley et al. 1981)).

**Température expérimentale**

Les relations de proportionnalités, $\mathbf{T0} \propto \mathcal{M}\mathbf{0}$ selon (6.5), $\mathcal{M}\mathbf{0} \propto \mathcal{M}\mathbf{1}$ selon (4.4) et $\mathcal{M}\mathbf{1} \propto \mathbf{T°}$ d'après (3.31), impliquent que $\mathbf{T0}$ augmente aussi linéairement avec la température; ce qui est bien observé (Buller, Kean et al. 1984; Ranatunga 1984; Edman 1988; Bottinelli, Canepari et al. 1996; Coupland and Ranatunga 2003; Piazzesi, Reconditi et al. 2003; Linari, Brunello et al. 2005).

Avec des conditions de température plus élevées lors de la mise en tension isométrique, il est envisageable que les sitA soient disponibles plus rapidement et que le nombre de hs et le nombre de WS contribuant aux raccourcissements de l'ordre 20 nm augmente. On peut donc tester les cas extrême où pour les températures élevées (supérieures à 15°C), la distribution des S1 se répartirait non plus sur $\boldsymbol{\delta_\theta}$ mais sur $\Delta\theta_{WS}^{Max}$ .

D'après (3.31), en passant au continu avec la définition d'une moyenne, on aurait :

$$\mathcal{M}\mathbf{0}_{T°>15°C} = \frac{\mathcal{M}\mathbf{1}_{T°>15°C}\cdot(\theta 1 + a_\theta)}{\Delta\theta_{WS}^{Max^2}} \cdot \int_{\theta 1}^{\theta_{stopWS}} \left(\frac{\theta_{stopWS} + a_\theta}{\theta + a_\theta} - 1\right)\cdot d\theta$$

Soit après intégration :

$$\frac{\mathcal{M}\mathbf{0}_{T°>15°C}}{\mathcal{M}\mathbf{1}_{T°>15°C}} = \frac{(\theta 1 + a_\theta)}{\Delta\theta_{WS}^{Max}} \cdot \left[\frac{(\theta_{stopWS} + a_\theta)}{\Delta\theta_{WS}^{Max}} \cdot \log\left(\frac{\theta_{stopWS} + a_\theta}{\theta 1 + a_\theta}\right) - 1\right] \tag{6.7}$$

***Application numérique***

Avec (3.10), (3.11) et $\mathbf{a_\theta = 316.9}$ , le calcul de (6.4) donne :

$$\frac{\mathcal{M}\mathbf{0}_{T°>15°C}}{\mathcal{M}\mathbf{1}_{T°>15°C}} = \frac{(-25 + 316.9)}{60} \cdot \left[\frac{(35 + 316.9)}{60} \cdot \log\left(\frac{65 + 316.9}{-25 + 316.9}\right) - 1\right] \approx 0.47 \tag{6.8}$$

On retrouve le calcul effectué au chapitre 4 pour (4.38)

Sur l'hyperbole (trait rouge ; Fig. 11), cette valeur correspond à la valeur angulaire de (4.40) :

$$\theta 0_{T°>15°C} = \theta 0^* \approx +4°$$

On devrait donc observer un décalage angulaire égal avec (4.17b) à:

$$\Delta\theta 0 = (\theta 0_{T°>15°C} - \theta 0) = (\theta 0^* - \theta 0) \approx +9°$$

Et consécutivement avec (4.41), un déplacement linéaire du hs vers la ligne M de :

$$\Delta X0 = X0^* \approx +1.5\text{nm}$$

Ces résultats sont comparables aux données de la littérature (Linari, Brunello et al. 2005) qui ne concernent, rappelons le, que les hs situés au milieu de la fmI. Pour les autres hs, le modèle reste satisfaisant.

# 7 Phase 1 d'un échelon de longueur (« *step length* »)

**Les résultats chiffrés concernant ce chapitre et les chapitres suivants sont extraits des expérimentations de G. Piazzesi, V. Lombardi et L. Lucii (Piazzesi and Lombardi 1995; Piazzesi, Lucii et al. 2002); ces travaux offrent l'avantage de produire des données recueillies sur des fibres musculaires appartenant à la même espèce de grenouille (Rana esculenta) et au même muscle (tibialis anterior) pour des raccourcissements réalisés avec des échelons de longueur et de force. Les différences observées avec les résultats d'autres chercheurs, notamment ceux de L.E. Ford, A.F. Huxley et R.M. Simmons, seront commentées dans un paragraphe intitulé « Discussion » à la fin des chapitres n° 7 à 15.**

## 7.1 Description de la phase 1 d'un échelon de longueur

Apres avoir été tétanisée isométriquement, la fmI est raccourcie continument et linéairement pendant une durée de temps inférieure à 0.2 ms (Huxley and Simmons 1971; Ford, Huxley et al. 1977). Cette phase transitoire est la phase 1 d'une perturbation par un échelon de longueur[1] $\Delta L_{p1}^{(k)} > 0$, k étant le numéro indiciel de l'échelon de la perturbation en longueur. La Fig. 13A offre 3 exemples d'échelons de longueur, (k-1), (k) et (k+1), respectivement; l'échelon (0) correspond au cas isométrique étudié au chapitre 6.

La durée de la phase 1 est appelée $\Delta t_{p1}$ ; dans l'exemple de la Fig. 13A, $\Delta t_{p1}$ = 150 µs

A l'instant t=0 de la perturbation, la fmI est en tension isométrique maximale dans les conditions décrites au paragraphe 6.1 et avec la valeur de l'éq. (6.5) imputée au paragraphe 6.3. Durant $\Delta t_{p1}$, aucune nouvelle tetM n'initie de WS, si bien que la répartition des $\Lambda 0$ S1 en cours de WS définie par (6.4) conserve la même homogénéité pour chacun des hs de la fmI.

Durant $\Delta t_{p1}$, le raccourcissement s'effectue à vitesse constante $V_{p1}^{(k)}$ , telle que (Fig. 13A) :

$$V_{p1}^{(k)} = \frac{\Delta L_{p1}^{(k)}}{\Delta t_{p1}} \qquad (7.1)$$

Ainsi, quelque soit la valeur $\Delta L_{p1}^{(k)}$ de l'échelon (k), toutes les équations précédentes et notamment celles du paragraphe 6.4 sont valables durant la phase 1.

A la fin de la phase 1 la longueur de la fmI (et de chacune des ses myofibrilles) devient constante et égale à $\left(L0 - \Delta L_{p1}^{(k)}\right)$ pout $t > \Delta t_{p1}$ , *i.e.* durant les phases suivantes.

A $t = \Delta t_{p1}$, la fin de la phase 1 se caractérise par 2 temps instantanés successifs (Fig. 13A et 13B) :

$$t_{stop_p1}^{1} = \Delta t_{p1} \qquad \text{où} \quad V = V_{p1}^{(k)}$$

$$t_{stop_p1}^{2} = \Delta t_{p1} \qquad \text{où} \quad V = 0$$

[1] *Voir note 1 de la page 47 (chap. 4).*

**Fig. 13 : Perturbation d'une fmi par 3 échelons de longueur respective, $\Delta L_{p1}^{(k-1)}, \Delta L_{p1}^{(k)}$, et $\Delta L_{p1}^{(k+1)}$**

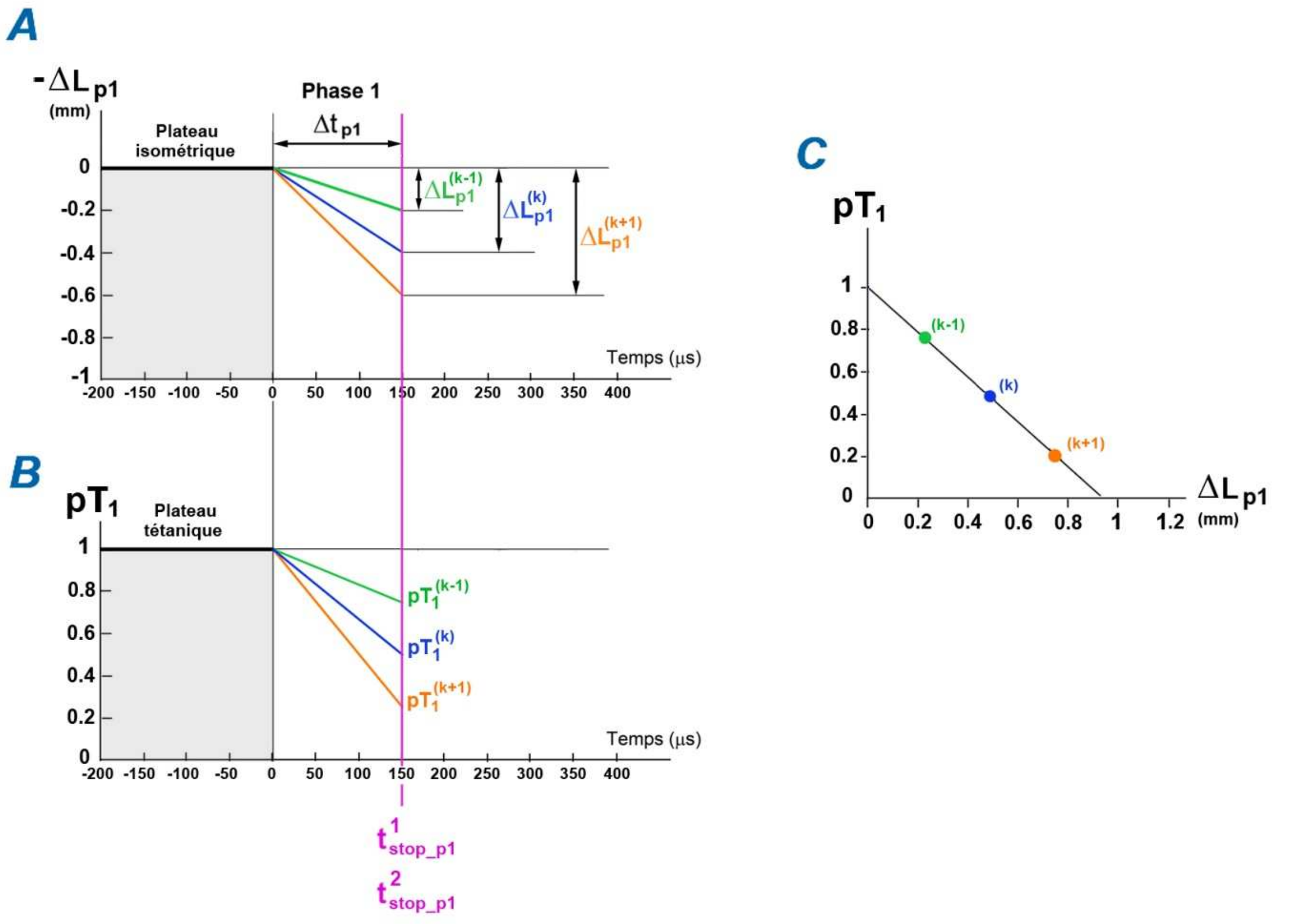

On nomme classiquement $\mathbf{T1}$ la tension minimale atteinte par la fmI à $\mathbf{t^{1}_{stop_p1}}$, et $\mathbf{pT1}$ la tension relative minimale égale au rapport de $\mathbf{T1}$ sur $\mathbf{T0^{fmI}}$ (Fig. 13B, 13C et 14).

Le raccourcissement $\mathbf{\Delta L_{p1}^{(k)}}$ de la fmI exprimée en mm (Fig. 14A et 14C) est ramené au raccourcissement $\mathbf{\Delta X_{p1}^{(k)}}$ de chaque[1] hs de la fmI (Huxley 1974).

La relation entre $\mathbf{T1}$ (ou $\mathbf{pT1}$) et $\mathbf{\Delta X_{p1}^{(k)}}$ est approximativement linéaire jusqu'à 3 nm environ dans chaque hs puis s'incurve légèrement (Huxley 1974; Ford, Huxley et al. 1977; Ford, Huxley et al. 1981; Piazzesi et Lombardi 1995; Piazzesi, Lucii et al. 2002 ).

La relation linéaire entre $\mathbf{pT1}$ et $\mathbf{\Delta X_{p1}^{(k)}}$ mène à la valeur $\mathbf{\Delta X_{p1}^{(k)} \approx 4\,nm}$ pour $\mathbf{pT1 = 0}$ (Piazzesi et Lombardi 1995; Piazzesi, Lucii et al. 2002). Cela implique une pente négative, égale en module à :

$$\mathbf{p \approx \frac{1}{4\,nm} = 0.25\,nm^{-1}} \qquad (7.2)$$

[1] *Cette proposition suppose que chaque hs de la fmI se raccourcit de la même valeur.*

## 7.2 Calcul de la tension de la fmI à la fin de la phase 1 pour V = $V_{p1}^{(k)}$

A $t^{1}_{stop_p1}$, la tension minimale atteinte par la fmI et correspondant à l'échelon de longueur $\Delta L^{(k)}_{p1}$ est notée $T^{1\ (k)}_{stop_p1}$ , *i.e.* $T1 = T^{1\ (k)}_{stop_p1}$ .

D'après l'Hyp. 12 et les valeurs fournies par (1.24) et (1.25), les tetM en fin de WS n'ont pas le temps de se détacher durant $\Delta t_{p1}$. Au paragraphe 4.3.4, il a été noté que ce cas particulier de la phase 1 correspondait au mode exagéré ; les éq. (4.25) et (4.32) sont alors modélisés par 2 relations linéaires d'éq. respectives (4.35a) et (4.36a), *i.e,* chaque hs se comporte comme un ressort linéaire de raideur commune $\chi^{hs}_{z1}$ et $\chi^{hs}_{z2}$ dans les zones 1 et 2, puisque chaque hs comporte le même nombre ( $\Lambda 0$ ) et la même répartition uniforme de S1 en cours de WS (voir chap. 6).

A $V = V^{(k)}_{p1}$, *i.e.* à $t^{1}_{stop_p1}$, il faut tenir compte des forces de viscosité s'exerçant sur les éléments composant chaque hs. On rappelle que les forces de viscosité sont proportionnelles à la vitesse absolue, et comme les vitesses absolues ou relatives sont constantes (voir chap. 2) et comme la durée de la phase 1 est constante quelque soit l'échelon (k), il devient possible de raisonner sur les distances. On démontre ainsi (Annexe D), que les valeurs relatives de la tension, nommées $pT1_{z1}$ dans la zone 1 et $pT1_{z2}$ dans la zone 2, exercées par la fmI[1] à $t^{1}_{stop_p1}$ sont égales à[2] :

**Si** $0 \le \Delta X^{(k)}_{p1} \le \Delta X^{min}_{WS}$ (Zone 1 ; trait bleu foncé ; Fig.14)

$$pT1_{z1} = \frac{T^{1\ (k)}_{stop_p1}}{T0^{fmI}} \approx \left[1-\left(\chi^{hs}_{z1} + 2\cdot Ns\cdot \nu^{hs}\right)\cdot \Delta X^{(k)}_{p1}\right]\cdot \mathbf{1}_{\left[0;\Delta X^{min}_{WS}\right]}\left(\Delta X^{(k)}_{p1}\right) \quad (7.3a)$$

$$\approx \left[1-\left(\chi^{fmI}_{z1} + \nu^{hs}\right)\cdot \Delta L^{(k)}_{p1}\right]\cdot \mathbf{1}_{\left[0;\Delta X^{min}_{WS}\right]}\left(\Delta X^{(k)}_{p1}\right) \quad (7.3b)$$

avec $$\chi^{fmI}_{z1} = \frac{\chi^{hs}_{z1}}{2\cdot Ns} \quad (7.3c)$$

et $\nu^{hs}$ est le coefficient de proportionnalité des forces de viscosité s'exerçant sur les ensembles $(Z_S + filA)$ et $(M_S + filM)$ défini à l'Annexe D par l'éq. (D.4c) à l'appui de l'Hyp. 10.

et $\Delta L^{(k)}_{p1} = 2\cdot Ns\cdot \Delta X^{(k)}_{p1}$ , d'après (D.8)

---

[1] *La fmI est composé de $N_{myof}$ myofibrilles identiques disposées en parallèle (chap. 2) ; comme les raideurs des ressorts en parallèle s'ajoutent et que les calculs de l'annexe D ont porté sur les valeurs relatives, ils sont aussi bien valides pour 1 myofibrille, que pour $N_{myof}$ myofibrilles identiques disposées en parallèle et donc pour la fmI.*

[2] *Rappel : voir Annexe A1 pour la définition d'une fonction indicatrice.*

**Fig. 14 : La tension relative en fonction de l'échelon (k) à la fin de la phase 1 suivant le raccourcissement de la fml par une perturbation en longueur**

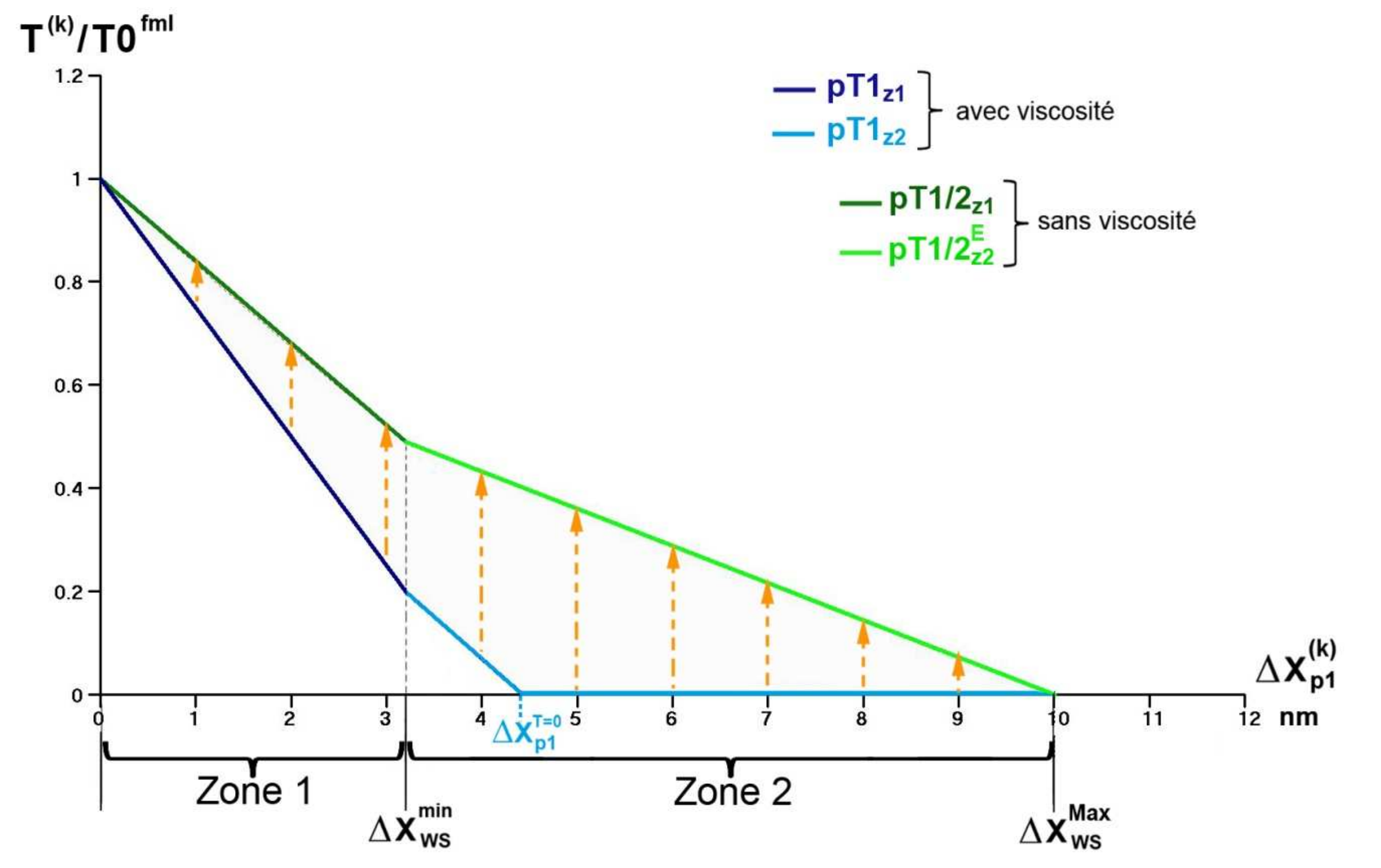

**Si** $\Delta X_{WS}^{min} \le \Delta X_{p1}^{(k)} \le \Delta X_{WS}^{Max}$ (Zone 2 ; trait bleu clair[1]; Fig.14)

$$pT1_{z2} = \frac{T_{stop_p1}^{1\ (k)}}{T0^{fmI}} \approx \left(\chi_{z2}^{hs} + 2 \cdot Ns \cdot \nu^{hs}\right) \cdot \left[\Delta X_{p1}^{T=0} - \Delta X_{p1}^{(k)}\right] \cdot \mathbb{1}_{\left[\Delta X_{WS}^{min};\Delta X_{p1}^{T=0}\right]}\left(\Delta X_{p1}^{(k)}\right) \quad (7.4a)$$

$$\approx \left(\chi_{z2}^{fmI} + \nu^{hs}\right) \cdot \left[\Delta L_{p1}^{T=0} - \Delta L_{p1}^{(k)}\right] \cdot \mathbb{1}_{\left[\Delta X_{WS}^{min};\Delta X_{p1}^{T=0}\right]}\left(\Delta X_{p1}^{(k)}\right) \quad (7.4b)$$

avec $$\chi_{z2}^{fmI} = \frac{\chi_{z2}^{hs}}{2 \cdot Ns} \quad (7.4c)$$

et $$\Delta X_{p1}^{T=0} = \Delta X_{WS}^{min} + \frac{\left[1 - \left(\chi_{z1}^{hs} + 2 \cdot Ns \cdot \nu^{hs}\right) \cdot \Delta X_{WS}^{min}\right]}{\left(\chi_{z2}^{hs} + 2 \cdot Ns \cdot \nu^{hs}\right)} \quad (7.4d)$$

et $$\Delta L_{p1}^{T=0} = 2 \cdot Ns \cdot \Delta X_{p1}^{T=0} \quad (7.4e)$$

La valeur d'abscisse $\Delta X_{p1}^{T=0}$ est indiquée sur la Fig. 14.

[1] *Dès que* $pT1_{z2}$ *s'annule, elle est considérée comme nulle jusqu'à la borne maximale de l'intervalle, même si des valeurs négatives expérimentales sont parfois relevées.*

Ainsi à $t^{1}_{stop_p1}$ où $V = V^{(k)}_{p1}$ , la fmI est assimilée dans chacune des 2 zones, à un ressort mécanique linéaire

La raideur de ce ressort est la somme de 2 termes :

- un terme provenant du comportement élastique des $\Lambda 0$ tetM en cours de WS dans chaque hs de la fmI
- un terme issu du comportement linéaire des forces de viscosité.

## 7.3 Calcul de la tension de la fmI à la fin de la phase 1 pour V= 0

A l'aide des équations (1.2) à (1.6), on vérifie que tous les solides (Z+2·filA) et (M+2·filM) de la fmI ont une vitesse linéaire nulle à $V = 0$ , *i.e.* à $t^{2}_{stop_p1}$ . On note que dans ces conditions, aucune des interactions mentionnées dans les éq. (5.1) à (5.3) n'est impliquée et à $V = 0$, la viscosité n'intervient pas.

Avec (6.2), par itération de (5.1) et (5.2) d'une part et de (3.1) et (3.3) d'autre part, on trouve (Fig. 4B et 12) :

$$\sum_{b=1}^{\Lambda 0} T_{A^{(b)}_{Ns,R}} = ... = \sum_{b=1}^{\Lambda 0} T_{A^{(b)}_{s,L}} = \sum_{b=1}^{\Lambda 0} T_{A^{(b)}_{s,R}} = \sum_{b=1}^{\Lambda 0} T_{D^{(b)}_{s,L}} = \sum_{b=1}^{\Lambda 0} T_{D^{(b)}_{s,R}} = \sum_{b=1}^{\Lambda 0} T_{A^{(b)}_{s-1,R}} = \sum_{b=1}^{\Lambda 0} T_{A^{(b)}_{s-1,R}} = ... = \sum_{b=1}^{\Lambda 0} T_{A^{(b)}_{0,L}} \tag{7.5}$$

Au chapitre 4, il a été constaté que chaque membre de cette égalité est décrit pour la zone 1 ( $0 \le \Delta X \le \Delta X^{min}_{WS}$ ) avec (4.25) et pour la zone 2 ( $\Delta X^{min}_{WS} \le \Delta X \le \Delta X^{Max}_{WS}$ ) avec (4.32) dont le choix a été justifié au paragraphe (4.3.4). Il s'agit bien des 2 mêmes fonctions pour chaque hs de la fmI puisque que le nombre de S1 en cours de WS est $\Lambda 0$ .

Comme ces 2 fonctions sont continument décroissantes sur les 2 intervalles successifs (traits verts foncés et clairs; Fig. 14), il y a bijection entre tension et raccourcissement, et consécutivement tous les hs de la fmi se raccourcissent de la même longueur $\Delta X^{(k)}_{p1}$ tel que :

$$\Delta X^{Ns,L}_{p1,(k)} = \Delta X^{Ns,R}_{p1,(k)} = \Delta X^{Ns-1,L}_{p1,(k)} = ... = \Delta X^{s,L}_{p1,(k)} = \Delta X^{s,R}_{p1,(k)} = ... = \Delta X^{1,L}_{p1,(k)} = \Delta X^{1,R}_{p1,(k)} = \Delta X^{(k)}_{p1} \tag{7.6}$$

avec $\Delta X^{s,R}_{p1,(k)}$ ( $\Delta X^{s,L}_{p1,(k)}$ ) la valeur du déplacement du hsR (hsL) du sarcomère n° s, respectivement, à $t = t^{2}_{stop_p1}$ dans hsR (hsL), lors de la perturbation d'échelon (k) et $\Delta X^{(k)}_{p1}$ la valeur commune du déplacement de toutes les S1 de la fmI soumise à une perturbation d'un échelon (k) en longueur

Il s'en déduit :

$$\Delta L^{(k)}_{p1} = 2 \cdot Ns \cdot \Delta X^{(k)}_{p1} \tag{7.7}$$

Après un échelon de longueur $\Delta L_{p1}^{(k)}$, la valeur de la tension exercée par la fmI à $t_{stop_p1}^{2}$ est notée $T_{stop_p1}^{2\ (k)}$ et ses valeurs relatives nommées $pT1/2_{z1}$ dans la zone 1 et $pT1/2_{z2}^{E}$ dans la zone 2[1]. D'après (4.35a), (4.36a), (5.3) et (7.6), celles-ci sont égales à :

Si $0 \le \Delta X_{p1}^{(k)} \le \Delta X_{WS}^{min}$ (zone1 ; trait vert foncé ; Fig.14) avec (7.3c)

$$pT1/2_{z1} = \frac{T_{stop_p1}^{2\ (k)}}{T0^{fmI}} \approx \left[1 - \chi_{z1}^{hs} \cdot \Delta X_{p1}^{(k)}\right] = \left[1 - \chi_{z1}^{fmI} \cdot \Delta L_{p1}^{(k)}\right] \quad (7.8)$$

Si $\Delta X_{WS}^{min} \le \Delta X_{p1}^{(k)} \le \Delta X_{WS}^{Max}$ (zone 2; trait vert clair; Fig.14) avec (7.4c)

$$pT1/2_{z2}^{E} = \frac{T_{stop_p1}^{2\ (k)}}{T0^{fmI}} \approx \chi_{z2}^{hs} \cdot \left[\Delta X_{WS}^{Max} - \Delta X_{p1}^{(k)}\right] = \chi_{z2}^{fmI} \cdot \left[2Ns \cdot \Delta X_{WS}^{Max} - \Delta L_{p1}^{(k)}\right] \quad (7.9)$$

Pour chaque hs de la fmI selon (7.5), on observe 2 relations similaires à celles présentées en trait rouge dans les Fig. 9 et 10 du chapitre 4.

A $t_{stop_p1}^{2}$ où $V = 0$, la fmI agit dans chacune des 2 zones, comme un ressort mécanique linéaire composé de $2Ns$ ressorts linéaires identiques disposés en série et synchronisés ; en effet, les inverses des raideurs s'ajoutent d'après les définitions respectives de $\chi_{z1}^{fmI}$ et $\chi_{z2}^{fmI}$ fournies par (7.3c) et (7.4c).

A $t_{stop_p1}^{2}$ où $V = 0$, il est notable que les forces de viscosité disparaissent. Dans ces conditions où $\nu^{hs} = 0$, les éq. (7.3a) et (7.4a) deviennent les éq. (7.8) et (7.9).

## 7.4 Application numérique à partir des observations expérimentales de la littérature

Les 2 termes $\nu^{hs}$ et $\nu^{fmI}$ ne sont pas accessibles au calcul. La donnée fournie avec (7.2) et introduite dans (7.3a) mène à :

$$\chi_{z1}^{hs} + 2 \cdot Ns \cdot \nu^{hs} \approx 0.25\ nm^{-1} \quad (7.10a)$$

Avec (4.35c), (7.10a) implique :

$$2 \cdot Ns \cdot \nu^{hs} \approx 0.09\ nm^{-1} \quad (7.10b)$$

[1] *E pour le mode exagéré ; en effet conformément aux indications du chap. 4, les S1 en fin de WS n'ont pas le temps de se détacher entre* $t_{stop_p1}^{1}$ *entre* $t_{stop_p1}^{2}$

La valeur moyenne de la longueur d'une fibre musculaire de Rana esculenta est égale à $\mathbf{L_0^{fmI} = 5\,mm}$ avec $\mathbf{L_0^{hs} = 1.1\,\mu m}$ d'après (6.2), soit :

$$\mathbf{2Ns \approx \frac{5mm}{1.1\,\mu m} \approx 4.54 \cdot 10^3}$$

$$\Rightarrow \qquad \mathbf{\nu^{hs} \approx 2 \cdot 10^{-5}\ nm^{-1}} \qquad (7.10c)$$

$$\Rightarrow \qquad \mathbf{\frac{\nu^{hs}}{\chi_{z1}^{hs}} \approx 1.25 \cdot 10^{-4}}$$

Ainsi les simplifications effectuées en Annexe D sont justifiées.

Les égalités (4.36c) et (7.10b) conduisent à la valeur de la pente de l'éq. (7.4a) :

$$\mathbf{\chi_{z2}^{hs} + 2 \cdot Ns \cdot \nu^{hs} \approx 0.16\ nm^{-1}} \qquad (7.10d)$$

Et le calcul de (7.4d) avec (3.23b), (7.10a) et (7.10d) donne :

$$\mathbf{\Delta X_{p1}^{T=0} \approx 4.45\ nm} \qquad (7.10e)$$

Les pentes des éq. linéaires (7.3a), (7.4a), (7.8) et (7.9) avec les valeurs respectives fournies par (7.10a), (7.10d), (4.39c) et (4.40c) apparaissent sur la Fig. 14 en bleu foncé, bleu ciel, vert foncé et vert clair, comme cela a déjà été mentionné.

## 7.5 Discussion

### *7.6.1 Questions soulevées par le modèle*

**Fin de la phase 1 et début de la phase 2**

Dans la présentation du début de chapitre (Fig. 13A et 13B), les 2 temps, $\mathbf{t_{stop_p1}^1}$ et $\mathbf{t_{stop_p1}^2}$ , coïncident en théorie, mais en pratique les moteurs électriques utilisés pour raccourcir les fibres musculaires sont asservis et nécessitent un délai de plusieurs µs pour s'arrêter (Ford, Huxley et al. 1977; Piazzesi, Lucii et al. 2002). Ainsi dès que la vitesse de raccourcissement diminue avant de s'annuler afin d'atteindre la valeur de l'échelon choisi, les forces de viscosité s'atténuent puis cessent leurs actions.

Consécutivement, la tension $\mathbf{T_{stop_p1}^{1\ (k)}}$ tend instantanément vers $\mathbf{T_{stop_p1}^{2\ (k)}}$ dans l'intervalle de temps séparant $\mathbf{t_{stop_p1}^1}$ et $\mathbf{t_{stop_p1}^2}$ qui n'est pas nul. Ce phénomène est illustré par les exemples de la Fig. 14 où pour chaque échelon multiple d'1 nm, les flèches en pointillé orange signalent la remontée de la tension relative.

Ainsi, dans notre modèle, la fin de la phase 1 à $\mathbf{t_{fin_p1}^2}$ où $\mathbf{V = 0}$ appartient déjà à la phase 2.

**Importance de la viscosité**

Dans nombre de travaux, les forces de viscosité sont tenues pour partie négligeable, phase 1 comprise (Ford, Huxley et al. 1977 ; Piazzesi, Francini et al. 1992 ; Linari, Dobbie et al. 1998 ; Piazzesi, Lucii et al. 2002). Tout au long de ce paragraphe et des chapitres suivants, la prise en compte de la viscosité durant la phase 1 puis au début de la phase 2 (nous négligerons les effets de la viscosité pour les phases suivantes) va s'avérer précieuse pour expliquer diverses observations expérimentales.

**Différences des valeurs de pente entre expérimentations**

Les pentes des relations linéaires entre $\mathbf{pT1}$ et $\mathbf{\Delta X_{p1}^{(k)}}$ dans les zones respectives 1 et 2 varient selon les publications.

Les données issues des travaux de G. Piazzesi et de ses coauteurs s'alignent sur $\mathbf{pT1_{z1}}$ (zone1; trait bleu foncé; Fig. 14) et $\mathbf{pT1_{z2}}$ (zone2; trait bleu clair; Fig. 14). Pour les autres chercheurs, toutes les pentes appartiennent à l'aire grisée (Fig.14) inscrite entre l'axe des abscisses et les droites définies par les éq. (7.3a), (7.4a), (7.7) et (7.8), soit respectivement, $\mathbf{pT1_{z1}}$ , $\mathbf{pT1_{z2}}$ , $\mathbf{pT1/2_{z1}}$ et $\mathbf{pT1/2_{z2}}$.

Ces différences de pentes sont motivées par diverses considérations :

- la durée $\mathbf{\Delta t_{p1}}$ varie selon les expérimentateurs de 110 µs (Linari, Dobbie et al. 1998) à 200 µs (Ford, Huxley et al. 1985) ; or la contribution des forces dues à la viscosité varie inversement avec $\mathbf{\Delta t_{p1}}$ modifiant *de facto* la valeur des pentes (Annexe D)
- l'espèce de grenouille et les muscles sont différents modifiant ainsi les caractéristiques qui ont servi à élaborer le modèle
- l'hétérogénéité des conditions expérimentales : préparation des fibres, température, pH, …
- la détermination de $\mathbf{pT1}$ est empirique ; en remarquant que lorsque la tension est mesurée instantanément durant la phase1, la pente observée dans ce cas est toujours inférieure à celle de $\mathbf{pT1_{z1}}$

En conséquence, les paramètres qui ont conduits aux données numériques s'appliquant aux éq. (7.3a), (7.4a), (7.7) et (7.8), variant d'une recherche à l'autre, il est normal d'observer de telles différences avec la précision que l'aire grise est modifiable selon les conditions des tests (notamment en fonction de $\mathbf{\Delta X_{WS}^{min}}$ et $\mathbf{\Delta X_{WS}^{Max}}$).

### *7.6.2 Faits observés prédits par le modèle*

**Changement de pentes entre zones 1 et 2**

Avec les éq. (7.3a) et (7.4a), le modèle conjecture une diminution en module de la pente reliant $\mathbf{pT_1}$ à $\mathbf{\Delta X_{p1}^{(k)}}$ dès que $\mathbf{\Delta X_{p1}^{(k)}}$ devient supérieur à $\mathbf{\Delta X_{WS}^{min}}$ , *i.e.* entre les zones 1 et 2 à partir de $\mathbf{3.2\ nm}$ (Fig. 14).

Cette prévision est observée dans toutes les expérimentations portées à notre connaissance concernant le muscle squelettique de la grenouille (Huxley 1974; Ford, Huxley et al. 1977; Ford, Huxley et al. 1981; Ford, Huxley et al. 1985; Linari and Woledge 1995; Piazzesi and Lombardi 1995 ; Linari, Lombardi et al. 1997; Piazzesi, Linari et al. 1997 ) ou celui d'autres espèces (Galler, Hilber et al. 1996).

Confirmée de plus avec l'élévation de la température expérimentale (voir paragraphe 12.3).

**Variation de la pente de $pT1_{z1}(\Delta X_{p1})$ selon la longueur initiale du hs**

Considérons 2 longueurs initiales de mise en tension isométrique maximale différentes avant raccourcissement, comme par exemple : $L0^{hs} = 2.2\ \mu m$ et $L0^{hs} = 3.2\ \mu m$

A même $V_{p1}^{(k)}$, les actions de la viscosité portant sur les mêmes éléments constitutifs de la fmI sont identiques pour les 2 types de tests. D'après (6.6), on observe $T0_{L0^{hs}=3.2\ \mu m} < T0_{L0^{hs}=2.2\ \mu m}$

Aussi, la contribution des forces de viscosité dans le calcul de la tension relative, où la tension isométrique maximale vaut $T0_{L0^{hs}=2.2\ \mu m}$ ou $T0_{L0^{hs}=3.2\ \mu m}$, est plus importante pour $L0^{hs} = 3.2\ \mu m$ La pente pour $L0^{hs} = 3.2\ \mu m$ doit donc être plus importante en module.

Ce pronostic est constaté (voir Fig. 11 dans (Ford, Huxley et al. 1981)).

Si les calculs des tensions relatives sont toutes calculée par rapport à la valeur de $T0^{fmI}$ de chaque expérience de longueur initiale de sarcomère réalisée sur la même fibre, la contribution relative des forces de viscosité sera identique mais celle des WS ayant initié un WS diminue si $L0^{hs}$ augmente[1] : la pente de $pT1_{z1}(\Delta X_{p1})$ doit maintenant diminuer en valeur absolue.

Ce qui est effectivement observé (voir Fig. 10 dans (Huxley 1974)).

**Variation de la pente de $pT1_{z1}(\Delta X_{p1})$ selon le nombre de WS ayant participé à la mise en tension isométrique maximale**

Une expérience n° 1 de raccourcissement est effectuée selon les critères définis et mène aux calculs précédents, illustrés par les 4 segments de droites de la Fig.14 où la tension relative est calculée relativement à $T0^{fmI}$ .

On réalise une expérience n° 2 dans des conditions identiques, hormis le fait que le nombre de WS ayant participé à la mise en tension isométrique maximale diminue. D'après (6.5) et les calculs appliqués dans ce chapitre, les tensions $T_{stop_p1}^{2\ (k)}$ mesurées à $t_{stop_p1}^{2}$ où V=0, vont diminuer proportionnellement à $\Lambda 0$ .

Si les calculs des tensions relatives sont toutes calculée par rapport à la valeur de $T0^{fmI}$ de la 1ère expérience, la contribution relative des forces de viscosité sera identique mais celle des WS ayant initié un WS sera moindre : la pente de $pT1_{z1}(\Delta X_{p1})$ doit diminuer en valeur absolue.

Cette conjecture est vérifiée en modulant le nombre de WS par le taux de calcium (Linari, Caremani et al. 2007), ou en recourant à un inhibiteur de déclenchement de WS (Linari, Piazzesi et al. 2009 ; Caremani, Lehman et al. 2011).

Il devrait être possible dans ces conditions de calculer la contribution exacte des forces de viscosité et les coefficients $\nu^{hs}$ et $\nu^{fmI}$ et tester la modélisation proposée.

---

[1] *Dès que* $L0^{hs} > 2.25\ \mu m$

**Variation de la pente de $pT1_{z1}(\Delta X_{p1})$ selon des variations de température expérimentale**

Les forces dues à la viscosité s'atténuent lorsque la température augmente. On doit donc selon le même raisonnement s'attendre à une diminution de leur contribution lorsque la tension relative est calculée par rapport à la tension maximale isométrique de chacun des tests en température.

Cette prédiction est caractérisée (Ford, Huxley et al. 1977 ; Piazzesi, Reconditi et al. 2003; Linari, Caremani et al. 2007).

**Variation de la pente de $pT1_{z1}(\Delta X_{p1})$ selon la durée de la phase 1**

Les forces dues à la viscosité s'atténuent lorsque la durée de la phase 1, $\Delta t_{p1}$, augmente (voir Appendix D) .

On doit donc selon le même raisonnement s'attendre à une diminution de leur contribution lorsque la tension relative est calculée par rapport à la tension maximale isométrique de chacun des tests en température.

Fait établi (voir Fig. 19 dans (Ford, Huxley et al. 1977))

Ce constat pose à nouveau le problème abordé au paragraphe précédent concernant les définitions relatives à la fin de la phase 1 et au début de la phase 2.

**Remarques**

On note l'importance de la prise en compte de la viscosité dans l'interprétation des résultats expérimentaux précédents.

Des prédictions supplémentaires seront apportées au chapitre 12 concernant la phase 1 d'un échelon de force, identique à celle d'un échelon de longueur.

# 8 Phase 2 d'un échelon de longueur

## 8.1 Description de la phase 2 d'un échelon de longueur

La phase 2 se caractérise par le rétablissement rapide de la tension, immédiatement après la fin de la phase 1; la durée de la phase 2, intitulée $\Delta t_{p2}$ , est brève et n'excède pas 5 ms (Huxley 1974).

Durant $\Delta t_{p2}$, la remontée de la tension en fonction du temps suit une courbe d'allure exponentielle (Huxley 1974; Ford, Huxley et al. 1977) pour atteindre une valeur de plateau nommée usuellement $\mathbf{T2}$ . On note $\mathbf{pT2}$ la valeur relative de $\mathbf{T2}$ à la tension isométrique maximale $\mathbf{T0}^{\mathbf{fmI}}$ .

La relation décroissante entre $\mathbf{pT2}$ et l'échelon de longueur $\Delta X_{p1}^{(k)}$ (traits fushsia et rouge; Fig. 15) prend une forme parabolique jusqu'à 5 nm environ, puis finit linéairement (Huxley 1974).

$\mathbf{pT2}$ s'annule pour une valeur comprise entre 10 et 16 nm selon les auteurs. Dans le travail référent (Piazzesi, Lucii et al. 2002), les auteurs précisent : $\mathbf{pT2 = 0}$ pour $\Delta X_{p2}^{T=0} \approx 10.7\,\mathbf{nm}$.

## 8.2 Calcul empirique de pT2 en fonction de la valeur de ΔXp1

A partir de l'ensemble des travaux portés à notre connaissance (Huxley 1974; Ford, Huxley et al. 1977; Ford, Huxley et al. 1981; Ford, Huxley et al. 1985; Lombardi, Piazzesi et al. 1992 ; Linari and Woledge 1995; Piazzesi and Lombardi 1995; Galler, Hilber et al. 1996 ; Linari, Lombardi et al. 1997; Piazzesi, Lucii et al. 2002 ; Linari, Piazzesi et al. 2009), l'examen de la relation entre $\Delta X_{p1}$ et $\mathbf{pT2}$ indique qu'elle passe par ou à proximité du point défini par les coordonnées[1] (rectangle noir ; Fig. 15) :

$$\left|\begin{array}{l}\Delta X_{p1} = d_{2sitA} = 5.4\,\text{nm} \\ pT2 = 0.8 = 80\%\end{array}\right. \tag{8.1}$$

On distingue ainsi à partir de ce point deux nouveaux domaines sur l'axe des abscisses, intitulés $\mathbf{d1}$ et $\mathbf{d2}$ :

**d1 où** $0 \le \Delta X_{p1} \le d_{2sitA}$ :

La valeur relative de la tension exercée par la fmI, notée $\mathbf{pT2_{d1}}$, est une fonction parabolique de $\Delta X_{p1}^{(k)}$ passant par le point de cordonnées $\left|\begin{array}{l}\Delta X_{p1} = 0 \\ pT2 = 1\end{array}\right.$ et le point défini avec (8.1), dont l'équation s'écrit :

$$pT2_{d1} = 1 - 0.015 \cdot \Delta X_{p1}^{(k)} - 0.004 \cdot {\Delta X_{p1}^{(k)}}^2 \tag{8.2a}$$

L'arc de parabole apparait en trait fuchsia sur la Fig. 15.

---

[1] $\mathbf{d_{2sitA}}$ *est la distance entre 2 sitA avec* $\mathbf{d_{2sitA}}$**=5.4nm** *; voir égalité (1.6) du paragraphe 1.3.*

**Fig. 15 : Evolution de la tension relative après le raccourcissement de la fml par un échelon (k) en longueur durant la phase 2**

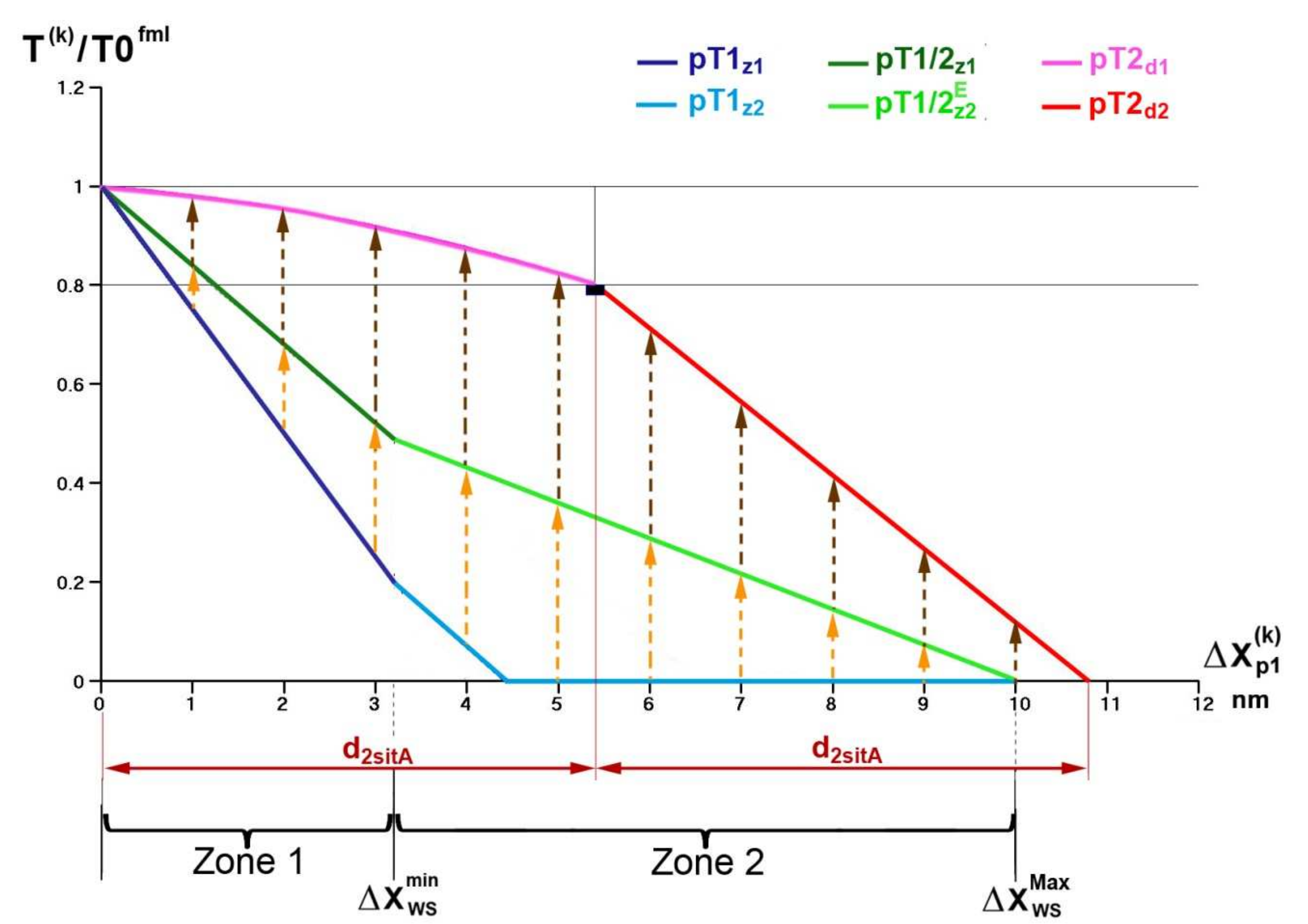

**d2 où** $\mathbf{d_{2sitA} < \Delta X_{p1} \le 2 \cdot d_{2sitA}}$ :

La valeur relative de la tension exercée par la fmI, notée $\mathbf{pT2_{d2}}$, se trouve sur une droite passant par le point défini par (8.1) et le point de cordonnées $\left|\begin{array}{l}\mathbf{\Delta X_{p1} = 2 \cdot d_{2sitA} = 10.8\ nm} \\ \mathbf{pT2 = 0}\end{array}\right.$ , dont l'équation se formule:

$$\mathbf{pT2_{d2} = 0.8 \cdot \left(2 - \frac{\Delta X^{(k)}_{p1}}{d_{2sitA}}\right)} \qquad (8.2b)$$

Le segment de droite apparait en trait rouge sur la Fig. 15.

## <u>8.3 Phénomènes participant à la remontée rapide de pT2</u>

### 8.3.1 Annulation des actions dues à la viscosité

Selon l'échelon (k), la remontée de la tension relative durant $\mathbf{\Delta t_{p2}}$ part d'un point situé sur des droites en bleu foncé ou clair d'abscisse $\mathbf{\Delta X^{(k)}_{p1}}$ pour arriver au point de même abscisse sur la parabole couleur fuchsia ou sur la droite en rouge d'équations respectives (8.2a) et (8.2b)

Voir les exemples de la Fig. 15 avec la succession de 2 flèches en pointillé de couleur orange puis brune, positionnées à chaque multiple du nm et dirigées selon l'axe des ordonnées.

Par définition, la phase 2 commence à $t^1_{stop_p1}$ ; au paragraphe 7.5 nous avons introduit $t^2_{stop_p1}$ lorsque les forces de viscosité ont totalement disparues avec $V = 0$. La valeur de $t^2_{stop_p1}$ est inconnue avec un ordre de grandeur de quelques centaine de µs. Arbitrairement a été choisi :

$$t^2_{stop_p1} = 250\,\mu s \tag{8.3a}$$

de manière que :

$$t^2_{stop_p1} - t^1_{stop_p1} = 100\,\mu s \tag{8.3b}$$

D'après les critères définis au paragraphe 8.1, $t^2_{stop_p1}$ appartient à $\Delta t_{p2}$. Au temps $t = t^2_{stop_p1}$, si aucun autre phénomène n'intervient, la tension relative est égale à $pT1/2$ (traits en vert foncé et clair; Fig. 15 à 17), calculée au paragraphe 7.4 avec les éq. (7.8) et (7.9) dans les zones respectives 1 et 2 correspondant à la condition $V = 0$.

### 8.3.2 Nouveaux SB $(SB_{fast})$

Dans notre modèle, la régénération rapide de la tension entre $pT1/2$ et $pT2$ (Flèches brunes verticales entre traits vert foncés et clairs et traits couleur fuchsia ou rouge; Fig. 15) provient de nouvelles tetM qui démarrent un WS suite à l'évt $SB_{fast}$ décrit au paragraphe 1.4.1. Sous l'effet du raccourcissement d'échelon $\Delta X^{(k)}_{p1}$, les tetM en mode $SB_{fast}$ remplissent désormais les condition de l'Hyp. 1 et sont à même d'initier un WS rapidement d'après les Hyp. 3 et 4.

Le calcul de $(pT2 - pT1/2)$ s'effectuent par différence des éq. (8.2a) et (8.2b) avec les éq. (7.8) et (7.9) dans les différents zones et domaines définis pour chacune des 4 équations. Le résultat apparait en trait marron sur la Fig. 16 ; formée d'une succession de 4 arcs ou segments de courbes, la courbe marron présente une allure concave avec un maximum atteint pour l'abscisse égale à $d_{2sitA}$ .

## <u>8.4 Calcul de la contribution des nouveaux SB en fonction de la valeur de $\Delta X_{p1}$</u>

Un premier calcul a été fait au paragraphe précédent à partir de la différence $(pT2 - pT1/2)$. Nous proposons une autre modélisation, similaire au niveau des résultats numériques, mais basée sur la manière dont l'évt $SB_{fast}$ s'accomplit selon la position d'une tetM par rapport aux molA du filA.

Par changement d'échelle apportée avec (4.9a), nous passons du déplacement angulaire au déplacement linéaire (Fig. 8; chap 4). Ainsi suite au déplacement d'échelon $\Delta X^{(k)}_{p1}$, les S1 de chaque hs de la fmI vont se répartir en 4 catégories (voir Fig.1 ; chap.1).

**Fig. 16 : Différence entre pT2 et pT1/2 après le raccourcissement de la fml par un échelon (k) en longueur**

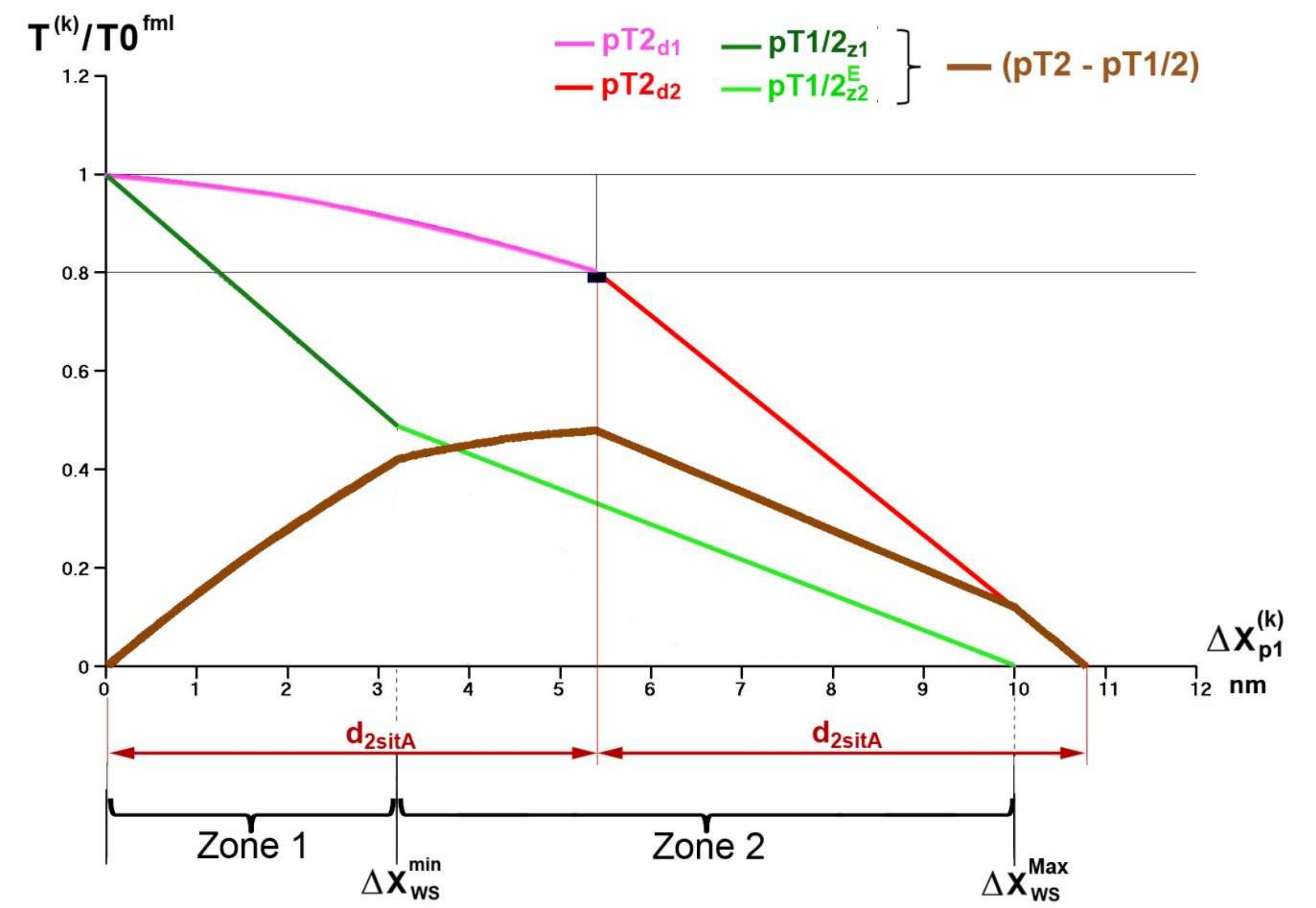

$\underline{\mathbf{cat1/S1_{iso}^{startWS} \rightarrow S1_{p2}^{WS}}}$ : les S1 initialement en isométrie et déplacés de $\mathbf{\Delta X_{p1}^{(k)}}$ sont en cours ou en fin de WS (état WS avec chemin 6 ; Fig. 1) et fournissent collectivement au temps $\mathbf{t_{stop_p1}^{2}}$ la valeur $\mathbf{pT1/2}$ dans les zones 1 et 2. Il est à noter que les S1 en fin de WS n'ont pas le temps de se détacher : pour la zone 2, la modélisation linéaire de $\mathbf{pT1/2_{z2}^{E}}$ est choisie selon (7.9), conformément aux indications du sous-paragraphe 4.3.4.3.

$\underline{\mathbf{cat2/S1_{iso}^{WB} \rightarrow S1_{p2}^{SBfast}}}$ : certains S1 en liaison faible durant le plateau isométrique transitent en liaison forte et initient simultanément un WS selon l'évt $\mathbf{SB_{fast}}$ car le déplacement $\mathbf{\Delta X_{p1}^{(k)}}$ leur offre les conditions imposées par l'Hyp. 1.

$\underline{\mathbf{cat3/S1_{iso}^{WB} \rightarrow S1_{p2}^{DE}}}$ : au contraire d'autres S1 ne remplissent plus les conditions de l'Hyp. 1 à cause du déplacement $\mathbf{\Delta X_{p1}^{(k)}}$ trop important car la position angulaire de S1b qui rendrait géométriquement S2 rigide devient supérieure à $\theta 2$. Durant les 2 phases transitoires suivantes, ces S1 seront considérés comme détachés (chemin 2 vers état DE; Fig. 1) et donc susceptibles de rentrer dans la catégorie suivante.

$\underline{cat4/S1_{iso}^{DE} \rightarrow S1_{p2}^{RS}}$ : les S1 restants sont détachés (état DE ; Fig. 1). Guidée par les protéines de maintien le long du filA associé d'après la géométrie hexagonale inter-filamentaire, chaque tetM détachée subit, en permanence et à la fois, l'action des forces thermiques et des forces d'attraction en $1/r^{\alpha}$ ($\alpha$ réel) d'une à 2 molA. Le raccourcissement brusque $\Delta X_{p1}^{(k)}$ entraine un déplacement linéaire mais aussi angulaire des segments S1 et S2 provoqué par l'attraction des molA. Ces déplacements sont non quantifiables à cause de la composante brownienne imposée par l'agitation thermique.

Cette situation nécessite pour un éventuel $SB_{fast}$ un redressement décrit au paragraphe 1.7 (évt RS vers état WB par chemin 9 ; Fig .1). Apres redressement certaines de ces tetM sont capables de remplir les conditions imposées pour un $SB_{fast}$ durant les phases transitoires suivantes, *i.e.* les phases 3 et 4.

Les nouveaux $SB_{fast}$ qui participent à la remontée de tension relative durant la seconde partie de la phase 2 correspondent à la catégorie n° 2 précédente $\left(S1_{p2}^{SB_{fast}}\right)$; leur contribution à la remontée de la tension relative jusqu'à $pT2$ sera intitulée $pT2_{SB_{fast}}$.

Rappelons que $d_{2sitA}$ est la distance entre 2 sitA , distance à partir de laquelle le modèle caractérise la liaison faible étudiée au paragraphe 1.3. Pour simplifier, nous admettons que la distance $d_{2sitA}$ est approximativement équivalente à $\delta_X$. Ainsi, par rapport à la position d'un sitA fixée à $0$ selon l'axe longitudinal $OX$, on distinguera l'intervalle $[-d_{2sitA}/2;0]$ favorable à la réalisation d'un $SB_{fast}$ et l'intervalle $[0;+d_{2sitA}/2]$ moins favorable à la réalisation d'un $SB_{fast}$ car plus propice à une modification de la positon angulaires de S1b supérieure à $\theta 2$.

Rappelons aussi que, quelque soit $\Delta X_{p1}^{(k)}$, une fois le WS initié, les positions angulaires des S1b se répartissent uniformément sur $\delta_\theta$ selon l'Hyp. 2 , et par changement d'échelle uniformément sur $\delta_X$.

Les considérations précédentes conduisent à distinguer 3 régions sur l'axe des abscisses, notées **r1**, **r2** et **r3**, respectivement, constituées de fractions ou de multiples de $d_{2sitA}$ (Fig. 17) :

**r1 où** $0 < \Delta X_{p1}^{(k)} \leq \left[\frac{d_{2sitA}}{2} \approx \frac{\delta_X}{2} \approx \Delta X_{WS}^{min}\right]$

Le nombre de S1 sujet à l'évt $SB_{fast}$ augmente linéairement avec $\Delta X_{p1}^{(k)}$, conformément à la nature linéaire d'une distribution uniforme et les S1 sont peu perturbés par $\Delta X_{p1}^{(k)}$ ; ceci se traduit par l'équation :

$$pT2_{SB_{fast}} \approx 0.135 \cdot \Delta X_{p1}^{(k)} \tag{8.4a}$$

**Fig. 17 : Contribution des nouveaux SB à la remontée de la tension relative durant la phase 2 suivant le raccourcissement de la fml après une perturbation en longueur d'échelon (k)**

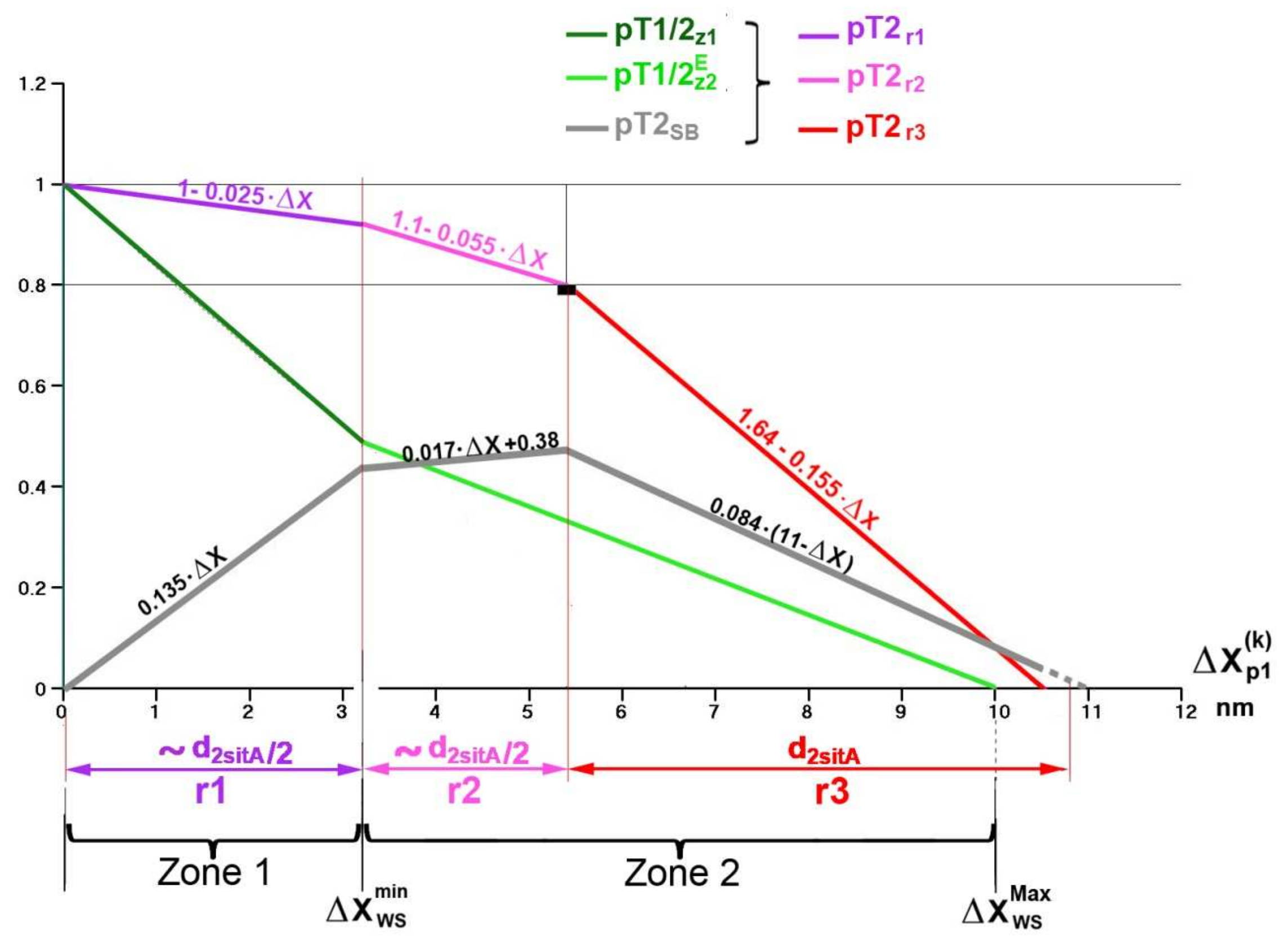

**r2 où** $\left[\frac{d_{2sitA}}{2} \approx \frac{\delta_X}{2} \approx \Delta X_{WS}^{min}\right] \leq \Delta X_{p1}^{(k)} \leq d_{2sitA}$

Le nombre de S1 susceptibles de réaliser l'évt $SB_{fast}$ augmente toujours linéairement avec $\Delta X_{p1}^{(k)}$, mais parmi ce nombre, certains rentrent dans la catégorie n° 3 $\left(S1_{p2}^{DE}\right)$ où le déplacement $\Delta X_{p1}^{(k)}$ entraine la position angulaire de S1b au delà de $\theta 2$.

Ceci se traduit par une droite de pente presque nulle d'équation :

$$pT2_{SB_{fast}} \approx 0.017 \cdot \Delta X_{p1}^{(k)} + 0.38 \tag{8.4b}$$

**r3 où** $d_{2sitA} < \Delta X_{p1} \leq 2 \cdot d_{2sitA}$

Les S1 sont perturbés suite au raccourcissement $\Delta X_{p1}^{(k)}$ et le nombre de S1 à même d'accomplir l'évt $SB_{fast}$ diminue linéairement avec $\Delta X_{p1}^{(k)}$ pour s'annuler à 11.2 nm, soit :

$$pT2_{SB_{fast}} = 0.9 - 0.08 \cdot \Delta X_{p1}^{(k)} \tag{8.4c}$$

Les valeurs des pentes et des ordonnées à l'origine des équations linéaires (8.4a), (8.4b) et (8.4c) sont calculées d'après la modélisation proposée au sous-paragraphe 8.3.2 (traits marrons ; Fig. 16).

On obtient $\mathbf{pT2}$ par sommation (Fig. 17) :

$$\mathbf{pT2 = pT1/2 + pT2_{SBfast}} \tag{8.5}$$

Soit dans les 3 régions :

**r1 où** $\mathbf{0 < \Delta X_{p1}^{(k)} \leq \left[ \frac{d_{2sitA}}{2} \approx \frac{\delta_X}{2} \approx \Delta X_{WS}^{min} \right]}$

En sommant l'éq. (7.8) dans laquelle on a introduit la valeur apportée par (4.35c) et l'éq. (8.4a), on obtient :

$$\mathbf{pT2 = 1 - 0.025 \cdot \Delta X_{p1}^{(k)}} \tag{8.6a}$$

**r2 où** $\mathbf{\left[ \frac{d_{2sitA}}{2} \approx \frac{\delta_X}{2} \approx \Delta X_{WS}^{min} \right] \leq \Delta X_{p1}^{(k)} \leq d_{2sitA}}$

En sommant l'éq. (7.9) dans laquelle on a introduit la valeur apportée par (4.36c) et l'éq. (8.4b), on obtient :

$$\mathbf{pT2 = 1.1 - 0.055 \cdot \Delta X_{p1}^{(k)}} \tag{8.6b}$$

**r3 où** $\mathbf{d_{2sitA} < \Delta X_{p1} \leq 2 \cdot d_{2sitA}}$

En sommant l'éq. (7.9) dans laquelle on a introduit la valeur apportée par (4.36c) et l'éq. (8.4c), on retrouve l'éq. (8.2b) :

$$\mathbf{pT2 = 1.64 - 0.155 \cdot \Delta X_{p1}^{(k)}} \tag{8.6c}$$

Les 3 segments de droites apparaissent en traits violet, fuchsia et rouge sur la Fig. 17 avec leurs équations respectives dans les régions 1, 2 et 3.

## 8.4 Modélisation de l'évolution temporelle de la tension relative durant la phase 2

Soit le début de la phase 1 à $\mathbf{t = 0}$. Pour chaque valeur de l'échelon de longueur $\mathbf{\Delta X_{p1}^{(k)}}$ , on a :

**Départ de la phase 2 classique** à $\mathbf{t = t_{stop_p1}^{1}}$ avec $\mathbf{t_{stop_p1}^{1} = 0.15\,ms}$

$$\mathbf{pT\left(t_{stop_p1}^{1}\right) = pT1} \tag{8.7a}$$

avec valeurs de $\mathbf{pT1}$ apportées par les éq. (7.3a) et (7.4a) dans les zones 1 et 2 respectivement (points bleus foncés et clairs ; Fig. 18)

**Remontée linéaire de la tension** de $t^1_{stop_p1}$ à $t^2_{stop_p1}$ avec $\left(t^2_{stop_p1} - t^1_{stop_p1}\right) = 0.1\,ms$

$$pT(t) = \left[pT1 + \left(t - t^1_{stop_p1}\right) \cdot \frac{pT1/2 - pT1}{t^2_{stop_p1} - t^1_{stop_p1}}\right] \cdot \mathbb{1}_{\left[t^1_{stop_p1};\, t^2_{stop_p1}\right]}(t) \tag{8.7b}$$

où les valeurs de $pT1/2$ sont apportées par les éq. (7.8) et (7.9) dans les zones 1 et 2, respectivement.

L'évolution temporelle linéaire apparait en traits bleus foncés et clairs sur la Fig. 18.

**Départ de la phase 2 du modèle** à $t = t^2_{stop_p1}$ avec $t^2_{stop_p1} = 0.25\,ms$

$$pT\left(t^2_{stop_p1}\right) = pT1/2 \tag{8.7c}$$

Dans notre modèle $t^2_{stop_p1}$ est la fin de la phase 1 et le début de la phase 2 (points verts foncés et clairs ; Fig. 18)

**Remontée exponentielle de la tension** de $t^2_{stop_p1}$ à $t_{stop_p2}$ avec $t_{stop_p2} \approx 3\,ms$

Dans cet intervalle de temps où $V = 0$, un raisonnement identique à ceux conduits aux paragraphes 6.3 et 7.3 mène à l'égalité des forces, des déplacements et du nombre de S1 en cours de WS et de SB dans chaque hs de la fmI.

La probabilité de réalisation de l'évt $SB_{fast}$ est donnée par (1.11). Dans chaque hs, le nombre moyen de S1 à l'instant t concernés par la réalisation d'un $SB_{fast}$ est noté $\Lambda^{hs}_{SBfast,p2}$ et est calculé avec (A.17) à l'Annexe A. Ces $\Lambda^{hs}_{SBfast,p2}$ S1 se répartissent uniformément sur $\delta_\theta$ selon l'Hyp. 2 et leur contribution relative à la remontée de la tension est égale au rapport de $\Lambda^{hs}_{SBfast,p2}$ sur $\Lambda 0$ ; on rappelle que $\Lambda 0$ est fourni par (6.4) et entre dans le calcul de $T0^{fmI}$ d'après (6.5).

Au niveau de la fmI, cette contribution relative est égale à la différence entre $pT2$ et $pT1/2$, où $pT2$ est la valeur plateau considérée comme la valeur stationnaire de l'exponentielle comme stipulé au 1[er] paragraphe du chapitre et dans le paragraphe 1.4.1.

On obtient :

$$pT(t) = \left[pT2 + \left(pT1/2 - pT2\right) \cdot e^{-\frac{t - t^2_{stop_p1}}{\Delta t_{SBfast}}}\right] \cdot \mathbb{1}_{\left\{t \geq t^2_{stop_p1}\right\}}(t) \tag{8.7d}$$

avec valeurs de $pT2$ apportées par (8.6a), (8.6b) et (8.6c) dans les régions appropriées.

et $\Delta t_{SBfast} \approx 0.7\,ms$ selon l'Hyp. 3 et (1.13)

L'évolution temporelle exponentielle apparait en traits verts foncés et clairs sur la Fig. 18.

**Fig. 18 :** **Remontée temporelle de la tension relative au cours de la phase 2 après le raccourcissement de la fml pour 7 échelons de longueur**

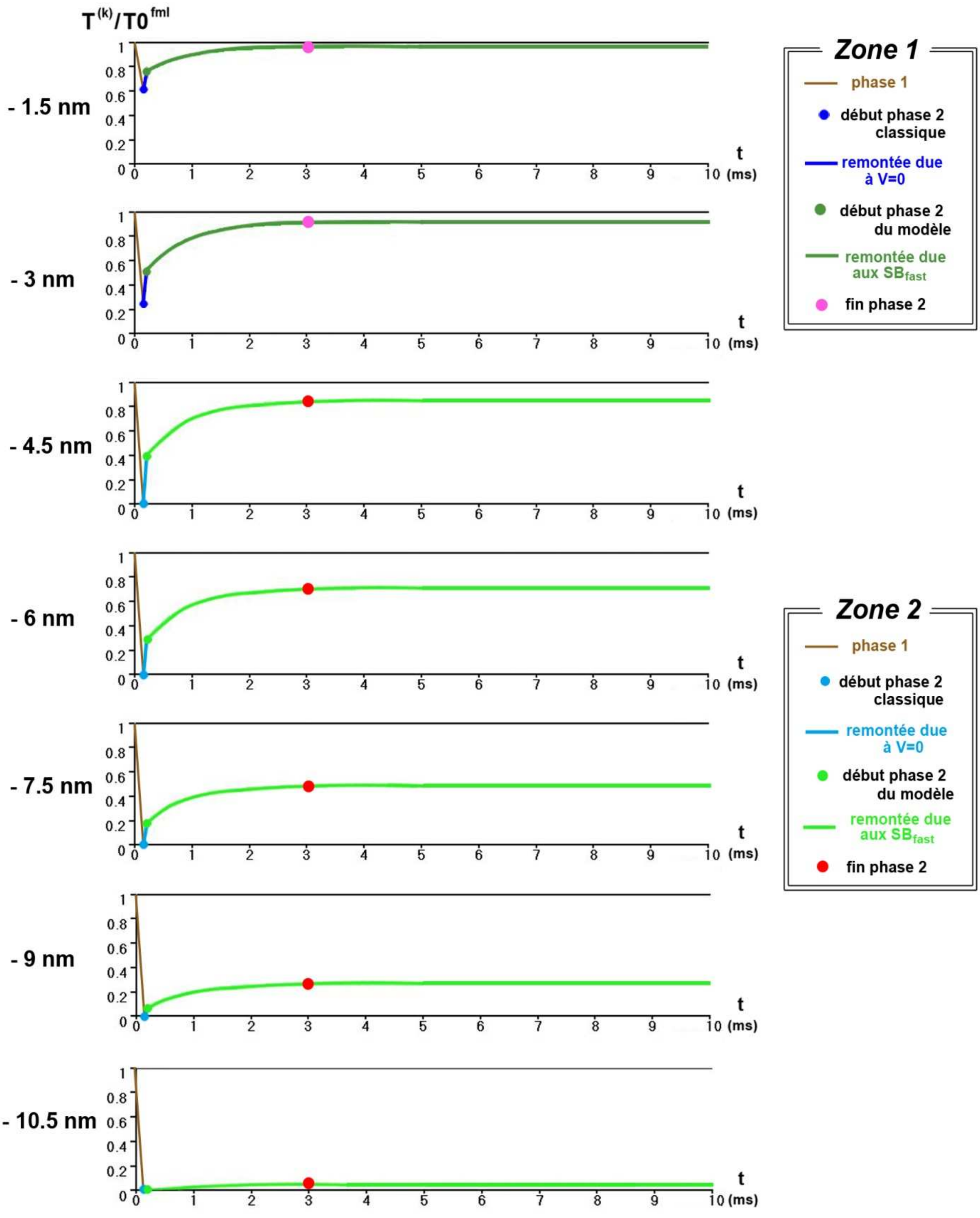

Notre modèle explique la remontée rapide de la tension pendant la phase 2 par la succession de 2 phénomènes : disparition des forces de viscosité dès que la fmI est immobilisée ( $\mathbf{V=0}$ ) puis nouvelles initiations de WS selon évt $\mathbf{SB_{fast}}$ .

Ce scénario permet de distinguer la participation de chacun des 2 phénomènes et de calculer leur contribution respective entre la fin de la phase 1 à $\mathbf{t^1_{stop_p1}}$ et la fin de la phase 2 à $\mathbf{t_{stop_p2}}$ (Fig. 15), telle que :

$$\mathbf{(pT2-pT1)} = \underbrace{\mathbf{(pT1/2-pT1)}}_{\mathbf{V=0}} + \underbrace{\mathbf{(pT2-pT1/2)}}_{\mathbf{SB_{fast}}} \quad (8.8)$$

Sur la Fig. 18, apparaissent les évolutions temporelles de ces 2 contributions (traits bleus pour l'annulation des forces de viscosité; traits verts pour les nouveaux WS dus à $\mathbf{SB_{fast}}$ ), définies par les équations établies au paragraphe 8.4, pour 7 échelons de longueur[1] multiples de 1.5 nm dans chaque hs de la fmI.

La durée de la phase 2 a été prise constante et égale à 3 ms, environ  pour s'ajuster au temps de début des plateaux des 7 exponentielles en traits verts, soit :

$$\mathbf{pT(t_{stop_p2}) \approx pT2} \quad (8.9)$$

Cette fin de la phase 2 correspond aux points rouges apparaissant sur les 7 cinétiques de la Fig. 18.

## 8.5 Discussion

### *8.5.1 Questions soulevées par le modèle*

**Fin de la phase 1 et début de la phase 2 (suite du paragraphe 7.6.1)**

Afin de distinguer les 2 phénomènes générateurs de force produisant la phase 2 classique d'un échelon de longueur (disparition des forces de viscosité quand $\mathbf{V=0}$, puis nouveaux déclenchements de WS consécutifs à l'évt $\mathbf{SB_{fast}}$ ), le modèle  propose une chronologie où leurs durées d'existence se succèdent. Dans la réalité, il est concevable que ces 2 phénomènes se manifestent selon une évolution temporelle concomitante, au moins dans les premières microsecondes.

Ainsi, si dans notre modèle $\mathbf{t^2_{stop_p1}}$ est défini comme la fin de la phase 1 et le début de la phase 2, il appartient dans les faits à la fois à la phase 1 à cause de la disparition des forces de viscosité et à la fois à la phase 2 avec l'apparition des nouveaux $\mathbf{SB_{fast}}$ .

La nature linéaire de l'éq. (8.7b)  entre  $\mathbf{t^1_{stop_p1}}$ à $\mathbf{t^2_{stop_p1}}$ précisant l'annulation des forces de viscosité est une simplification de la remontée brusque de la tension relative car il existe la possibilité que le phénomène présente une phase d'amortissement nécessitant une composante exponentielle.

De plus, la durée de cette remontée est considérée comme constante et égale à $\mathbf{100\,\mu s}$ , quelle que soit la valeur de l'échelon et quel que soit le différentiel entre $\mathbf{pT1}$ et $\mathbf{pT1/2}$ ; la réalité doit se révéler plus complexe.

[1] *Rappel : les raccourcissements sont mesurés négativement comme indiqué sur la Fig. 19, mais présentés positivement dans les calculs de notre modèle.*

**Fig. 19 : variabilité de la pente selon la modélisation linéaire de pT2 pour $\Delta X_{p1}^{(k)} > d_{2sitA}$**

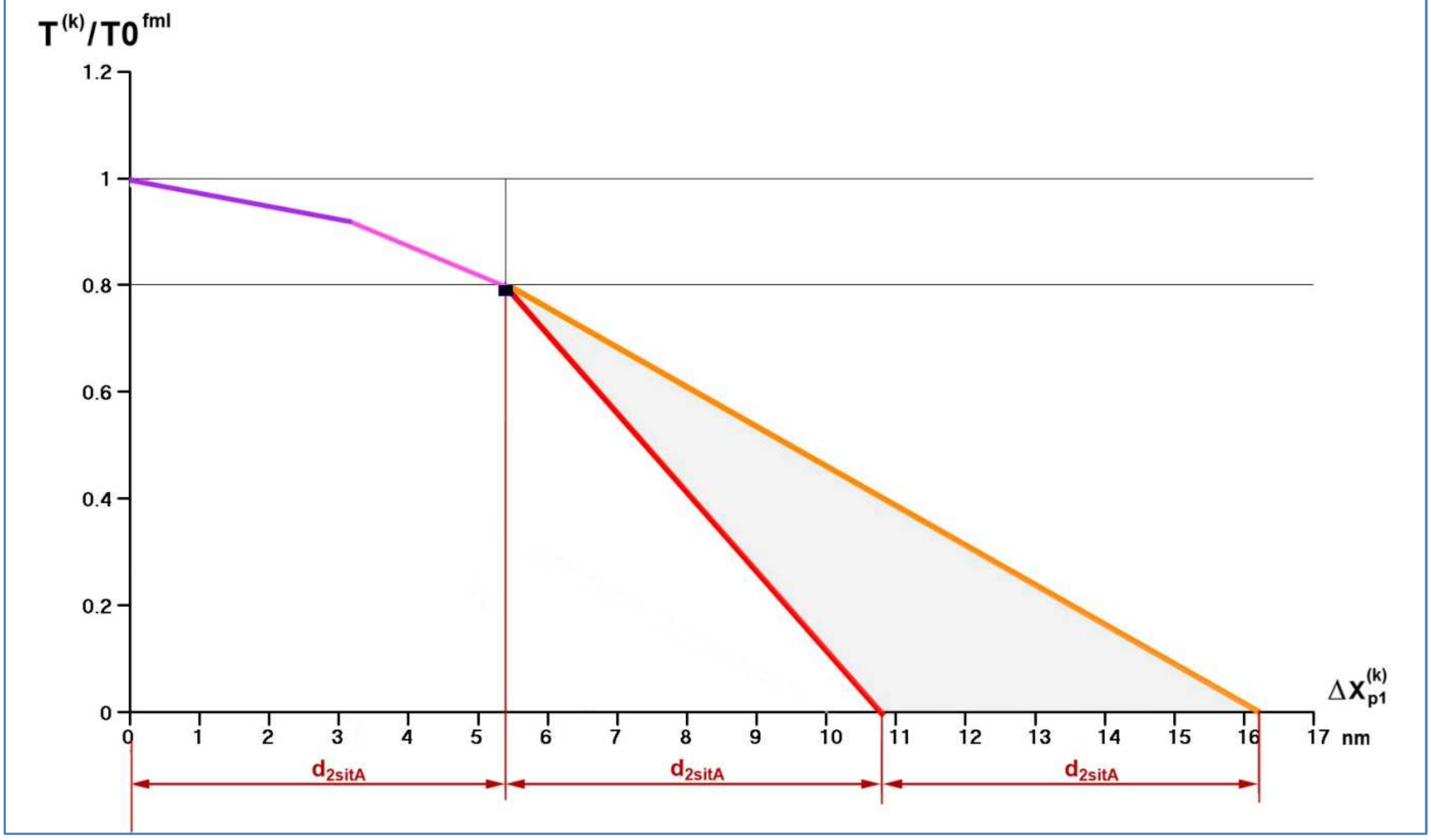

Toutes ces remarques conduisent à observer des évolutions de courbes plus lissées dans les graphiques issus de tests expérimentaux (Huxley 1974; Ford, Huxley et al. 1977) que celles proposées par la modélisation à la Fig. 18.

Nonobstant ces approximations, le modèle rend correctement compte des faits observés et met en lumière la combinaison de ces 2 phénomènes générateurs de force durant la phase 2 pour un échelon en longueur.

**Remontée de la tension relative jusqu'à $\mathbf{pT2}$**

De nouvelles expérimentations ont démontré que la remontée de la tension était engendré par le recrutement de nouveaux S1 (Huxley et Kress 1985 ; Huxley, Reconditi et al. 2006a). Ainsi comme cela a été suggéré dans le paragraphe 8.4, la différence $\mathbf{(pT2 - pT1/2)}$ qui explicite cette remontée rend compte des possibilités offertes à des S1 d'initier de nouveaux WS, possibilités dépendant grandement de la géométrie inter-filamentaire après la perturbation, *i.e.* de la position angulaire de S1b par rapport à la molA à laquelle S1a est liée faiblement.

**Différences entre expérimentations**

Le calcul empirique de $\mathbf{pT2}$ a été effectué dans 2 zones séparées :

Si $\mathbf{\Delta X_{p1} \le d_{2sitA}}$

La modélisation de $\mathbf{pT2}$ , soit par la parabole d'éq. 8.2a, soit par la succession des 2 segments de droites d'éq. (8.6a) et (8.6b) s'ajuste en 1ère approximation avec les mesures de l'ensemble des travaux portés à notre connaissance concernant le muscle squelettique de la grenouille ou celui d'autres espèces (voir références mentionnées au début du paragraphe 8.2).

Si $\Delta X_{p1} > d_{2sitA}$

Les données issues des travaux de G. Piazzesi et de ses coauteurs s'alignent sur le segment de droite d'éq. (8.2b) ou (8.6c). Tous les autres publications produisent aussi une modélisation de type linéaire mais présentent une différence de pente, les segments de droites appartenant au triangle grisé de la Fig. 19.

Les coordonnées des 3 sommets sont définies par (8.1) et :

$$\left| \begin{array}{l} \Delta X = 2 \cdot d_{2sitA} = 10.8\,nm \\ pT2 = 0 \end{array} \right. \quad \text{et} \quad \left| \begin{array}{l} \Delta X = 3 \cdot d_{2sitA} \approx 16 nm \\ pT2 = 0 \end{array} \right. \qquad (8.10)$$

Les explications de ces différences de pente sont similaires à celles détaillées au paragraphe 7.6.1.

### *8.5.2 Faits observés prédits par le modèle*

**Stabilité des valeurs de pT2**

La nature géométrique des conditions de réalisation de $SB_{fast}$ qui conduisent à la valeur de $pT2$ suggère que les facteurs modifiant les valeurs de la tension relative lors de la phase 1 à cause de la viscosité ne doivent plus exercer d'influence à la fin de la phase 2 et donc sur $pT2$.

On observe des valeurs similaires pour $pT2$ selon :

- la longueur initiale du hs (voir Fig. 11 dans (Ford, Huxley et al. 1981))
- le nombre de WS ayant participé à la mise en tension isométrique maximale (voir Fig. 3D dans (Linari, Piazzesi et al. 2009)).
- des variations de température expérimentale (Ford, Huxley et al. 1977).
- la durée de la phase 1 (voir Fig. 19 dans (Ford, Huxley et al. 1977)).

**Temps moyen de survenue de $SB_{fast}$ identique pour tous les échelons en longueur**

Une fois éliminé les effets dus à la viscosité, le temps moyen $\Delta t_{SBfast}$ apparaissant dans l'éq. (1.11) suffit pour modéliser les remontées de la tension relative pour n'importe quel échelon en longueur.

La valeur de $\Delta t_{SBfast}$ prise à 0.7 ms d'après l'égalité (1.13) est conforme à celles des différents modèles proposés dans la littérature (Huxley 1969 ; Huxley et Simmons 1971 ; Irving, Lombardi et al. 1992 ; Piazzesi and Lombardi 1995 ; Piazzesi, Reconditi et al. 1999 ; Huxley 2000; Reconditi, Linari et al. 2004 ; Huxley, Reconditi et al. 2006).

On comprend aussi la difficulté de mener des raccourcissements où les forces de viscosité n'influeraient pas en augmentant la durée de la phase 1. A cause de la valeur de $\Delta t_{SBfast}$, inférieure à la ms, de nouveaux S1 commenceraient à être recrutés avant la fin de la phase 1, comme c'est probablement le cas si $\Delta t_{p1} = 1\,ms$ (voir Fig. 19 dans (Ford, Huxley et al. 1977)).

# 9 Phase 3 d'un échelon de longueur

## 9.1 Description

Après la phase 2, il est constaté que le taux de rétablissement de la tension se réduit notablement avec possibilité de s'annuler et parfois même de devenir négatif sur une période qui varie entre 2 et 20 ms. Cette phase transitoire est nommée phase 3 (Huxley 1974; Ford, Huxley et al. 1977).

## 9.2 Interprétation du modèle

Durant la phase 3 avec. $\mathbf{t > t_{stop_p2}}$ , les 4 catégories proposées au paragraphe 8.4 deviennent:

$\mathbf{1/S1_{iso}^{startWS} \rightarrow S1_{p2}^{WS} \rightarrow S1_{p3}^{WS}\,ou\,S1_{p3}^{DE}}$ : les S1 initialement en isométrie sont toujours en cours ou en fin de WS. Les S1 en fin de WS ont le temps de se détacher et ne participent plus à la valeur de $\mathbf{pT1/2}$ . Ainsi dans la zone 2, la contribution durant la phase 3 n'est plus $\mathbf{pT1/2_{z2}^{E}}$ (trait en pointillé vert clair; Fig. 20) mais tend à devenir $\mathbf{pT1/2_{z2}^{A}}$ (trait plein vert clair; Fig. 20).

$\mathbf{2/S1_{iso}^{WB} \rightarrow S1_{p2}^{SB_{fast}}}$ : les S1 ayant initié leur WS suite à la survenue de l'évt $\mathbf{SB_{fast}}$ fournissent collectivement la valeur $\mathbf{(pT2 - pT1/2)}$ dans les régions 1, 2 et 3; ils maintiennent cette contribution constante durant les phases 3 et 4 selon (8.9).

$\mathbf{3/S1_{iso}^{WB} \rightarrow S1_{p2}^{DE} \rightarrow S1_{p3}^{DE}}$ : ces S1 sont considérés, soit comme détachés, soit comme rentrant dans la catégorie suivante.

$\mathbf{4/S1_{iso_p2_p3}^{DE} \rightarrow S1_{p2_p3}^{RS} \rightarrow S1_{p3\&p4}^{WB} \rightarrow S1_{p3\&p4}^{SBfast}}$: les S1 après RS se lient faiblement ; parmi ceux-ci, certains se retrouvent en position de réaliser l'évt $\mathbf{SB_{fast}}$ et d'initier un WS. Ces S1 sont désignés par le sigle : $\mathbf{SB_{RS}}$.

Les S1 participant à la remontée de la tension relative sont séparés en 2 entités temporelles :

- **Après $\mathbf{t_{stop_p1}^{2}}$, *i.e.* durant les phase 2, 3 et 4**, les S1 qui initient un WS après avoir réalisé l'évt $\mathbf{SB_{fast}}$
- **Après $\mathbf{t_{stop_p2}}$, *i.e.* durant les phases 3 et 4**, les S1 qui initient un WS après avoir accompli l'évt **RS** suivi de l'évt $\mathbf{SB_{fast}}$. L'enchainement de ces 2 évts sera intitulé $\mathbf{SB_{RS}}$.

Fig. 20 : Relaxation de la tension relative durant la phase 3 après le raccourcissement de la fml pour un échelon (k) en longueur

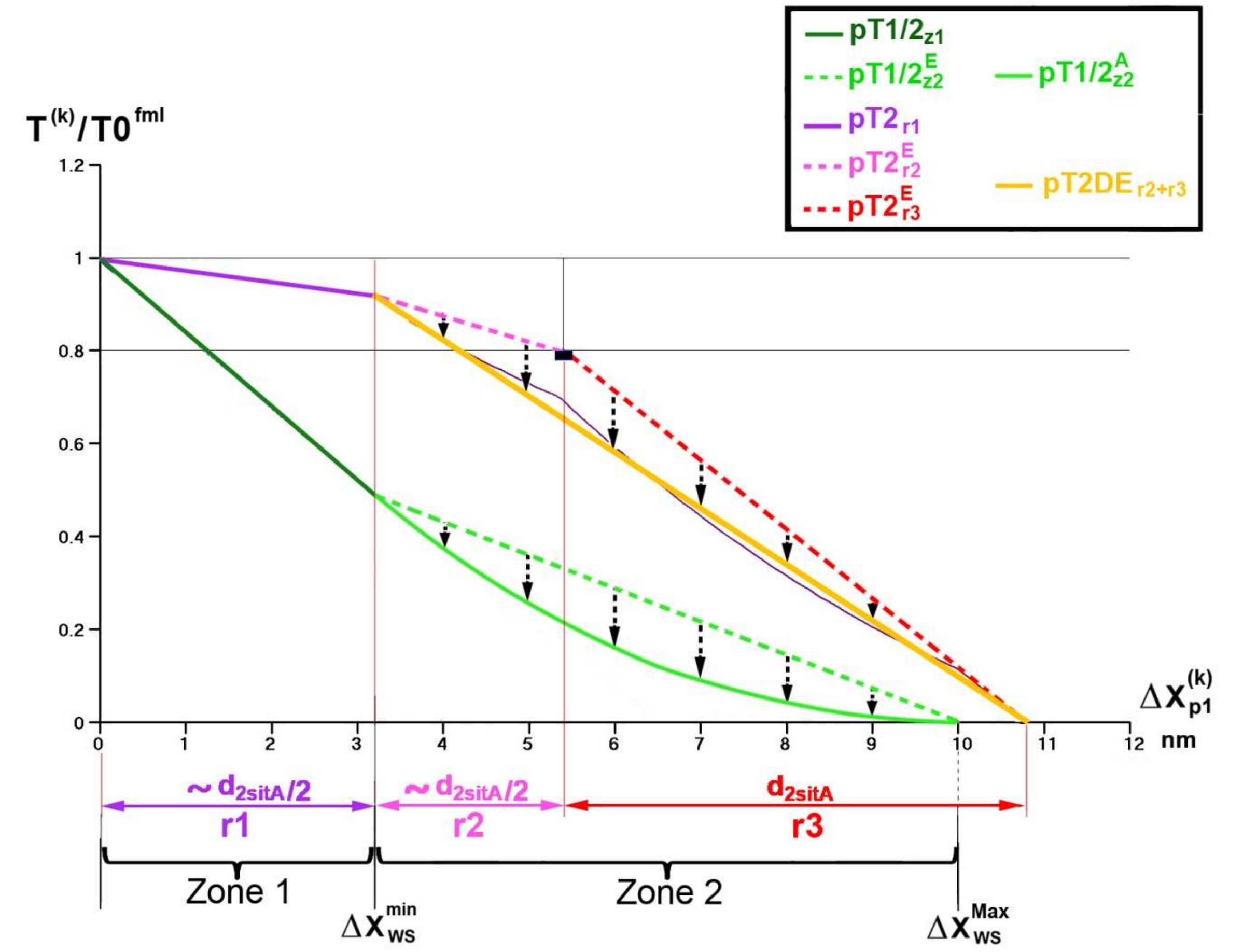

## 9.3 Modélisation de l'évolution temporelle de la tension relative durant la phase 3

### *9.3.1 Diminution de la tension relative due au détachement des S1 en fin de WS*

Les S1 en fin de WS se détachent et ne participent plus à la valeur de $\mathbf{pT1/2}$ . Ce phénomène n'a pas d'incidence dans la zone 1 où par définition aucun S1 n'est en fin de WS et concerne uniquement la zone 2 ; dans ce cas, la contribution des S1 qui ont participé à la mise en place du plateau tétanique n'est plus $\mathbf{pT1/2_{z2}^{E}}$ mais devient $\mathbf{pT1/2_{z2}^{A}}$ , conformément aux indications du paragraphe 4.3.4.4.

La différence $\left(\mathbf{pT1/2_{z2}^{E}}-\mathbf{pT1/2_{z2}^{A}}\right)$ est calculable dans la zone 2 à partir des équations (4.30) et (4.32).

Conséquemment il faut soustraire $\left(\mathbf{pT1/2_{z2}^{E}}-\mathbf{pT1/2_{z2}^{A}}\right)$ à $\mathbf{pT2}$. Nommons $\mathbf{pT2DE}$ la nouvelle valeur de la tension relative qui tient compte du détachement des S1 en fin de WS, soit (Fig. 20) :

$$\mathbf{pT2DE}=\mathbf{pT2}-\left(\mathbf{pT1/2_{z2}^{E}}-\mathbf{pT1/2_{z2}^{A}}\right) \tag{9.1}$$

Le calcul opéré dans les équations respectives (8.6b) et (8.6c) correspondant, respectivement, aux régions 2 et 3 mène à une courbe approximativement linéaire (courbe grisée figurant derrière la droite en jaune ; Fig. 20).

D'après (8.6a) et la définition des régions **r1**, **r2** et **r3**, la droite de couleur jaune de la Fig. 20 passe par les points: $\begin{vmatrix} \Delta X_{WS}^{min} \\ \left(1 - 0.025 \cdot \Delta X_{WS}^{min}\right) \end{vmatrix}$ et $\begin{vmatrix} 2 \cdot d_{2sitA} \\ 0 \end{vmatrix}$

Avec les données du modèle, elle adopte pour équation dans les régions 2 et 3 :

$$pT2DE_{r2+r3} = 0.121 \cdot \left(10.8 - \Delta X_{p1}^{(k)}\right) \tag{9.2}$$

A l'appui de (8.6a), on vérifie dans la région 1 :

$$pT2DE_{r1} = pT2_{r1} = 1 - 0.025 \cdot \Delta X_{p1}^{(k)} \tag{9.3}$$

On obtient ainsi dans les régions 1, 2 et 3, la contribution de la relaxation de la tension relative induite par le détachement des S1 en fin de WS, intitulée $pT_{DE}$ en recourant à l'éq. (A.17) de l'annexe A :

$$pT_{DE}(t) = (pT2DE - pT2) \cdot \left[1 - e^{-\frac{t - t_{fin_p2}}{\Delta t_{DE}}}\right] \cdot \mathbf{1}_{\left\{t \geq t_{stop_p2}\right\}}(t) \tag{9.4}$$

avec valeurs de $pT2$ apportées par (8.6a), (8.6b) et (8.6c) et celles de $pT2DE$ issues de (9.2) et (9.3) dans chaque région appropriée et $\Delta t_{DE} = 8\,ms$ et $\Delta t_{P_{DE}} \approx t_{fin_p2} \approx 3\,ms$ d'après l'Hyp. 12 et selon (1.23), (1.24) et (1.25)

### *9.3.2 Augmentation de la tension relative due aux nouveaux démarrages de WS*

Au paragraphe 1.7, l'évènement 7 intitulé RS a été décrit par une succession d'évènements et apprécié comme un évènement global.

La contribution de l'augmentation de la tension relative engendrée par l'évènement RS suivi de l'évènement caractérisant le déclenchement d'un WS, intitulée $pT_{RS}$ , est égale a l'appui de (1.2) et (A.17)[1] :

$$pT_{RS}(t) = (1 - pT2DE) \cdot \left[1 - \frac{\Delta t_{RS} \cdot e^{-\frac{t - t_{fin_p2}}{\Delta t_{RS}}} - \Delta t_{SBfast} \cdot e^{-\frac{t - t_{fin_p2}}{\Delta t_{SBfast}}}}{\Delta t_{RS} - \Delta t_{SBfast}}\right] \cdot \mathbf{1}_{\left\{t \geq t_{stop_p2}\right\}}(t) \tag{9.5}$$

avec $\Delta t_{RS} = 30\,ms$ et $\Delta t_{RS} \approx t_{fin_p2} \approx 3\,ms$ selon l'Hyp. 14 et les égalités (1.28), (1.29) et (1.30)

et $\Delta t_{SBfast} = 0.7\,ms$ selon l'Hyp. 3 et les égalités (1.11) et (1.13)

[1] *La valeur finale de la tension relative est donnée pour égale à 1 dans (9.5) ; c'est en effet ce qui est observé mais cette valeur sera justifiée au prochain chapitre.*

**Fig. 21 :** Plateaux temporels de la tension relative lors la phase 3 après le raccourcissement de la fml pour 7 échelons de longueur

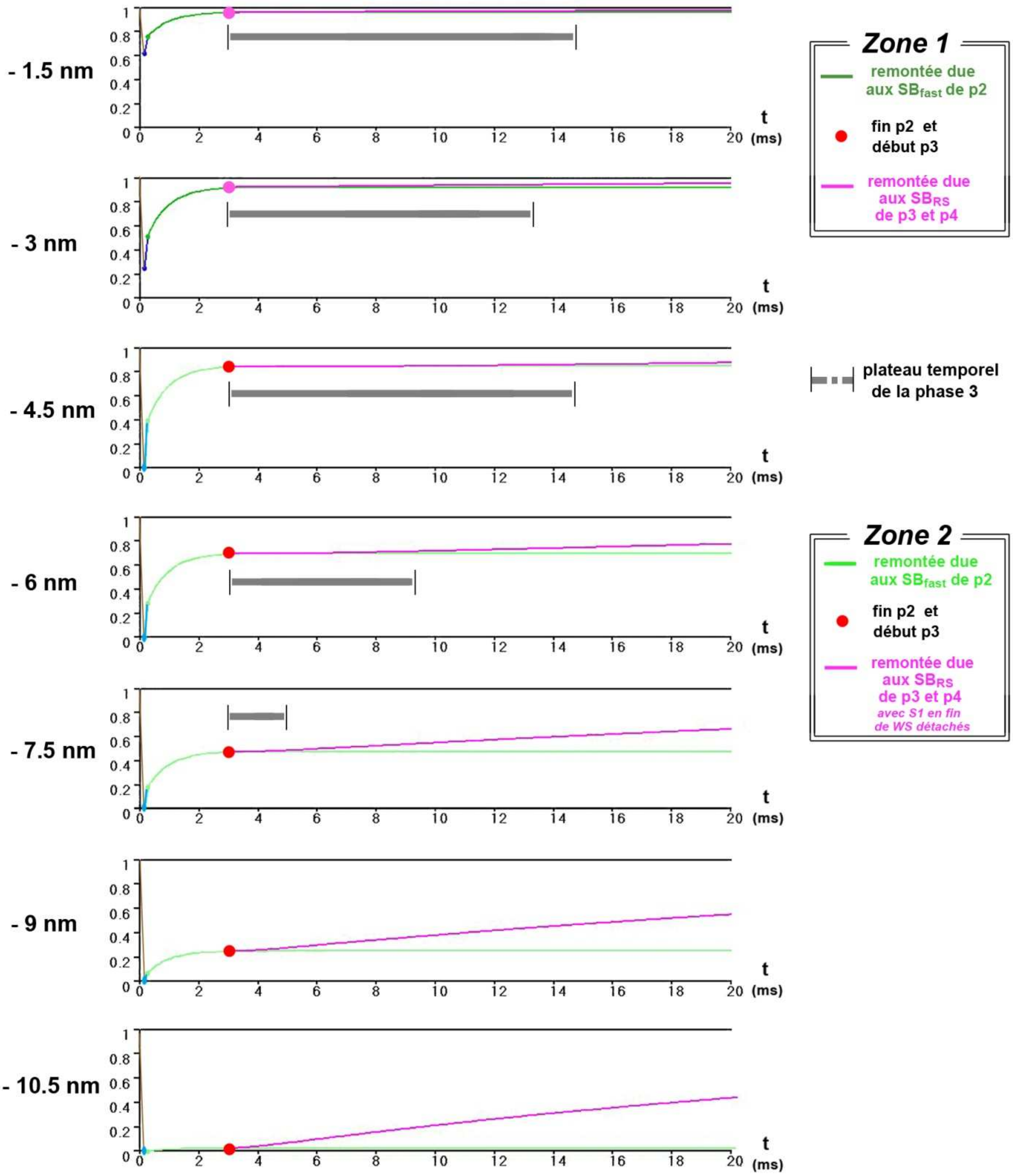

### *9.3.3 Calcul de la tension relative durant les phases 3 et 4 d'un échelon de longueur*

En sommant les contributions apportées par (9.4) et (9.5), on obtient au total l'évolution temporelle de la tension relative de la fimI après $\mathbf{t_{stop_p2}}$ , *i.e.* durant les phases 3 et 4, soit :

$$\mathbf{pT(t) = pT2 + pT_{DE}(t) + pT_{RS}(t)} \qquad (9.6)$$

avec valeurs de $\mathbf{pT_{DE}(t)}$ et $\mathbf{pT_{RS}(t)}$ apportées, respectivement, par (9.4) et (9.5)

Le début de l'évolution temporelle de $\mathbf{pT(t)}$ apparait en trait mauve sur la Fig. 21 entre 3 et 20ms pour 7 échelons de longueurs. La modélisation présente des plateaux de longueur variable selon l'échelon; il est relevé que pour les valeurs de raccourcissement supérieur à 8 nm, les courbes ne présentent pas de plateau.

On remarque que les départs des courbes en trait mauve de la Fig. 21, tracés à partir de l'éq. (9.6), sont particulièrement plats. En effet la convolution qui a menée à (9.5) concoure au phénomène d'aplatissement de la phase 3 à l'appui des approximations de l'annexe A. Pour des valeurs de x proches de 0, les approximations (A.3) et (A.4) mènent, proportionnellement et respectivement à $\mathbf{x}$ et $\mathbf{x^2}$ , *i.e.* la 1ere bissectrice du plan et une parabole tangente en 0.

## <u>9.4 Discussion</u>

### *Faits prédits par le modèle*

Par l'entremise de l'évt DE et grâce à la convolution mise en place pour l'évt RS, le modèle pronostique correctement le phénomène de plateau caractéristique de la phase 3.

# 10 Phase 4 d'un échelon de longueur

## 10.1 Description

La phase 4 est caractérisée par le rétablissement progressif de la tension exercée par la fmI avec approche asymptotique jusqu'à la valeur de la tension isométrique $\mathbf{T0^{fmi}}$ (Huxley 1974).

La phase 4 se termine lorsqu'une diminution de $\mathbf{T0^{fmi}}$ est observée, soit environ entre 200 et 300ms après le début de la phase 1 (Ford, Huxley et al. 1977).

## 10.2 Interprétation et modélisation de la phase 4

### *10.2.1 Remontée lente de pT vers 1*

Les événements qui concourent à l'évolution temporelle de la tension relative ont commencé à la phase 3 et leur description a été apportée au paragraphe 9.2.

La modélisation de la remontée temporelle de $\mathbf{pT}$ a été fournie au paragraphe 9.3 avec les éq. (9.4) à (9.6). Dans l'éq. (9.5), la valeur finale a été posée à 1 pour les raisons qui suivent.

$\mathbf{pT2DE}$ est la valeur de $\mathbf{pT}$ à la fin de la phase 3 qui correspond à la répartition des S1 ayant initié un WS durant la phase 2, compte tenu des S1 encore en cours de WS, les S1 en fin de WS ayant cessé toute influence; dans les régions 2 et 3, $\mathbf{pT2DE}$ appartient à la droite (trait jaune ; Fig. 22) dont l'équation est formulée dans (9.2).

$\mathbf{pT2DE}$ étant proportionnelle à la valeur de l'échelon (k) avec un taux négatif, on peut supposer que par définition d'une distribution uniforme, la répartition des S1 suit cette proportionnalité. Comme la structure géométrique inter-filamentaire est répétée à l'identique dans chaque hs de la fmI, le nombre d'occurrences de $\mathbf{SB_{fast}}$ est une constante pour la gamme de longueurs étudiées avec $\mathbf{L0^{hs}}$ proche de $\mathbf{1.1\,\mu m}$. Les S1 qui contribuent à la remontée de $\mathbf{pT}$ en initiant un WS durant les phases 3 et 4 vont se répartir uniformément conformément à l'Hyp. 2 pour compléter proportionnellement (avec un taux positif) le nombre total d'occurrences et donc mener à une valeur finale identique[1] de $\mathbf{pT}$ pour tous les échelons appartenant aux régions 2 et 3.

Or la valeur $\mathbf{\Delta X_{p1}^{(k)} = 2 \cdot d_{sitA}}$ où $\mathbf{pT = 0}$ correspond à la mise en conditions isométriques maximales de la fmI. Au 1$^{er}$ paragraphe du chapitre 6, trois conditions d'obtention de $\mathbf{T0^{fmi}}$ étaient posées dont l'une stipulait que tous les hs de la fmI possèdent la même longueur de référence $\mathbf{L0^{hs}}$.

La géométrie inter-filamentaire permet d'expliquer que $\mathbf{T0^{fmi}}$ reste constante lorsque les expérimentations sont réalisées avec $\mathbf{L0^{hs}}$ variant entre $\mathbf{1\,\mu m}$ et $\mathbf{1.125\,\mu m}$ (Gordon, Huxley et al. 1966).

---

[1] *i.e valeur appartenant à un intervalle de confiance inhérent à la variabilité biologique, mais intervalle de confiance d'étendue réduite puisque cette valeur est reproductible.*

**Fig. 22 : Remontée de la tension relative durant la phase 4 suivant le raccourcissement de la fml par une perturbation en longueur d'échelon (k)**

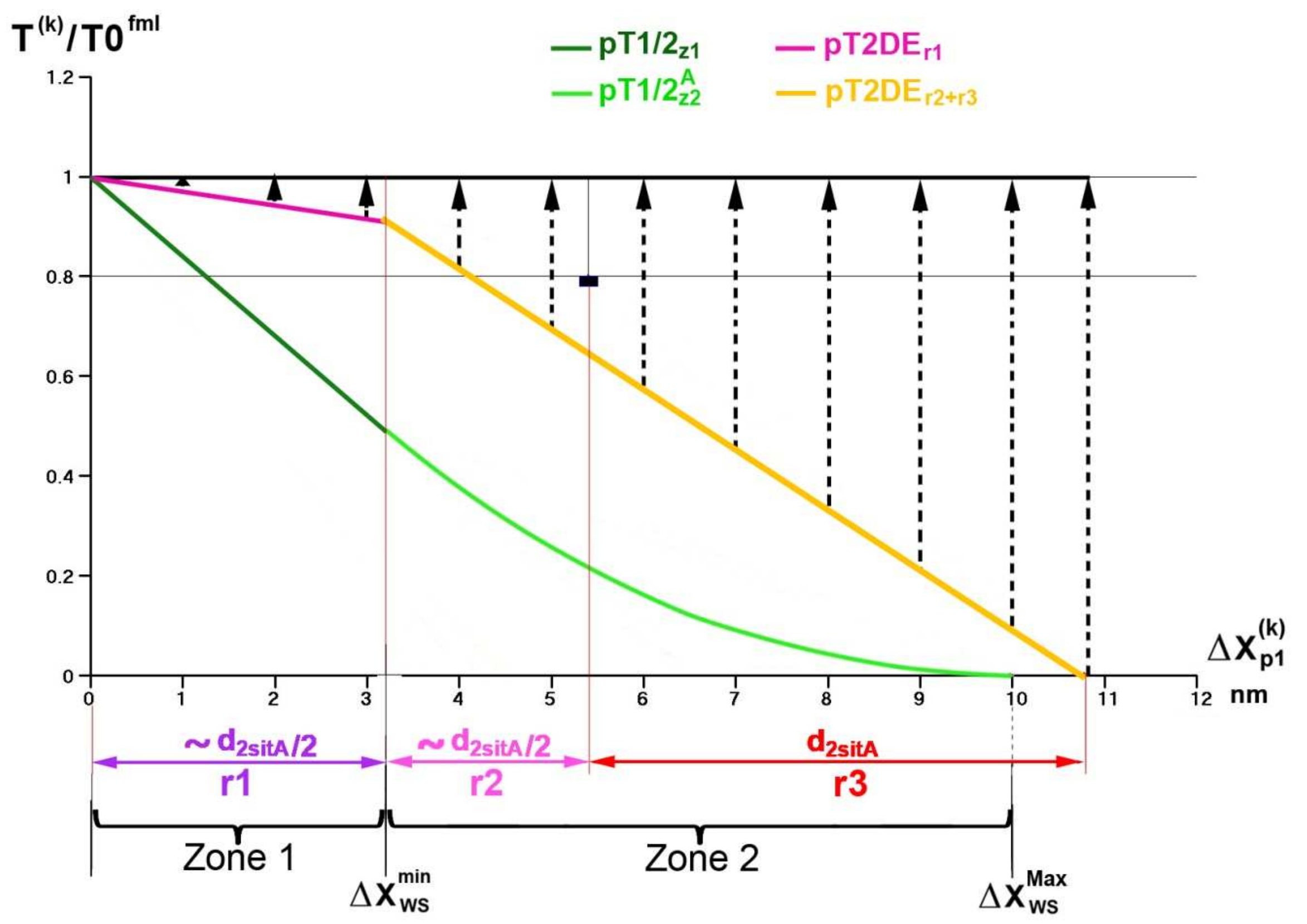

Dans le modèle, la valeur retenue est $\mathbf{L0^{hs} = 1.1\,\mu m}$ selon l'égalité (6.2). Ainsi si on avait choisi avant la mise en isométrie de la fmI une longueur égale à $\left(\mathbf{L0^{hs} - \Delta X_{p1}^{(k)}}\right)$, on aurait obtenu à nouveau $\mathbf{T0^{fmi}}$ quelque soit la valeur de l'échelon k, et notamment pour $\mathbf{\Delta X_{p1}^{(k)} = 2 \cdot d_{sitA}}$.

On doit retrouver $\mathbf{T0^{fmi}}$ pour cette valeur et d'après le raisonnement précédent pour toutes les valeurs de raccourcissement comprise dans les régions 2 et 3.

L'explication donnée pour la remontée de $\mathbf{pT}$ tendant exponentiellement vers $\mathbf{1}$ revient à conclure que la distribution des positions angulaires des S1b appartenant aux tetM démarrant un WS et en cours de WS à la fin de la phase 4 se rapproche de la répartition uniforme sur $\boldsymbol{\delta_\theta}$ en condition isométrique maximale (voir chap. 6); cette assertion est notablement vérifiée pour les régions $\mathbf{r1}$ où le déplacement est peu conséquent, et pour la 2ème partie de $\mathbf{r3}$ où les S1 ayant participé à la mise en isométrie maximale ont presque tous terminés leur WS, laissant de nouveaux S1 initier un WS.

Ce phénomène a été reconnu auparavant avec la formulation de l'Hyp. 11.

La répartition uniforme de $\Lambda 0$ S1b sur $\delta_\theta$ appartenant à $\Lambda 0$ tetM ayant initié un WS dans chaque hs de la fmI apparait ainsi comme la distribution de référence aboutissant à la force maximale de la fibre. Cette conjecture sera utilisée pour la phase 4 d'un échelon de force dont la valeur est proche de $T0^{fmI}$.

### *10.2.1 Equation générale de l'évolution temporelle de pT après la phase 1*

A l'appui des éq. (8.7a), (8.7b),(8.7c),(8.7d), (9.1), (9.4), (9.5) et (9.6), l'évolution temporelle de la remontée de la tension relative se formule[1] durant les phases 2, 3 et 4 d'un échelon de longueur :

$$
\begin{aligned}
pT(t) = & \left[ pT1 + \left( t - t^{1}_{stop_p1} \right) \cdot \frac{pT1/2 - pT1}{t^{2}_{stop_p1} - t^{1}_{stop_p1}} \right] \cdot \mathbf{1}_{\left[ t^{1}_{stop_p1} ; t^{2}_{stop_p1} \right]}(t) \\
& + \left[ pT2 + (pT1/2 - pT2) \cdot e^{-\frac{t - t^{2}_{stop_p1}}{\Delta t_{SBfast}}} \right] \cdot \mathbf{1}_{\left\{ t \geq t^{2}_{stop_p1} \right\}}(t) \\
& + pT2 + (pT2DE - pT2) \cdot \left[ 1 - e^{-\frac{t - t_{stop_p2}}{\Delta t_{DE}}} \right] \cdot \mathbf{1}_{\left\{ t \geq t_{stop_p2} \right\}}(t) \\
& + (1 - pT2DE) \cdot \left[ 1 - \frac{\Delta t_{RS} \cdot e^{-\frac{t - t_{stop_p2}}{\Delta t_{RS}}} - \Delta t_{SBfast} \cdot e^{-\frac{t - t_{stop_p2}}{\Delta t_{SBfast}}}}{\Delta t_{RS} - \Delta t_{SBfast}} \right] \cdot \mathbf{1}_{\left\{ t \geq t_{stop_p2} \right\}}(t)
\end{aligned}
\tag{10.1}
$$

Les évolutions de $pT$ selon l'éq. (10.1) pour 7 échelons multiples de 1.5 nm sont reproduites dans la Fig. 23, où sont représentées les 4 phases transitoires qui suivent un raccourcissement en longueur.

## 10.3 Discussion

Les évolutions de $pT$ calculées d'après (10.1) et représentées dans la Fig. 23 sont conformes aux allures des courbes de la littérature (voir Fig. 11 dans (Ford, Huxley et al. 1977) ; Fig. 4 dans (Ford, Huxley et al. 1981)) Elles s'ajustent correctement sur les tracés de la Fig. 3 dans (Piazzesi et Lombardi 1995).

---

[1] *La définition d'une fonction indicatrice est donnée à l'Annexe A1.*

**Fig. 23 :** Remontée temporelle de la tension relative au cours de la phase 4 lors du raccourcissement de la fml pour 7 échelons de longueur

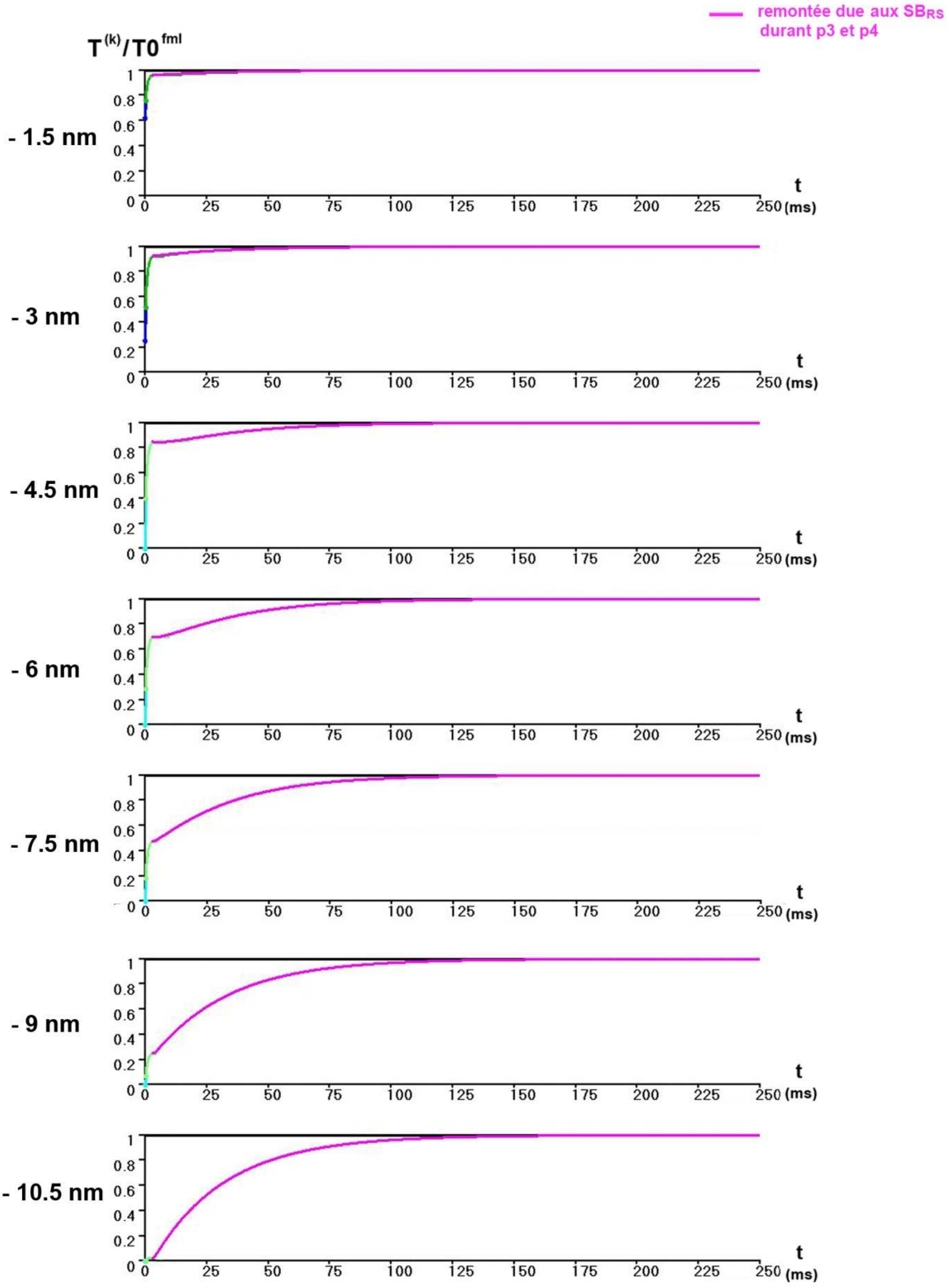

# 11 Perturbation en escalier (« *staircase shortening* »)

## 11.1 Description

Apres avoir été tétanisée isométriquement, la fmI est raccourcie selon une série d'échelons de longueur identique, espacés dans le temps par un intervalle constant (Irving, Lombardi et al. 1992 ; Lombardi, Piazzesi et al. 1992; Piazzesi et Lombardi 1995). La cinétique temporelle de cette série de raccourcissements ressemble à un escalier, d'où la dénomination. Dans un souci de simplification, un échelon est appelé « marche » et une perturbation de la fmI composée de n marches ou échelons identiques successifs est intitulée « escalier ».

La longueur de la marche rapportée à un hs et l'intervalle de temps entre 2 marches successives sont nommés, respectivement, $\Delta X_{step}^{(e)}$ et $\Delta t_{step}^{(e)}$, où $(e)$ est le numéro indiciel de l'escalier.

Les 9 escaliers mesurés et étudiés par M. Linari, V. Lombardi et G. Piazzesi (Linari, Lombardi et al. 1997) serviront de support expérimental référent aux calculs de ce chapitre.

Soit d'après la Fig. 1 de l'article[1] pour $e = 1,2,..,9$ :

$$\Delta X_{step}^{(1)} = \Delta X_{step}^{(2)} = \Delta X_{step}^{(3)} = 1.5nm \quad \text{avec} \quad \Delta t_{step}^{(1)} = 3ms,\ \Delta t_{step}^{(2)} = 2.1ms,\ \Delta t_{step}^{(3)} = 1.4ms$$

$$\Delta X_{step}^{(4)} = \Delta X_{step}^{(5)} = \Delta X_{step}^{(6)} = 3.4nm \quad \text{avec} \quad \Delta t_{step}^{(4)} = 6.1ms,\ \Delta t_{step}^{(5)} = 4.1ms,\ \Delta t_{step}^{(6)} = 3ms$$

$$\Delta X_{step}^{(7)} = \Delta X_{step}^{(8)} = \Delta X_{step}^{(9)} = 4.3nm \quad \text{avec} \quad \Delta t_{step}^{(7)} = 10ms,\ \Delta t_{step}^{(8)} = 6.1ms,\ \Delta t_{step}^{(9)} = 4.1ms$$

Les fibres musculaires utilisées lors des expériences relatives à cet article appartiennent à la même espèce de grenouille (Rana esculenta) et au même muscle (tibialis anterior) que celles des 2 publications (Piazzesi et Lombardi 1995; Piazzesi, Lucii et al. 2002) servant de base expérimentale aux chapitres 7 à 10. D'autre part, les conditions imposées par l'évt $SB_{fast}$ précédant l'initiation d'un WS sont vérifiées pour toutes les marches des 9 escaliers. Il est ainsi possible de recourir aux équations chiffrées des chapitres précédents, consacrés à la description des phases 1 à 4 d'une perturbation de la fmI par un échelon de longueur.

## 11.2 Interprétation du modèle

A l'appui de la modélisation proposée pour un échelon de longueur, la cinétique temporelle de la tension relative $pT(t)$ se décompose pour chaque marche d'indice $(m)$ de l'escalier d'indice $(e)$ en plusieurs phases successives :

- Diminution brutale de $pT$ lors de la phase 1 de raccourcissement entre $t_{deb_p1}^{(m)}$, le temps de début de la marche, et $t_{stop_p1}^{1,(m)}$, la fin de la phase 1 avec présence des forces de viscosité.
- Puis augmentation rapide de $pT$ au début de la phase 2 entre $t_{stop_p1}^{1,(m)}$ et $t_{stop_p1}^{2,(m)}$, induite par l'arrêt des actions de la viscosité car V=0.

[1] *Comme dans les chapitres précédents les raccourcissements sont considérés comme positifs contrairement à l'usage.*

- Suivie d'une augmentation exponentielle de $\mathbf{pT}$ durant le reste de la phase 2 entre $\mathbf{t^{2,(m)}_{stop_p1}}$ et $\mathbf{t^{(m)}_{stop_p2}}$ due à des S1 réalisant l'évt $\mathbf{SB_{fast}}$. les conduisant à l'initiation d'un nouveau WS
- Enfin si $\mathbf{\Delta t^{(e)}_{step} > 3ms}$, un plateau ou une augmentation modérée de $\mathbf{pT}$ durant les phases 3 et 4, due à l'initiation de nouveaux WS avec des S1 réalisant l'évt $\mathbf{RS}$ suivi de l'évt $\mathbf{SB_{fast}}$ avec initiation d'un nouveau WS

La durée entre le début de la marche $\mathbf{(m)}$ de l'escalier $\mathbf{(e)}$ jusqu'à l'arrêt des actions dues à la viscosité est égale à :

$$\mathbf{\Delta t^{(e)}_{p1/2} = t^{2,(m)}_{stop_p1} - t^{(m)}_{deb_p1}} \qquad (11.1)$$

La durée de la phase 2 consacrée à l'augmentation de $\mathbf{pT}$ due à l'initiation de nouveaux WS concernant la marche $\mathbf{(m)}$ de l'escalier $\mathbf{(e)}$ vérifie :

$$\mathbf{\Delta t^{(e)}_{p2} = t^{(m)}_{stop_p2} - t^{2,(m)}_{stop_p1} = min\left(\Delta t^{(e)}_{step}, 3ms\right) - t^{2,(m)}_{stop_p1}} \qquad (11.2)$$

où 3 ms correspond à la durée totale des phases 1 et 2 d'un échelon de longueur ; voir paragraphes 1.4.1 et 8.4 avec égalité (8.9)

On calcule $\mathbf{m_{STA}}$, le nombre de marches nécessaires pour que tous les S1 ayant participé à la mise en tension isométrique maximale initiale aient terminé leur WS, soit :

$$\mathbf{m_{STA} = int\left(\frac{\Delta X^{Max}_{WS}}{\Delta X^{(e)}_{step}}\right) + 1} \qquad (11.3)$$

où le sigle « int » signifie partie entière

Soit d'après les données de l'article :

pour e = 1 à 3 $\mathbf{m_{STA} = int(10/1.5) + 1 = 7}$

et pour e = 4 à 9 $\mathbf{m_{STA} = int(10/3.4) + 1 = int(10/4.3) + 1 = 3}$

Pour chacun des 9 escaliers (voir Fig. 1 dans (Linari, Lombardi et al. 1997)), on observe à partir de la 1[ère] marche une diminution linéaire de la cinétique de la tension pour les marches successives tant que la somme des raccourcissements imposés reste inférieur à $\mathbf{\Delta X^{Max}_{WS}}$, puis cette cinétique devient quasiment reproductible à l'identique lors des échelons postérieurs, *i.e* pour $\mathbf{m > m_{STA}}$. Autrement formulé, une fois que tous les S1 ayant participé à la mise en tension isométrique ont achevé leur WS, une cinétique répétitive est atteinte (Fig. 24).

Avec (1.11) et (11.3), la mise en équation de cet état stationnaire pour les marches d'indice (m) tel que $\mathbf{m > m_{STA}}$ mène à :

$$\Delta pT^{(e)}_{SB_{fast};STA} = \left[ \left( pT2^{(e)}_{STA} - pT1/2^{(e)}_{STA} \right) \cdot \left( 1 - e^{-\frac{\Delta t^{(e)}_{p2}}{\Delta t_{SBfast}}} \right) \right] \cdot \mathbf{1}_{\left[ t^{2,(m)}_{stop_p1};\, t^{(m)}_{stop_p2} \right]}(t) \tag{11.4}$$

où

$\Delta pT^{(e)}_{SB_{fast};STA}$ caractérise l'augmentation de $\mathbf{pT}$ due à l'initiation de nouveaux WS durant la phase 2 de chaque marche

$pT1/2^{(e)}_{STA}$ est une constante qui caractérise la valeur de $\mathbf{pT}$ au temps $t = \left( t^{(m)}_{deb_p1} + \Delta t^{(e)}_{p1/2} \right)$

$pT2^{(e)}_{STA}$ est une constante qui caractérise la valeur de $\mathbf{pT}$ au temps $t = \left( t^{(m)}_{deb_p1} + 3\,ms \right)$

$\Delta t^{(e)}_{p2}$ et $\Delta t^{(e)}_{p1/2}$ sont apportée par (11.1) et (11.2)

$\Delta t_{SBfast} \approx 0.7\ ms$ selon l'Hyp. 3 et (1.13)

L'équation (11.4) n'est correcte que pour les escaliers dont le laps de temps entre 2 marches, $\Delta t^{(e)}_{step}$, permet d'atteindre l'état stationnaire de la phase 2 ; ainsi selon (8.9), $\Delta t^{(e)}_{step}$ doit égale où supérieure à 3 ms ; dans l'exemple référent, cette condition est vérifiée pour 7 des 9 escaliers.

D'après l'Hyp. 2, les S1 initiant un WS à chaque marche $\mathbf{(m)}$ de l'escalier $\mathbf{(e)}$ se répartissent uniformément sur $\delta_X$ puis, à chaque marche suivante, leur contribution est fournie par les éq. (4.35a) et (4.36a) pour les zone 1 et 2, respectivement (les S1 en fin de WS n'ayant pas le temps de se détacher, le mode requis est « Exagéré »).
En divisant chaque membre de ces équations par $\mathbf{T0^{fmI}}$, on obtient pour l'état stationnaire :

pour la zone 1 $$pT^{(m)}_{z1}(\Delta X) = \Delta pT^{(e)}_{SB_{fast};STA} \cdot \left( 1 - \Delta X \cdot \chi^{hs}_{z1} \right) \tag{11.5a}$$

pour la zone 2 $$pT^{(m)}_{z2}(\Delta X) = \Delta pT^{(e)}_{SB_{fast};STA} \cdot \left( \Delta X^{Max}_{WS} - \Delta X \right) \tag{11.5b}$$

Ainsi dès que $\mathbf{m > m_{STA}}$, le terme $pT1/2^{(e)}_{STA}$ est égal avec (11.5a) et (11.5b) à :

$$pT1/2^{(e)}_{STA} = \Delta pT^{(e)}_{SB_{fast};STA} \cdot C^{(e)}_{STA} \tag{11.6}$$

où $C^{(e)}_{STA}$ est une constante propre à chaque escalier $\mathbf{(e)}$ telle que :

$$C^{(e)}_{STA} = \sum_{m=1}^{(m_{STA}-1)} \left[ si\left( m \cdot \Delta X^{(e)}_{step} < \Delta X^{min}_{WS} \right), \left( 1 - m \cdot \Delta X^{(e)}_{step} \cdot \chi^{hs}_{z1} \right) et\, sin\, on \left( \chi^{hs}_{z2} \cdot \left( \Delta X^{Max}_{WS} - m \cdot \Delta X^{(e)}_{step} \right) \right) \right] \tag{11.7}$$

En introduisant (11.6) et (11.7) dans (11.4), on trouve :

$$\Delta pT^{(e)}_{SB_{fast};STA} = \left(pT2^{(e)}_{STA} - C^{(e)}_{STA} \cdot \Delta pT^{(e)}_{SB_{fast};STA}\right) \cdot \left(1 - e^{-\frac{\Delta t^{(e)}_{p2}}{\Delta t_{SBfast}}}\right)$$

On en déduit :

$$\Delta pT^{(e)}_{SB_{fast};STA} = \frac{pT2^{(e)}_{STA} \cdot \left(1 - e^{-\frac{\Delta t^{(e)}_{p2}}{\Delta t_{SBfast}}}\right)}{1 + C^{(e)}_{STA} \cdot \left(1 - e^{-\frac{\Delta t^{(e)}_{p2}}{\Delta t_{SBfast}}}\right)} \qquad (11.8)$$

***<u>Application numérique</u>***

Avec les données des chapitres précédents, tous les termes de (11.8) sont calculables à l'exception de $pT2^{(e)}_{STA}$, valeur relevée sur la dernière marche de chaque escalier $(e)$ de la Fig. 1 (Linari, Lombardi et al. 1997); la valeur est prise à la fin de la phase 2, soit 3 ms, excepté pour les escaliers n° 2 et 3.

Les calculs de $\Delta pT^{(e)}_{SB_{fast};STA}$ effectués à partir de 11.8 apparaissent dans la colonne 7 du Tab. 4.

En appliquant l'éq. (10.1) avec les valeurs du tableau précédent, il est possible de tracer les évolutions temporelles pour chacun des 9 escaliers (Fig. 24).

La contribution due à l'arrêt des actions de viscosité est calculée par différence entre $pT1/2$ et $pT1$ pour la 1ère marche. Cette contribution est ensuite reportée dans tous les calculs relatifs aux marches suivantes de l'escalier $(e)$ (col 3; Tab. 4), puisque les actions de viscosité portent sur les mêmes éléments constitutifs de la myofibrille (disqM et disqZ) et sont donc identiques d'une marche à l'autre.

La 1ère marche est calculée à partir des valeurs de $pT1/2$ et $pT2$, apportées par les équations des chapitres 7 et 8. Ces valeurs sont identiques pour les mêmes valeurs de $\Delta X^{(e)}_{step}$ (col 4; Tab. 4).

Pour la 2ème marche, on calcule le terme $pT1/2$ sur le même principe que pour le calcul de l'éq. (11.7) : on somme la contribution des SI ayant initiés un WS en isométrie maximale déplacés de $2 \cdot \Delta X^{(e)}_{step}$ et la contribution des $\Delta pT^{(e)}_{SB_{fast}}$ de la 1ère marche (col 4 ; Tab. 4) déplacées de $\Delta X^{(e)}_{step}$.

**TAB. 4 : Valeurs introduites dans le calcul de l'évolution de pT pour 9 escaliers d'après la Fig. 1 de l'article de M. Linari (1997)**

| | 1 | 2 | 3 | 4 | 5 | 6 | 7 |
|---|---|---|---|---|---|---|---|
| (e) | $\Delta X^{(e)}_{step}$ (nm) | $\Delta t^{(e)}_{step}$ (ms) | $\Delta pT^{(e)}_{Visco}$ ∓ | $\Delta pT^{(e)}_{SBfast}$ ∓ 1ère marche | $\Delta pT^{(e)}_{SBfast}$ ⊗ 2ème marche | $\Delta pT^{(e)}_{SBfast}$ ⊗ Marches suivantes | $\Delta pT^{(e)}_{SBfast;STA}$ ∓ |
| **1** | 1.5 | 3.0 | 0.09 | 0.2 | 0.22 | 0.22 | 0.22 |
| **2** | 1.5 | 2.1 | 0.09 | 0.2 | 0.21 | 0.18 | |
| **3** | 1.5 | 1.4 | 0.09 | 0.2 | 0.21 | 0.195 | |
| **4** | 3.4 | 6.1 | 0.21 | 0.43 | 0.385 | 0.40 | 0.37 |
| **5** | 3.4 | 4.1 | 0.21 | 0.43 | 0.33 | 0.33 | 0.3 |
| **6** | 3.4 | 3.0 | 0.21 | 0.43 | 0.3 | 0.3 | 0.25 |
| **7** | 4.3 | 10.0 | 0.28 | 0.45 | 0.47 | 0.47 | 0.46 |
| **8** | 4.3 | 6.1 | 0.28 | 0.45 | 0.38 | 0.38 | 0.35 |
| **9** | 4.3 | 4.1 | 0.28 | 0.45 | 0.35 | 0.29 | 0.28 |

∓ ***Valeurs calculées d'après modèle***

⊗ ***Valeurs adoptées pour tracés de la Fig. 24***

Idem pour la 3ème marche : on somme la contribution des SI ayant initiés un WS en isométrie déplacées de $3 \cdot \Delta X^{(e)}_{step}$ (comptabilisée nulle si $3 \cdot \Delta X^{(e)}_{step} > \Delta X^{Max}_{WS}$), la contribution des $\Delta pT^{(e)}_{SBfast}$ de la 1ère marche (col 4 ; Tab. 4) déplacées de $2 \cdot \Delta X^{(e)}_{step}$, la contribution des $\Delta pT^{(e)}_{SBfast}$ de la 2ème marche (col 5 ; Tab. 4) déplacées de $\Delta X^{(e)}_{step}$.

… etc.

Pour toutes les marches d'indice supérieur à 3, on prend la valeur de la colonne 6 du Tab. 4.

La modélisation des 9 escaliers a nécessité des adaptations pour les valeurs de $\Delta pT^{(e)}_{SBfast}$ des marches n° 1 et n° 2, comme signalé dans les colonnes respectives 4 et 5 du Tab. 4.

Concernant les autres marches, la valeur stationnaire calculée à l'aide de (11.8) a été remplacée par une valeur empirique proche (col. 6 ; Tab. 4).

Pour les échelons de durée supérieurs à 3 ms, la contribution des $SB_{RS}$ est tenue pour quantité négligeable dans les calculs. Les pentes en mauves qui apparaissent sur le graphique sont tracées artificiellement après la fin de la phase 2.

## Fig. 24 : Modélisation de 9 perturbations en escalier *(d'après Linari, 1997)*

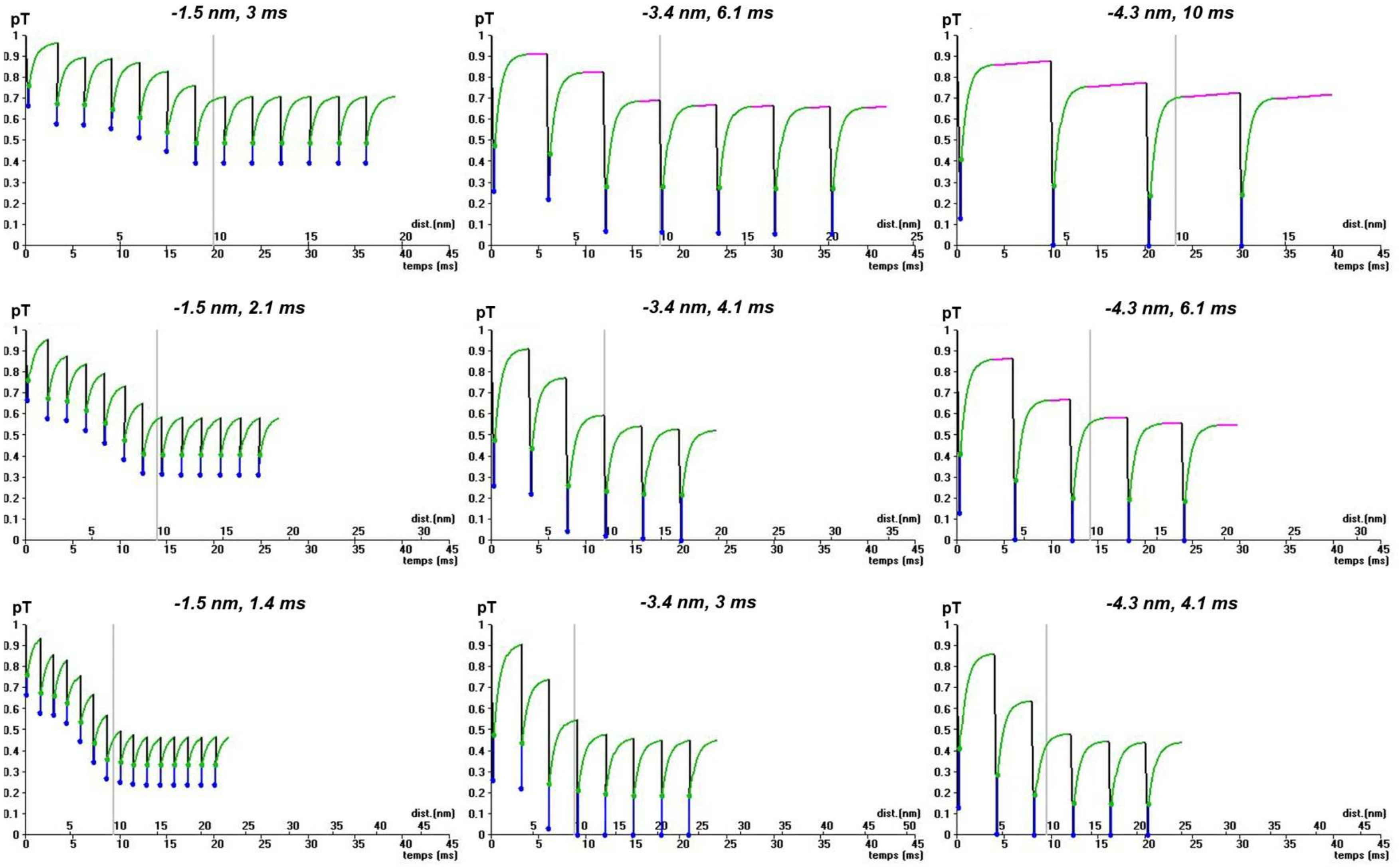

## 11.3 Discussion

A nouveau, l'importance de la viscosité est relevée, car sa prise en compte permet d'interpréter l'ensemble des 9 perturbations avec une seule constante de temps, $\Delta t_{SBfast}$ .

Des divergences apparaissent entre les tracés du modèle et les mesures expérimentales :

- concernant les valeurs de pT1, les auteurs précisent que la durée de la phase 1, $\Delta t_{p1}$, n'est pas identique pour tous les échelons alors qu'elle est constante $(\Delta t_{p1} = 150\ \mu s)$ pour l'ensemble des marches du modèle.
- la contribution due à l'évènement DE, qui doit intervenir dès que la durée de la perturbation devient supérieure à une dizaine de ms, n'a pas été comptabilisée.
- la contribution due à l'évènement RS été négligée

Les valeurs adoptées dans la col. 6 du Tab. 4 sont comparables aux valeurs calculées dans la col. 7 à l'aide de (10.8), validant ainsi la modélisation.

Dans le tableau 4, on note :

<u>$\mathbf{pour\Delta X_{step}^{(e)} = 1.5nm}$</u> , les valeurs calculées dans les col. 4 et 7 sont proches des valeurs empiriques des col 5 et 6, indiquant que la répétition des échelons perturbent modérément les positions angulaires des S1b; ainsi les distributions diffèrent peu, autorisant un taux d'initiation de WS à peu près identique pour chacune des marches. Les valeurs de pT2 restent similaires à celles calculées au chap. 8.

<u>$\mathbf{pour\Delta X_{step}^{(e)} = 3.4nm\ \&\ 4.3nm}$</u>, il y a diminution importante du taux d'initiation de WS, jusqu'à 70% entre la 1$^{ère}$ (col. 4) et la 3$^{ème}$ marche (col. 6), *i.e.* dès que $\mathbf{m > m_{STA}}$ , signalant que la répétition des marches perturbent les postions angulaires des S1b et consécutivement diminue les possibilités de d'enclenchement pour de nouveaux WS ; globalement, les valeurs de $\mathbf{pT2}$ sont inférieures à celles présentées au chap. 8.

L'augmentation de la durée des échelons semblent favoriser le repositionnement angulaire de S1b; par exemple, pour l'escalier n° 7 où $\mathbf{\Delta t_{step}^{(7)} = 10ms}$, on retrouve des valeurs calculées dans les col. 4 et 7 presque égales à celles des valeurs empiriques des col. 5 et 6, indiquant des taux de renouvellement de WS proches.

On retrouvera ce problème de modification des valeurs de $\mathbf{pT2}$ en fonction de la vitesse de raccourcissement continu lors de la phase 4 d'une perturbation selon un échelon de force (voir chap. 15).

# 12 Phase 1 d'un échelon de force (« *step force* »)

## 12.1 Description

Apres avoir été tétanisée isométriquement, la tension appliquée à la fmI est diminuée continument et linéairement jusqu'à une valeur $\mathbf{T^{(j)}}$ pendant une durée de temps inférieure à 0.2 ms (Huxley 1974; Piazzesi, Lucii et al. 2002 ); cette période transitoire est la phase 1 d'une perturbation par un échelon de force $\mathbf{T^{(j)}}$, j étant le numéro indiciel de l'échelon, tel que :

$$\mathbf{0 < pT^{(j)} \leq 1} \tag{12.1}$$

avec $\mathbf{pT^{(j)} = T^{(j)} / T0^{fmI}}$

La Fig. 13B (chap. 7) offre 3 exemples d'échelons de forces; l'échelon (0) correspond au cas isométrique étudié au chapitre 6.

La durée de la phase 1 est une constante appelée $\mathbf{\Delta t_{p1}^{F}}$, commune à tous les échelons. Dans l'exemple de la Fig. 13B, $\mathbf{\Delta t_{p1}^{F} = \Delta t_{p1}}$ = 150 µs.

La valeur $\mathbf{T^{(j)}}$ est maintenue constante durant les phases suivantes.

## 12.2 Interprétation du modèle : équivalence avec la phase 1 d'un échelon de longueur

Au chapitre 7, la modélisation de la phase 1 d'une perturbation de la fmI par un échelon de longueur a mené à 2 relations linéaires entre tension relative et longueur de raccourcissement de la fmI, avec pour équations (7.3b) et (7.4b) respectivement dans les zones 1 et 2. La modélisation linéaire indique que le comportement de la fmI est analogue à celui d'un ressort mécanique linéaire dans chacune de ces 2 zones.

La description du paragraphe précédent et les relations (7.3b) et (7.4b) impliquent que le raccourcissement s'effectue durant $\mathbf{\Delta t_{p1}^{F}}$ avec une vitesse constante nommée $\mathbf{V_{p1}^{(j)}}$ ; ainsi les équations des chap. 2 à 6 sont valables durant la phase 1, quelque soit la valeur de l'échelon (j). Il s'en déduit que les phases 1 d'un échelon de longueur et d'un échelon de force, réalisées dans des conditions similaires, sont identiques jusqu'à $\mathbf{t_{fin_p1}^{1} = \Delta t_{p1}^{F} = \Delta t_{p1}}$ où $\mathbf{V = V_{p1}^{(j)}}$ (Fig. 13A, 13B et 13C; chap. 7).

Ainsi, les relations linéaires (7.3a) et (7.4a) entre tension relative et longueur de raccourcissement d'un hs de la fmI, respectivement dans les zones 1 et 2, sont utilisables pour un échelon de force (traits bleus ; Fig. 14 ; chap. 7).

## 12.3 Discussion

### *Faits observés prédits par le modèle:*

**Superposition des valeurs entre échelon de force et échelon de longueur**

Pour des expériences menées dans des conditions similaires, les relations entre longueur de raccourcissement d'un hs et tension relative étant identiques entre les 2 types de perturbation, les points mesurés pour les échelons de longueur doivent se superposer avec ceux mesurés pour les échelons de force; ce phénomène est souligné par G. Piazzesi et ses coauteurs : voir Fig. 4 dans (Piazzesi, Lucii et al. 2002).

**Variation de la pente de $pT1_{z1}(\Delta X_{p1})$ selon des variations de température expérimentale**

Pour les mêmes raisons développées au paragraphe 7.6.2, on doit noter une diminution de la contribution relative des forces de viscosité lorsque la tension relative est calculée par rapport à la tension maximale isométrique de chacun des tests en température.

Ce pronostic est confirmé durant les phases 1 de différents échelons en force pour une gamme de température expérimentale variant de 2 à 17°C (Decostre, Bianco et al. 2005).

Au paragraphe 6.4.3, suite à une augmentation de la température expérimentale, il avait été prédit une augmentation de $\delta_\theta$ , observée expérimentalement (Linari, Brunello et al. 2005). On devrait dans ce cas noter une diminution de la valeur de $\Delta X_{WS}^{min}$ et donc de la zone 1, et en conséquence observer une cassure entre les 2 pentes définies dans (7.3a) et (7.4a) à une valeur d'abscisse légèrement inférieure à $\Delta X_{WS}^{min}$.

Cette prévision est observée[1] (voir Fig. 1B dans (Decostre, Bianco et al. 2005)) avec $\Delta X_{WS}^{min}$ = 3.2 nm.

[1] *Il est possible d'utiliser les données du chapitre 7, puisque les expérimentations dans cet article ont été réalisées avec des fibres prélevées chez la même espèce de grenouille et sur le même muscle que ceux utilisés pour les calculs du modèle.*

# 13 Phase 2 d’un échelon de force

## 13.1 Description

A la fin de la phase 1, la tension exercée sur la fmI est maintenue constante à $\mathbf{T^{(j)}}$ ; il est observé sur une durée inférieure à 5 ms un raccourcissement rapide qui caractérise la phase 2 (Huxley 1974). Des données plus récentes suggèrent que cette durée peut s’étendre au-delà de 5 ms pour les valeurs de tension supérieures à $\mathbf{0.7 \cdot T0^{fmi}}$, jusqu'à 11 ms pour $\mathbf{0.8 \cdot T0^{fmi}}$ (Piazzesi, Lucii et al. 2002).

Plusieurs équipes de chercheurs ont relevé que les valeurs de la tension relative se répartissent en 2 secteurs, notés **s1** et **s2** (Fig. 25), où s’établissent des règles différentes pour les phases 2, 3 et 4 d’un échelon de force :

**s1 où $\mathbf{0.8 < pT^{(j)} < 1}$**

Les phases 2 et 3 n’apparaissent pas (Edman 1988; Piazzesi, Lucii et al. 2002 ), la phase 4 succédant directement à la phase 1 ; ou autrement dit, les phases 2, 3 et 4 sont indistinctes.

**s2 où $\mathbf{0 < pT^{(j)} \leq 0.8}$**

a) Plus l’échelon j est important, plus $\mathbf{pT^{(j)}}$ diminue, plus $\mathbf{\Delta t_{p2}^{(j)}}$ la durée de la phase 2 à l’échelon n° j est courte, et plus la distance de raccourcissement est importante.
b) Durant la phase 2, la vitesse de raccourcissement nommée $\mathbf{V_{p2}^{(j)}}$ n’est pas constante. Elle s’amortit progressivement jusqu’à une réduction conséquente qui caractérisera la phase 3.
c) A la fin de la phase 2, les points ayant pour abscisses les longueurs de raccourcissements mesurées à cet instant et pour ordonnées les échelons de force[1] $\mathbf{pT^{(j)}}$ respectifs, s’alignent sur la droite[2] d’éq. (8.6c) qui relie $\mathbf{\Delta X_{p1}^{(k)}}$ et $\mathbf{pT2^{(k)}}$, la tension relative à la fin de la phase 2 d’un échelon de longueur[1] n° k (voir Fig. 4 dans (Piazzesi, Lucii et al. 2002)).

## 13.2 Phénomènes participant au raccourcissement de la fmI durant la phase 2

Au chapitre 4, la modélisation de la tension produite par un ensemble de S1, ayant initié leurs WS puis étant déplacés collectivement, dégageait 2 types d’allure (Fig. 10 ; chap. 4), et donc 2 zones respectives, concernant les raccourcissements associées à 2 zones relatives aux tensions relatives; séparées par la valeur $\mathbf{pT = 0.49}$ d’après (4.35b). Pour simplifier, cette valeur est arrondie à $\mathbf{0.5}$.

---

[1] *Les indices (j) et (k) caractérisent, respectivement, un échelon de force et un échelon de longueur.*
[2] *Les points de la droite sont des valeurs moyennes tirées d’expériences réalisées avec plusieurs fibres musculaires.*

**Fig. 25 : Raccourcissement d'un hs de la fml au début de la phase 2 après une perturbation en force d'échelon (j)**

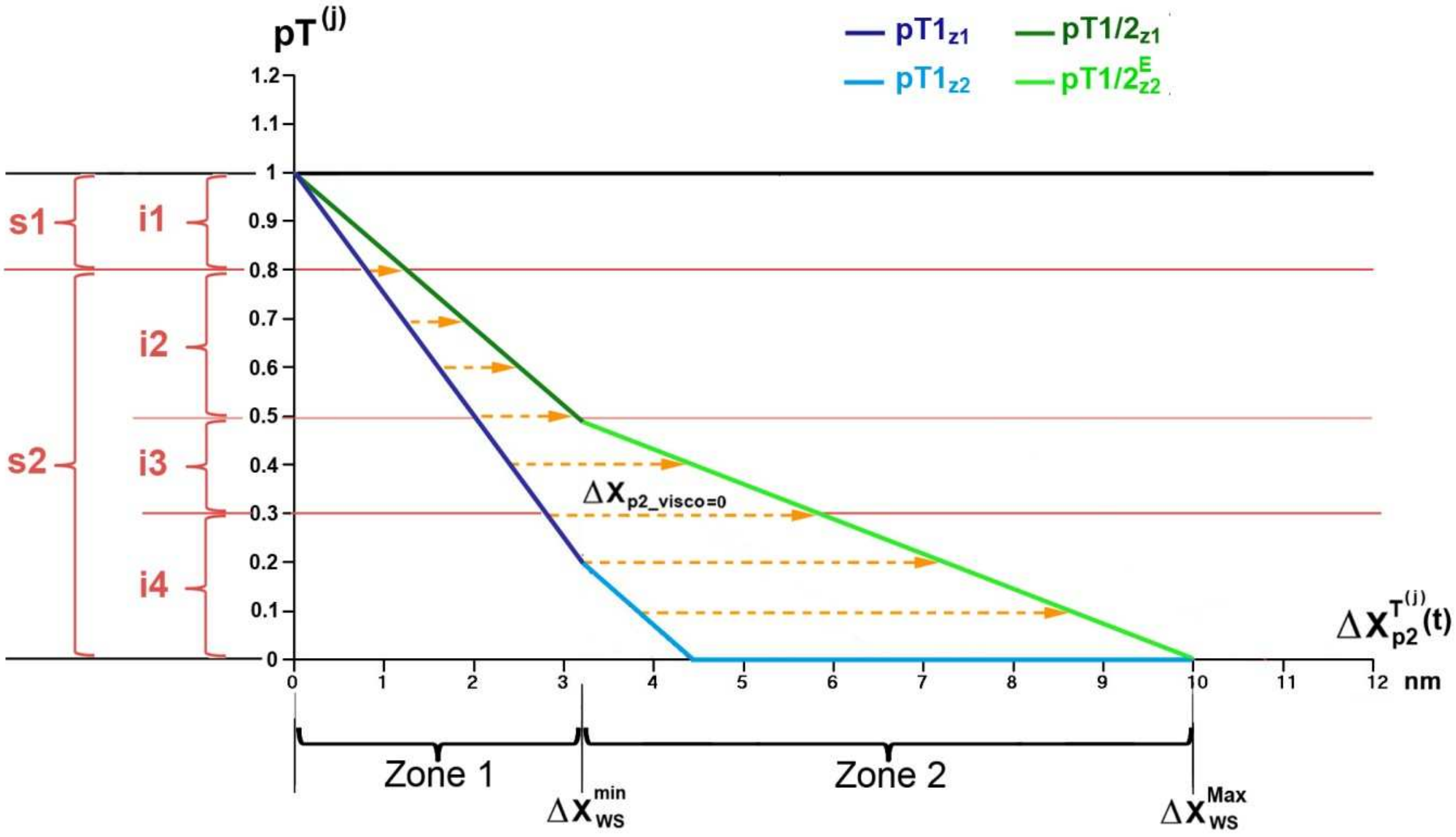

Sur cette base, les 2 secteurs **s1** et **s2** sont transformés en 4 intervalles de tension relative, notés **i1**, **i2**, **i3** et **i4** (Fig. 25), tels que :

$$\mathbf{i1\ où\ 0.8 \leq pT^{(j)} \leq 1} \qquad (13.1a)$$

$$\mathbf{i2\ où\ 0.5 \leq pT^{(j)} < 0.8} \qquad (13.1b)$$

$$\mathbf{i3\ où\ 0.3 \leq pT^{(j)} < 0.5} \qquad (13.1c)$$

$$\mathbf{i4\ où\ 0 < pT^{(j)} < 0.3} \qquad (13.1d)$$

On désigne $\mathbf{\Delta X_{p2}^{T^j}(t)}$ le déplacement linéaire instantané, commun[1] à tous les hs de la fmI, durant la phase 2 après une perturbation en force d'échelon (j) à partir de la position initiale en isométrie $\mathbf{L_0^{hs}}$.

Durant la phase 2, une difficulté se présente car la vitesse n'étant plus constante, les équations des chapitres 2 à 5, ne sont plus valides.

1 *Ce fait est admis par cohérence avec les résultats précédents, même si cela n'a pas été démontré pour une vitesse non constante.*

### *13.2.1 Annulation des actions dues à la viscosité au début de la phase 2*

Au chapitre précédent, la fmI, les myofibrilles et les hs ont été assimilées durant la phase 1 à des ressorts mécaniques. Si la tension d'un ressort mécanique est modifiée rapidement, celui-ci s'immobilise après quelques oscillations dans la nouvelle position d'équilibre (V=0) décrite par la relation linéaire entre force et variation de longueur selon la loi classique de Hooke, la force exercée sur le ressort étant compensée par la tension du ressort.

Au début de la phase 2 d'un échelon de force, la fmI ne s'immobilise pas mais poursuit immédiatement son raccourcissement.

En effet pour chaque hs de la fmI, la position d'équilibre (V=0) ne situe pas sur les droites paramétrant la fin de la phase 1 classique (traits bleu foncé et clair dans les zones 1 et 2 ; Fig. 25) mais sur les 2 droites représentant l'interruption des actions due à la viscosité. Cette position d'équilibre a été modélisée au chapitre 7 par 2 segments de droite dans les zones 1 et 2, d'équations respectives (7.8) et (7.9) (trait vert foncé et clair; Fig. 25) ; l'éq. (7.9) dans la zone 2 se réfère au mode E (Exagéré) car les S1 en fin de WS n'ayant pas le temps de se détacher exercent une influence exagérée selon l'Hyp. 13, les égalités (1.26) et (1.27), et les considérations du paragraphe 4.3.4.3.

### *13.2.2 Variation de la quantité de mouvement de la fmI et amortissement d'origine visqueuse*

Considérons une masse $\mathbf{m}$ attachée à un ressort qui se déplace horizontalement avec frottement visqueux. Lorsqu'on relâche la masse $\mathbf{m}$ hors de sa position d'équilibre avec une vitesse initiale non nulle, on observe un phénomène d'amortissement autour de la position d'équilibre puis immobilisation de la masse. Ce problème classique de mécanique est décrit à l'aide du théorème de la résultante cinétique appliquée à la masse $\mathbf{m}$, *i.e.* en égalant la force engendrée par la variation temporelle de la quantité de mouvement de $\mathbf{m}$ avec la somme algébrique de la force de rappel du ressort et de la force de frottement visqueux, proportionnelle à la vitesse.

Lorsque l'expérience se déroule sans masse $\mathbf{m}$, on observe aussi ce phénomène où la variation de la quantité de mouvement provient de la masse propre du ressort. Ce fait s'applique à la fmI à la fin la phase 1 lorsque les forces de viscosité vont s'annuler au fur et à mesure de la diminution de la vitesse de raccourcissement : la force créée par la variation temporelle de la quantité de mouvement de la fmI[1] est égale à la somme algébrique de la force motrice totale des S1 en cours de WS, des forces de viscosité et de $\mathbf{T^{(j)}}$, la force exercée sur la fmI.

Il en résulte un mouvement d'amortissement où la vitesse n'est pas constante, passant de la vitesse de début de phase 2 notée $\mathbf{V^{(j)}_{start_p2}}$ à $\mathbf{V=0}$ qui correspondant à la position d'équilibre (traits vert foncé et clair ; Fig. 25) ; conséquemment, chaque hs entrainé par la force décélératrice d'amortissement se raccourcit vers la position d'équilibre.

A l'aide des équations définissant la relation entre $\mathbf{pT}$ et $\mathbf{\Delta X}$ pour $\mathbf{pT1_{z1}}$, $\mathbf{pT1_{z2}}$, $\mathbf{pT1/2_{z1}}$ et $\mathbf{pT1/2^{E}_{z2}}$ soit, respectivement, (7.3a), (7.4a), (7.8) et (7.9), il est possible de calculer la différence entre la valeur du raccourcissement à la fin de la phase 1 et la valeur du raccourcissement correspondant à la position d'équilibre.

---

[1] *Voir calcul de la quantité de mouvement de la fmI au paragraphe E1 de l'annexe E. On rappelle que la masse de la fmI est constituée des masses des (disqM+2·filM) et (disqZ+2·filA), les masses des tetM comprises.*

Cet écart est appelé $\Delta X^{(j)}_{p2_visco=0}$ pour chaque valeur de $pT^{(j)}$ (Fig. 25). En rapprochant ces calculs avec les données de G. Piazzesi et de ses coauteurs (voir Fig. 1 dans (Piazzesi, Lucii et al. 2002)), on constate que la durée du raccourcissement associée à $\Delta X^{(j)}_{p2_visco=0}$ est approximativement constante et égale à $0.7\,ms$ .quelque soit l'échelon j.

A ce stade de la phase 2, deux évènements peuvent se manifester :

- La force motrice des S1 en cours de WS n'est pas suffisante pour s'opposer à $T^{(j)}$ et la fmI cesse de se raccourcir et s'immobilise ou s'allonge
- La force motrice des S1 en cours de WS est suffisante pour compenser à $T^{(j)}$ et la fmI continue de se raccourcir au delà de la position d'équilibre: cela suppose l'apparition de nouvelles forces motrices, phénomène décrit au paragraphe suivant

Une instabilité peut naître entre ces 2 évènements et engendrer une bifurcation avec phénomène critique et apparition d'oscillations perdurant lors de la phase 4 (Huxley 1974).

### *13.2.3 Nouveaux WS au cours de la phase 2*

On rappelle qu'en conditions de raccourcissement continu, les S1 ne transitent pas de l'état WB vers l'état WS par l'évt $SB_{fast}$ mais par l'évt $SB_{slow}$ d'après les Hyp. 4 et 6, et que les phénomènes décrits dans ce chapitre ne sont établis que pour $0 < pT^{(j)} \le 0.8$.

Suite au raccourcissement provoqué par un échelon de force, les S1 de chaque hs de la fmI se répartissent durant la phase 2 en 4 catégories :

$\underline{1/S1^{startWS}_{iso} \rightarrow S1^{WS}_{p2}}$ : les S1 ayant contribué à la tension du plateau tétanique initial sont en cours ou en fin de WS (état WS avec chemin 6 ; Fig. 1). Leurs forces motrices est la principale contribution à la force opposée à $T^{(j)}$ à la fin de la phase 2, lorsque la vitesse de raccourcissement après avoir fortement diminuée devient modérée ou faible.

$\underline{2/S1^{WB}_{iso} \rightarrow S1^{DE}_{p2}}$ : les S1 initialement en WB sont déplacés durant $\Delta t^{F}_{p1}$ d'approximativement $\Delta X^{T^j}_{p2}(t)$ et sont susceptibles de se détacher, poussés par S2 ou sous l'action des forces thermiques. Durant les phases transitoires suivantes, ces S1 seront considérés comme détachés (chemin 2 vers état DE; Fig. 1).

$\underline{3/S1^{WB}_{iso} \rightarrow S1^{SB_{slow}}_{p2}}$ : lors d'un raccourcissement continu, des S1 peuvent transiter par l'évt $SB_{slow}$ conformément aux Hyp. 5 et 6, et produire un WS. La probabilité de survenue de cet évènement est donnée par (1.14) avec la valeur de la constante de temps fournie par l'intervalle de valeurs défini dans (1.17).

On vérifie que plus $\mathbf{pT^{(j)}}$ baisse, plus la durée de la phase 2 décroit, et plus la probabilité de survenue d'un $\mathbf{SB_{slow}}$ avec initiation d'un WS diminue.

<u>$\mathbf{4/S1_{iso}^{DE} \rightarrow S1_{p2}^{DE}\ ou\ S1_{p2}^{RS}}$</u> : les S1 restants sont détachés (état DE ; Fig.1). Le raccourcissement continu ne favorise pas le redressement. Ces S1 seront considérés comme inopérants ou détachés.

## 13.3 Interprétation du modèle :

### *13.3.1 Le raccourcissement est décomposé en 2 temps et analysé dans 3 intervalles*

G. Piazzesi et ces coauteurs ont caractérisée une valeur du raccourcissement du hs correspondant à la fin de la phase 2 pour chaque valeur de $\mathbf{pT^{(j)}}$ (voir Fig. 1 dans (Piazzesi, Lucii et al. 2002)); cette valeur est nommée $\mathbf{\Delta X_{stop_p2}^{T^j}}$.

Les points d'abscisse $\mathbf{\Delta X_{stop_p2}^{T^j}}$ et d'ordonnée $\mathbf{pT^{(j)}}$ s'alignent sur une droite qui se superpose à la droite reliant les points d'abscisse $\mathbf{\Delta X_{p1}^{(k)}}$ et d'ordonnée $\mathbf{pT2_{d2}^{(k)}}$ (Piazzesi, Lucii et al. 2002), ce segment de droite apparait en trait rouge dans **r3** **(**Fig. 17 ; chap. 8), où $\mathbf{pT2_{d2}^{(k)}}$ est la tension relative à la fin de la phase 2 de l'échelon de longueur n° k.

Nous avons reproduit la droite des points relatifs à $\mathbf{pT2_{d2}^{(k)}}$ modélisée par (8.2b) ou (8.6c) en trait rouge sur la Fig. 26.

A l'aide des équations définissant la relation entre $\mathbf{pT}$ et $\mathbf{\Delta X}$ pour $\mathbf{pT1/2_{z1}}$, $\mathbf{pT1/2_{z2}^{E}}$ et $\mathbf{pT2_{d2}}$ soit, respectivement, (7.8), (7.9) et (8.2b), il est possible de calculer la différence entre la valeur de raccourcissement correspondant à la position d'équilibre $\mathbf{V = 0}$ et $\mathbf{\Delta X_{stop_p2}^{T^j}}$. Cet écart est appelé $\mathbf{\Delta X_{p2_amort}^{(j)}}$ pour chaque valeur de $\mathbf{pT^{(j)}}$ (Fig. 26), tel que :

$$\mathbf{\Delta X_{stop_p2}^{T^j} = \Delta X_{p1}^{(j)} + \Delta X_{p2_visco=0}^{(j)} + \Delta X_{p2_amort}^{(j)}} \qquad (13.2)$$

A la lumière du paragraphe précédent, il est nécessaire que de nouveaux WS soient produits par l'entremise de l'évt $\mathbf{SB_{slow}}$ pour que le raccourcissement de chaque hs de la fmI perdure à partir de la position d'équilibre.

On observe un comportement différent selon l'intervalle (Fig. 26) :

**Dans i2 où $\mathbf{0.5 \leq pT^{(j)} \leq 0.8}$**

La distance pour atteindre la position d'équilibre est très inférieure à la distance à parcourir entre la position d'équilibre et la valeur de fin de phase 2, soit :

$$\mathbf{\Delta X_{p2_visco=0}^{(j)} << \Delta X_{p2_amort}^{(j)}}$$

**Fig. 26 : Tension relative en fonction du raccourcissement d'un hs durant la seconde partie de phase 2 suivant une perturbation en force d'échelon (j)**

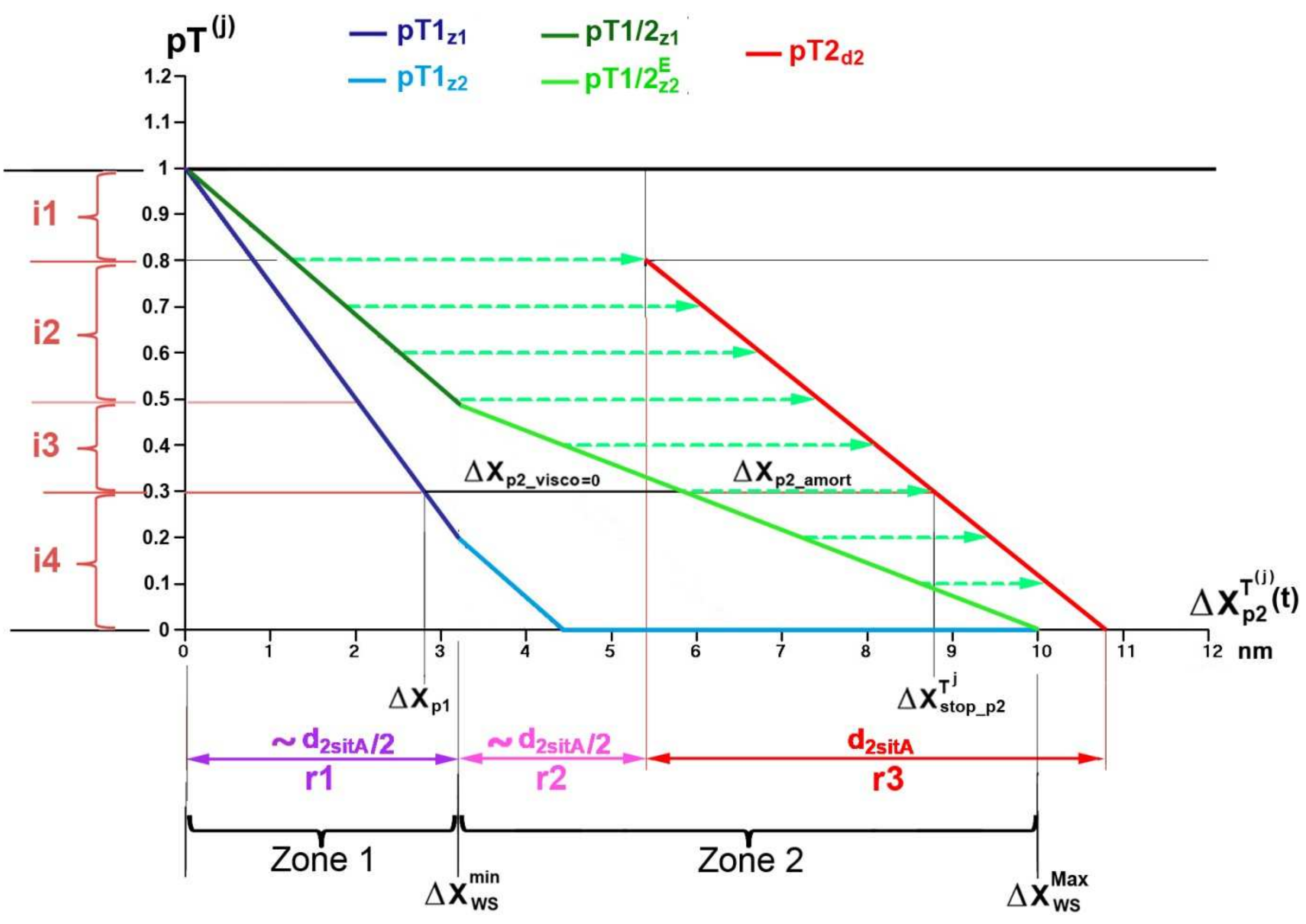

On remarque que la distance $\Delta X^{(j)}_{p2_amort}$ est à peu près identique quelque soit la valeur de l'échelon ( j) dans $\mathbf{i2}$, ce qui est logique puisque selon nos calculs ces 2 segments de droites ont des pentes quasi identiques.

Pour maintenir la valeur de l'échelon de force, le nombre de S1 en $\mathbf{SB_{slow}}$ doit augmenter en conséquence et d'après (1.14), le temps nécessaire à la génération de nouveaux WS doit s'accroitre également en fonction de $\mathbf{pT^{(j)}}$. Dans $\mathbf{i2}$, on observe effectivement que la durée de raccourcissement relative à $\Delta X^{(j)}_{p2_amort}$ décroit de 11 ms pour $\mathbf{pT=0.8}$ à 2.5 ms pour $\mathbf{pT=0.5}$ (Piazzesi, Lucii et al. 2002).

**Dans i3 où $0.3<pT^{(j)}\leq 0.5$**

La réduction de la valeur de $\mathbf{pT^{(j)}}$ coïncide avec une augmentation de $\Delta X^{(j)}_{p2_visco=0}$ et une diminution de $\Delta X^{(j)}_{p2_amort}$ qui reste néanmoins inférieur à $\Delta X^{(j)}_{p2_visco=0}$ (Fig. 26).

Dans $\mathbf{i3}$, la probabilité de survenue de l'évt $\mathbf{SB_{slow}}$ s'amoindrit, et la durée de raccourcissement concernant $\Delta X^{(j)}_{p2_amort}$ décroit de 2.5 ms à 1 ms (Piazzesi, Lucii et al. 2002).

**Dans i4 où $0 < pT^{(j)} \leq 0.3$**

La baisse de $pT^{(j)}$ est suivie d'une forte augmentation de $\Delta X^{(j)}_{p2_visco=0}$ alliée à une forte diminution de $\Delta X^{(j)}_{p2_amort}$ (Fig. 26) , telles que :

$$\Delta X^{(j)}_{p2_visco=0} >> \Delta X^{(j)}_{p2_amort}$$

Le taux de nouveaux WS nécessaire au maintien de l'échelon de force tend vers zéro, ce qui correspond à une chute de la probabilité de finalisation de l'évt $SB_{slow}$ avec une durée pour $\Delta X^{(j)}_{p2_amort}$ inférieure à la ms (Piazzesi, Lucii et al. 2002).

**Remarque** : il faut une valeur minimale de temps de recrutement et donc de tension pour que l'échelon de force soit maintenu et que le raccourcissement continu soit effectif. Il ne peut y avoir de raccourcissement pour $pT = 0$ (voir chap. 16).

### *13.3.2 Première partie de la phase 2 : raccourcissement vers la position d'équilibre V=0*

Plus $pT^{(j)}$ diminue, plus la distance $\Delta X^{(j)}_{p2_visco=0}$ augmente pour une durée identique ; *de facto*, la vitesse de raccourcissement augmente lorsque $pT^{(j)}$ décroit.

### *13.3.3 Seconde partie de la phase 2 : raccourcissement du à l'initiation de WS avec évt $SB_{slow}$*

Plus $pT^{(j)}$ diminue, plus la durée de raccourcissement relative à $\Delta X^{(j)}_{p2_amort}$ se contracte :

Dans $i2$, la distance $\Delta X^{(j)}_{p2_amort}$ étant constante, la vitesse moyenne de raccourcissement augmente passant de $0.4\,nm.ms^{-1}\ (pT = 0.8)$ à $1.7\,nm.ms^{-1}\ (pT = 0.5)$

Dans $i3$ la distance $\Delta X^{(j)}_{p2_amort}$ diminue, mais la vitesse moyenne de raccourcissement continue d'augmenter de $1.7\,nm.ms^{-1}\ (pT = 0.5)$ à $3\,nm.ms^{-1}\ (pT = 0.3)$

Cette dernière valeur est la vitesse moyenne maximale de la seconde partie de la phase 2, d'après nos calculs effectués à partir des données de la Fig. 1 de l'article de G. Piazzesi (Piazzesi, Lucii et al. 2002).

Dans $i4$ la distance $\Delta X^{(j)}_{p2_amort}$ continue de diminuer, mais on note que la vitesse diminue passant de $3\,nm.ms^{-1}\ (pT = 0.3)$ à $1.9\,nm.ms^{-1}\ (pT = 0.1)$

Cette dernière valeur est proche de celle trouvée pour $pT = 0.5$. Ce phénomène s'explique par la difficulté de réalisation de l'évt $SB_{slow}$ pour des durées inférieures à la ms, concevable selon les valeurs de $\Delta t_{SBslow}$ données dans (1.17).

.

### *13.3.4 Fin de la phase 2 : commencement des détachements de S1 en fin de WS*

La fin de la phase 2 correspond au détachement des WS en fin de WS, selon (1.26) et (1.27) en nombre suffisant pour influer sur la valeur de $\mathbf{pT^{(j)}}$ ; par conséquence, la valeur de la force motrice totale des S1 en cours de WS baisse, passant du mode Exagéré au mode Amorti vus au chap. 4. Ceci se traduit par un ralentissement de la vitesse de raccourcissement qui caractérise la phase 3.

## 13.4 Discussion

### *13.4.1 Faits prédits par le modèle*

La présence des forces de viscosité et leur annulation durant la phase 2 d'un échelon de force permet d'expliquer le comportement de la fibre musculaire lors de la phase 2 d'un échelon en force, comme cela avait été le cas pour la phase 2 d'un échelon en longueur.

Le modèle explique qualitativement les variations des valeurs de la vitesse de raccourcissement avec $\mathbf{pT^{(j)}}$.

Le fait que de nouveaux WS apparaissent est vérifié (voir Fig. 2A dans (Piazzesi, Reconditi et al. 2007)) où l'intensité de réflexion $\mathbf{I_{M3}}$, qui est un indicateur du nombre de S1 en cours de WS, augmente entre la phase 2 et la phase 3 à tous les échelons de force

### *13.4.2 Question soulevée par la fin de la phase 2*

La concordance exacte entre fin de phase 2 pour un échelon de longueur (droite en rouge ; Fig. 15 ; chap.8) et pour un échelon de force (droite en rouge ; Fig. 26) repose sur les résultats d'une seule publication.

Selon le modèle, la fin de la phase 2 pour les échelons de longueur dépend de la durée de survenue de l'évt DE, dont la durée moyenne est fournie par l'égalité (1.24).

Concernant les échelons de force, la fin de la phase 2 semble aussi provenir du temps de survenue de l'évt DE, avec les inégalités (1.26) et (1.27) dans $\mathbf{i2}$ et $\mathbf{i3}$, mais dans $\mathbf{i4}$ la distance totale parcourue se rapprochant de $\mathbf{\Delta X_{WS}^{Max}}$, cas de figure où la majorité des S1 ayant contribué à $\mathbf{T0^{fmI}}$ sont en fin de WS, semble plutôt être le facteur influent.

Le modèle n'est donc pas en mesure d'expliquer ce phénomène s'il est avéré.

Au chap. 8 se rapportant à un échelon de longueur, il avait été relevé que les pentes relatives à $\mathbf{pT2_{d2}}$ sont plus élevées chez les autres auteurs (Fig. 19 ; chap. 8). Il est donc important de vérifier par de nouvelles expérimentations s'il y a superposition ou non de ces 2 droites.

### *13.4.3 Analogie entre phase 2 et phase 4 d'un échelon de force*

La mise en place du raccourcissement lors de la seconde partie de la phase 2 (recrutement de S1 initiant un WS avec évt $\mathbf{SB_{slow}}$) est un prélude des mécanismes mis en jeu lors de la phase 4 (voir chap. 15). En effet, dans le modèle, la position d'équilibre (V=0) illustrée par les 2 segments de droites en vert des Fig. 25 et 26 a été élaborée pour expliquer le déroulement de la phase 2 d'un échelon de longueur (chap. 4 et 8); cette position d'équilibre servira aussi à expliciter la phase 4 d'un échelon de force (voir chap. 15 et 17).

Cette analogie avait été relevée précédemment (Piazzesi, Lucii et al. 2002 ; Piazzesi, Reconditi et al. 2007).

# 14 Phase 3 d'un échelon de force

## 14.1 Description

La vitesse de la phase 3 est considérablement réduite par rapport à celle de la phase 2 avec possibilité de s'annuler (Huxley 1974). La durée de la phase 3 est d'autant plus longue que l'échelon de force est faible (Piazzesi, Lucii et al. 2002).

Comme pour la phase 2, la phase 3 ne concerne que les valeurs de $\mathbf{T}^{(j)}$ inférieures à $\mathbf{0.8 \cdot T0^{fmi}}$ , appartenant aux intervalles de tension relative $\mathbf{i2}$, $\mathbf{i3}$ et $\mathbf{i4}$ (Fig. 26 ; chap.13).

La phase 3 se décompose en 2 périodes de durées comparables (Piazzesi, Lucii et al. 2002) :

- une 1ère période où la fmI se raccourcit à vitesse constante lente variant graduellement de $\mathbf{0.1\,nm.ms^{-1}}$ $(\mathbf{pT = 0.8})$ à $\mathbf{1\ nm.ms^{-1}}$ $(\mathbf{pT = 0.1})$; on note que plus la valeur de $\mathbf{pT^{(j)}}$ est faible, plus la durée de cette période est courte.
- puis une période durant la quelle la vitesse de raccourcissement n'est plus constante, passant de la valeur relative à la période précédente jusqu'à la vitesse de croisière constante qui caractérise la phase 4.

La fin de la phase 3 survient après un raccourcissement égal ou supérieur à 10 nm, 15 nm au maximum (Piazzesi, Lucii et al. 2002; Reconditi, Linari et al. 2004; Piazzesi, Reconditi et al. 2007 ).

## 14.2 Interprétation du modèle

Comme indiqué à la fin du chapitre précédent, le ralentissement de la vitesse durant la phase 3 d'un échelon de force résulte du détachement des S1 en fin de WS, exactement comme pour la phase 3 d'un échelon de longueur (chap. 9).

La contribution de ses S1 cesse après détachement, et la tension instantanée tend à baisser, passant de la valeur en mode Exagéré vers la valeur inférieure du mode Amorti (Fig. 10 ; chap. 4). Il faut donc qu'un nombre suffisant de S1 démarrent un WS suite à l'évt $\mathbf{SB_{slow}}$ pour compenser cette différence et revenir à l'équilibre, *i.e.* à la valeur de l'échelon $\mathbf{pT^{(j)}}$.

La phase 3 est marquée par cet effet de balance entre diminution de la tension relative due au détachement des S1 en fin de WS et augmentation provoquée par la survenue de nouveaux WS, jusqu'à obtention de l'état stationnaire caractérisée par la vitesse de croisière de la phase 4.

La durée de la phase 3 est donc déterminée :

- par le détachement complet des S1 qui ont participé à la mise en isométrie maximale précédant la perturbation en force : à la fin de la phase 3, le raccourcissement total depuis $\mathbf{L0_0^{hs}}$ est compris entre 10 et 13 nm, *i.e* supérieure à $\mathbf{\Delta X_{Ws}^{Max}}$ quelque soit la valeur de l'échelon de force. On retrouve ainsi un constat relevé au chap. 11 : la distance du raccourcissement dans chaque hs doit nécessairement être supérieure à $\mathbf{\Delta X_{Ws}^{Max}}$ pour que l'état stationnaire s'instaure.
- par le nombre de nouveaux S1 démarrant un WS : ainsi logiquement, d'après (1.14), plus $\mathbf{pT^{(j)}}$ est élevé, plus la durée de la phase 3, *i.e.* le temps mis à retrouver l'équilibre s'allonge.

# 15 Phase 4 d'un échelon de force

## 15.1 Description

Après un plateau tétanique suivi d'un raccourcissement soudain d'un échelon de force (j), la fmI est maintenue à la tension invariante $\mathbf{T}^{(j)}$, *i.e.* la fmI se contracte isotoniquement. Postérieurement à cette phase 1, et aux phases 2 et 3 pour les valeurs de $\mathbf{T}^{(j)}$ inférieures à $\mathbf{0.8 \cdot T0^{fmI}}$ (voir chap. 13 et 14), la fmI se raccourcit à une vitesse constante (« *steady velocity* »), notée $\mathbf{V}^{(j)}$, durant plusieurs dizaines de ms (Edman, Mulieri et al. 1976; Edman and Hwang 1977); cet état stationnaire caractérise la phase 4 d'un échelon de force (Hill 1938; Podolsky 1960; Huxley 1974).

Pour la gamme d'échelons de force (j) appartenant à l'intervalle $\left]\mathbf{0 ; T0^{fmI}}\right]$, la relation force-vitesse entre $\mathbf{T}^{(j)}$ et $\mathbf{V}^{(j)}$ a été étudiée par de nombreux chercheurs.

En 1935, W. Fenn et B. Marsh ont modélisé les données expérimentales avec une équation de type exponentielle (Fenn and Marsh 1935) :

$$\mathbf{T}^{(j)} = \mathbf{0.95 \cdot T0 \cdot e^{-a \cdot V^{(j)}} - k \cdot V^{(j)}} \tag{15.1}$$

où $\mathbf{a}$ et $\mathbf{k}$ sont 2 constantes représentatives de la fibre musculaire

Trois années plus tard, A. Hill a proposé une équation de nature hyperbolique (Hill 1938) :

$$\left(\mathbf{T}^{(j)} + \mathbf{a}\right) \cdot \left(\mathbf{V}^{(j)} + \mathbf{b}\right) = \mathbf{b} \cdot \left(\mathbf{T0} + \mathbf{a}\right) \tag{15.2}$$

où $\mathbf{a}$ et $\mathbf{b}$ sont 2 constantes caractéristiques de la fibre musculaire homogènes à une tension et une vitesse, respectivement

En adaptant les coefficients des vitesses d'attachement et de détachement, A.F. Huxley (Huxley 1957) a trouvé un bon accord entre son modèle et l'hyperbole de Hill, qui sert d'équation de référence depuis.

En 1973, P. Allen et W. Stainsby (Allen 1973) ont noté que, pour les faibles valeurs de vitesse, les points se situent en dessous de l'hyperbole de Hill. K. Edman (Edman, Mulieri et al. 1976) a repris cette observation en distinguant 2 intervalles $\left]\mathbf{0 ; 0.78 \cdot T0}\right]$ et $\left[\mathbf{0.78 ; T0}\right]$, développant ainsi le concept de nature biphasique de la relation Force-Vitesse. Pour la caractériser, K. Edman (Edman 1988) a suggéré de remplacer l'équation (15.2) par :

$$\mathbf{V}^{(j)} = \frac{\left(\mathbf{T0^{*}} - \mathbf{T}^{(j)}\right) \cdot \mathbf{b}}{\mathbf{T}^{(j)} + \mathbf{a}} \cdot \left(\mathbf{1} - \frac{\mathbf{1}}{\mathbf{e}^{-\mathbf{k1}\left(\mathbf{T}^{(j)} - \mathbf{k2 \cdot T0}\right)}}\right) \tag{15.3}$$

où $\mathbf{a}$ et $\mathbf{b}$ sont les 2 constantes de l'éq. de Hill (15.2)
$\mathbf{T0^{*}}$ est la valeur de la tension isométrique prédite par l'équation de Hill
et $\mathbf{k1}$ et $\mathbf{k2}$ sont 2 constantes

Il est important de noter la répétitivité des mesures qui dans des conditions identiques s'ajustent sur la même courbe (Edman, Mulieri et al. 1976; Edman 1988), soulignant ainsi l'aspect déterministe de la relation bijective entre $\mathbf{T^{(j)}}$ et $\mathbf{V^{(j)}}$.

L'arrangement quasi-cristallin des molM d'une fmI, sur lequel une source de rayons X est projetée, produit différentes réflexions, dont l'une est désignée par l'acronyme $\mathbf{M3}$ ; l'intensité de cette réflexion, nommée $\mathbf{I_{M3}}$ est considérée comme un bon indicateur du nombre de tetM en cours de WS (Irving, Piazzesi et al. 2000). Il est constaté que $\mathbf{I_{M3}}$ décroit en fonction de l'échelon $\mathbf{pT^{(j)}}$ durant la phase 4 (voir Tab. 1 dans (Huxley, Reconditi et al. 2006b) ; voir Fig. 2A et 2B dans (Piazzesi, Reconditi et al. 2007)). Comparativement, les valeurs sont plus élevées pour H.E. Huxley.

Par l'entremise de cette technique utilisée sur des fibres intactes, H.E. Huxley et ses coauteurs (Huxley, Reconditi et al. 2006b) observent différentes configurations selon l'échelon de force durant la phase 4 :

- les distributions des positions angulaires de S1b apparaissent plutôt uniformes que gaussiennes aux différents échelons de force $\mathbf{T^{(j)}}$ vérifiant $\mathbf{pT \geq 0.3}$.

- à faibles vitesses, *i.e.* $\mathbf{pT \approx 0.9}$, la distribution des positions angulaires des S1b appartenant aux S1 en cours de WS au sein d'un hs est uniforme avec une dispersion légèrement plus grande que celle de $\boldsymbol{\delta_X}$, un déplacement angulaire moyen de 7° et un déplacement linéaire relatif moyen de 1 nm par rapport aux positions isométriques maximales respectives, $\boldsymbol{\theta 0}$ et $\mathbf{X0}$

- à vitesses modérées, *i.e.* $\mathbf{0.5 \leq pT \leq 0.6}$, la distribution des positions angulaires des S1b appartenant aux S1 en cours de WS au sein d'un hs est uniforme avec une valeur de dispersion proche de $\boldsymbol{\Delta\theta_{WS}^{Max}}$ associée à un déplacement angulaire moyen de 28° par rapport à $\boldsymbol{\theta 0}$

- à vitesses rapides, *i.e.* $\mathbf{0.3 \leq pT \leq 0.38}$, la distribution des positions angulaires des S1b appartenant aux S1 en cours de WS au sein d'un hs est uniforme avec une valeur de dispersion supérieure à $\boldsymbol{\Delta\theta_{WS}^{Max}}$ qui accompagne un déplacement angulaire moyen de 38° et un déplacement linéaire relatif moyen de 6 nm par rapport à $\boldsymbol{\theta 0}$ et $\mathbf{X0}$

- la dispersion angulaire des S1b augmente globalement d'un facteur 1.5 entre vitesses réduites et vitesses élevées, *i.e.* entre $\mathbf{pT \approx 0.9}$ et $\mathbf{pT \approx 0.3}$, respectivement

## 15.2 Vitesse relative et répartition identiques au sein de chaque hs de la fmi

A $\mathbf{V}$ cste, et avec les Hyp. 1 à 15 énoncées au chap. 1, il est possible d'utiliser les équations des chapitres 2 à 5.

Les vitesses de raccourcissement durant la phase 4, sont très inférieures à celles observées lors des phases 1 et 2 (Piazzesi, Lucii et al. 2002); en conséquence, il est supposé que la viscosité n'interfère pas (une justification de cette hypothèse sera apportée dans le paragraphe « Discussion » en fin de chapitre).

Les forces et moments de liaison ainsi que les moments moteurs des S1 en cours de WS sont seuls considérés, toutes les autres interactions sont négligées.

Les éq. (3.1) à (3.5), (3.13), (5.1) et (5.2), conduisent à la valeur en module de $\mathbf{T}^{(j)}_{s,R_L}$, la tension instantanée à l'échelon de force n° j exercée au niveau du disque Z du hsR ou hsL n° s d'une des myofibrille de la fmI :

$$\mathbf{T}^{(j)}_{s,R_L} = \frac{1}{L_{S1b} \cdot S_{WS}} \cdot \sum_{b=1}^{\Lambda_{WS}^{j,s,R_L}} \left| \mathcal{M}^{(b)}_{s,R_L} \right| \tag{15.4}$$

où $\mathcal{M}^{(b)}_{s,R_L}$ est le moment moteur instantané de la tête n° b appartenant au hsR ou hsL n°s et $\Lambda_{WS}^{j,s,R_L}$ est le nombre instantané de tetM en cours de WS dans le hsR ou hsL n° s à l'échelon n° j

Le théorème de l'énergie cinétique appliqué à la fmI, considérée comme un système de solides articulés entre eux, apporte l'égalité suivante (voir Annexe E2) relativement à une myofibrille :

$$\mathbf{T}^{(j)} = \frac{N_{myof}}{V^{(j)}} \cdot \sum_{s=1}^{Ns} \sum_{b=1}^{\Lambda_{WS}^{j,s,R_L}} \left( \dot{\theta}^{(b)}_{s,R} \cdot \mathcal{M}^{(b)}_{s,R} + \dot{\theta}^{(b)}_{s,L} \cdot \mathcal{M}^{(b)}_{s,L} \right) = \mathbf{cste} \tag{15.5}$$

où $\dot{\theta}^{(b)}_{s,R_L}$ est la vitesse angulaire inter-segmentaire (entre S1a et S1b) instantanée de la tête n° b appartenant au hsR ou hsL n° s

A vitesse constante, par itération de (5.1) à (5.3), l'égalité des tensions appliquées aux disques Z se traduit avec (15.4) par :

$$\forall s,\ \mathbf{T}^{(j)}_{s,R_L} = \frac{\mathbf{T}^{(j)}}{N_{myof}} = \frac{1}{L_{S1b} \cdot S_{WS}} \cdot \sum_{b=1}^{\Lambda_{WS}^{j,s,R_L}} \left| \mathcal{M}^{(b)}_{s,R_L} \right| = \mathbf{cste} \tag{15.6}$$

Avec (3.15) et (15.4), l'éq. (15.5) se reformule :

$$\mathbf{T}^{(j)}_{s,R_L} = \frac{-1}{L_{S1b} \cdot R_{WS} \cdot V^{(j)}} \cdot \sum_{b=1}^{\Lambda_{WS}^{j,s,R_L}} \left( u_{s,R} \cdot \mathcal{M}^{(b)}_{s,R} + u_{s,L} \cdot \mathcal{M}^{(b)}_{s,L} \right) = \mathbf{cste} \tag{15.7}$$

où $u_{s,R_L}$ est la vitesse linéaire relative du hsR ou hsL n° s

A l'appui de (2.5) et (3.14), l'égalité des 2 expressions (15.6) et (15.7) induit comme solution :

$$\mathbf{V^{(j)} = 2 \cdot Ns \cdot u^{(j)}} \qquad (15.8)$$

où $\mathbf{u^{(j)}}$ est la valeur en module de la vitesse de déplacement relatif commun à tous les hs de la fmI durant la phase 4 d'un échelon de force n° j.

On note à nouveau à ce stade de l'article toute l'importance de (3.14) et son implication sur la géométrie inter-segmentaire d'une tetM en cours de WS.
L'égalité (15.6) prédit que le nombre, noté $\mathbf{\Lambda^{(j)}}$, et la répartition des tetM en cours de WS sont identiques et constants dans chaque hs de la fmI à tout instant durant la phase 4 d'un échelon de force (j) ; l'égalité (15.8) prédit l'égalité des vitesses de déplacements relatifs au sein de chaque hs. Il est donc possible de raisonner sur un hs de la fmI pour la suite du chapitre en cherchant la relation entre la valeur relative de la tension $\mathbf{pT^{(j)}}$ et $\mathbf{u^{(j)}}$.

Par contre, l'égalité (15.6) ne renseigne ni sur le nombre de tetM en cours de WS, ni sur leur répartition en fonction de l'échelon $\mathbf{pT^{(j)}}$.

## 15.3 Equation générale reliant force et vitesse durant la phase 4

A l'appui de la conclusion du paragraphe précédent et d'après les indications apportées par les travaux de H. Huxley (Huxley, Reconditi et al. 2006b), on déduit qu'il existe une bijection entre $\mathbf{pT^{(j)}}$ et $\mathbf{\Delta X^{(j)}}$, *i.e.* il existe une relation notée $\mathbf{G}$ définissant la valeur de l'échelon de force n° j exprimée en valeur relative en fonction de la position moyenne angulaire des S1b appartenant aux tetM en cours de WS et, par changement d'échelle, en fonction du déplacement linéaire moyen relatif de ces tetM, telle que :

$$\mathbf{pT^{(j)} = \frac{T^{(j)}}{T0^{fmI}} = G(\Delta X^{(j)})} \qquad (15.9)$$

On rappelle que, grâce aux égalités (2.1) et (15.6), à V cste, la tension (ou force) relative de la fmI est égale à celle d'une myofibrille ou d'un hs.

Soit une durée élémentaire notée $\mathbf{t_e}$. Pendant $\mathbf{t_e}$, le hs et donc l'ensemble des tetM en cours de WS se raccourcissent d'une longueur élémentaire notée $\mathbf{dX_e}$.

$\mathbf{dX_e}$ entraine une diminution de la tension relative, intitulée $\mathbf{dpT1}$, égale selon (15.9) à :

$$\mathbf{dpT1 = \dot{G}(\Delta X^{(j)}) \cdot dX_e} \qquad (15.10)$$

Pendant $\mathbf{t_e}$, la probabilité de survenue d'un nouveau WS est apportée par (1.14) selon l'Hyp. 6. La contribution de ces nouveaux $\mathbf{SB_{slow}}$, intitulée $\mathbf{dpT2}$, vaut en moyenne d'après (A.17) de l'annexe A :

$$\mathbf{dpT2} = \left(\mathbf{pT^{(j)}_{Lim}} - \mathbf{pT^{(j)}}\right) \cdot \left(1 - e^{-\frac{t_e}{\Delta t_{SBslow}}}\right) \qquad (15.11)$$

où $\mathbf{pT^{(j)}_{Lim}} \in \left]\mathbf{pT^{(j)}};1\right]$, tel que $\left(\mathbf{pT^{(j)}_{Lim}} - \mathbf{pT^{(j)}}\right)$ définit la contribution maximale possible engendrée par les initiations de WS suite à l'évt $\mathbf{SB_{slow}}$, relative à cet échelon de force, contribution dans laquelle intervient la nature géométrique inter-filamentaire.

L'approximation apportée par l'égalité (A.4) de l'annexe A appliquée à (15.11) mène à :

$$\mathbf{dpT2} \approx \left(\mathbf{pT^{(j)}_{Lim}} - \mathbf{pT^{(j)}}\right) \cdot \frac{t_e}{\Delta t_{SBslow}} \qquad (15.12)$$

En considérant qu'il n'y pas d'autres contributions au maintien de $\mathbf{pT^{(j)}}$ pendant $\mathbf{t_e}$, et notamment qu'il n'y a pas d'implication créée par le détachement des S1 en fin de WS, l'état stationnaire conduit à :

$$\mathbf{dpT1 + dpT2 = 0} \qquad (15.13)$$

De (15.10), (15.12) et (15.13), on déduit l'équation reliant force et vitesse relative pour un hs durant la phase 4 :

$$\mathbf{u^{(j)}} = \frac{dX_e}{t_e} \approx \frac{\mathbf{pT^{(j)}} - \mathbf{pT^{(j)}_{Lim}}}{\Delta t_{SBslow} \cdot \dot{G}(\Delta X^{(j)})} \qquad (15.14a)$$

Autrement formulée :

$$\mathbf{pT^{(j)}} = \mathbf{pT^{(j)}_{Lim}} + \left(\Delta t_{SBslow} \cdot \dot{G}(\Delta X^{(j)})\right) \cdot \mathbf{u^{(j)}} \qquad (15.14b)$$

Avec (15.8) et (15.9), on déduit la relation bijective entre $\mathbf{V^{(j)}}$ et $\mathbf{T^{(j)}}$ :

$$\mathbf{V^{(j)}} = \left(\frac{2 \cdot Ns}{T0^{fmI} \cdot \Delta t_{SBslow} \cdot \dot{G}(\Delta X^{(j)})}\right) \cdot \left(\mathbf{T^{(j)}} - \mathbf{T^{(j)}_{Lim}}\right) \qquad (15.14c)$$

Les éq. (15.14a) à (15.14c) comportent 3 types d'inconnue :

$\mathbf{G}$, la fonction reliant $\mathbf{pT^{(j)}}$ et $\mathbf{\Delta X^{(j)}}$ caractérisant l'état stationnaire de la phase 4 pour chaque échelon n° j

$\mathbf{\Delta t_{SBslow}}$ exprime la durée moyenne de survenue de l'évt $\mathbf{SB_{slow}}$

$\mathbf{pT^{(j)}_{Lim}} = \mathbf{T^{(j)}_{Lim}} / \mathbf{T0^{fmI}}$ renseigne sur la potentialité maximale de contribution à la valeur de la tension avec prise en compte de la nature géométrique inter-filamentaire

$\mathbf{\Delta t_{SBslow}}$ et $\mathbf{T^{(j)}_{Lim}}$ sont des fonctions potentielles de l'échelon de force $\mathbf{T^{(j)}}$.

## 15.4 Interprétation du modèle

### *15.4.1 Relation « linéaire en 1ère approximation » entre T et V dans les intervalles i1 et i2*

Dans **i1** et **i2** (Fig. 27), *i.e.* $\mathbf{0.5 \le pT^{(j)} \le 1}$, la vitesse relative $\mathbf{u^{(j)}}$ est inférieure à $\mathbf{0.8\,nm \cdot ms^{-1}}$ (Edman 1988; Huxley, Reconditi et al. 2006b); aussi plus de $\mathbf{12\,ms}$ sont nécessaires pour qu'un S1 effectue un pas maximal de 10 nm si la position angulaire de départ de S1b est égale à $\boldsymbol{\theta 1}$ , *i.e.* selon les valeurs de $\boldsymbol{\Delta t_{DE}^{stepF}}$ consignées dans (1.17), les S1 en fin de WS ont majoritairement le temps de se détacher sans exercer d'action. Rappelons que l'Hyp. 13 autorise le détachement d'une tetM avant que la position angulaire de S1b n'atteigne la position finale $\boldsymbol{\theta_{stopWS}}$, proposition qui figurait déjà dans (Reconditi, Linari et al. 2004; Huxley, Reconditi et al. 2006b).

En utilisant la définition d'une moyenne et en suivant le même raisonnement que celui mené au paragraphe 4.1, l'égalité (4.39) mène à la valeur correspondant dans le modèle à $\boldsymbol{\Lambda 0}$ positions angulaires de S1b appartenant à $\boldsymbol{\Lambda 0}$ tetM en cours de WS uniformément distribuées dans $\boldsymbol{\Delta\theta_{WS}^{Max}}$ nommée $\mathbf{pT0^*}$ , telle que :

$$\mathbf{pT0^* = 0.75} \tag{15.15}$$

En croisant les considérations précédentes avec les informations apportées par (Huxley, Reconditi et al. 2006b; Piazzesi, Reconditi et al. 2007), nous soumettons le scénario suivant :

$\mathbf{pT^{(j)}}$ décroit linéairement en fonction dc $\boldsymbol{\Delta X^{(j)}}$ à partir du point de coordonnées $\left|\begin{array}{l}\mathbf{\Delta X = 0}\\ \mathbf{pT = 1}\end{array}\right.$ jusqu'au point $\left|\begin{array}{l}\mathbf{\Delta X = \Delta X_{WS}^{min}}\\ \mathbf{pT = 0.5}\end{array}\right.$ , de telle manière que :

> $\boldsymbol{\Lambda_{WS}^{(j)}}$, le nombre de tetM en cours de WS dans chaque hs de la fmI décroit de $\boldsymbol{\Lambda 0}$ à $\mathbf{(0.75 \cdot \Lambda 0)}$.
>
> Les positions angulaires des $\boldsymbol{\Lambda^{(j)}}$ S1b se répartissent uniformément sur un intervalle $\boldsymbol{\Delta\theta_{WS}^{(j)}}$ qui varie progressivement de $\boldsymbol{\delta_\theta}$ à $\left(\mathbf{1.1 \cdot \Delta\theta_{WS}^{Max}}\right)$ . Tant que $\boldsymbol{\Delta\theta_{WS}^{(j)} < \Delta\theta_{WS}^{Max}}$, on conjecture avec l'Hyp. 13 que les tetM en cours de WS se détachent lorsque que la position angulaire $\boldsymbol{\theta}$ de S1b atteint la borne $\left(\left|\theta 1\right| + \Delta\theta_{WS}^{(j)}\right)$ inférieure à $\left|\boldsymbol{\theta_{stopWS}}\right|$. Plus $\mathbf{pT^{(j)}}$ se rapproche de $\mathbf{0.5}$, plus les positions angulaires se décalent majoritairement vers $\boldsymbol{\theta_{stopWS}}$ (la distribution n'étant plus rigoureusement uniforme) afin que le déplacement relatif moyen atteigne $\mathbf{\Delta X_{WS}^{min} = 3.2\,nm}$ et que $\mathbf{pT^{(j)}}$ adopte la valeur $\mathbf{0.5}$ inférieure à $\mathbf{0.56 = (0.75 \cdot 0.75)}$.

Selon les conditions décrites ci-dessus, la relation entre $\mathbf{pT^{(j)}}$ et $\mathbf{\Delta X^{(j)}}$ s'apparente à un segment de droite dont l'équation est apportée par les éq. (4.35a) et (7.8) dans la **zone 1** correspondant à **i1** et **i2** (Fig. 10 ; chap. 4), soit :

$$\mathbf{pT^{(j)} = G_{z1}\left(\Delta X^{(j)}\right) \approx 1 - \Delta X^{(j)} \cdot \chi_{z1}^{hs}} \tag{15.16}$$

La dérivation de (15.16) donne :

$$\mathbf{\dot{G}_{z1}\left(\Delta X^{(j)}\right) \approx -\chi_{z1}^{hs}} \tag{15.17}$$

A l'instar des modélisations proposées aux chapitres 8 et 9 pour $\mathbf{pT2}$, $\mathbf{pT_{Lim}^{(j)}}$ est formulé par une équation linéaire telle que :

$$\mathbf{pT_{Lim}^{(j)} = 1 - \sigma_{z1} \cdot \Delta X^{(j)}} \tag{15.18}$$

où $\mathbf{\sigma_{z1}}$ est une constante positive vérifiant :

$$\mathbf{\sigma_{z1} < \chi_{z1}^{hs}} \tag{15.19}$$

On déduit de (15.16) et (15.18) :

$$\mathbf{pT_{Lim}^{(j)} = \left(1 - \frac{\sigma_{z1}}{\chi_{z1}^{hs}}\right) + \frac{\sigma_{z1}}{\chi_{z1}^{hs}} \cdot pT^{(j)}} \tag{15.20}$$

Avec (15.17) et (15.20), l'éq. (15.14b) se reformule pour $\mathbf{0.5 \le pT^{(j)} \le 1}$ :

$$\mathbf{pT^{(j)} = 1 - \left[\Delta t_{SBslow} \cdot \frac{\left(\chi_{z1}^{hs}\right)^2}{\left(\chi_{z1}^{hs} - \sigma_{z1}\right)}\right] \cdot u^{(j)}} \tag{15.21a}$$

où $\mathbf{\Delta t_{SBslow}}$ est considérée constante et vérifie (1.17) d'après l'Hyp. 5

Conventionnellement, on exprime la vitesse en fonction de la force :

$$\mathbf{u^{(j)} = \left[\frac{\left(\chi_{z1}^{hs} - \sigma_{z1}\right)}{\Delta t_{SBslow} \cdot \left(\chi_{z1}^{hs}\right)^2}\right] \cdot \left(1 - pT^{(j)}\right)} \tag{15.21b}$$

Soit avec (15.7) la relation de $\mathbf{V^{(j)}}$ en fonction de $\mathbf{T^{(j)}}$ :

$$\mathbf{V^{(j)} = \left[\frac{1}{\Delta t_{SBslow} \cdot \chi_{z1}^{fmI}} \cdot \frac{\left(\chi_{z1}^{hs} - \sigma_{z1}\right)}{\chi_{z1}^{hs}}\right] \cdot \left(1 - \frac{T^{(j)}}{T0^{fmI}}\right)} \tag{15.21c}$$

où selon (7.3c), $\mathbf{\chi_{z1}^{fmI} = \frac{\chi_{z1}^{hs}}{2 \cdot Ns}}$

Pour $\mathbf{0.5 \le pT^{(j)} \le 1}$, la relation force-vitesse entre $\mathbf{V^{(j)}}$ et $\mathbf{T^{(j)}}$ est un segment de droite dont la pente négative est en module :

1/ inversement proportionnelle au coefficient de raideur de la fmI, assimilée à un ressort mécanique linéaire et, par conséquence, proportionnelle à la longueur de la fibre musculaire et à sa compliance.

2/ inversement proportionnelle à la valeur de la tension isométrique, $\mathbf{T0^{fmI}}$ et à la constante de temps, $\mathbf{\Delta t_{SBslow}}$, caractérisant l'évènement $\mathbf{SB_{slow}}$ décrit au paragraphe 1.4.2.

3/ proportionnelle à la constante $\left(\chi_{z1}^{hs} - \sigma_{z1}\right) / \chi_{z1}^{hs}$, calculée à partir de coefficients élaborés lors d'un échelon de longueur.

## ***Application numérique***

D'après les données expérimentales (Piazzesi and Lombardi 1995), la valeur de $\chi_{z1}^{hs}$ a été calculée selon (4.36c) :

$$\chi_{z1}^{hs} = 0.16\ \mathrm{nm}^{-1}$$

Et en prenant pour $\sigma_{z1}$, la valeur de la pente de l'éq. (8.6a) qui se rapporte à l'évt $\mathbf{SB_{fast}}$ caractéristique de la remontée de la tension de la phase 2 d'un échelon de longueur, voir Fig. 17 du chap. 8, soit :

$$\sigma_{z1} = 0.025\ \mathrm{nm}^{-1}$$

Avec ses 2 valeurs, (15.20b) s'écrit pour $\mathbf{0.5 \le pT^{(j)} \le 1}$ :

$$u^{(j)} = \frac{5.27}{\Delta t_{SBslow}} \cdot \left(1 - pT^{(j)}\right) \tag{15.21d}$$

### *15.4.2 Relation de « nature hyperbolique » entre T et V dans les intervalles i3 et i4*

Comme la bijection $\mathbf{G}$ entre $\mathbf{pT^{(j)}}$ et $\mathbf{\Delta X^{(j)}}$ dans **i1** et **i2** se superpose à la relation entre $\mathbf{pT^{(k)}}$ et $\mathbf{\Delta X^{(k)}}$ caractéristique d'un échelon de longueur, nous admettons par extension qu'il en est de même pour les intervalles **i3** et **i4**, *i.e.* $\mathbf{0 < T^{(j)} \le 0.5}$ ; l'expression de G est fournie par l'éq. (4.30) dans la zone 2 avec le mode Amorti où les détachements des tetM en fin de WS n'influent pas, soit :

$$pT^{(j)} = G_{z2}\left(\Delta X^{(j)}\right) \approx \Omega \cdot \left[ \left(X3 + a_X\right) \cdot \log\left( \frac{X3 + a_X}{\Delta X^{(j)} + X1 + a_X} \right) - \left( \Delta X_{WS}^{Max} - \Delta X^{(j)} \right) \right] \tag{15.22}$$

avec la constante $\Omega = \dfrac{\mathcal{M}1 \cdot \left(X1 + a_X\right)}{\mathcal{M}0 \cdot \Delta X_{WS}^{Max} \cdot \delta_X}$

La dérivation de (15.22) donne :

$$\dot{G}_{z2}\left(\Delta X^{(j)}\right) \approx \Omega \cdot \left[1 - \frac{\left(X3 + a_X\right)}{\left(\Delta X^{(j)} + X1 + a_X\right)}\right] \tag{15.23}$$

A l'exemple des extrapolations expérimentales adoptées aux chapitres 8 et 9 pour $pT2$, $pT^{(j)}_{Lim}$ est modélisé ous une forme linéaire telle que :

$$pT^{(j)}_{Lim} = \eta_{z2} - \sigma_{z2} \cdot \Delta X^{(j)} \tag{15.24}$$

où $\eta_{z2}$ et $\sigma_{z2}$ sont 2 constantes positives avec $\eta_{z2} > 1$

En prenant pour (15.24), les coefficients de la droite donnée par (8.6b), voir Fig. 17 du chap. 8, soit :

$$\nu_{z2} = 1.1 \tag{15.25a}$$

$$\sigma_{z2} = 0.055\ nm^{-1} \tag{15.25b}$$

Il n'est pas possible d'expliciter une formule analytique simple entre $pT^{(j)}$ et $u^{(j)}$, ou entre $T^{(j)}$ et $V^{(j)}$ comme au paragraphe précédent pour les intervalles $i1$ et $i2$.

On procède selon la méthode suivante :

1/ Une longueur fixe élémentaire de raccourcissement est choisie tel que :

$$dX_e = 0.1\,nm \tag{15.26}$$

2/ A partir d'une valeur de $pT^{(j)}$ comprise ente 0.05 et 0.5, la relation bijective fournie par (15.22) permet d'interpoler $\Delta X^{(j)}$ qui est ensuite introduite dans (15.23) ; la diminution élémentaire de la tension relative, $dpT1$ est déterminée d'après (15.10) et (15.26).

3/ Avec l'égalité (15.13), $dpT1$ est rapportée dans (15.12)

4/ La valeur de $\Delta t_{SBslow}$ est prise constante et identique à celles utilisées dans les calculs de (15.21a) à (15.21d), et autorise le calcul du temps élémentaire $t_e$ dans (15.12)

5/ on en déduit d'après (15.26), la valeur de la vitesse relative pour un hs, $u^{(j)} = dX_e / t_e$

En incrémentant $pT^{(j)}$ entre 0.05 et 0.5, cette méthode est utilisée par simulation numérique pour calculer et tracer la relation force-vitesse entre $u^{(j)}$ et $pT^{(j)}$.

### *15.4.3 Relation avec G explicité par le modèle entropique avec mode Amorti (Fig. 10 ; chap. 4)*

La bijection $pT^{(j)} = G(\Delta X^{(j)})$ est exprimée à partir des 2 relations développées au chap. 4, reliant la tension relative exercée par un ensemble de tetM ayant initié leur WS dans un hsL après un raccourcissement discret en longueur de la fmI, *i.e.* les 2 éq. (4.25) et (4.30) du modèle entropique défini dans les zones 1 et 2 (Fig. 10 ; chap.4). La nature hyperbolique de la relation entre $u^{(j)}$ et $pT^{(j)}$ ou entre $V^{(j)}$ et $T^{(j)}$, issue elle-même de l'équation entropique/hyperbolique du nanomoteur d'une tetM (3.31), est soulignée.

**Fig. 27 : Relations entre $u^{(j)}$ et $pT^{(j)}$**

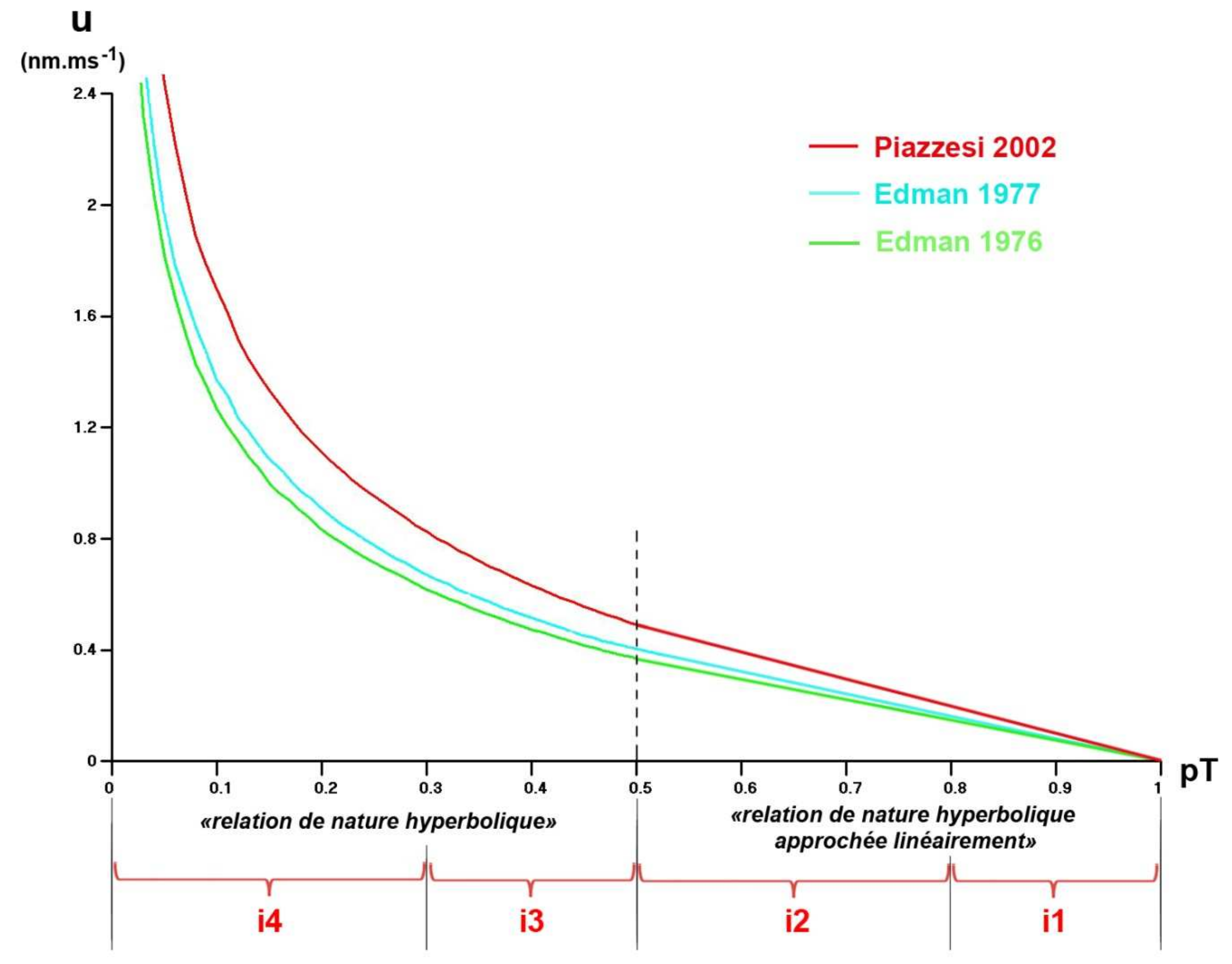

Dans la **zone1**, *i.e.* les intervalles **i1** et **i2** pour la variable $\mathbf{pT^{(j)}}$ , l'éq. (4.25) est approché par la relation linéaire donnée par (4.35a) et (7.8).

La confrontation de cette modélisation avec les points de la Fig. 3A dans (Piazzesi, Lucii et al. 2002) aboutit à un accord convenable en posant $\mathbf{\Delta t_{SBslow} \approx 5.7\,ms}$ (Fig. 27).

Bien qu'il s'agisse de fibres recueillies sur un autre muscle (semitendinosus) d'une autre espèce de grenouille (Rana tempororia), un ajustement correct est trouvé (Fig. 27) :

- pour les points correspondants de la Fig. 3 dans (Edman, Mulieri et al. 1976) relativement aux 2 fibres de longueur respective 8.6 mm et 11 mm avec une valeur identique : $\mathbf{\Delta t_{SBslow} = 7.6ms}$
- pour les points de la Fig. 2 dans (Edman et Hwang 1977) avec $\mathbf{\Delta t_{SBslow} = 7ms}$

On note cependant, pour $\mathbf{pT^{(j)} < 0.1}$ ou $\mathbf{u^{(j)} > 1.5\,nm \cdot ms^{-1}}$, un décalage des points expérimentaux vers la gauche indiquant les limites du modèle. D'où les modifications apportées dans le paragraphe suivant.

## 15.5 Rectifications dans les intervalles i1 et i4

### *15.5.1 Changement d'équation dans i1*

Selon l'expression (15.4), la diminution de $\mathbf{pT^{(j)}}$ est due, soit à la diminution de $\mathbf{\Lambda_{WS}^{(j)}}$ (le nombre de tetM par hs en cours de WS), soit à l'élargissement de $\mathbf{\Delta\theta_{WS}^{(j)}}$ (l'intervalle de répartition des positions angulaires des $\mathbf{\Lambda_{WS}^{(j)}}$ S1b), soit aux deux. En conséquence, $\mathbf{\Lambda_{WS}^{(j)}}$ et $\mathbf{\Delta\theta_{WS}^{(j)}}$ vérifient les 2 inégalités suivantes dans **i1** :

$$\mathbf{\Lambda_{WS}^{(j)} \le \Lambda 0} \tag{15.27a}$$

$$\mathbf{\delta_\theta \le \Delta\theta_{WS}^{(j)} \le \Delta\theta_{WS}^{Max}} \tag{15.27b}$$

A l'appui des observations apportées dans (Huxley, Reconditi et al. 2006b), nous proposons de remplacer dans **i1** le modèle « linéaire » défini par l'éq. (15.21b) par une nouvelle équation après les justifications ci-dessous :

pour $\mathbf{0.9 \le pT^{(j)} \le 1}$

$\mathbf{\Lambda_{WS}^{(j)}}$ diminue très peu tout en restant proche de **1** et $\mathbf{\Delta\theta_{WS}^{(j)}}$ augmente très légèrement tout en restant voisin de $\mathbf{\delta_\theta}$. Cette conjoncture s'interprète par la difficulté accrue de débuter de nouveaux WS et se traduit par la diminution de la valeur de $\mathbf{pT_{Lim}^{(j)}}$. Ainsi le paramètre $\mathbf{\sigma_{z1}}$ de l'éq. (15.21b) est remplacé par une valeur 4 fois plus élevée notée $\mathbf{\sigma_{i1}}$, tel que :

$$\mathbf{\sigma_{i1} \approx 0.1\ nm^{-1}} \tag{15.28}$$

Le nouveau segment de droite s'apparente à une tangente de l'axe des abscisses (Fig. 28)

***Rappel*** : durant le raccourcissement continu caractérisant l'état stable de la phase 4, les conditions pour lesquelles $\mathbf{\Delta\theta_{WS}^{(j)}}$ est proche de $\mathbf{\delta_\theta}$ et donc inférieur à $\mathbf{\Delta\theta_{WS}^{Max}}$, imposent que parmi les $\mathbf{\Lambda_{WS}^{(j)}}$ tetM en cours de WS, celles dont le sous-segment S1b prend une position angulaire $\mathbf{\theta}$ dépassant la borne supérieure de $\mathbf{\delta_\theta}$ , *i.e.* $\mathbf{\theta 2}$, se détachent de la molA. On observe que pour $\mathbf{0.9 \le pT^{(j)} \le 1}$ les vitesses relatives dans un hs sont très faibles, *i.e.* $\mathbf{0 \le u^{(j)} \le 0.05 nm \cdot ms^{-1}}$ selon les données de la Fig. 2C dans (Edman 1988) ; il appert qu'à ces faibles vitesses, la durée du WS ($\mathbf{\Delta t_{WS}}$) est très supérieure à $\mathbf{\Delta t_{DE}^{stepF}}$, même pour des WS incomplets. Ce scénario est étayé par l'étude du relâchement de la fmI après tetanisation (Brunello, Fusi et al. 2009).

pour $\mathbf{0.8 \le pT^{(j)} \le 0.9}$

Il s'opère un retour vers le modèle déterminé par (15.21b) où $\mathbf{\Lambda_{WS}^{(j)}}$ diminue toujours très peu tout en restant à proximité de **1** et $\mathbf{\Delta\theta_{WS}^{(j)}}$ s'élargit pour se rapprocher de $\mathbf{\Delta\theta_{WS}^{Max}}$.

Ainsi lorsque $\mathbf{pT^{(j)}}$ diminue vers **0.8**, on retrouve les préalables menant au calcul de $\mathbf{pT0^*}$ avec (15.15), *i.e.* la valeur **0.75** avoisinant **0.78**, la donnée signalée par K. Edman (voir paragraphe 15.1).

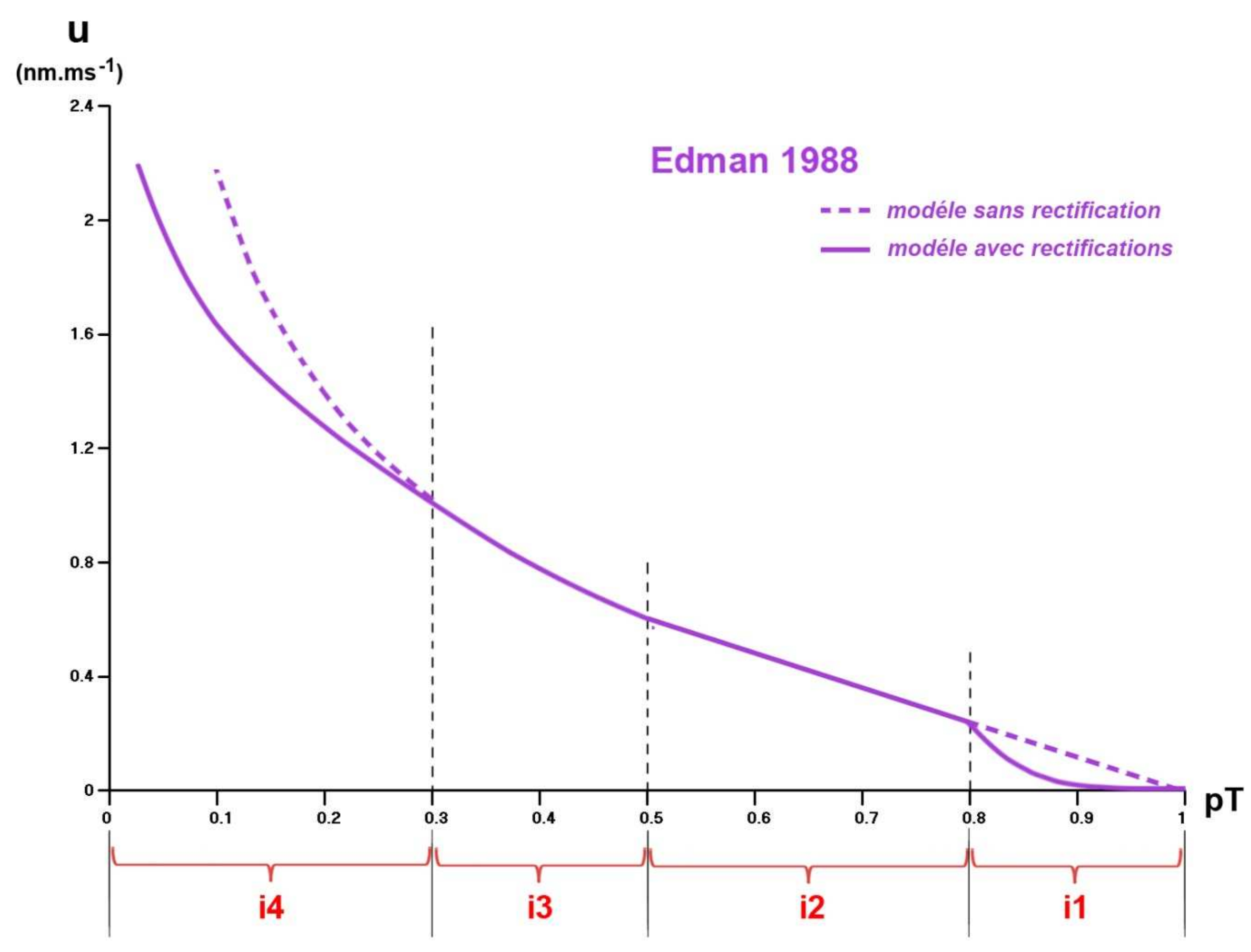

**Fig. 28 : Relations entre $u^{(j)}$ et $pT^{(j)}$ après rectifications dans i1 et i4**

L'examen des 2 sous-intervalles de **i1** est schématisé par l'introduction d'une fonction empirique de type puissance en remplacement de (15.21d), soit :

$$\mathbf{u^{(j)}} = \frac{\mathbf{5.27}}{\mathbf{\Delta t_{SBslow}}} \cdot \frac{\left(\mathbf{1 - pT^{(j)}}\right)^{\mathbf{n}}}{\left(\mathbf{1 - 0.8}\right)^{\mathbf{n-1}}} \qquad (15.29)$$

avec $\mathbf{n \approx 4}$

***Remarque*** : pour n=1, (15.29) redevient (15.21d)

### *15.5.2 Changement de la valeur de* $\mathbf{pT_{Lim}^{(j)}}$ *dans i4* $\left(\mathbf{0 < pT^{(j)} \leq 0.3}\right)$

Les calculs d'intégration effectués dans le chapitre 4, dont certains servent à modéliser la fonction **G**, reposaient sur l'hypothèse que le nombre de S1 initiant un WS pendant un instant t de l'ordre de la ms était suffisant pour autoriser le passage au continu et justifier le recours à l'Hyp. 2 d'uniformité.

Dans **i4** où $\mathbf{u^{(j)} > 0.8nm \cdot ms^{-1}}$ les conditions posées deviennent caduques et les calculs non conformes aux résultats expérimentaux ; dans **i4** les vitesses théoriques sont plus élevées que les vitesses mesurées.

Ce phénomène s'interprète par un empêchement accru de réunir les conditions de déclenchement d'un WS, avec pour conséquence une diminution de $\mathbf{pT_{Lim}^{(j)}}$ . Dans **i4**, nous proposons de remplacer les coefficients des égalités (15.25a) et (15.25b) par les coefficients de la droite d'éq. (8.6c), voir Fig. 17 du chap. 8, soit :

$$\nu_{i4} = 1.64 \qquad (15.30a)$$

$$\sigma_{i4} = 0.155\ \mathrm{nm}^{-1} \qquad (15.30b)$$

Cette difficulté se traduit expérimentalement par un moins bon alignement des points sur les courbes issues de modélisations théoriques, comme cela apparait sur les Fig. 2A et 2B dans (Edman 1988) pour les très faibles valeurs de forces.

### *15.5.3 Comparaison avec les données de K. Edman (1988)*

Sur la base du modèle développé au paragraphe 15.4 et pour les raisons développées dans le paragraphe précédent, il est procédé aux aménagements suivants :

- dans **i1** , (15.29) est substituée à (15.21b)
- dans **i2** , (15.21b) est conservée : la linéarité dans cet intervalle provient de la baisse proportionnelle de $\Lambda_{WS}^{(j)}$ jusqu'à la valeur **0.75** et d'un faible élargissement[1] progressif de $\Delta\theta_{WS}^{(j)}$, correspondant à un déplacement moyen relatif variant d'une valeur supérieure à $|X0^* - X0| = 1.5\ \mathrm{nm}$ jusqu'à $\Delta X_{WS}^{min} = 3.2\ \mathrm{nm}$ ; ainsi $\mathbf{pT^{(j)}}$ tend vers la valeur **0.5** inférieure à $0.56 = (0.75 \cdot 0.75)$
- dans **i3** , (15.24) s'écrit avec les 2 valeurs identiques du paragraphe précédent apportées aux 2 coefficients par (15.25a) et (15.25b)
- dans **i4** , (15.24) s'écrit avec 2 nouvelles valeurs fournies aux 2 coefficients avec (15.30a) et (15.30b)

En introduisant la valeur commune, $\Delta t_{SBslow} \approx 4.5\ \mathrm{ms}$ , dans les équations, cette nouvelle modélisation présentée à la Fig. 28, après rectifications dans **i1** et **i4** , s'accorde convenablement avec les points des Fig. 2A, 2B, 3A et 3B dans (Edman 1988).

On observe une relation entre $\mathbf{u^{(j)}}$ et $\mathbf{pT^{(j)}}$ qui se module non pas dans 2 mais 4 intervalles.

---

[1] *La répartition des postions angulaires $\theta^{(b)}$ des $\Lambda_{WS}^{(j)}$ S1b n'est plus tout à fait uniforme dans $\Delta\theta_{WS}^{(j)}$ car les nouvelles tetM initiant un WS ne sont pas en proportion suffisante pour compenser le décalage créé par le raccourcissement continu et la position moyenne se déporte légèrement vers la borne supérieure.*

## 15.6 Discussion

### *15.6.1 Questions soulevées par le modèle*

**Paramètre $a_\theta$**

Au paragraphe 4.5, la constante $a_\theta$ a été donnée égale à $316.9$ afin de parvenir, selon (4.39) et (15.15), à l'identité : $pT0^* = 0.75$ .

Existe-t-il des facteurs qui influent sur ce paramètre déterminant ? Si la réponse est affirmative, la typologie, la température, le taux de calcium ou le taux d'ATP se trouvent-ils parmi ces facteurs ?

Autre interrogation : est-il nécessaire d'adapter la valeur de $a_\theta$ de telle sorte que $pT0^* = 0.8$ ?

**Evènement $SB_{slow}$**

Dans le modèle, la durée de l'évt $SB_{fast}$ lors de la phase 2 d'un échelon de longueur se révèle en moyenne inférieure à la ms (Hyp. 3), car le déplacement relatif imposé et la condition isométrique (Hyp. 4) favoriseraient la réalisation des modalités préalables à l'initiation du WS pour de nouvelles tetM.

Durant les phases 2 à 4 d'un échelon de force, la durée de l'évt $SB_{slow}$ se révèlerait plus élevée en moyenne (Hyp. 5 et inégalité (1.15)) car le raccourcissement continu perturberait la survenue de ces mêmes modalités, probablement celle qui concerne la contrainte de rigidité de S2: il serait plus difficile de remplir cette condition lorsque le filA (auquel S1a est fixé) et le filM (auquel S2 est accroché) se déplacent l'un relativement à l'autre (Hyp. 6) ; par comparaison, l'immobilité des filA et filM (Hyp.4) faciliterait cette même condition.

Pour le modèle, la valeur de $SB_{slow}$ a été déterminée comme constante et donc indépendante de la vitesse de raccourcissement de la phase 4 ; ce point reste à être vérifié.

**Evènement DE**

L'équation générale de la relation Force/Vitesse est fondée sur l'égalité (14.13), dans laquelle seuls les évènements $SB_{slow}$ et $WS$ interviennent. Avec notre modèle, l'évt $DE$ n'exercerait pas d'influence durant la phase 4 d'un échelon de force ; cette conjecture est en désaccord avec les prérequis de nombreux modèles précédents, notamment celui de A. F. Huxley.

L'Hyp. 13 a été spécifiée afin d'harmoniser les différentes données expérimentales présentées au paragraphe 15.1:

Dans $i1$, les tetM en cours de WS se détachent avant que la position angulaire de S1b n'atteigne la position finale, conformément au complément apporté à l'Hyp. 13 (voir paragraphe 1.6). Ce postulat conduit aux inégalités de (15.27b) et a été mentionnée auparavant par M. Reconditi et H.E. Huxley; il a été vérifié expérimentalement pour l'isométrie maximale (Brunello, Fusi et al. 2009).

Dans **i2** et **i3**, où les vitesses relatives sont inférieures à 1 $nm.ms^{-1}$, les durées d'un WS autorisent un détachement sans conséquence d'action sur le filM , d'après les égalités (1.26) et (1.27).

Dans **i4**, où les vitesses relatives sont supérieures à 1 $nm.ms^{-1}$, les durées moyennes de détachement doivent être inférieures à 5 ms : l'Hyp. 13 intègre le concept qu'un mécanisme autorise ce détachement rapide.

Dans leur article (Reconditi, Brunello et al. 2011), consacré à la mise en place du plateau tétanique, M. Reconditi et ses coauteurs concluaient par ces mots :

> ***«The time course of force generation in physiological conditions is not limited by the working stroke, or by that of structural changes in the thin or thick filaments, but by the rate of strong myosin head attachment to actin.*** »

Nous postulons que cette assertion peut s'appliquer également lors d'un raccourcissement continu, notamment pour la phase 4 d'un échelon de force, en ajoutant les 2 autres spécifications de l'Hyp. 1 (rigidité de S2 et déclenchement du moment moteur).

**Nécessité de nouveaux travaux expérimentaux à des fins de validation**

L'équation générale, déclinée selon les éq. (15.14a) à (15.14c), dépend de 3 inconnues :

- $\Delta t_{SBslow}$ donné pour constant quelque soit la valeur de l'échelon de force $pT^{(j)}$ ; les valeurs[1] de $\Delta t_{SBslow}$ varient entre 4 et 8 ms, conformément à l'Hyp. 5 et à (1.17).
- $G$ , la bijection entre $pT^{(j)}$ et $\Delta X^{(j)}$ qui, suite à un raisonnement s'appuyant sur des observations, se confond avec la bijection entre $pT^{(k)}$ et $\Delta X^{(k)}$ ; cette bijection a été mise en équation dans les zones 1 et 2 au chapitre 4 à partir du modèle entropique avec mode Amorti (pas d'interférences dues à l'évt DE) et utilisée pour expliquer et décrire les 4 phases d'un échelon (k) de longueur (voir chap. 7 à 10).
- $pT^{(j)}_{Lim}$ , fonction de $pT^{(j)}$ ou $\Delta X^{(j)}$, dont les valeurs sont alignées sur les équations entre $pT2$ et $\Delta X^{(k)}$ régissant la possibilité de survenue de l'évt $SB_{fast}$ durant la phase 2 d'un échelon (k) de longueur (voir chap. 8).

La construction de la bijection $G$ provient d'une unique source (Piazzesi, Lucii et al. 2002) qui, seule à notre connaissance, a marié les 2 types de perturbations, échelons de longueur et de force.

La superposition des 2 bijections a été élaborée à partir des observations de 2 publications (Huxley, Reconditi et al. 2006; Piazzesi, Reconditi et al. 2007b) dont les données ont du être modifiées pour rester compatibles avec les calculs du modèle. Rappelons d'abord qu'il apparait des divergences chiffrées entre ces 2 publications pour les valeurs de $I_{M3}$ en fonction de $pT^{(j)}$ .

Les répartitions angulaires durant la phase 4 d'un échelon de force sont évaluées dans (Huxley, Reconditi et al. 2006a) sur la base de la distribution des positions angulaires isométriques de la Fig. 15 de cet article, soit :

$$\theta 1 = -60°$$
$$\theta 2 = -14°$$
$$\theta_{stopWS} = +30°$$
$$\Delta\theta_{WS} = 90°$$

[1] *Pour des fibres de grenouilles, dans des conditions physiologiques standard et avec des températures expérimentales proches de 0°C.*

**Fig. 29 : Evolution de 5 paramètres pour 4 valeurs de $\beta$ dans un hsL entre $\theta 1 = -60°$ et $\theta_{stopWS} = +30°$**

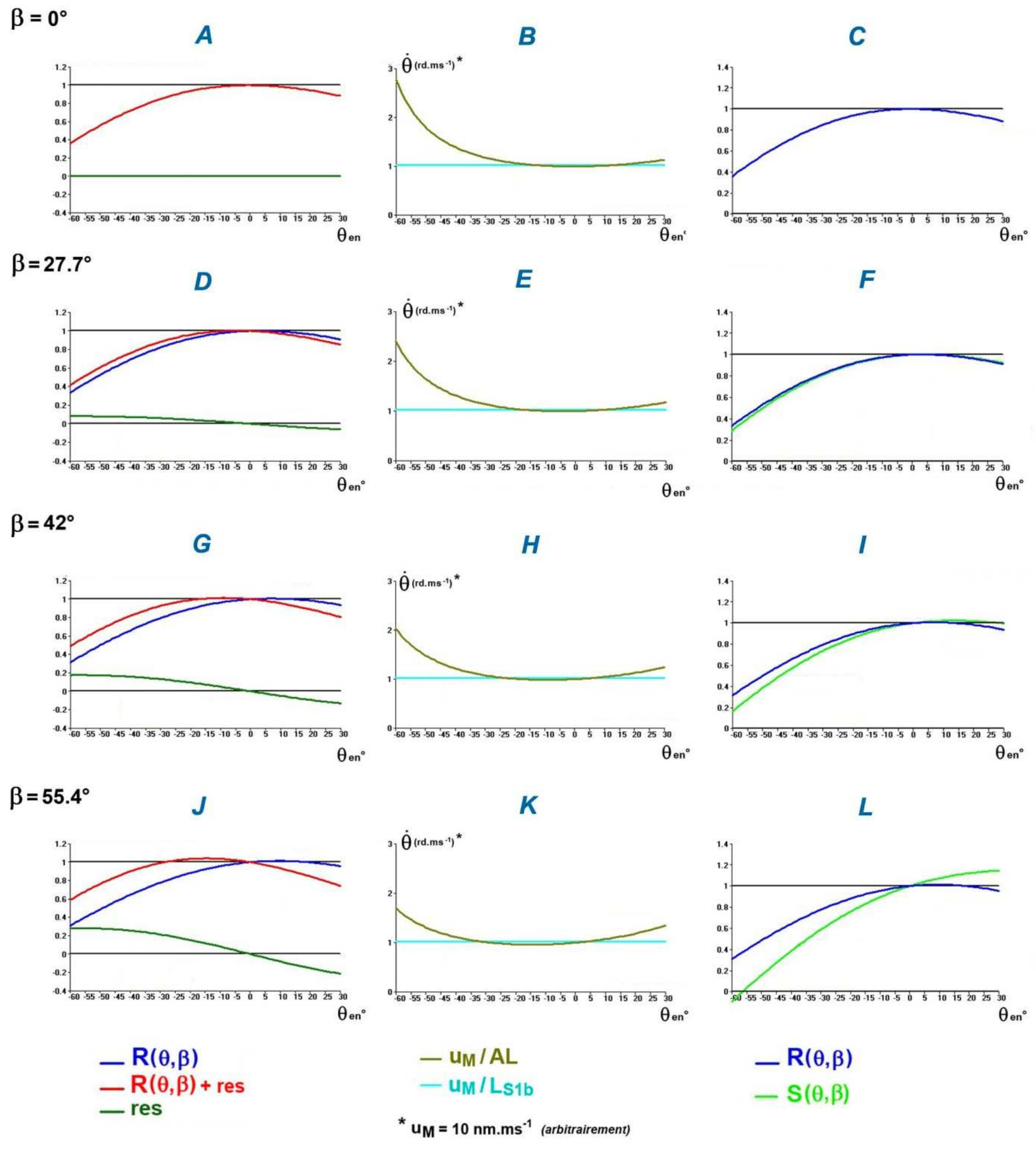

Avec ces données, nous avons repris les calculs effectués dans le chap. 3, et il apparait une grande hétérogénéité des valeurs de $\mathbf{R(\theta,\beta)}$, $\mathbf{S(\theta,\beta)}$, $\dot{\theta}^{(b)}_{s,R_L}$ et $u_{Ms,R_L}$ en fonction de $\theta$ pour 4 valeurs de $\beta$ (Fig. 29), à comparer avec les cinétiques de la Fig. 5 au chap. 3.

Ces disparités n'autorisent plus les approximations indispensables aux développements effectués dans les chapitres 4 à 15, particulièrement en ce qui concerne les égalités (3.12), (3.13) et (3.16).

Toutes ces remarques nous amènent à souhaiter que de nouvelles expérimentations soient conduites afin de confirmer ou d'infirmer les conjectures et les choix qui ont présidé à la conception de notre modèle.

### *15.6.2 Faits observés prédits par le modèle*

**i1 borné par 1 : l'état pseudo-absorbant de l'isométrie maximale**

Ce cas particulier décrit[1] pour $\mathbf{0.9 \leq pT^{(j)} \leq 1}$, où $\Lambda_{\mathbf{WS}}^{(\mathbf{j})}$ diminue très peu tout en restant proche de $\mathbf{1}$ et $\Delta\theta_{\mathbf{WS}}^{(\mathbf{j})}$ augmente très légèrement tout en restant voisin de $\boldsymbol{\delta_\theta}$, s'interprète en considérant l'isométrie maximale comme un état pseudo-absorbant ou comme le point selle d'un attracteur.

Cela avait déjà été souligné à la rubrique « Répétitions de la mesure de $\mathbf{T0}$ » du paragraphe 6.4.3, au paragraphe 10.2.1 intitulé « Remontée lente de pT vers 1 », et indiqué avec l'énoncé de l'Hyp. 11.

Ce phénomène dénote la capacité de la fibre musculaire à maintenir une contraction isométrique maximale, *i.e.* à rigidifier les Ns sarcomères disposés en série.

Concernant le détachement des tetM en isométrie maximale, où $\mathbf{pT^{(j=0)} = pT0 = 1}$, notre modèle conjecture les conclusions formulées dans (Brunello, Fusi et al. 2009) : les positions angulaires des S1b appartenant aux tetM qui se détachent sont à proximité de la borne maximale de $\Delta\theta_{\mathbf{WS}}^{(\mathbf{j})}$, on doit donc observer un déplacement relatif moyen des tetM en cours de WS, *i.e.* non détachées, vers $\mathbf{X0 = 0}$ et une augmentation de la force par tetM.

**i1 borné par 0. 8 : valeur proche de 0.78 ≈ 0.75 ≈ pT0***

Le calcul de $\mathbf{pT0^*}$ procédé dans (15.15) est issu de (4.39) où $\Delta\theta_{\mathbf{WS}}^{(\mathbf{j})}$ devient voisin de $\Delta\theta_{\mathbf{WS}}^{\mathbf{Max}}$, configuration prédite par le modèle lorsque $\mathbf{pT^{(j)}}$ se rapproche de $\mathbf{0.8}$.

Ainsi l'origine de la valeur $\mathbf{0.78}$ qui est à la base de la nature « biphasique » étudiée à plusieurs reprises par K. Edman (Edman, Mulieri et al. 1976; Edman 1979; Edman 1988) résulterait de $\mathbf{pT0^*}$.

**Phases 2 et 3 absentes dans i1**

Il n'apparait pas d'effets dus à la viscosité, ni à l'évt DE, influences principales interférant durant les phases 2 et 3 (voir chap. 13 et 14) à cause de la faiblesse des valeurs de vitesse de raccourcissement; en conséquence ces 2 phases ne se distinguent pas de la phase 4 dans **i1** (Piazzesi, Lucii et al. 2002).

**Alignement des points expérimentaux dans i2 sur une droite passant par le point (pT = 1 ; V = 0)**

Toutes les relations F/V dont nous avons pris connaissance vérifient cette propriété (Fig. 27 et 28); soit tous les travaux relatifs à la phase 4 d'un échelon de force déjà cités, et les articles suivants : (Edman 1979; Ford, Huxley et al. 1985) .

Les pentes des relations linéaires, exprimées par exemple avec (15.21a), (15.21b) ou (15.21c), sont proportionnelles :

- au coefficient de raideur de la fmI si **T** est fonction de **V**
- au coefficient de raideur d'un hs de la fmI si **pT** est fonction de **u**
- à la longueur ou au nombre de sarcomères si **V** est fonction de **T**

---

[1] *Mais aussi valable pour* $\mathbf{1 \leq pT^{(j)} \leq 1.1}$ *; l'allongement de la fibre ou «stretching» fera l'objet d'un autre article.*

- à la tension maximale isométrique si T est fonction de **V**

- à la durée moyenne de survenue de l'évt $SB_{slow}$ si **T** est fonction de **V**, ou si **pT** est fonction de **u**

Ainsi ces relations linéaires permettent de caractériser la nature élastique du comportement collectif des tetM pour un hs, une myofibrille ou la fibre musculaire.

La baisse proportionnelle de $\Lambda_{WS}^{(j)}$ et de l'élargissement pratiquement constant de $\Delta\theta_{WS}^{(j)}$ autour de $\Delta\theta_{WS}^{Max}$ permet de prédire que $R_{M3}$, paramètre qui rend compte de ce phénomène doit rester constant dans **i2**, ce qui est effectivement observé avec la Fig. 3D dans (Piazzesi, Reconditi et al. 2007).

**Alignement des points expérimentaux dans i3 sur une courbe « de nature hyperbolique » dans i3**

Toutes les relations F/V dont nous avons pris connaissance vérifient cette propriété (Fig. 27 et 28); soit tous les travaux relatifs déjà mentionnés.

Cet arc de courbe est obtenu à partir de paramètres identiques à ceux adoptés dans i2 :

- la même constante de temps $\Delta t_{SBslow}$
- les mêmes équations pour $pT_{Lim}^{(j)}$ provenant des données obtenues d'après les études réalisées les échelons de longueur adoptées dans **i2** pour l'évt $SB_{fast}$

Ces résultats reflètent une homogénéité du modèle.

**Limite du modèle dans i4 et détermination difficile de $V^{Max}$**

Les vitesses élevées imposent des temps courts impropres aux calculs d'intégration.

Ce hiatus est représenté expérimentalement par des mesures plus dispersées pour les très faibles valeurs de forces, et par des données variant d'une publication à l'autre. On retrouve la complexité analysée au paragraphe 8.2 (Fig. 19) avec des valeurs de pentes différentes entre résultats de différents chercheurs.

Malgré cela, une modélisation a été obtenue avec la même constante de temps $\Delta t_{SBslow}$ et les mêmes équations pour $pT_{Lim}^{(j)}$ adoptées dans les intervalles **i2** et **i3**, indiquant à nouveau la cohérence du modèle.

L'évaluation correcte de $V^{Max}$ est rendue difficile car le raccourcissement nécessite une tension minimale (voir chapitre 16).

**Déplacement relatif moyen et force moyenne par tetM selon les 4 intervalles**

L'étude des 4 intervalles a souligné le rôle de la dispersion $\Delta\theta_{WS}^{(j)}$ ; les évolutions de $\Delta\theta_{WS}^{(j)}$ expliquent que, dans **i1** et **i2** comparativement à **i3** et **i4,** le pas moyen d'une tetM et la force moyenne par tetM doivent être, pour le premier, inférieur et, pour la seconde, supérieure (Reconditi, Linari et al. 2004).

**Nombre et répartition des tetM en cours de WS identiques dans chaque hs de la fmI**

La validité de l'éq. (15.4) s'appuie sur la répétitivité des mesures réalisées avec différentes fibres durant la phase 4 d'un échelon de force par des techniques de diffractométrie des rayons X (Linari, Piazzesi et al. 2000; Huxley, Reconditi et al. 2003; Reconditi, Linari et al. 2004; Huxley, Reconditi et al. 2006b; Piazzesi, Reconditi et al. 2007 ; Linari, Piazzesi et al. 2009; Reconditi, Brunello et al. 2011).

**Vitesse proportionnelle à la longueur du levier S1b**

Durant la phase 4 d'un échelon de force n° j, l'éq. (3.15) se réécrit :

$$\mathbf{u^{(j)} = L_{S1b} \cdot R_{WS} \cdot \dot{\theta}^{(j)}} \qquad (15.31a)$$

où $\dot{\theta}^{(j)}$ est la valeur en module de la vitesse de rotation constante commune à tous les S1b appartenant aux $\Lambda_{WS}^{(j)}$ tetM en cours de WS dans n'importe quel hs de la fmI

En combinant (15.31a) avec (15.8), on obtient en module :

$$\mathbf{V^{(j)} = 2 \cdot Ns \cdot L_{S1b} \cdot R_{WS} \cdot \dot{\theta}^{(j)}} \qquad (15.31b)$$

On déduit de ces 2 formulations que la vitesse de raccourcissement d'un hs ou de la fmI varie proportionnellement avec le bras de levier, *i.e.* la longueur du levier S1b dans notre modèle, ce qui est vérifié expérimentalement (Uyeda, Abramson et al. 1996).

**In fluence de la longueur de la fibre sur V et rôle négligeable de la viscosité**

Les relations (15.8) et (15.31b) indiquent que la vitesse de raccourcissement de la fmI varie en proportion avec le nombre de sarcomères disposés en série, et consécutivement avec une longueur de référence comme celle de la fmI en conditions isométriques maximales $\mathbf{L0_{fmi}}$.

Si on choisit 2 fibres de longueurs de référence différentes mais de mêmes types (prélevées sur le même muscle de la même grenouille), en rapportant la vitesse des 2 fibres à sa longueur initiale, on doit observer une superposition des points ; ce phénomène est vérifié :

- voir Fig. 3 dans (Edman, Mulieri et al. 1976) relativement aux 2 fibres de longueur respective 8.6 mm et 11 mm avec une valeur identique : $\mathbf{\Delta t_{SBslow} = 7.6\ ms}$
- voir Fig 2A et 2B dans (Edman 1988) avec une fibre de 6.3 mm et un segment de 0.6 mm extrait de cette même fibre avec une valeur identique : $\mathbf{\Delta t_{SBslow} = 4.5\ ms}$

Cette superposition des points a une autre conséquence :

Avec (7.3a) et (7.3b), il a été constaté qu'en présence de forces de viscosité, le nombre de sarcomères influe sur la raideur de la fmI modélisée comme un ressort mécanique (voir Annexe D). En conséquence, si la viscosité se manifeste pendant la phase 4, la relation force-vitesse de la fmI devrait varier en fonction de la longueur de la fibre étudiée. La superposition quasi parfaite des points des Fig. 2A et 2 B dans (Edman 1988), même pour les vitesses les plus élevées, avec une fibre de 6.3 mm et un segment de 0.6 mm, suppose que la viscosité n'intervient pas durant la phase 4 d'un échelon de force.

**Rôle de la température**

L'accroissement de la température augmente le nombre de chocs d'origine thermique et doit ainsi améliorer les possibilités de réalisation de l'évt $\mathbf{SB_{slow}}$ en occasionnant une diminution de $\mathbf{\Delta t_{SBslow}}$.

Suite à des expérimentations effectuées sur la même fibre avec des températures expérimentales de 1.8° c et 11°C, (voir Fig. 5 dans (Edman 1988)) nos calculs basés sur les points de cette figure donnent les valeurs respectives, $\mathbf{\Delta t_{SBslow}^{t=1.8°C} = 6.5ms}$ et $\mathbf{\Delta t_{SBslow}^{t=11°C} = 3ms}$ ; la baisse de plus de la moitié de la constante de temps relativement à une augmentation de 9°C, confirme la prédiction.

On observe un résultat similaire avec d'autres travaux (Ranatunga 1984 ; Piazzesi, Reconditi et al. 2003; Elangovan, Capitanio et al. 2012).

**Influence de la distance inter-filamentaire**

La réduction de la distance entre le milieu du diamètre du filA et le milieu du diamètre du filM ($\mathbf{d_{filAM}}$) entraîne *de facto* une difficulté accrue pour les tetM de remplir les conditions favorables au déclenchement d'un WS (voir paragraphe 3.5.1); il apparait par conséquent une restriction des chances de réalisation de l'évt $\mathbf{SB_{slow}}$ se traduisant par une diminution de $\mathbf{\Delta t_{SBslow}}$, ce qui apparait effectivement avec la Fig. 3 dans (Edman and Hwang 1977) et avec la Fig. 4 dans (Edman, Reggiani et al. 1988).

# 16 Valeurs minimale et maximale du nombre de WS

Rappel du corolaire 1 se rapportant aux Hyp. 1 et 7 (voir paragraphe 1.5) : seul un des 2 S1 reliés à S2 a la possibilité de se lier à un sitA pour y effectuer un WS.

Pour chaque hs de la fmI, on nomme $\Lambda_{TOT}$ le nombre total de S1, $\Lambda_{CB}$ le nombre de CB et $\Lambda_{WS}$ le nombre de WS.

D'après la définition de ces entités, on vérifie :

$$\frac{\Lambda_{TOT}}{2} \geq \Lambda_{CB} \geq \Lambda_{WS} \tag{16.1}$$

$\Lambda 0$ a été introduit dans (6.4) comme le nombre maximal de tetM initiant un WS dans un hs de la fmI, *i.e.* la valeur maximale de WS par hs. Le nombre de CB par hs est le nombre de possibilités de réalisation des états WB, SB et WS (Fig.1 ; chap. 1), tel que :

$$\Lambda_{CB} \geq \Lambda 0 \tag{16.2}$$

Lors de la phase 4 d'un échelon de force, chaque ensemble $(Z_S + filA)$ est animé d'un mouvement linéaire continu dans le temps à vitesse relative $u^{(j)} = cste$ (voir paragraphe 15.2). Ce mouvement est réalisé par une succession de WS non synchronisés car l'instant de début de WS dépend de facteurs aléatoires, comme la position spatiale de tel S1 par rapport à une molA pour un possible CB, ou de tel autre S1 par rapport à un sitA pour un éventuel WS.

Durant la phase 4 d'un échelon de force, l'égalité (15.6) prédit que le nombre et la répartition des tetM en cours de WS sont identiques et constants dans chaque hs de la fmI ; associée à l'égalité (15.13), (15.6) souligne l'état stable en considérant que le nombre de S1 qui entament leur WS est équivalent à celui des S1 qui terminent leur WS. Il faut que la succession des WS non synchronisés se répartissent (approximativment) de manière uniforme sur le cycle complet d'un mouvement (collectif) de rotation, *i.e.* une répartition uniforme sur le cercle de rayon unitaire 1. Le nombre minimal de la répartition de WS durant la phase 4 correspond au cas où un S1 commence son WS quand un S1 finit le sien, soit le cas extrême $2\pi / \Delta\theta_{WS}^{Max}$ .

Il s'en déduit :

$$\Lambda_{WS} > \frac{2\pi}{\Delta\theta_{WS}^{Max}} \tag{16.3}$$

L'éq. (16.3) signifie qu'il faut une valeur minimale de tension pour que la fmI se raccourcisse et que conséquemment, il n'y a pas de raccourcissement possible pour la fmI à tension nulle (Hill 1949).

D’après ce qui précède, avec (16.2) et (16.3), les inégalités de (16.1) se réécrivent :

$$\frac{\Lambda_{TOT}}{2} \geq \Lambda_{CB} \geq \Lambda 0 \geq \Lambda_{WS} > \frac{2\pi}{\Delta\theta_{WS}^{Max}} \tag{16.4}$$

Rappel de l’éq. (1.19) avec $p_{WS}^{Max}$ la probabilité maximale de WS :

$$p_{WS}^{Max} = \frac{1}{2} \cdot \frac{3 \cdot 5.4}{36} \approx 22.5\%$$

Soit en divisant chaque membre des inégalités de (16.4) par $\Lambda_{TOT}$ et en introduisant la donnée précédente :

$$50\% \geq \frac{\Lambda_{CB}}{\Lambda_{TOT}} \geq 22.5\% \geq \frac{\Lambda 0}{\Lambda_{TOT}} \geq \frac{\Lambda_{WS}}{\Lambda_{TOT}} > \frac{2\pi}{\Lambda_{TOT} \cdot \Delta\theta_{WS}^{Max}} \tag{16.5}$$

# 17 Signature de la fibre musculaire en contraction

Par souci de simplification, un échelon de longueur sera noté **stepL** et un échelon de force **stepF** .

## 17.1 Synthèse des écritures

Avec les données de la littérature et les calculs des chap. 3 et 4, la tension relative se formule en fonction du raccourcissement relatif d'un hs

dans la **zone 1**, d'après (4.25) :

$$\mathbf{pT_{z1}(\Delta X) = 67.26 \cdot \log\left(\frac{\Delta X + 55.5}{\Delta X + 48.7}\right) - 7.79} \tag{17.1}$$

dans la **zone 2**, d'après (4.30) :

$$\mathbf{pT_{z2}^{E}(\Delta X) = \frac{457.4}{(10 - \Delta X)} \cdot \log\left(\frac{58.7}{\Delta X + 48.7}\right) - 7.79} \tag{17.2a}$$

dans la **zone 2**, d'après (4.32) :

$$\mathbf{pT_{z2}^{A}(\Delta X) = 67.26 \cdot \log\left(\frac{58.7}{\Delta X + 48.7}\right) + 1.146 \cdot \Delta X - 11.46} \tag{17.2b}$$

A l'aide de (17.1) et (17.2a), la fonction $\mathbf{G^{E}}$ (traits bleu et rouge ; Fig. 30) est définie[1] par :

$$\mathbf{G^{E}(\Delta X) = pT_{z1}(\Delta X) \cdot 1\!\!1_{\left[0;\Delta X_{WS}^{min}\right]}(\Delta X) + pT_{z2}^{E}(\Delta X) \cdot 1\!\!1_{\left]\Delta X_{WS}^{min};\Delta X_{WS}^{Max}\right]}(\Delta X)} \tag{17.3a}$$

$\mathbf{G^{E}}$ fournit la valeur de $\mathbf{pT}$ lorsqu'après le déplacement relatif $\mathbf{\Delta X}$ du disqM par rapport au disqZ, les tetM en fin de WS sont attachées et exercent une action collective sur le filM (mode Exagéré).

Grâce à (17.1) et (17.2b), la fonction $\mathbf{G^{A}}$ (traits bleu et vert ; Fig. 30) est définie[1] par :

$$\mathbf{G^{A}(\Delta X) = pT_{z1}(\Delta X) \cdot 1\!\!1_{\left[0;\Delta X_{WS}^{min}\right]}(\Delta X) + pT_{z2}^{A}(\Delta X) \cdot 1\!\!1_{\left]\Delta X_{WS}^{min};\Delta X_{WS}^{Max}\right]}(\Delta X)} \tag{17.3b}$$

avec $\mathbf{\Delta X_{WS}^{min} = 3.2\ nm}$

et $\mathbf{\Delta X_{WS}^{Max} = 10\ nm}$

$\mathbf{G^{A}}$ fournit la valeur de $\mathbf{pT}$ lorsqu'après le déplacement relatif $\mathbf{\Delta X}$, les tetM en fin de WS sont détachées et n'exercent pas d'action collective sur le filM (mode Amorti).

[1] *Voir la définition d'une fonction indicatrice à l'Annexe A1.*

**Fig. 30 : Relation bimodale de la tension relative en fonction du raccourcissement relatif d'un hs de la fmI**

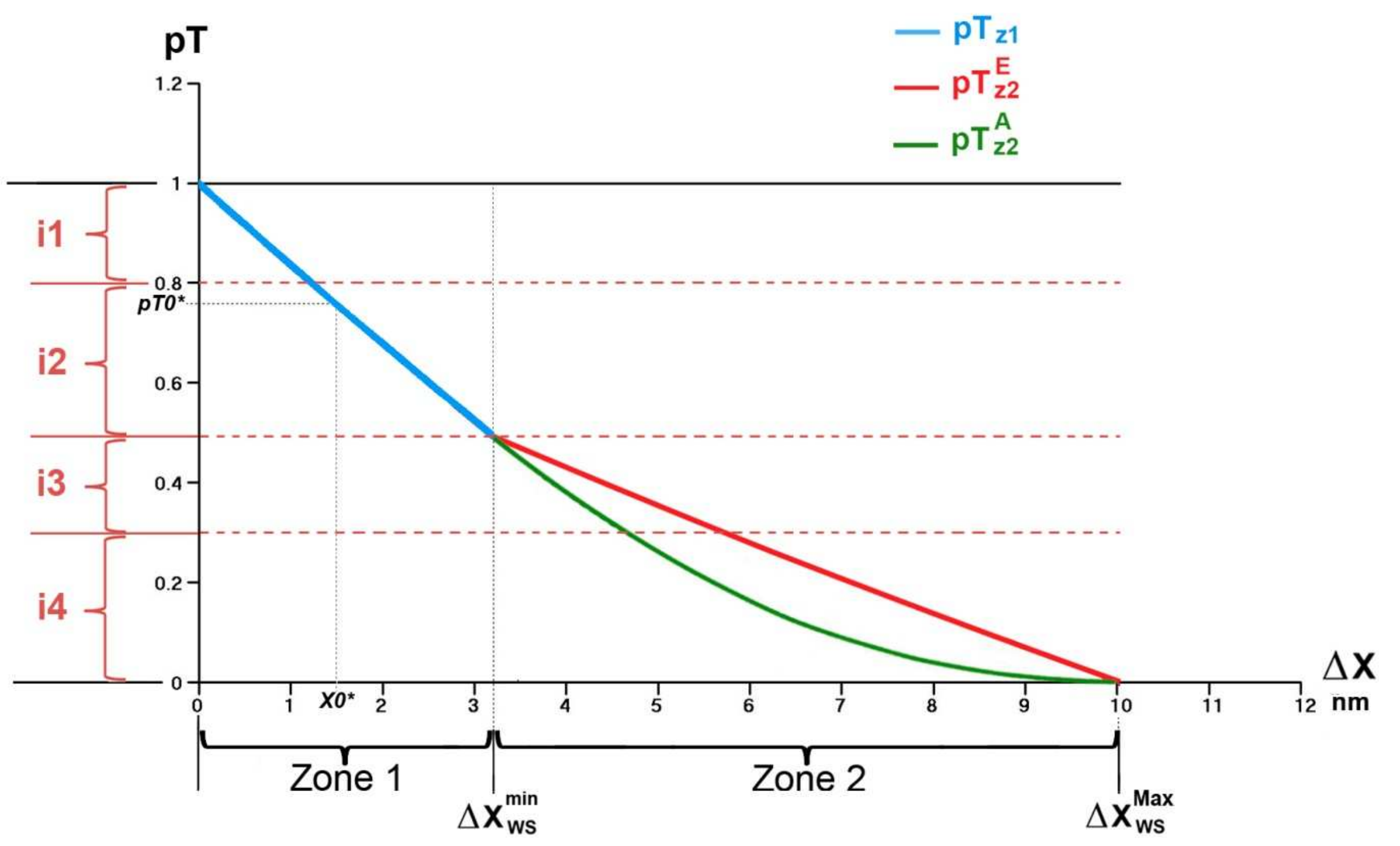

## 17.2 Convergences entre les 4 phases transitoires pour les 2 types de perturbation

**Phase 1**

Pour **stepL** n° k ou **stepF** n° j : phase identique dans les 2 cas avec action essentielle des forces de viscosité lorsque la durée de la phase 1 est inferieure à 0.2 ms. La fmI se comporte comme un ressort mécanique linéaire.

**Phase 2**

Phase qui se décompose en 2 périodes

***1ère période de la phase 2*** : annulation des forces de viscosité

Pour **stepL** n° k : remontée de $\mathbf{pT^{(k)}}$ vers $\mathbf{G^{E}\left(\Delta X^{(k)}\right)}$

Pour **stepF** n° j : maintien de $\mathbf{pT^{(j)}}$ avec raccourcissement jusqu'à $\mathbf{\Delta X^{(j)}_{p2_visco=0}}$ tel que

$$\mathbf{pT^{(j)} = G^{E}\left(\Delta X^{(j)}_{p2_visco=0}\right)}$$

***2ère période de la phase 2*** :

Pour **stepL** n° k : remontée de $\mathbf{pT^{(k)}}$ par le déclenchement de nouveaux WS suite à l'évt $\mathbf{SB_{fast}}$

Pour **stepF** n° j : maintien de $\mathbf{pT^{(j)}}$ engendré par l'initiation de nouveaux **WS** après réalisation de l'évt $\mathbf{SB_{slow}}$ ; le nombre et la répartition des tetM concernées par cet évènement varient selon l'intervalle $\mathbf{i1, i2, i3}$ **ou** $\mathbf{i4}$.

**Phase 3**

Phase de ralentissement dans les 2 cas, du en grande partie à l'évt $\mathbf{DE}$

Pour $\mathbf{stepL}$ n° k : plateau temporel avec ralentissement ou arrêt de la remontée de $\mathbf{pT^{(k)}}$ causé par la baisse de $\mathbf{G^E\left(\Delta X^{(k)}\right)}$ vers $\mathbf{G^A\left(\Delta X^{(k)}\right)}$ par suite du détachement des tetM en fin de WS .

Pour $\mathbf{stepF}$ n° j : maintien de $\mathbf{pT^{(j)}}$ avec ralentissement net de $\mathbf{V^{(j)}}$, provoqué par la réduction de $\mathbf{pT^{(j)}}$ à cause du tassement occasionné par le détachement des tetM en fin de WS entrainant $\mathbf{G^E\left(\Delta X^{(j)}(t)\right)}$ vers $\mathbf{G^A\left(\Delta X^{(j)}(t)\right)}$.

**Phase 4**

Pour $\mathbf{stepL}$ n° k : remontée de $\mathbf{pT^{(k)}}$ à $\mathbf{\Delta X^{(k)} = cste}$ vers l'état pseudo-absorbant $\mathbf{pT0 = 1}$

Pour $\mathbf{stepF}$ n° j : maintien de $\mathbf{pT^{(j)}}$ avec vitesse de raccourcissement $\mathbf{V^{(j)} = cste}$ , maintien engendré par l'initiation de nouveaux $\mathbf{WS}$ après réalisation de l'évt $\mathbf{SB_{slow}}$ ; le nombre et la répartition des tetM concernées par cet évènement varient selon l'intervalle $\mathbf{i1, i2, i3}$ **ou** $\mathbf{i4}$ .

La position angulaire moyenne des S1b appartenant aux tetM en cours de WS correspond à un déplacement relatif moyen $\mathbf{\Delta X^{(j)}}$ tel que $\mathbf{pT^{(j)} = G^A\left(\Delta X^{(j)}\right)}$ dans $\mathbf{i2}$ **et** $\mathbf{i3}$ ; des adaptations sont nécessaires dans $\mathbf{i1}$ **et** $\mathbf{i4}$ .

La formulation analytique de $\mathbf{G^A}$ autorise le calcul de la relation entre $\mathbf{pT^{(j)}}$ et $\mathbf{V^{(j)}}$.

***Remarques :***

- similitudes entre la 2ème période de la phase 2 et la phase 4 (fait déjà souligné par G. Piazzesi)
- l'état stable de la phase 4 débute lorsque $\mathbf{\Delta X \geq \Delta X_{WS}^{Max}}$ , *i.e.* lorsque toutes les tetM qui ont initié un WS lors de la mise en isométrie maximale sont détachées et ne contribuent plus à $\mathbf{pT^{(j)}}$. L'état stable d'une perturbation en escalier (répétition d'1 $\mathbf{stepL}$ ) apparait avec le même préalable.

## 17.3 Discussion

La nature bimodale du comportement collectif des tetM, formulée grâce aux 2 fonctions $\mathbf{G^E}$ et $\mathbf{G^A}$ , permet d'expliciter, directement ou indirectement, les 4 phases transitoires qui se succèdent après une perturbation par un échelon de longueur ou de force.

On peut donc présenter ces 2 fonctions comme la signature de la fmI.

Il reste à découvrir les caractéristiques de cette nature bimodale : possèdent-elles des composantes communes à toutes les fibres musculaires, ou, tout au contraire, varient-elles selon la structure de la tetM ou selon d'autres facteurs ?

# 18 Limites du modèle

## 18.1 Passage de l'état WB à l'état WS

L'Hyp. 1 stipule trois conditions pour l'initiation du WS : rigidité de S2, liaison forte et démarrage du moment moteur. Ainsi, l'Hyp. 1 requiert la présence de S2[1] , tout en soulignant son importance dans le modèle.

.Le modus operandi qui permet le passage de l'état WB à l'état WS reste à découvrir.

## 18.2 Dénombrement des tetM subissant les évts SBfast ou SBslow

Par définition, $\left(\mathbf{pT2}^{(k)} - \mathbf{pT1/2}^{(k)}\right)$ comptabilise le nombre maximal de possibilités de survenue de l'évt $\mathbf{SB_{fast}}$ pour la phase 2 d'un échelon de longueur n° k.

Toutes les expressions concernant $\mathbf{pT2}^{(k)}$ ont été fournies empiriquement (voir paragraphes 8.2, 8.4, 8.5.1 et 11.2). L'évolution de $\left(\mathbf{pT2}^{(k)} - \mathbf{pT1/2}^{(k)}\right)$ a été décrit au paragraphe 6.3.2 par une courbe concave dont le maximum se situait à l'abscisse $\mathbf{\Delta X = d_{2sitA}}$ (Fig. 16 et 17; chap. 8).

La modélisation théorique de $\mathbf{pT2}$ reste à être entreprise.

Par définition, $\left(\mathbf{pT}_{\mathbf{Lim}}^{(\mathbf{j})} - \mathbf{pT}^{(\mathbf{j})}\right)$ comptabilise le nombre maximal de possibilités de survenue de l'évt $\mathbf{SB_{slow}}$ pour la phase 4 d'un échelon de force n° j.

Les équations de $\mathbf{pT}_{\mathbf{Lim}}^{(\mathbf{j})}$ ont été alignées sur celles de $\mathbf{pT2}^{(k)}$ (voir paragraphe 15.4).

Or les formulations de $\mathbf{pT2}^{(k)}$ ont été agencées relativement au point de coordonnées (Fig. 15 ; chap. 8) :

$$\left|\begin{array}{l}\mathbf{\Delta X_{p1} = d_{2sitA} = 5.4\,nm}\\ \mathbf{pT2 = 0.8 = 80\%}\end{array}\right.$$

Quel est le lien entre ces 2 données, la distance entre 2 sitA qui détermine la manière dont les tetM dans les états WB ou SB sont perturbées après un échelon de longueur, et la borne inférieure de l'intervalle $\mathbf{i1}$ qui concerne l'étalement des positions angulaires des S1b de $\mathbf{\delta_\theta}$ à $\mathbf{\Delta\theta_{WS}^{Max}}$ pour un échelon de force compris entre $\mathbf{0.8 \cdot T0^{fmI}}$ et $\mathbf{T0^{fmI}}$ ?

Est-ce fortuit ou existe –il une relation de causalité ? Y-a-t-il un rapprochement à effectuer avec la superposition des 2 droites schématisant la fin des phases 2 pour les 2 types de perturbation ?

---

[1] *De notre point de vue, le déplacement d'un S1 seul sur un filA serait concerné uniquement par les états WB et DE (Fig.1, chap.1), et non par les états SB et WS qui nécessitent la présence de S2. Les sauts d'un S1 isolé, de sitA en sitA, s'expliqueraient par les ruptures successives et aléatoires de la liaison faible provoquées par les forces thermiques, puis par l'attraction exercée par le sitA voisin (d'où possibilité de saut dans la direction opposée à celle de la polarisation des molA); les forces mesurées lors de la variation de la quantité de mouvement du filA après chaque saut appartiennent, en règle générale, à l'intervalle déterminé par les inégalités de (1.9) ; ceci expliquerait pourquoi plusieurs sauts sont effectués avec une seule molécule d'ATP puisque celle-ci n'est pas utilisée pour déclencher un WS.*

## CONCLUSION

De futures expérimentations valideront ou réfuteront les thèses avancées dans les chapitres précédents. Toutefois, notre modèle a démontré que le recours aux lois de la mécanique classique permettait d'appréhender certains des problèmes soulevés par le comportement collectif des têtes de myosine, et ceci en considérant la fibre musculaire squelettique pour ce qu'elle est intrinsèquement au quotidien[1] :

> un objet mécanique générateur de forces et de mouvements déterministes.

[1] *à commande motrice identique.*

# ANNEXE A

Les notions mathématiques, présentées succinctement dans les paragraphes qui suivent, figurent dans un grand nombre d'ouvrages de mathématiques consacrés aux probabilités. On les trouve par exemple, en ordre dispersé, dans (Engel 1976; Roddier 1988; Combrouze 1993; Suquet 2013).

## A1 Fonction indicatrice

Soit $\omega$ un évènement élémentaire d'une expérience et $\Omega$ l'ensemble de tous les évènements élémentaires de cette expérience.

Soit l'évènement $\mathbf{A}$ , un sous-ensemble d'événements élémentaires de $\Omega$ .

On note $\mathbf{1}_A$ la fonction indicatrice relative à $\mathbf{A}$ telle que :

$$\mathbf{1}_A(\omega) = \begin{cases} 1 & \text{si } \omega \in A \\ 0 & \text{sinon} \end{cases}$$

## A2 Rappels

$$\int_0^t \lambda \cdot e^{-\lambda \cdot z} . dz = \left[ -e^{-\lambda \cdot z} \right]_0^t = 1 - e^{-\lambda \cdot t} \qquad \text{(A.1)}$$

avec $\lambda \in \mathbf{R}_+^*$

Le développement en série entière de l'exponentielle donne :

$$e^x = 1 + \frac{x}{1!} + \frac{x^2}{2!} + \frac{x^3}{3!} + \ldots + \frac{x^n}{n!} + \ldots \qquad \text{(A.2)}$$

où $x \in \mathbf{R}$

On en déduit que si $\mathbf{x}$ tend vers $\mathbf{0}$ :

$$1 - e^{-\frac{x}{a}} \approx \frac{x}{a} \qquad \text{(A.3)}$$

où $\mathbf{a}$ est une constante avec $a \in \mathbf{R}_+^*$

$$1 - \frac{a \cdot e^{-\frac{x}{a}} - b \cdot e^{-\frac{x}{b}}}{a - b} \approx \frac{x^2}{a \cdot b} \qquad \text{(A.4)}$$

où $\mathbf{a}$ et $\mathbf{b}$ sont 2 constantes avec $a, b \in \mathbf{R}_+^*$ vérifiant $a \neq b$

## A3 Loi de probabilité de la somme de 2 v.a.[1] indépendantes suivant chacune une loi exponentielle

Soit $T_1$ et $T_2$ , 2 v.a. suivant chacune une loi exponentielle de paramètre respectif $\lambda_1$ et $\lambda_2$, telles que :

$$\lambda_1 \neq \lambda_2 \tag{A.5}$$

Les densités de probabilités de $T_1$ et $T_2$ , notées respectivement $f_1$ et $f_2$, s'écrivent :

$$f_1(t) = \lambda_1 \cdot e^{-\lambda_1 \cdot t} \cdot \mathbb{1}_{R_+}(t) \tag{A.6}$$

$$f_2(t) = \lambda_2 \cdot e^{-\lambda_2 \cdot t} \cdot \mathbb{1}_{R_+}(t) \tag{A.7}$$

où $t$ représente le temps instantané,

Considérons la somme de $T_1$ et $T_2$, *i.e.* $(T_1 + T_2)$ et la densité de cette v.a. notée $f_{1+2}$ . On démontre classiquement que, si $T_1$ et $T_2$ sont 2 v.a. indépendantes, $f_{1+2}$ est égale au produit de convolution de $f_1$ et $f_2$, soit :

$$f_{1+2}(t) = (f_1 * f_2)(t) = \int_{-\infty}^{+\infty} f_1(t-z) \cdot f_2(z) \cdot dz$$

Qui se réécrit avec (A.6) et (A.7) :

$$f_{1+2}(t) = \int_{-\infty}^{+\infty} \left[\lambda_1 \cdot e^{-\lambda_1 \cdot (t-z)} \cdot \mathbb{1}_{R_+}(t-z) \cdot \lambda_2 \cdot e^{-\lambda_2 \cdot z} \cdot \mathbb{1}_{R_+}(z)\right] \cdot dz \tag{A.8}$$

Le produit des 2 indicatrices se réécrit :

$$\mathbb{1}_{R_+}(t-z) \cdot \mathbb{1}_{R_+}(z) = \begin{cases} 1 & \text{si } z \geq 0 \text{ et } t - z \geq 0 \\ 0 & \text{sinon} \end{cases}$$

$$= \begin{cases} \mathbb{1}_{[0;t]}(z) \text{ si } t \geq 0 \\ 0 \text{ sinon} \end{cases}$$

$$= \mathbb{1}_{R_+}(t) \cdot \mathbb{1}_{[0;t]}(z) \tag{A.9}$$

En introduisant (A.9) dans (A.8), on obtient :

$$f_{1+2}(t) = \mathbb{1}_{R_+}(t) \cdot \lambda_1 \cdot \lambda_2 \cdot e^{-\lambda_1 \cdot t} \cdot \int_{-\infty}^{+\infty} e^{-z \cdot (\lambda_2 - \lambda_1)} \cdot \mathbb{1}_{[0;t]}(z) \cdot dz$$

Soit avec la définition d'une fonction indicatrice donnée au 1er paragraphe et après intégration avec (A.1) :

$$f_{1+2}(t) = \quad = \frac{\lambda_1 \cdot \lambda_2 \cdot e^{-\lambda_1 \cdot t}}{\lambda_2 - \lambda_1} \cdot \left[1 - e^{-t \cdot (\lambda_2 - \lambda_1)}\right]$$

---

[1] *v.a. : acronyme de « variable aléatoire ».*

La densité de probabilité de la somme de 2 v.a. suivant 2 loi exponentielles de paramètres respectifs $\lambda_1$ et $\lambda_2$ (avec $\lambda_1 \neq \lambda_2$ ) se formule :

$$f_{1+2}(t) = \frac{\lambda_1 \cdot \lambda_2}{(\lambda_2 - \lambda_1)} \cdot e^{-\lambda_1 \cdot t} + \frac{\lambda_1 \cdot \lambda_2}{(\lambda_1 - \lambda_2)} \cdot e^{-\lambda_2 \cdot t} \qquad \text{(A.10)}$$

On obtient directement ce résultat avec le produit des fonctions caractéristiques des densités $f_1$ et $f_2$ (voir paragraphe suivant). Cette démonstration a été choisie car la réécriture du produit d'indicatrices en (A.9) montre que le temps de survenue d'un des 2 évènements précède celui de l'autre, et s'applique bien à 2 évènements qui se succèdent dans le temps.

## A4 Loi de probabilité de la somme de 3 v.a. indépendantes suivant chacune une loi exponentielle

On définit la fonction caractéristique d'une v.a. par la transformée de Fourier de la densité de cette v.a., soit :

$$\mathcal{F}[f(x)] = \int f(t) \cdot e^{i \cdot x \cdot t} \cdot dt \qquad \text{(A.11)}$$

où $i$ représente l'unité des nombres imaginaires purs.

Soit $T_1$, $T_2$ et $T_3$ , 3 v.a. suivant chacune une loi exponentielle de paramètre respectif $\lambda_1$, $\lambda_2$ et $\lambda_3$ telles que :

$$\lambda_1 \neq \lambda_2 \ , \lambda_2 \neq \lambda_3, \ \lambda_1 \neq \lambda_3 \qquad \text{(A.12)}$$

On démontre que la fonction caractéristique de la densité de la somme de 2 v.a. indépendantes est égale au produit des fonctions caractéristiques des 2 densités de ces 2 v.a. (Roddier 1988).

Considérons la somme de $T_1$, $T_2$ et $T_3$ , *i.e.* $(T_1 + T_2 + T_3)$ et la densité de cette v.a. notée $f_{1+2+3}$ .
D'après ce qui précède, avec (A.11) et par distributivité de la somme, du produit et du produit de convolution, la fonction caractéristique de $f_{1+2+3}$ s'écrit :

$$\mathcal{F}[f_{1+2+3}(x)] = \left(\int_0^\infty \lambda_1 \cdot e^{-(\lambda_1 - i \cdot x) \cdot t} \cdot dt\right) \cdot \left(\int_0^\infty \lambda_2 \cdot e^{-(\lambda_2 - i \cdot x) \cdot t} \cdot dt\right) \cdot \left(\int_0^\infty \lambda_3 \cdot e^{-(\lambda_3 - i \cdot x) \cdot t} \cdot dt\right)$$

Après intégration avec (A.1), on obtient :

$$\mathcal{F}[f_{1+2+3}(x)] = \frac{\lambda_1}{(\lambda_1 - i \cdot x)} \cdot \frac{\lambda_2}{(\lambda_2 - i \cdot x)} \cdot \frac{\lambda_3}{(\lambda_3 - i \cdot x)}$$

Et après développement :

$$\mathcal{F}\left[f_{1+2+3}(x)\right]=$$
$$\frac{\lambda_1\cdot\lambda_2\cdot\lambda_3}{(\lambda_2-\lambda_1)}\cdot\left[\frac{1}{(\lambda_3-\lambda_1)}\cdot\left(\frac{1}{\lambda_1-i\cdot x}-\frac{1}{\lambda_3-i\cdot x}\right)-\frac{1}{(\lambda_3-\lambda_2)}\cdot\left(\frac{1}{\lambda_2-i\cdot x}-\frac{1}{\lambda_3-i\cdot x}\right)\right]$$

En passant par la transformation inverse de la transformée de Fourier, on obtient la densité de la somme de 3 v.a. indépendantes d'après (A.11) et avec (A.12) :

$$f_{1+2+3}(t)=\frac{\lambda_1\cdot\lambda_2\cdot\lambda_3}{(\lambda_2-\lambda_1)\cdot(\lambda_3-\lambda_1)}\cdot e^{-\lambda_1\cdot t}+\frac{\lambda_1\cdot\lambda_2\cdot\lambda_3}{(\lambda_1-\lambda_2)\cdot(\lambda_3-\lambda_2)}\cdot e^{-\lambda_2\cdot t}+\frac{\lambda_1\cdot\lambda_2\cdot\lambda_3}{(\lambda_1-\lambda_3)\cdot(\lambda_2-\lambda_3)}\cdot e^{-\lambda_3\cdot t} \qquad \text{(A.13)}$$

## A5 Dénombrement

Soit $\mathbf{N}$ tetM concernées par un événement $\mathbf{A}$ dont la probabilité de survenue entre les instants $\mathbf{0}$ et $\mathbf{t}$ est notée $p_t$ .

Considérons $\mathbf{N}$ v.a. indépendantes, nommées génériquement $K_b$ (avec b = 1, 2, …, N) qui correspondent au temps de survenue de $\mathbf{A}$ entre les instants $\mathbf{0}$ et $\mathbf{t}$ pour chacune des $\mathbf{N}$ tetM.

Soit $L_b$, la v.a. de Bernouilli égale à $\mathbb{1}_A(t)$ pour la tetM n° $\mathbf{b}$ . Les $\mathbf{N}$ $L_b$ (avec b = 1, 2,…, N) sont des v.a. de Bernouilli indépendantes de paramètre $p_t$ .

Par définition, la somme des $\mathbf{N}$ $L_b$ est une v.a. nommée $N_A$ qui suit une loi binomiale de paramètres $\mathbf{N}$ et $p_t$ . On en déduit que le nombre moyen de tetM pour lesquelles $\mathbf{A}$ est survenu entre $\mathbf{0}$ et $\mathbf{t}$ est égal à :

$$E(N_A)=N\cdot p_t \qquad \text{(A.14)}$$

Il en découle que $N_{\overline{A}}$ le nombre de tetM pour lesquelles $\mathbf{A}$ n'est pas survenu entre $\mathbf{0}$ et $\mathbf{t}$ suit une loi binomiale de paramètres $\mathbf{N}$ et $(1-p_t)$, et est en moyenne égal à :

$$E(N_{\overline{A}})=N\cdot(1-p_t) \qquad \text{(A.15)}$$

***Cas particuliers*** :

**Cas 1** : la densité de probabilité est une loi exponentielle de paramètre $\lambda$, soit $f(t) = \lambda \cdot e^{-\lambda \cdot t}$ .

Les $N$ v.a. $K_b$ (avec b = 1, 2,…, N) suivent indépendamment la même loi exponentielle de paramètre $\lambda$.

La probabilité $p$ de survenue de $A$ avant $t$ pour la tête n° $b$ est égale par définition et avec (A.1), à:

$$p_t = P(K_b \leq t) = \int_0^t \lambda \cdot e^{-\lambda \cdot z}.dz = 1 - e^{-\lambda \cdot t} \tag{A.16}$$

En introduisant (A.16) dans (A.14), on obtient :

$$E(N_A) = N \cdot \left(1 - e^{-\lambda \cdot t}\right) \tag{A.17}$$

**Cas 2** : la densité de probabilité est $f_{1+2}(t)$, la densité de la sommé de 2 v.a. suivant chacune une loi exponentielle de paramètres différents (voir paragraphe A3) :

$$E(N_A) = N \cdot p_{1+2,t} \tag{A.18}$$

avec $p_{1+2,t} = \int_0^t f_{1+2}(z).dz$

Soit après intégration de (A.10) avec (A.1), en introduisant (A.18) dans (A.14), on obtient :

$$E(N_A) = N \cdot \left[1 - \frac{\lambda_2}{\lambda_2 - \lambda_1} \cdot e^{-\lambda_1 \cdot t} - \frac{\lambda_1}{\lambda_1 - \lambda_2} \cdot e^{-\lambda_2 \cdot t}\right] \tag{A.19}$$

**Cas 3** : la densité de probabilité est $f_{1+2+3}(t)$, la densité de la sommé de 3 v.a. suivant chacune une loi exponentielle de paramètres différents 2 à 2 (voir paragraphe A4) :

$$E(N_A) = N \cdot p_{1+2+3,t} \tag{A.20}$$

avec $p_{1+2+3,t} = \int_0^t f_{1+2+3}(z).dz$

Soit après intégration de (A.13) avec (A.1), en introduisant (A.20) dans (A.14), on obtient :

$$E(N_A) =$$

$$N \cdot \left[1 - \frac{\lambda_2 \cdot \lambda_3 \cdot e^{-\lambda_1 \cdot t}}{(\lambda_2 - \lambda_1) \cdot (\lambda_3 - \lambda_1)} - \frac{\lambda_1 \cdot \lambda_3 \cdot e^{-\lambda_2 \cdot t}}{(\lambda_1 - \lambda_2) \cdot (\lambda_3 - \lambda_2)} - \frac{\lambda_1 \cdot \lambda_2 \cdot e^{-\lambda_3 \cdot t}}{(\lambda_2 - \lambda_3) \cdot (\lambda_1 - \lambda_3)}\right]$$

(A.21)

## ANNEXE B

Dans le modèle, tous les filA sont identiques, tous les filM sont identiques et leur répartitions transversales respectives à l'intérieur de la fmI sont parfaitement homogènes de telle manière que la distance entre le milieu du diamètre du filA et le milieu du diamètre du filM, $\mathbf{d_{filAM}}$, est une constante (Fig. 3E ; chap. 2).
Les données qui suivent sont propres à chaque tête de myosine étudiée en cours de WS.

## S1 n° b dans le demi-sarcomère droit (hsR)

$\mathbf{A^{o(b)}_{s-1,L}}$, $\mathbf{B^{o(b)}_{s,R}}$ et $\mathbf{C^{o(b)}_{s,R}}$ points du plan $\mathbf{OXY^{o}}$ sont les projections orthogonales de $\mathbf{A^{(b)}_{s-1,L}}$, $\mathbf{B^{(b)}_{s,R}}$ et $\mathbf{C^{(b)}_{s,R}}$ points du plan $\mathbf{OXY^{(b)}}$ propre à S1 n° b dans hsR (Fig. 3D et 3E).

La trigonométrie dans le plan $\mathbf{OXY^{o}}$ donne à l'aide des coordonnées (X;Y) des points $\mathbf{A^{o(b)}_{s-1,L}}$, $\mathbf{B^{o(b)}_{s,R}}$ et $\mathbf{C^{o(b)}_{s,R}}$ et $\mathbf{D^{(b)}_{s,R}}$ de la Fig. 3E :

$$XA^{o(b)}_{s-1,L} - XD^{(b)}_{s,R} = L_{S1b} \cdot \sin\theta^{(b)}_{s,R} + L^{\circ}_{S2,R} \cdot \cos\varphi^{(b)}_{s,R} \tag{B.1}$$

$$\begin{aligned} YA^{o(b)}_{s-1,L} - YD^{(b)}_{s,R} &= -d_{AB} \cdot \cos\beta^{(b)}_{s,R} - L_{S1b} \cdot \cos\theta^{(b)}_{s,R} \cdot \cos\beta^{(b)}_{s,R} + L^{\circ}_{S2,R} \cdot \sin\varphi^{(b)}_{s,R} \\ &= -\left(d_{filAM} - r_{filM} - r_{filA} \cos\beta^{(b)}_{s,R}\right) \end{aligned} \tag{B.2}$$

où $\mathbf{\theta^{(b)}_{s,R}}$ est l'angle instantané de rotation selon $\mathbf{OZ^{(b)}}$ du segment S1b par rapport à l'axe $\mathbf{C^{(b)}_{s,R}Y}$ dans le plan $\mathbf{OXY^{(b)}}$ (Fig. 3C) où $\mathbf{\varphi^{(b)}_{s,L}}$ est l'angle instantané de rotation selon $\mathbf{OZ^{o}}$ de la projection orthogonale du segment S2 dans le plan $\mathbf{OXY^{o}}$ par rapport à l'axe $\mathbf{D^{(b)}_{s,L}X}$ (Fig. 3E) où $\mathbf{\beta^{(b)}_{s,R}}$ est l'angle constant durant le WS entre $\mathbf{OXY^{o}}$ et $\mathbf{OXY^{(b)}}$ selon l'axe longitudinal du filA dans hsR (Fig. 3B et 3D) où $\mathbf{L^{o}_{S2,R}}$ est la projection orthogonale de $\mathbf{L_{S2}}$ sur le plan $\mathbf{OXY^{o}}$ telle que

$$L^{o}_{S2,R} = \sqrt{L^{2}_{S2} - \left[\left(r_{filA} + d_{AB} + L_{S1b} \cdot \cos\theta^{(b)}_{s,R}\right) \cdot \sin\beta^{(b)}_{s,R}\right]^2} \tag{B.3}$$

et où $\mathbf{r_{filA}}$ (rayon du filA), $\mathbf{r_{filM}}$ (rayon du filM), $\mathbf{d_{AB}}$ (distance entre A et B) et $\mathbf{d_{filAM}}$ sont des constantes

Durant le WS, on vérifie en dérivant (B.3) :

$$\frac{dL^{o}_{S2,R}}{dt} = \dot{\theta}^{(b)}_{s,R} \cdot \frac{L_{S1b} \cdot \left(\sin\beta^{(b)}_{s,R}\right)^2 \cdot \sin\theta^{(b)}_{s,R} \cdot \left(r_{filA} + d_{AB} + L_{S1b} \cdot \cos\theta^{(b)}_{s,R}\right)}{L^{\circ}_{S2,R}} \tag{B.4}$$

On obtient par dérivation des équations (B.1) et (B.2) avec (B.4) :

$$\dot{X}A^{o(b)}_{s-1,L} - \dot{X}D^{(b)}_{s,R} = \dot{\theta}^{(b)}_{s,R} \cdot \left( L_{S1b} \cdot \cos\theta^{(b)}_{s,R} \right) - \dot{\varphi}^{(b)}_{s,R} \cdot \left( \overset{\circ}{L}_{S2,R} \cdot \sin\varphi^{(b)}_{s,R} \right)$$
$$+ \dot{\theta}^{(b)}_{s,R} \cdot \cos\varphi^{(b)}_{s,R} \cdot \frac{L_{S1b} \cdot \left( \sin\beta^{(b)}_{s,R} \right)^2 \cdot \sin\theta^{(b)}_{s,R} \cdot \left( r_{filA} + d_{AB} + L_{S1b} \cdot \cos\theta^{(b)}_{s,R} \right)}{\overset{\circ}{L}_{S2,R}}$$

(B.5)

$$0 = \dot{\theta}^{(b)}_{s,R} \cdot \left( L_{S1b} \cdot \cos\beta^{(b)}_{s,R} \cdot \sin\theta^{(b)}_{s,R} \right) + \dot{\varphi}^{(b)}_{s,R} \cdot \left( \overset{\circ}{L}_{S2} \cdot \cos\varphi^{(b)}_{s,R} \right)$$
$$+ \dot{\theta}^{(b)}_{s,R} \cdot \sin\varphi^{(b)}_{s,R} \cdot \frac{L_{S1b} \cdot \left( \sin\beta^{(b)}_{s,R} \right)^2 \cdot \sin\theta^{(b)}_{s,R} \cdot \left( r_{filA} + d_{AB} + L_{S1b} \cdot \cos\theta^{(b)}_{s,R} \right)}{\overset{\circ}{L}_{S2,R}}$$

(B.6)

De (B.6), on tire :

$$\dot{\varphi}^{(b)}_{s,R} = -\dot{\theta}^{(b)}_{s,R} \cdot \frac{L_{S1b} \cdot \cos\beta^{(b)}_{s,R} \cdot \sin\theta^{(b)}_{s,R}}{\overset{\circ}{L}_{S2,R} \cdot \cos\varphi^{(b)}_{s,R}}$$
$$- \dot{\theta}^{(b)}_{s,R} \cdot \tan\varphi^{(b)}_{s,R} \cdot \frac{L_{S1b} \cdot \left( \sin\beta^{(b)}_{s,R} \right)^2 \cdot \sin\theta^{(b)}_{s,R} \cdot \left( r_{filA} + d_{AB} + L_{S1b} \cdot \cos\theta^{(b)}_{s,R} \right)}{\left( \overset{\circ}{L}_{S2,R} \right)^2}$$

(B.7)

Puisque les points $A^{(b)}_{s-1,L}$ et $A^{o(b)}_{s-1,L}$ ont la même vitesse que le disque $Z_{s-1}$ et le point $D^{(b)}_{s,R}$ a la même vitesse que le disque $M_s$, d'après (1.3), on a :

$$\dot{X}A^{o(b)}_{s-1,L} - \dot{X}D^{(b)}_{s,R} = \dot{X}A^{(b)}_{s-1,L} - \dot{X}D^{(b)}_{s,R} = \frac{dM_sZ_{s-1}}{dt} = -u_{Ms,R} \qquad \text{(B.8)}$$

Et (B.7) et (B.8) introduits dans (B.5) donnent :

$$u_{Ms,R} = -\dot{\theta}^{(b)}_{s,R} \cdot L_{S1b} \cdot \left[ \left( \cos\theta^{(b)}_{s,R} + \sin\theta^{(b)}_{s,R} \cdot \tan\varphi^{(b)}_{s,R} \cdot \cos\beta^{(b)}_{s,R} \cdot \right) + res^{(b)}_{s,R} \right] \qquad \text{(B.9)}$$

où $$res^{(b)}_{s,R} = \left[ \frac{\left( \sin\beta^{(b)}_{s,R} \right)^2 \cdot \sin\theta^{(b)}_{s,R} \cdot \left( r_{filA} + d_{AB} + L_{S1b} \cdot \cos\theta^{(b)}_{s,R} \right)}{\overset{\circ}{L}_{S2,R} \cdot \cos\varphi^{(b)}_{s,R}} \right]$$

On note de plus que $\varphi_{s,R}^{(b)}$ dépend de $\theta_{s,R}^{(b)}$ et $\beta_{s,R}^{(b)}$ puisque (B.2) donne :

$$\varphi_{s,R}^{(b)} = \mathrm{Arcsin}\left( \frac{\left(r_{filA} + d_{AB} + L_{S1b} \cdot \cos\theta_{s,R}^{(b)}\right) \cdot \cos\beta_{s,R}^{(b)} + r_{filM} - d_{filAM}}{\sqrt{L_{S2}^{2} - \left[\left(r_{filA} + d_{AB} + L_{S1b} \cdot \cos\theta_{s,R}^{(b)}\right) \cdot \sin\beta_{s,R}^{(b)}\right]^{2}}} \right) \tag{B.10}$$

## S1 n° b dans le demi-sarcomère gauche (hsL)

Par symétrie, on obtient dans hsL :

$$u_{Ms,L} = -\dot{\theta}_{s,L}^{(b)} \cdot L_{S1b} \cdot \left[\left(\cos\theta_{s,L}^{(b)} + \sin\theta_{s,L}^{(b)} \cdot \tan\varphi_{s,L}^{(b)} \cdot \cos\beta_{s,L}^{(b)} \cdot\right) + res_{s,L}^{(b)}\right] \tag{B.11}$$

où $\beta_{s,L}^{(b)}$ est l'angle constant durant le WS entre $OXY^{o}$ et $OXY^{(b)}$ selon l'axe longitudinal du filA dans hsL

où $\varphi_{s,L}^{(b)}$ est l'angle instantané de rotation selon $OZ^{o}$ de la projection orthogonale du segment S2 dans le plan $OXY^{o}$ par rapport à l'axe $D_{s,L}^{(b)}X$ (Fig. 2E) où $\theta_{s,L}^{(b)}$ est l'angle instantané de rotation selon $OZ^{(b)}$ du segment S1b par rapport à l'axe $C_{s,L}^{(b)}Y$ dans le plan $OXY^{(b)}$

où $L_{S2,L}^{o}$ est la projection orthogonale de $L_{S2}$ sur le plan $OXY^{o}$ telle que

$$L_{S2,L}^{o} = \sqrt{L_{S2}^{2} - \left[\left(r_{filA} + d_{AB} + L_{S1b} \cdot \cos\theta_{s,L}^{(b)}\right) \cdot \sin\beta_{s,L}^{(b)}\right]^{2}} \tag{B.12}$$

et où
$$res_{s,L}^{(b)} = -\left[ \frac{\left(\sin\beta_{s,L}^{(b)}\right)^{2} \cdot \sin\theta_{s,L}^{(b)} \cdot \left(r_{filA} + d_{AB} + L_{S1b} \cdot \cos\theta_{s,L}^{(b)}\right)}{L_{S2,L}^{o} \cdot \cos\varphi_{s,L}^{(b)}} \right] \tag{B.13}$$

On note de plus que $\varphi_{s,L}^{(b)}$ dépend de $\theta_{s,L}^{(b)}$ et $\beta_{s,L}^{(b)}$ :

$$\varphi_{s,L}^{(b)} = -\mathrm{Arcsin}\left( \frac{\left(r_{filA} + d_{AB} + L_{S1b} \cdot \cos\theta_{s,L}^{(b)}\right) \cdot \cos\beta_{s,L}^{(b)} + r_{filM} - d_{filAM}}{\sqrt{L_{S2}^{2} - \left[\left(r_{filA} + d_{AB} + L_{S1b} \cdot \cos\theta_{s,L}^{(b)}\right) \cdot \sin\beta_{s,L}^{(b)}\right]^{2}}} \right) \tag{B.14}$$

## ANNEXE C

On note $\boldsymbol{\gamma}$ et $\boldsymbol{\delta}$, respectivement, l'accélération linéaire et la quantité d'accélération angulaire d'un système matériel,. Appliquons dans le repère galiléen OXY°Z° le théorème de la résultante cinétique (*TRC*) et le théorème du moment cinétique (*TMC*) aux centres de gravité des 3 solides indéformables composant la tête de myosine effectuant son WS (voir Fig. 4A et 4B ; chap. 3) :

### Demi sarcomère droit (hsR)

La tête de myosine n° b localisée dans hsR est composée des segments $S1a_R$, $S1b_R$ et $S2_R$ .

Pour $S1a_R$ (centre de gravité en $\mathbf{G_{S1a_R}}$ ; masse $\mathbf{m_{S1a}}$) les actions de liaisons sont modélisées par un torseur dont les éléments de réduction sont en $\mathbf{A^{(b)}_{s-1,L}}$ [ $\overrightarrow{\mathbf{F}_{A^{(b)}_{s-1,L}}}$ , $-\overrightarrow{\mathcal{M}_{A^{(b)}_{s-1,L}}}$ ] et en $\mathbf{B^{(b)}_{s,R}}$ [ $-\overrightarrow{\mathbf{F}_{B^{(b)}_{s,R}}}$ , $\overrightarrow{\mathcal{M}_{B^{(b)}_{s,R}}}$ ]

*TRC*

$$\mathbf{mS_{1a}} \cdot \overrightarrow{\gamma_{G^{(b)}_{S1a_R}}} = \vec{0} = -\overrightarrow{F_{B^{(b)}_{s,R}}}\left|\begin{matrix} -T_B \\ N_B \\ Q_B \end{matrix}\right. + \overrightarrow{F_{A^{(b)}_{s-1,L}}}\left|\begin{matrix} T_A \\ -N_A \\ -Q_A \end{matrix}\right.$$

Pour $\mathbf{S1b_R}$ (centre de gravité en $\mathbf{G_{S1b_R}}$ situé au milieu de $S1b_R$ ; masse $\mathbf{m_{S1b}}$ ), les actions de liaisons sont modélisées par un torseur dont les éléments de réduction sont en $\mathbf{B^{(b)}_{s,R}}$ [ $\overrightarrow{\mathbf{F}_{B^{(b)}_{s,R}}}$ , $-\overrightarrow{\mathcal{M}_{B^{(b)}_{s,R}}}$ ] et en $\mathbf{C^{(b)}_{s,R}}$ [ $-\overrightarrow{\mathbf{F}_{C^{(b)}_{s,R}}}$ , $\vec{\mathbf{0}}$ ]

*TRC*

$$\mathbf{m_{S1b}} \cdot \overrightarrow{\gamma_{G^{(b)}_{S1b_R}}} \approx \vec{0} = -\overrightarrow{F_{C^{(b)}_{s,R}}}\left|\begin{matrix} -T_C \\ N_C \\ Q_C \end{matrix}\right. + \overrightarrow{F_{B^{(b)}_{s,R}}}\left|\begin{matrix} T_B \\ -N_B \\ -Q_B \end{matrix}\right.$$

*TMC*

$$\overrightarrow{\delta_{G^{(b)}_{S1b_R}}} = \vec{0} = -\overrightarrow{\mathcal{M}_{B^{(b)}_{s,R}}}\left|\begin{matrix} 0 \\ -M_B \cdot \sin\beta \\ -M_B \cdot \cos\beta \end{matrix}\right.$$

$$+\overrightarrow{G_{S1b_R}B^{(b)}_{s,R}}\left|\begin{matrix} L_{S1b}/2 \, \sin\theta \\ -L_{S1b}/2 \, \cos\theta \cdot \cos\beta \\ -L_{S1b}/2 \, \cos\theta \cdot \sin\beta \end{matrix}\right. \wedge \overrightarrow{F_{B^{(b)}_{s,R}}}\left|\begin{matrix} T_B \\ -N_B \\ -Q_B \end{matrix}\right.$$

$$+\overrightarrow{G_{S1b_R}C^{(b)}_{s,R}}\left|\begin{matrix} -L_{S1b}/2 \, \sin\theta \\ L_{S1b}/2 \, \cos\theta \cdot \cos\beta \\ L_{S1b}/2 \, \cos\theta \cdot \sin\beta \end{matrix}\right. \wedge -\overrightarrow{F_{C^{(b)}_{s,R}}}\left|\begin{matrix} -T_C \\ N_C \\ Q_C \end{matrix}\right.$$

Pour $\mathbf{S2_R}$ (centre de gravité en $\mathbf{G_{S2_R}}$ situé au milieu de $S2_R$ ; masse $\mathbf{m_{S2}}$), les actions de liaisons sont modélisées par un torseur dont les éléments de réduction sont en $\mathbf{C^{(b)}_{s,R}}$ [ $\overrightarrow{\mathbf{F}_{C^{(b)}_{s,R}}}$ , $\vec{\mathbf{0}}$ ] et en $\mathbf{D^{(b)}_{s,R}}$ [ $\overrightarrow{-\mathbf{F}_{D^{(b)}_{s,R}}}$ , $\vec{\mathbf{0}}$ ]

*TRC*

$$\mathbf{m_{S2}} \cdot \overrightarrow{\gamma_{G^{(b)}_{S2_R}}} \approx \vec{\mathbf{0}} = -\overrightarrow{\mathbf{F}_{D^{(b)}_{s,R}}}\begin{vmatrix}-\mathbf{T_D}\\ \mathbf{N_D}\\ \mathbf{Q_D}\end{vmatrix} + \overrightarrow{\mathbf{F}_{C^{(b)}_{s,R}}}\begin{vmatrix}\mathbf{T_C}\\ -\mathbf{N_C}\\ -\mathbf{Q_C}\end{vmatrix}$$

*TMC*

$$\overrightarrow{\delta_{G^{(b)}_{S2_R}}} = \vec{\mathbf{0}} = \overrightarrow{\mathbf{G_{S2_R}D^{(b)}_{s,R}}}\begin{vmatrix}-\mathbf{L^{o}_{S2}}/2\ \cos\varphi\\ -\mathbf{L^{o}_{S2}}/2\ \sin\varphi\\ -\mathbf{L^{z}_{S2}}/2\end{vmatrix} \wedge -\overrightarrow{\mathbf{F}_{D^{(b)}_{s,R}}}\begin{vmatrix}-\mathbf{T_D}\\ \mathbf{N_D}\\ \mathbf{Q_D}\end{vmatrix}$$

$$+\overrightarrow{\mathbf{G_{S2_R}C^{(b)}_{s,R}}}\begin{vmatrix}\mathbf{L^{o}_{S2}}/2\ \cos\varphi\\ \mathbf{L^{o}_{S2}}/2\ \sin\varphi\\ \mathbf{L^{z}_{S2}}/2\end{vmatrix} \wedge \overrightarrow{\mathbf{F}_{C^{(b)}_{s,R}}}\begin{vmatrix}\mathbf{T_C}\\ -\mathbf{N_C}\\ -\mathbf{Q_C}\end{vmatrix}$$

où $\mathbf{L^{o}_{S2,R}}$ est la projection orthogonale de $\mathbf{L_{S2}}$ selon $\mathbf{OZ^{o}}$ sur le plan $\mathbf{OXY^{o}}$ (voir équation A3 de l'annexe A) et $\mathbf{L^{z}_{S2,R}}$ est la projection orthogonale de $\mathbf{L_{S2}}$ selon $\mathbf{OY^{o}}$ sur l'axe $\mathbf{OZ^{o}}$

On projette ces 5 équations vectorielles sur les axes $\mathbf{OX}$, $\mathbf{OY^{o}}$ et $\mathbf{OZ^{o}}$ (Fig. 4B). En notant que les forces et moments sont exprimés en module, et avec les angles exprimés en valeurs algébriques définis sur la Fig. 3E du chap. 2, on obtient :

Pour $\mathbf{S1a_R}$

*En X* : $\quad \mathbf{0 = T_{A^{(b)}_{s-1,L}} - T_{B^{(b)}_{s,R}}}$ (C.1)

*En Y* : $\quad \mathbf{0 = -N_{A^{(b)}_{s-1,L}} + N_{B^{(b)}_{s,R}}}$ (C.2)

Pour $\mathbf{S1b_R}$

*En X* : $\quad \mathbf{0 = T_{B^{(b)}_{s,R}} - T_{C^{(b)}_{s,R}}}$ (C.3)

*En Y* : $\quad \mathbf{0 = -N_{B^{(b)}_{s,R}} + N_{C^{(b)}_{s,R}}}$ (C.4)

*En Z* :

$$\mathcal{M}_{B^{(b)}_{s,R}} \cdot \cos\beta^{(b)} = \frac{\mathbf{L_{S1b}}}{\mathbf{2}} \cdot \left[\left(\mathbf{T_{B^{(b)}_{s,R}} + T_{C^{(b)}_{s,R}}}\right) \cdot \cos\beta^{(b)}_{s,R} \cdot \cos\theta^{(b)}_{s,R} - \left(\mathbf{N_{B^{(b)}_{s,R}} + N_{C^{(b)}_{s,R}}}\right) \cdot \sin\theta^{(b)}_{s,R}\right]$$

(C5)

Pour **S2**$_{\mathbf{R}}$

*En X* : $$\mathbf{0 = T_{C_{s,R}^{(b)}} - T_{D_{s,R}^{(b)}}} \qquad \text{(C.6)}$$

*En Y* : $$\mathbf{0 = -N_{C_{s,R}^{(b)}} + N_{D_{s,R}^{(b)}}} \qquad \text{(C.7)}$$

*En Z* : $$\mathbf{0 = \left(T_{C_{s,R}^{(b)}} + T_{D_{s,R}^{(b)}}\right) \cdot \sin\varphi_{s,R}^{(b)} + \left(N_{C_{s,R}^{(b)}} + N_{D_{s,R}^{(b)}}\right) \cdot \cos\varphi_{s,R}^{(b)}} \qquad \text{(C.8)}$$

Des éq. C.1, C.3 et C.6, on tire : $\mathbf{T_{A_{s-1,L}^{(b)}} = T_{B_{s,R}^{(b)}} = T_{C_{s,R}^{(b)}} = T_{D_{s,R}^{(b)}}}$

Avec les éq. C.2, C.4 et C.7, on a : $\mathbf{N_{A_{s-1,L}^{(b)}} = N_{B_{s,R}^{(b)}} = N_{C_{s,R}^{(b)}} = N_{D_{s,R}^{(b)}}}$

Les équations C.1 à C.8 mènent à : $\mathbf{\mathcal{M}_{B_{s,R}^{(b)}} = L_{S1b} \cdot T_{A_{s-1,L}^{(b)}} \cdot \left(\cos\theta_{s,R}^{(b)} + \frac{\sin\theta_{s,R}^{(b)} \cdot \tan\varphi_{s,R}^{(b)}}{\cos\beta_{s,R}^{(b)}}\right)}$

## Demi sarcomère gauche (hsL)

Le même raisonnement appliqué aux actions de liaisons de la tête de myosine n° b positionnée dans hsL, et constituée des segments **S1a**$_{\mathrm{L}}$, **S1b**$_{\mathrm{L}}$ et **S2**$_{\mathrm{L}}$, donne :

$$\mathbf{T_{A_{s,R}^{(b)}} = T_{B_{s,L}^{(b)}} = T_{C_{s,L}^{(b)}} = T_{D_{s,L}^{(b)}}}$$

$$\mathbf{N_{A_{s,R}^{(b)}} = N_{B_{s,L}^{(b)}} = N_{C_{s,L}^{(b)}} = N_{D_{s,L}^{(b)}}}$$

$$\mathbf{\mathcal{M}_{B_{s,L}^{(b)}} = L_{S1b} \cdot T_{A_{s,R}^{(b)}} \cdot \left(\cos\theta_{s,L}^{(b)} + \frac{\sin\theta_{s,L}^{(b)} \cdot \tan\varphi_{s,L}^{(b)}}{\cos\beta_{s,L}^{(b)}}\right)}$$

## ANNEXE D

### D.1 Forces de viscosité durant la phase 1 d'un échelon en longueur

La formule de Stokes-Einstein fournit une expression de la force de viscosité, notée $\mathbf{F_{visc}}$ , s'appliquant à une sphère se déplaçant dans un fluide à la vitesse absolue V.

$\mathbf{F_{visc}}$ est une force s'opposant au mouvement et colinéaire à V, soit classiquement :

$$\mathbf{F_{visc} = -6 \cdot \pi \cdot r \cdot \eta \cdot V} \tag{D.1}$$

avec $\mathbf{r}$ le rayon de la sphère

*et* $\eta$ le coefficient de viscosité du liquide dans lequel se meut la molécule sphérique

(par ex : $\eta_eau(20°)=10^{-3}\ Pa$)

Rappelons que le coefficient $\eta$ diminue lorsque la température augmente (et inversement).

De manière générale, un corps se déplaçant dans un fluide visqueux à la vitesse absolue $\mathbf{V}$ est soumis à une force $\mathbf{F_{visc}}$ opposée à $\mathbf{V}$ et variant en 1[ère] approximation linéairement avec $\mathbf{V}$ , telle que :

$$\mathbf{F_{visc} = -\alpha \cdot V} \tag{D.2}$$

où $\alpha$ est le coefficient de proportionnalité entre $\mathbf{F_{visc}}$ et $\mathbf{V}$

Considérons le hsL du sarcomère n° s en cours de raccourcissement durant la phase 1. Les forces de viscosité freinant les ensembles $(\mathbf{Z_S + filA})$ et $(\mathbf{M_S + filM})$ dans le hsL n° s sont proportionnelles à la vitesse absolue de ceux-ci , d'après (D.2).

Comme la phase 1 est réalisée à vitesse constante et à durée identique $\mathbf{\Delta t_{p1}}$ quelque soit l'échelon (k), les forces de viscosité s'appliquant aux éléments du hsL sont proportionnelles aux raccourcissements globaux $\mathbf{\Delta L_{p1}^{Zs,L}}$ et $\mathbf{\Delta L_{p1}^{Ms,L}}$ tels que [1] :

Pour $(\mathbf{Z_S + filA})$
$$\mathbf{V_{p1}^{Zs,L} = \frac{\Delta L_{p1}^{Zs,L}}{\Delta t_{p1}} = \frac{1}{\Delta t_{p1}} \cdot \left( \sum_{i=1}^{s} \Delta X_{p1}^{i,R} + \sum_{i=1}^{s} \Delta X_{p1}^{i,L} \right)} \tag{D.3a}$$

Pour $(\mathbf{M_S + filM})$
$$\mathbf{V_{p1}^{Ms,L} = \frac{\Delta L_{p1}^{Ms,L}}{\Delta t_{p1}} = \frac{1}{\Delta t_{p1}} \cdot \left( \sum_{i=1}^{s} \Delta X_{p1}^{i,R} + \sum_{i=1}^{s-1} \Delta X_{p1}^{i,L} \right)} \tag{D.3b}$$

avec $\mathbf{\Delta X_{p1,(k)}^{i,R}} \left( \mathbf{\Delta X_{p1,(k)}^{i,L}} \right)$ la valeur du raccourcissement interne du hsR (hsL) n° i, respectivement, à $\mathbf{t = \Delta t_{p1} = t_{fin_p1}^{1}}$ lors de la perturbation en longueur d'échelon (k)

[1] *Voir les éq. (1.2) à (1.6) du chapitre 1. Rappelons que tous les raccourcissements sont mesurés négativement et introduits positivement dans les équations du modèle (voir note 1 page 47 du chapitre 4).*

Rappel de **Hypothèse 9** : les constantes de proportionnalités entre force de viscosité et vitesse de déplacement du disque Z (avec les filA et autres protéines associées) et du disque M (avec ses filM et autres protéines associées situés) sont égales.

Nommons $\nu^{Zs}$ et $\nu^{Ms}$ les coefficients de proportionnalité caractérisant, respectivement, $(Z_S + filA)$ et $(M_S + filM)$ en présence de viscosité.

Ainsi les force dues à la viscosité s'exerçant sur $(M_S + filM)$ et $(Z_S + filA)$ vérifient à $t^1_{fin_p1}$ dans hsL :

$$T^{Zs}_{visco} / T0^{hsL} = -\nu^{hs} \cdot \Delta L^{Zs,L}_{p1} \qquad \text{(D.4a)}$$

$$T^{Ms}_{visco} / T0^{hsL} = -\nu^{hs} \cdot \Delta L^{Zs,L}_{p1} \qquad \text{(D.4b)}$$

$$\text{avec } \nu^{hs} = \nu^{Zs} = \nu^{Ms} \qquad \text{(D.4c)}$$

On déduit les équations similaires pour hsR.

Remarquons que :

$\nu^{hs}$ diminue lorsque la température augmente (et inversement).

$\nu^{hs} \propto 1/\Delta t_{p1}$ d'après (D.2), (D.3a), (D3b), (D.4a), (D.4b) et (D.4c). Ainsi $\nu^{hs}$ diminue lorsque $\Delta t_{p1}$ augmente (et inversement).

## D.2 Calcul de la tension à la fin de la phase 1 en présence des forces de viscosité

Considérons pour une myofibrille de la fmI un raccourcissement $\Delta L$ et examinons le comportement de n hs en série, n variant de 1 à Ns, en suivant un raisonnement par récurrence.

Pour simplifier :

$\chi^{hs}_{z1}$ , le coefficient de l'éq. (4.35a)[1] sera noté $\chi$

$\nu^{hs}$ sera noté $\nu$

$T0^{hsL} = T0^{hsR}$ sera noté $T0$

La tension exercée sur l'élément le plus à gauche du dernier hs de la myofibrille sera noté $T$

### D.2.1 n = 1 et $\Delta L = \Delta X_1$

A $t^1_{fin_p1}$, le hs n°1 se raccourcit de $\Delta X_1$ et la somme des forces appliquées sur $(M_1 + filM)$ sont $T$, la tension exercée par les S1 en cours de dans le hsR n° 1 , et la force due à la viscosité . Le ***TRC*** est appliqué à $(M_1 + filM)$ (Fig. D.1) et (4.35a) avec (D.4b) mène à :

$$0 = -T/T0 + (1 - \chi \cdot \Delta X_1) - \nu \cdot \Delta L$$

$$\Rightarrow \; T/T0 = 1 - \Delta L \cdot [\chi + \nu]$$

La myofibrille composée de 1 hs se comporte comme un ressort linéaire.

[1] *Rappel : avec l'Hyp .2, les forces de viscosité s'exerçant sur les tetM liées sont négligeables, même à vitesse élevée. Ainsi les équations de l'Annexe C et des chapitres 4 et 5 restent valides.*

**Fig. D1 : modélisation d'une myofibrille composée de 3 hs et se raccourcissant à V cste durant la fin de la phase 1 en présence de forces dues à la viscosité**

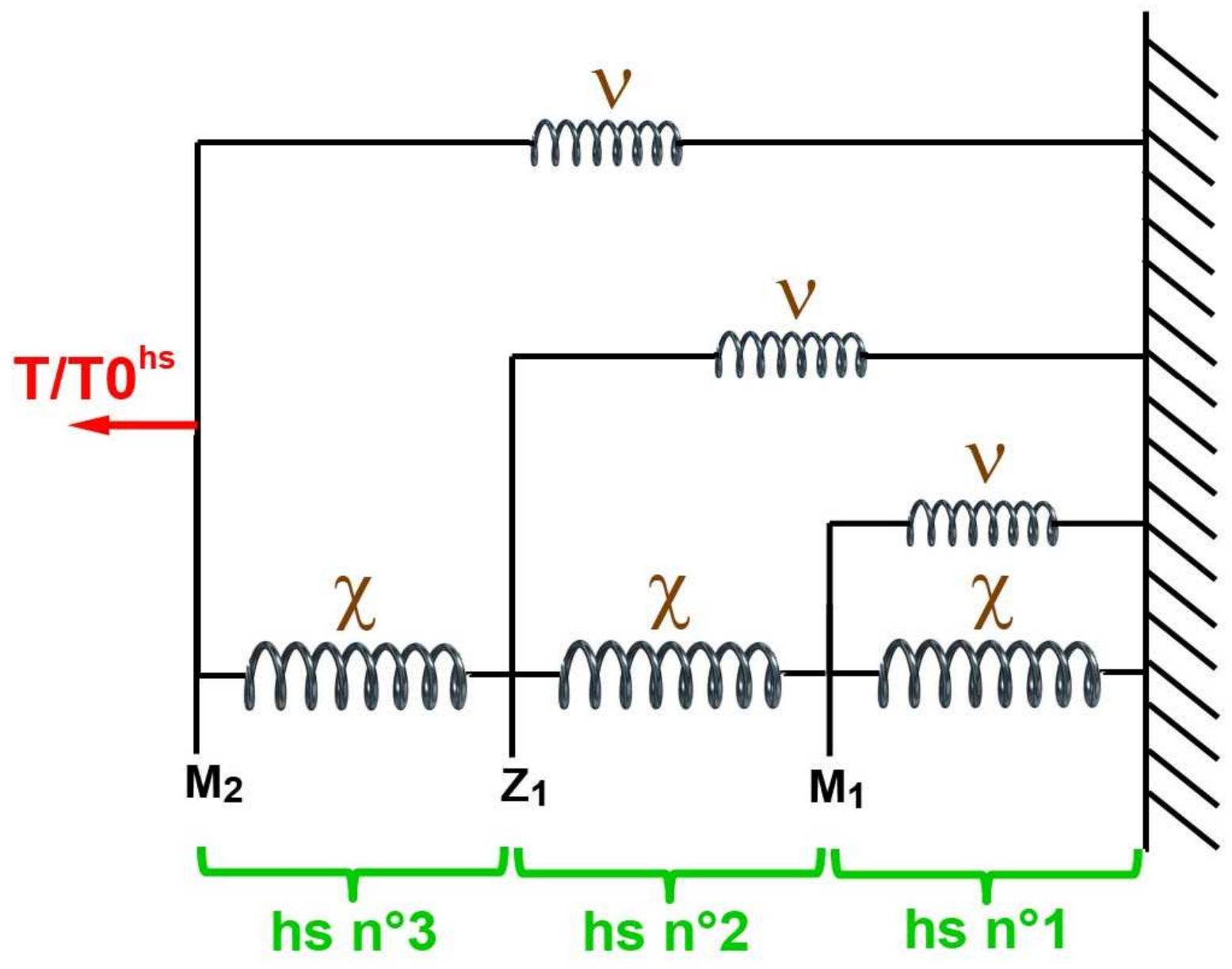

**D.2.2 n = 2 et $\Delta L = \Delta X_1 + \Delta X_2$**

A $t^{1}_{fin_p1}$, les hs n°1 et 2 se raccourcissent de $\Delta X_1$ et $\Delta X_2$.

*TRC* à $(M_1 + filM)$ : $0 = (1 - \chi \cdot \Delta X_1) - (1 - \chi \cdot \Delta X_2) - \nu \cdot \Delta X_1$

$$\Rightarrow \Delta X_2 = \Delta X_1 \cdot \left(1 + \frac{\nu}{\chi}\right)$$

$$\Rightarrow \Delta L = \Delta X_1 + \Delta X_2 = \Delta X_1 \cdot \left(2 + \frac{\nu}{\chi}\right)$$

$$\Rightarrow \Delta X_2 = \Delta L \cdot \frac{1 + \frac{\nu}{\chi}}{2 + \frac{\nu}{\chi}}$$

*TRC* à $(Z_1 + filA)$ : $0 = -T/T0 + (1 - \chi \cdot \Delta X_2) - \nu \cdot \Delta L$

$$\Rightarrow \quad T/T0 = 1 - \chi \cdot \Delta X_2 - \nu \cdot \Delta L = 1 - \Delta L \cdot \left[\chi \cdot \frac{1 + \frac{\nu}{\chi}}{2 + \frac{\nu}{\chi}} + \nu\right]$$

La myofibrille composée de 2 hs en série se comporte comme un ressort linéaire.

### D.2.3 n= 3 $\Delta L = \Delta X_1 + \Delta X_2 + \Delta X_3$

A $t^1_{fin_p1}$, les hs n°1, 2 et 3 se raccourcissent de $\Delta X_1$, $\Delta X_2$ et $\Delta X_3$.

*TRC* à $(M_1 + filM)$ : $0 = (1 - \chi \cdot \Delta X_1) - (1 - \chi \cdot \Delta X_2) - \nu \cdot \Delta X_1$

$$\Rightarrow \Delta X_2 = \Delta X_1 \cdot \left(1 + \frac{\nu}{\chi}\right)$$

$$\Rightarrow \Delta X_1 + \Delta X_2 = \Delta X_1 \cdot \left(2 + \frac{\nu}{\chi}\right)$$

*TRC* à $(Z_1 + filA)$ : $0 = (1 - \chi \cdot \Delta X_2) - (1 - \chi \cdot \Delta X_3) - \nu \cdot (\Delta X_1 + \Delta X_2)$

$$\Rightarrow \Delta X_3 = \Delta X_2 + \frac{\nu}{\chi} \cdot (\Delta X_1 + \Delta X_2) = \Delta X_1 \cdot \left(1 + \frac{3\nu}{\chi} + \frac{\nu^2}{\chi^2}\right)$$

$$\Rightarrow \Delta L = \Delta X_1 + \Delta X_2 + \Delta X_3 = \Delta X_1 \cdot \left[3 + \frac{4\nu}{\chi} + \frac{\nu^2}{\chi^2}\right]$$

*TRC* à $(M_2 + filM)$ : $0 = -T/T0 + (1 - \chi \cdot \Delta X_3) - \nu \cdot \Delta L$

$$\Rightarrow \quad T/T0 = 1 - \chi \cdot \Delta X_3 - \nu \cdot \Delta L = 1 - \Delta L \cdot \left[\chi \cdot \frac{1 + \frac{3\nu}{\chi} + \frac{\nu^2}{\chi^2}}{3 + \frac{4\nu}{\chi} + \frac{\nu^2}{\chi^2}} + \nu\right]$$

La myofibrille composée de 3 hs en série se comporte comme un ressort linéaire (Fig. D.1).

### D.2.4 terme général n avec $\Delta L = \Delta X_1 + \Delta X_2 + \Delta X_3 + ... + \Delta X_n$

A $t^1_{fin_p1}$, les n hs se raccourcissent de $\Delta X_1$, $\Delta X_2$, $\Delta X_3$ .,... $\Delta X_n$, respectivement. Par récurrence, on calcule :

$$T/T0 = 1 - \Delta L \cdot \left[\chi \cdot \frac{1 + N_{1,n} \cdot \frac{\nu}{\chi} + N_{2,n} \cdot \frac{\nu^2}{\chi^2} + N_{3,n} \cdot \frac{\nu^3}{\chi^3} + ... + \frac{\nu^{n-1}}{\chi^{n-1}}}{n + \Delta_{1,n} \cdot \frac{\nu}{\chi} + \Delta_{2,n} \cdot \frac{\nu^2}{\chi^2} + \Delta_{3,n} \cdot \frac{\nu^3}{\chi^3} + ... + \frac{\nu^{n-1}}{\chi^{n-1}}} + \nu\right] \qquad \text{(D.5)}$$

avec $N_{1,n} = \frac{n \cdot (n-1)}{2}$ , $\Delta_{1,n} = \Delta_{1,n-1} + N_{1,n}$

$N_{2,n} = N_{2,n-1} + \Delta_{1,n-1}$ , $\Delta_{2,n} = \Delta_{2,n-1} + N_{2,n}$

$N_{3,n} = N_{3,n-1} + \Delta_{2,n-1}$ , $\Delta_{3,n} = \Delta_{3,n-1} + N_{3,n}$

...

$N_{k,n} = N_{k,n-1} + \Delta_{k-1,n-1}$ , $\Delta_{k,n} = \Delta_{k,n-1} + N_{k,n}$

... etc

La myofibrille composée de n hs en série se comporte comme un ressort linéaire.

### D.2.5 n = 2·Ns

On remarque avec les formules précédentes qu'à $t^{1}_{fin_p1}$ :

$$\Delta X_{2Ns} \le \Delta X_{2Ns-1} \le \ldots \le \Delta X_{n+1} \le \Delta X_{n} \le X_{n-1} \ldots \le \Delta X_{2} \le \Delta X_{1}$$

*i.e.*, en revenant aux notations apportées à l'écriture de (7.3) :

$$\Delta X^{Ns,L}_{p1,(k)} \le \Delta X^{Ns,R}_{p1,(k)} \le \Delta X^{Ns-1,L}_{p1,(k)} \le \ldots \le \Delta X^{s,L}_{p1,(k)} \le \Delta X^{s,R}_{p1,(k)} \le \ldots \le \Delta X^{1,L}_{p1,(k)} \le \Delta X^{1,R}_{p1,(k)} \quad \text{(D.6)}$$

Ainsi lorsque la viscosité est prise ne compte, les valeurs des raccourcissements et des vitesses relatives de raccourcissements ne sont pas rigoureusement égales entre chaque hs de la fmI.

L'ordre de grandeur de $\nu/\chi$ est 1/2Ns d'après les résultats expérimentaux de la phase d'un échelon de longueur; ainsi pour une myofibrille avec $L^{fmI}_{0} = 5\,mm$ et $L^{hs}_{0} = 1.05\,\mu m$, le calcul mène à : $\nu/\chi \approx 2 \cdot 10^{-4}$ (voir développements de l'éq. 7.10a dans le paragraphe 7.4); on en déduit que tous les termes en $\nu/\chi$ sont négligeables.
Ainsi (D.5) se réécrit pour $n = 2 \cdot Ns$ dans la zone 1 :

$$T/T0^{hs} \approx 1 - \Delta L \cdot \left[ \frac{\chi^{hs}_{z1}}{2 \cdot Ns} + \nu \right] \quad \text{(D.7)}$$

De même avec cette approximation, les raccourcissements des 2·Ns hs de la myofibrille sont pratiquement identiques, et (D.6) redevient :

$$\Delta X^{Ns,L}_{p1,(k)} \approx \Delta X^{Ns,R}_{p1,(k)} \approx \ldots \approx \Delta X^{s,L}_{p1,(k)} \approx \Delta X^{s,R}_{p1,(k)} \approx \ldots \approx \Delta X^{1,L}_{p1,(k)} \approx \Delta X^{(k)}_{p1} \approx \frac{\Delta L^{(k)}_{p1}}{2 \cdot Ns} \quad \text{(D.8)}$$

Avec (D.8), (D.7) se reformule à $t^{1}_{fin_p1}$ :

$$T/T0^{hs} \approx 1 - \Delta X^{(k)}_{p1} \cdot \left[ \chi^{hs}_{z1} + 2 \cdot Ns \cdot \nu \right] \quad \text{(D.9)}$$

Pour la zone 2, on obtient de manière analogue à partir de l'éq. (4.36a) :

$$T/T0^{hs} \approx \left[ \chi^{hs}_{z2} + 2 \cdot Ns \cdot \nu \right] \cdot \left( \Delta X^{Max}_{WS} - \Delta X^{(k)}_{p1} \right) \quad \text{(D.10)}$$

Lors d'un raccourcissement réalisé à vitesse élevée, la myofibrille composée de 2·Ns hs en série se comporte comme un ressort linéaire dont la raideur est la somme de 2 termes : l'un du à l'élasticité des $\Lambda 0$ tetM en cours de WS dans chaque hs de la myofibrille et l'autre (égal à $2 \cdot Ns \cdot \nu$) provient des forces de viscosité.

# ANNEXE E

## E.1 Quantité de mouvement d'une fmI se raccourcissant à V cste

Rappel de **Hypothèse 9** : les masses du disque Z (avec les filA et autres protéines associées, à droite et à gauche) et du disque M (avec ses filM et autres protéines associées, à droite et à gauche) sont égales.

Nommons $\mathbf{m}^{\mathbf{Zs}}$ et $\mathbf{m}^{\mathbf{Ms}}$ les masses respectives de $(\mathbf{Z_S + 2 \cdot filA})$ et $(\mathbf{M_S + 2 \cdot filM})$ telles que :

. $$\mathbf{m}^{\mathbf{Zs}} = \mathbf{m}^{\mathbf{Ms}} = \mathbf{m} \qquad \text{(E.1a)}$$

On en déduit que la masse d'une myofibrille, composée de $\mathbf{Ns}$ sarcomères identiques disposés en série, est égale à :

$$\mathbf{m_{myof} = Ns \cdot \left(m^{Zs} + m^{Ms}\right) = 2Ns \cdot m} \qquad \text{(E.1b)}$$

Et celui d'une fmI constituée de $\mathbf{N_{myof}}$ myofibrilles identiques disposées en parallèle :

$$\mathbf{m_{fmI} = 2 \cdot N_{myof} \cdot Ns \cdot m} \qquad \text{(E.1c)}$$

Par définition, la quantité de mouvement d'un système déformable composé de segments rigides est égale à la somme des quantités de mouvement de chacun des segments rigides. En négligeant les quantités de mouvement linéaire et angulaire des 3 segments composant les tetM, on obtient la quantité de mouvement d'une myofibrille notée $\mathbf{Q_{myof}}$ en sommant de droite à gauche (Fig. 2 ; chap. 2) :

$$\mathbf{Q_{myof} = \sum_{s=1}^{Ns} \left(m \cdot V_{Ms}\right) + \sum_{s=1}^{Ns-1} \left(m \cdot V_{Zs}\right) + \frac{m}{2} \cdot V_{ZNs}} \qquad \text{(E.2)}$$

avec $\mathbf{V_{Zs}}$ et $\mathbf{V_{Ms}}$, les vitesses absolues respectives de $(\mathbf{Z_S + filA})$ et $(\mathbf{M_S + filM})$, exprimées en module, dans le repère du laboratoire.

Si la fmI et donc les $\mathbf{N_{myof}}$ myofibrilles se raccourcissent à vitesse constante (à l'exemple de la phase 1 d'un échelon de longueur ou de force), il a été démontré que tous les hs de la myofibrille se raccourcissaient en module à une même vitesse relative, notée $\mathbf{u}$, que ce soit en présence (éq. D.8 ; Annexe D) ou en absence (éq. 7.6 ; paragraphe 7.3) des forces de viscosité.

Avec les formules (2.2) et (2.5) établies au chap.2, l'éq. (E.2) se réécrit :

$$\mathbf{Q_{myof} = m \cdot u \cdot \left[\sum_{s=1}^{Ns} (2s-1) + \sum_{s=1}^{Ns-1} (2s) + \frac{2Ns}{2}\right]} \qquad \text{(E.3)}$$

Avec l'égalité $\mathbf{\sum_{j=1}^{n} j = \frac{j \cdot (j+1)}{2}}$, on obtient en développant (E.3) :

$$\mathbf{Q_{myof} = 2 \cdot m \cdot u \cdot Ns^2} \qquad \text{(E.4)}$$

$\mathbf{V}$, la vitesse de raccourcissement de la fmI et des ses $\mathbf{N_{myof}}$ myofibrilles, est égale en module selon (2.5) à :

$$\mathbf{V = V_{Z_{Ns}} = 2 \cdot Ns \cdot u} \tag{E.5}$$

De (E.4) et (E.5), on déduit avec (E.1c) la quantité de mouvement en module de la fmi à Vcste, soit :

$$\mathbf{Q_{fmI} = N_{myof} \cdot Q_{myof} = \frac{1}{2} \cdot m_{fmI} \cdot V} \tag{E.6}$$

## E.2 Théorème de l'énergie cinétique appliqué à la fmI se raccourcissant à V cste

Le théorème de l'énergie cinétique s'écrit en puissance, *i.e.* sous forme différentielle :

$$\mathbf{\frac{dEC^{fmI}}{dt} = P_{Fext} + P_{Fint}} \tag{E.7}$$

où $\mathbf{EC^{fmI}}$ est l'énergie cinétique du système mécanique étudié, *i.e.* la fmI

$\mathbf{P_{Fext}}$ est la puissance des forces extérieures

$\mathbf{P_{Fint}}$ est la puissance des forces intérieures

L'énergie cinétique d'un système mécanique est égale à la somme des énergies cinétiques de translation et de rotation de tous les éléments rigides qui le compose, soit dans le cas de la fmI, tous les $\mathbf{(Z_S + 2 \cdot filA)}$, $\mathbf{(M_S + 2 \cdot filM)}$, S1a, S1b et S2. Ces différents segments rigides sont tous en translation ou rotation uniforme, excepté les segments S1b et S2 des différents S1 en cours de WS mais leurs termes apportent une contribution négligeable. Ainsi à V constante, comme par exemple durant la phase 4 d'un échelon de force, on vérifie :

$$\mathbf{\frac{dEC^{fmI}}{dt} = 0} \tag{E.8}$$

Par définition la fmI est soumise à une extrémité à une tension noté $\mathbf{T}$ cste qui travaille à $\mathbf{V}$ cste et à l'autre extrémité fixe à une force qui ne travaille pas (Fig. 2 ; chap.2). La puissance des forces extérieures s'écrit donc :

$$\mathbf{P_{Fext} = T \cdot V = -|T| \cdot V} \tag{E.9}$$

En biomécanique ou en robotique, on constate qu'au niveau d'une articulation qui relie 2 segments rigides en mouvement, seul le moment articulaire inter-segmentaire travaille. Aussi la puissance d'un système mécanique déformable, qui est constitué de segments rigides articulés entre eux et qui se déplace dans l'espace, est égale à la somme des puissances des moments articulaires. En appliquant ce principe à la fmI en cours de raccourcissement à V cste, la puissance des forces intérieures de la fmI est égale à la somme des puissances des moments articulaires des S1 en cours de WS.

Comme d'après les Hyp. 7 et 8, les vecteurs des moments moteurs $\mathcal{M}$ et des vitesses angulaires $\dot{\theta}$ sont colinéaires selon l'axe OZ (Fig. 3B et 3C ; chap. 2), la puissance des forces intérieures se formule :

$$\mathbf{P_{F\,int}} = \mathbf{N_{myof}} \cdot \sum_{s=1}^{Ns} \sum_{b=1}^{\Lambda_{WS}^{s,R_L}} \left( \dot{\theta}_{s,R}^{(b)} \cdot \mathcal{M}_{s,R}^{(b)} + \dot{\theta}_{s,L}^{(b)} \cdot \mathcal{M}_{s,L}^{(b)} \right) \qquad \text{(E.10)}$$

où $\Lambda_{WS}^{s,R_L}$ est le nombre de S1 en cours de WS dans le hsR ou hsL n° s d'1 des myofibrilles de la fmI.
s est l'indice du sarcomère n° s
et b le numéro indiciel du S1 en cours de WS dans le hsR ou dans le hsL du sarcomère n° s

En introduisant (E.8), (E.9) et (E.10) dans (E.7), on obtient :

$$|\mathbf{T}| = \frac{\mathbf{N_{myof}}}{\mathbf{V}} \cdot \sum_{s=1}^{Ns} \sum_{b=1}^{\Lambda_{WS}^{s,R_L}} \left( \dot{\theta}_{s,R}^{(b)} \cdot \mathcal{M}_{Bs,R}^{(b)} + \dot{\theta}_{s,L}^{(b)} \cdot \mathcal{M}_{Bs,L}^{(b)} \right) \qquad \text{(E.11)}$$

# BIBLIOGRAPHIE

Ait-Haddou, R. and W. Herzog (2002). "Force and motion generation of myosin motors: muscle contraction." J Electromyogr Kinesiol **12**(6): 435-45.

Allen P. D. a. S., W. N. (1973). "A 5 parameter curve : The best fit for the force : velocity relationship of in situ dog skeletal muscle." Physiologist **16**: 252.

Astumian, R. D. and M. Bier (1996). "Mechanochemical coupling of the motion of molecular motors to ATP hydrolysis." Biophys J **70**(2): 637-53.

Baumketner, A. and Y. Nesmelov (2011). "Early stages of the recovery stroke in myosin II studied by molecular dynamics simulations." Protein Sci **20**(12): 2013-22.

Bottinelli, R., M. Canepari, et al. (1996). "Force-velocity properties of human skeletal muscle fibres: myosin heavy chain isoform and temperature dependence." J Physiol **495 ( Pt 2)**: 573-86.

Brunello, E., P. Bianco, et al. (2006). "Structural changes in the myosin filament and cross-bridges during active force development in single intact frog muscle fibres: stiffness and X-ray diffraction measurements." J Physiol **577**(Pt 3): 971-84.

Brunello, E., L. Fusi, et al. (2009). "Structural changes in myosin motors and filaments during relaxation of skeletal muscle." J Physiol **587**(Pt 18): 4509-21.

Buller, A. J., C. J. Kean, et al. (1984). "Temperature dependence of isometric contractions of cat fast and slow skeletal muscles." J Physiol **355**: 25-31.

Caremani, M., J. Dantzig, et al. (2008). "Effect of inorganic phosphate on the force and number of myosin cross-bridges during the isometric contraction of permeabilized muscle fibers from rabbit psoas." Biophys J **95**(12): 5798-808.

Caremani, M., S. Lehman, et al. (2011). "Orthovanadate and orthophosphate inhibit muscle force via two different pathways of the myosin ATPase cycle." Biophys J **100**(3): 665-74.

Chin, L., P. Yue, et al. (2006). "Mathematical simulation of muscle cross-bridge cycle and force-velocity relationship." Biophys J **91**(10): 3653-63.

Combrouze, A. (1993). Probabilités et statistiques. Paris, Presses Uinversitaires de France.

Cooke, R. and W. Bialek (1979). "Contraction of glycerinated muscle fibers as a function of the ATP concentration." Biophys J **28**(2): 241-58.

Cooke, R., H. White, et al. (1994). "A model of the release of myosin heads from actin in rapidly contracting muscle fibers." Biophys J **66**(3 Pt 1): 778-88

Cooke, R. (1997). "Actomyosin interaction in striated muscle." Physiol Rev **77**(3): 671-97.

Corrie, J. E., B. D. Brandmeier, et al. (1999). "Dynamic measurement of myosin light-chain-domain tilt and twist in muscle contraction." Nature **400**(6743): 425-30.

Coupland, M. E. and K. W. Ranatunga (2003). "Force generation induced by rapid temperature jumps in intact mammalian (rat) skeletal muscle fibres." J Physiol **548**(Pt 2): 439-49.

Decostre, V., P. Bianco, et al. (2005). "Effect of temperature on the working stroke of muscle myosin." Proc Natl Acad Sci U S A **102**(39): 13927-32.

Duke, T. A. (1999). "Molecular model of muscle contraction." Proc Natl Acad Sci U S A **96**(6): 2770-5.

Edman, K. A., L. A. Mulieri, et al. (1976). "Non-hyperbolic force-velocity relationship in single muscle fibres." Acta Physiol Scand **98**(2): 143-56.

Edman, K. A. and J. C. Hwang (1977). "The force-velocity relationship in vertebrate muscle fibres at varied tonicity of the extracellular medium." J Physiol **269**(2): 255-72.

Edman, K. A. (1979). "The velocity of unloaded shortening and its relation to sarcomere length and isometric force in vertebrate muscle fibres." J Physiol **291**: 143-59.

Edman, K. A. (1988). "Double-hyperbolic force-velocity relation in frog muscle fibres." J Physiol **404**: 301-21.

Edman, K. A., C. Reggiani, et al. (1988). "Maximum velocity of shortening related to myosin isoform composition in frog skeletal muscle fibres." J Physiol **395**: 679-94.

Elangovan, R., M. Capitanio, et al. (2012). "An integrated in vitro and in situ study of kinetics of myosin II from frog skeletal muscle." J Physiol **590**(Pt 5): 1227-42.

Engel, A. (1976). Processus aléatoires pour les débutants. Paris, Cassini.

Fenn, W. O. and B. S. Marsh (1935). "Muscular force at different speeds of shortening." J Physiol **85**(3): 277-97.

Feynman, R. P., R. B. Leighton, et al., Eds. (1963). The Feynman lectures on physics. Reading MA, Addison-Wesley.

Fischer, S., B. Windshugel, et al. (2005). "Structural mechanism of the recovery stroke in the myosin molecular motor." Proc Natl Acad Sci U S A **102**(19): 6873-8.

Ford, L. E., A. F. Huxley, et al. (1974). "Proceedings: Mechanism of early tension recovery after a quick release in tetanized muscle fibres." J Physiol **240**(2): 42P-43P.

Ford, L. E., A. F. Huxley, et al. (1977). "Tension responses to sudden length change in stimulated frog muscle fibres near slack length." J Physiol **269**(2): 441-515.

Ford, L. E., A. F. Huxley, et al. (1981). "The relation between stiffness and filament overlap in stimulated frog muscle fibres." J Physiol **311**: 219-49.

Ford, L. E., A. F. Huxley, et al. (1985). "Tension transients during steady shortening of frog muscle fibres." J Physiol **361**: 131-50.

Galler, S., K. Hilber, et al. (1996). "Force responses following stepwise length changes of rat skeletal muscle fibre types." J Physiol **493 ( Pt 1)**: 219-27.

Geeves, M. A. and K. C. Holmes (1999). "Structural mechanism of muscle contraction." Annu Rev Biochem **68**: 687-728.

Geeves, M. A. and K. C. Holmes (2005). "The molecular mechanism of muscle contraction." Adv Protein Chem **71**: 161-93.

Gordon, A. M., A. F. Huxley, et al. (1966). "The variation in isometric tension with sarcomere length in vertebrate muscle fibres." J Physiol **184**(1): 170-92.
Hanson, J. and H. E. Huxley (1953). "Structural basis of the cross-striations in muscle." Nature **172**(4377): 530-2.
Highsmith, S. (1999). "Lever arm model of force generation by actin-myosin-ATP." Biochemistry **38**(31): 9791-7.
Hill, A. V. (1938). "The heat of shortening and dynamic constants of muscle." Proc R Soc Lond B Biol Sci **126**: 136-195.
Hill, A. V. (1949). "The absence of lengthening during relaxation in a completely unloaded muscle." J Physiol **109**(1-2): Proc, 8.
Holmes, K. C. (1997). "The swinging lever-arm hypothesis of muscle contraction." Curr Biol **7**(2): R112-8.
Holmes, K. C., I. Angert, et al. (2003). "Electron cryo-microscopy shows how strong binding of myosin to actin releases nucleotide." Nature **425**(6956): 423-7.
Hopkins, S. C., C. Sabido-David, et al. (2002). "Orientation changes of the myosin light chain domain during filament sliding in active and rigor muscle." J Mol Biol **318**(5): 1275-91.
Houdusse, A., V. N. Kalabokis, et al. (1999). "Atomic structure of scallop myosin subfragment S1 complexed with MgADP: a novel conformation of the myosin head." Cell **97**(4): 459-70.
Houdusse, A., A. G. Szent-Gyorgyi, et al. (2000). "Three conformational states of scallop myosin S1." Proc Natl Acad Sci U S A 97(21): 11238-43.
Houdusse, A. and H. L. Sweeney (2001). "Myosin motors: missing structures and hidden springs." Curr Opin Struct Biol **11**(2): 182-94.
Huxley, A. F. (1957). "Muscle structure and theories of contraction." Prog Biophys Biophys Chem **7**: 255-318.
Huxley, A. F. and R. M. Simmons (1971). "Proposed mechanism of force generation in striated muscle." Nature **233**(5321): 533-8.
Huxley, A. F. (1974). "Muscular contraction." J Physiol **243**(1): 1-43.
Huxley, A. F. (2000). "Mechanics and models of the myosin motor." Philos Trans R Soc Lond B Biol Sci **355**(1396): 433-40.
Huxley, H. E. (1969). "The mechanism of muscular contraction." Science **164**(3886): 1356-65.
Huxley, H.E., M. Reconditi, et al. (2006a). "X-ray interference studies of crossbridge action in muscle contraction: evidence from quick releases." J Mol Biol **363**(4): 743-61.
Huxley, H.E., M. Reconditi, et al. (2006b). "X-ray interference studies of crossbridge action in muscle contraction: evidence from muscles during steady shortening." J Mol Biol **363**(4): 762-72.
Huxley, H. E. and M. Kress (1985). "Crossbridge behaviour during muscle contraction." J Muscle Res Cell Motil **6**(2): 153-61.
Irving, M., V. Lombardi, et al. (1992). "Myosin head movements are synchronous with the elementary force-generating process in muscle." Nature **357**(6374): 156-8.
Irving, M., G. Piazzesi, et al. (2000). "Conformation of the myosin motor during force generation in skeletal muscle." Nat Struct Biol **7**(6): 482-5.
Juanhuix, J., J. Bordas, et al. (2001). "Axial disposition of myosin heads in isometrically contracting muscles." Biophys J **80**(3): 1429-41.
Karatzaferi, C., M. K. Chinn, et al. (2004). "The force exerted by a muscle cross-bridge depends directly on the strength of the actomyosin bond." Biophys J **87**(4): 2532-44.
Koppole, S., J. C. Smith, et al. (2007). "The structural coupling between ATPase activation and recovery stroke in the myosin II motor." Structure **15**(7): 825-37.
Kuhner, S. and S. Fischer (2011). "Structural mechanism of the ATP-induced dissociation of rigor myosin from actin." Proc Natl Acad Sci U S A **108**(19): 7793-8.
Lan, G. and S. X. Sun (2005). "Dynamics of myosin-driven skeletal muscle contraction: I. Steady-state force generation." Biophys J **88**(6): 4107-17.
Larsson, L. and R. L. Moss (1993). "Maximum velocity of shortening in relation to myosin isoform composition in single fibres from human skeletal muscles." J Physiol **472**: 595-614.
Linari, M., I. Dobbie, et al. (1995). "Comparison between tension transients during isometric contraction and in rigor in isolated fibers from frog skeletal muscle." Biophys J **68**(4 Suppl): 218S.
Linari, M., V. Lombardi, et al. (1997). "Cross-bridge kinetics studied with staircase shortening in single fibres from frog skeletal muscle." J Muscle Res Cell Motil **18**(1): 91-101.
Linari, M., I. Dobbie, et al. (1998). "The stiffness of skeletal muscle in isometric contraction and rigor: the fraction of myosin heads bound to actin." Biophys J **74**(5): 2459-73.
Linari, M., G. Piazzesi, et al. (2000). "Interference fine structure and sarcomere length dependence of the axial x-ray pattern from active single muscle fibers." Proc Natl Acad Sci U S A **97**(13): 7226-31.
Linari, M., E. Brunello, et al. (2005). "The structural basis of the increase in isometric force production with temperature in frog skeletal muscle." J Physiol **567**(Pt 2): 459-69.
Linari, M., G. Piazzesi, et al. (2009). "The effect of myofilament compliance on kinetics of force generation by myosin motors in muscle." Biophys J **96**(2): 583-92.
Llinas, P., O. Pylypenko, et al. (2012). "How myosin motors power cellular functions: an exciting journey from structure to function: based on a lecture delivered at the 34th FEBS Congress in Prague, Czech Republic, July 2009." FEBS J **279**(4): 551-62.
Lombardi, V., G. Piazzesi, et al. (1992). "Rapid regeneration of the actin-myosin power stroke in contracting muscle." Nature **355**(6361): 638-41.
Lowey, S. and K. M. Trybus (2010). "Common structural motifs for the regulation of divergent class II myosins." J Biol Chem **285**(22): 16403-7.
Lymn, R. W. and E. W. Taylor (1971). "Mechanism of adenosine triphosphate hydrolysis by actomyosin." Biochemistry **10**(25): 4617-24.

Nielsen, B. G. (2002). "Entropic elasticity in the generation of muscle force--a theoretical model." J Theor Biol **219**(1): 99-119.
Nyitrai, M., R. Rossi, et al. (2006). "What limits the velocity of fast-skeletal muscle contraction in mammals?" J Mol Biol **355**(3): 432-42.
Offer, G. and K. W. Ranatunga (2013). "A cross-bridge cycle with two tension-generating steps simulates skeletal muscle mechanics." Biophys J **105**(4): 928-40.
Piazzesi, G. and V. Lombardi (1995). "A cross-bridge model that is able to explain mechanical and energetic properties of shortening muscle." Biophys J **68**(5): 1966-79.
Piazzesi, G., M. Reconditi, et al. (1999). "Changes in conformation of myosin heads during the development of isometric contraction and rapid shortening in single frog muscle fibres." J Physiol **514 ( Pt 2)**: 305-12.
Piazzesi, G., L. Lucii, et al. (2002). "The size and the speed of the working stroke of muscle myosin and its dependence on the force." J Physiol **545**(Pt 1): 145-51.
Piazzesi, G., M. Reconditi, et al. (2002). "Mechanism of force generation by myosin heads in skeletal muscle." Nature **415**(6872): 659-62.
Piazzesi, G., M. Reconditi, et al. (2003). "Temperature dependence of the force-generating process in single fibres from frog skeletal muscle." J Physiol **549**(Pt 1): 93-106.
Piazzesi, G., M. Reconditi, et al. (2007). "Skeletal muscle performance determined by modulation of number of myosin motors rather than motor force or stroke size." Cell **131**(4): 784-95.
Podolsky, R. J. (1960). "Kinetics of muscular contraction: the approach to the steady state." Nature **188**: 666-8.
Prochniewicz, E., P. Guhathakurta, et al. (2013). "The structural dynamics of actin during active interaction with myosin depends on the isoform of the essential light chain." Biochemistry **52**(9): 1622-30.
Ranatunga, K. W. (1984). "The force-velocity relation of rat fast- and slow-twitch muscles examined at different temperatures." J Physiol **351**: 517-29.
Rayment, I., H. M. Holden, et al. (1993). "Structure of the actin-myosin complex and its implications for muscle contraction." Science **261**(5117): 58-65.
Rayment, I., W. R. Rypniewski, et al. (1993). "Three-dimensional structure of myosin subfragment-1: a molecular motor." Science **261**(5117): 50-8.
Reconditi, M., M. Linari, et al. (2004). "The myosin motor in muscle generates a smaller and slower working stroke at higher load." Nature **428**(6982): 578-81.
Reconditi, M., E. Brunello, et al. (2011). "Motion of myosin head domains during activation and force development in skeletal muscle." Proc Natl Acad Sci U S A **108**(17): 7236-40.
Roddier, F. (1988). Distributions et transformations de Fourier. Paris, McGRAW-HILL.
Rome, L. C., A. A. Sosnicki, et al. (1990). "Maximum velocity of shortening of three fibre types from horse soleus muscle: implications for scaling with body size." J Physiol **431**: 173-85.
Smith, D. A. and M. A. Geeves (1995). "Strain-dependent cross-bridge cycle for muscle." Biophys J **69**(2): 524-37.
Squire, J. M., H. A. Al-Khayat, et al. (2005). "Molecular architecture in muscle contractile assemblies." Adv Protein Chem **71**: 17-87.
Squire, J. M. and C. Knupp (2005). "X-ray diffraction studies of muscle and the crossbridge cycle." Adv Protein Chem **71**: 195-255.
Sugi, H., H. Minoda, et al. (2008). "Direct demonstration of the cross-bridge recovery stroke in muscle thick filaments in aqueous solution by using the hydration chamber." Proc Natl Acad Sci U S A **105**(45): 17396-401.
Suquet, C. (2013). Probabilitès via l'intégrale de Riemann. ellipses. Paris.
Uyeda, T. Q., P. D. Abramson, et al. (1996). "The neck region of the myosin motor domain acts as a lever arm to generate movement." Proc Natl Acad Sci U S A **93**(9): 4459-64.
Yanagida, T., K. Kitamura, et al. (2000). "Single molecule analysis of the actomyosin motor." Curr Opin Cell Biol **12**(1): 20-5.
Yu, L. C. and B. Brenner (1989). "Structures of actomyosin crossbridges in relaxed and rigor muscle fibers." Biophys J **55**(3): 441-53.